\documentclass[10pt,conference,letterpaper]{IEEEtran}                    
\usepackage{graphicx}
 \usepackage{algorithmic}
 \usepackage{algorithm}
\usepackage{amsmath}
\usepackage{multicol}
\usepackage{cases}
\usepackage[final]{changes}
\usepackage{mathptmx}
\usepackage{setspace}
\usepackage{cite}
\usepackage{url}
\usepackage{graphicx}
\usepackage{tfrupee}
\usepackage{soul}
\usepackage{subfig}
\usepackage{array,etoolbox}
\usepackage{spreadtab}
\usepackage{multirow}
\usepackage{hyperref}

\preto\tabular{\setcounter{magicrownumbers}{0}}
\newcounter{magicrownumbers}
\def\rownumber{}
\IEEEoverridecommandlockouts

\def\hb{\hbox to 10.7 cm{}}
\newcommand{\fref}[1]{Fig.~\ref{#1}}

\newcommand{\cref}[1]{Chapter~\ref{#1}}

\newcommand{\junk}[1]{}

\begin{document}

\title{Analyzing HC-NJDG Data to Understand the Pendency in High Courts in India}

\author{\IEEEauthorblockN{Kshitiz Verma\thanks{{\bf Disclaimer: } The data is taken from HC-NJDG and may not reflect the actual status of the statistics in the Hon'ble High Courts. Due care has been taken in the analysis but some errors may still be there. Kindly notify the author in case of discovery of such errors. The reference to the Hon'ble High Courts should be construed as the reference being made for the department responsible for updates on HC-NJDG.}}
\IEEEauthorblockA{LNMIIT Jaipur, India\\
Email: vermasharp@gmail.com}
}

\maketitle
\begin{abstract}

Indian Judiciary is suffering from burden of millions of cases that are lying pending in its courts at all the levels. Hon'ble Supreme Court of India has initiated e-Courts project to deploy Information and Communication Technology in the judiciary so as to efficiently impart justice without compromising on its quality. The National Judicial Data Grid (NJDG) is an important outcome of this project that indexes all the cases pending in the courts and publishes the data publicly. The launch of NJDG has also resulted in a jump of 30 ranks in the World Bank's Ease Of Doing Business Report. 

In this paper, we analyze the data that we have collected on the pendency of 24 high courts in the Republic of India as they were made available on High Court NJDG (HC-NJDG). We collected data on 73 days beginning August 31, 2017 to December 26, 2018, including these days. Thus, the data collected by us spans a period of almost sixteen months. We have analyzed various statistics available on the NJDG portal for High Courts, including but not limited to the number of judges in each high court, the number of cases pending in each high court, cases that have been pending for more than 10 years, cases filed, listed and disposed, cases filed by women and senior citizens, etc. Our results show that:
\begin{enumerate}
\item statistics as important as the number of judges in high courts have serious errors on NJDG (Fig. \ref{fig:hcjudges}, \ref{fig:jud_date_j}, \ref{fig:njudgeshc}, \ref{fig:njudgeshc1}, Table \ref{tab:njdg_doj}).
\item pending cases in most of the high courts are increasing rather than decreasing (Fig. \ref{fig:hcpendency}, \ref{fig:np_hc1}).
\item regular update of HC-NJDG is required for it to be useful. Data related to some high courts is not being updated regularly or is updated erroneously on the portal (\fref{fig:np_hc2}).
\item there is a huge difference in terms of average load of cases on judges of different high courts (\fref{fig:pend_date_r}).
\item if all the high courts operate at their approved strength of judges, then for most of the high courts pendency can be nullified within 20 years from now (Fig. \ref{fig:years_combat_real}, \ref{fig:years_combat_national_avg_lowest}).
\item the pending cases filed by women and senior citizens are disproportionately low, they together constitute less than 10\% of the total pending cases (\fref{fig:womsen_hc_date} - \ref{fig:wom_hc})
\item a better scheduling process for preparing causelists in courts can help reducing the number of pending cases in the High Courts (\fref{fig:fdl_hc}).
\item some statistics are not well defined (\fref{fig:po_hc3}).
\end{enumerate}


\end{abstract}

\section{Introduction}
\label{sec:intro}

\par More than 29 million cases are pending in all the levels in Indian courts as of December, 2018 \cite{pendingcases_ht,NJDG,njdg_hc}. A substantial percentage of them have been around for more than ten years. It has been realized that more scientific ways of dealing with such a mammoth backlog is required without compromising on the quality of justice. Thus, Indian Judiciary has started, on the initiatives of the Hon'ble Supreme Court of India, e-Courts project to take help of the Information and Communications Technologies (ICT) in the judicial sector through its e-Committee \cite{sci_actionplan2005,sci_actionplan2014}. The e-Courts project has been carried out in two phases. This has lead to the digitization of court records and many services are being provided online. The Chief Justice of India has recently launched such services as a part of e-Court project \cite{CJI_efiling}. More government projects are yet to be implemented to bring connectivity to approximately 3000 courts by the year end \cite{more_courts,digital_push}. A great leap in providing free access to the judicial information was provided by the implementation of the National Judicial Data Grid (NJDG) \cite{NJDG}. The National Judicial Data Grid (NJDG), an important outcome of the e-Courts projects, has data of more than 29 million cases pending in Indian courts at all the levels. Many, possibly different and independent parameters have been considered in estimating the number of years required to nullify the backlog. According to Justice V. V. Rao, a prediction made in 2010 \cite{justicerao320}, it would take 320 years to clear the backlog. On the other hand, a commitment to curb pendency in five years was made by the then CJI Justice H.L. Dattu in 2015\cite{fiveYearsJDattu}. However, at both these moments, NJDG did not exist and such prediction was largely based on the wisdom and foresight of the learned judges. Today when NJDG exists, we have concrete data that can be studied to form better predictions and methods to reduce the backlog.

\par In the same spirit, National Judicial Data Grid for High Courts (HC-NJDG) was also launched and was visioned to be a game changer\cite{njdg_gamechg}. In this paper, we exclusively study HC-NJDG and report results related to High Courts only as the data on high courts is relatively easy to verify. As of December 26, 2018, there were more than 4.9 million cases pending in the High Courts of India. We have chosen to study pendency in high courts only because they are generally well equipped with resources to implement the suggested measures effectively. Also, since the number of high courts is only 24 (the 25$^{th}$ started on January 1, 2019) analysis is easy to interpret. Most of the high courts also publish pendency statistics on their website, so it is easy to cross verify as well.

\par As one can imagine, the sheer idea of the existence of NJDG appears to be a mammoth task, implementing it properly is going to be daunting. Indexing millions of cases to the level of details envisaged in the e-Court project is a task that has never been carried out by any government anywhere. Hence, the mere existence of NJDG is a miracle in itself which also gets reflected from a sudden jump in India's rank in the Ease of Doing Business Report by the World Bank. There is an improvement of 30 ranks from the year 2016 to 2017 \cite{WB_lauds}. However, as mentioned before, it is a non-trivial task and it is not yet time to celebrate its success. This effort can be called successful only after the judges, court staff, advocates and litigants find it useful in reducing their pain and NJDG helps improving the efficiency of the whole system. It is still very far from that stage. Note that the success of portals like NJDG depends immensely on individual high courts updating their data regularly on the portal. Recently more than 1000 lower courts were reported not updating their data regularly on NJDG\cite{lc_not_updating}. The same is true for some of the high courts as well.

\subsection{Results in the paper}

Even though the importance of NJDG cannot be ruled out, its usefulness remains a mirage until it is implemented flawlessly. Our work is an attempt in the direction of helping NJDG to become more useful. We summarize some of our feedback for NJDG and present some results below: 

\begin{enumerate}
\item {\bf Define terms used on the portal:} As a first improvement, there is a need to explain the terms that are used on the NJDG portal for high courts. For example, there are statistics about \emph{Cases-Under Objection} or \emph{Cases Pending Registration} but these terms lack a clear definition. There are many such terms which are never defined on the portal. 

\item {\bf Lack of proper documentation:} Portal must have a well defined documentation so that the observer does not feel isolated. Right now, the portal is not accompanied with any easy to locate documentation to help the reader to interpret the data. 

\item {\bf Erroneous data:} The portal should be as error free as possible. For example, the number of judges in some of the high courts on HC-NJDG is incorrect. Some of the high courts have even reported the number of judges to be more than the approved strength of the respective high court. 

\item {\bf Absence of frequent updates:} Some high courts are not updating the data regularly. Such failures in updating the data are crucial as it impacts the data at the national level, which is simply the aggregation of the data received from all the high courts. Hence, errors are aggregated in the national data on pending cases. Ideally, the data on HC-NJDG should be updated on a daily basis. 

\item {\bf Need for different parameter:} All the statistics provided are on daily basis apart from the \emph{Cases Disposed} and \emph{Cases Filed}. If these two parameters can also be made consistent with the remaining parameters then comparisons will become easier. 

\item {\bf Time to Combat Pendency:} We do a detailed analysis of how long will it take to get rid of pendency in high courts without assuming anything unrealistic. We have also claimed that the data from NJDG may not be reliable as it is, hence, we have done another level of careful analysis to consider only those numbers that can be considered realistic. 

\item {\bf Preparing causelists more scientifically:} Daksh report has a crucial finding that the high court judges have very little time -- of the order of two to fifteen minutes -- to devote to a case as the number of cases listed on daily causelists are very high \cite{daksh_report}. The report also mentions that the financial loss to the nation due to non-hearing of the listed cases is a significant portion of the Gross Domestic Product (GDP). Our first inference from the data analyzed is that the high number of cases on causelists should be avoided because the number of cases listed in a day are equivalent to the number of cases disposed in a month for many high courts. 
A reduction in the size of causelists will help all the stakeholders to ease their life because the judges, court staff and advocates will have lesser load. The litigants will also be better off if their cases are heard for a longer period by the judges and if the chances of their cases being heard increase.
\end{enumerate}

\subsection{Organization of the paper}
\par The rest of the paper is organized as follows: Section \ref{sec:rw} encompasses the related studies and the scope of the work. Section \ref{sec:njdg} discusses the methodology for data collection and also summarizes some of the main results in this paper. Section \ref{sec:judges_hc} presents analysis of data on the number of the judges in high courts in India. Section \ref{sec:pending_hc} elaborates on pending cases in the high courts. Section \ref{sec:combat} is home to the most important result of the paper in which we estimate the time required to nullify the pendency in the high courts. Section \ref{sec:misc} focuses on the cases filed by senior citizens and women, inter-relationship between the cases filed, disposed and listed in the high courts, cases under objection and cases pending registration. Section \ref{sec:conclusion} concludes the paper.


\section{Related work}
\label{sec:rw}
In this section, we state the scope of our work. We also compare our work with major studies on pendency in courts in India. 

Before we proceed any further to discuss the technical findings of the paper, we first understand the scope and the limitations of our work. There exist many articles and reports on pendency of court cases in India. In this paper, instead of studying the pending cases in all the courts of India, we decided to limit ourselves to only the high courts. This limits the number of court complexes in our study and the improvements suggested in the study are relatively easier to implement in high courts than in lower courts. Thus, we would concretely know where things can be improved. Hence, throughout this study, we have concentrated on the pendency in high courts rather than the subordinate courts. 

In our study, we have not taken input from any real person. No judge, advocate, litigant or court staff was interviewed. This may have both the impacts, positive and negative. Intervention of humans, who are involved in updating NJDG may have provided more insights to interpret our results. On the other side of it, their views might have biased our results. So we decided to leave it for future because we wanted our assessment to be purely technical and statistical based only on the observations made from the data that we have collected from NJDG. This study, to the best of our knowledge, is the first one to analyze NJDG data over such a long period of time. Most of the existing studies consider the data from only one day on NJDG. Hence, we differ from the other studies in this basic premises itself. Our work also tries to find out the answer to the question, "How reliable is single day analysis of NJDG?". We answer this as negative, i.e., the NJDG data collected on just one day may not be taken as reliable for any reasonable analysis. There have been instances when the data on NJDG was very erroneous and such days are not rare. For example, a recent article on the pendency statistics of Bombay High Court claimed that 4.64 lakh cases are pending \cite{bombay_hc464}. Our finding is that the number has stayed the same, i.e., 4,64,074 ever since the data related to Bombay High Court was published on NJDG, throughout our data collection period.

There have been many news reports, articles and studies on pendency in Indian courts \cite{arrears}  \cite{over_22lakh} \cite{jkpile} \cite{verma2018courts}. A study by Alok Prasanna Kumar has used number of the District and Magistrate courts, collected from the National Judicial Data Grid as of 18 March, 2016 \cite{comparative}. In our study, we show that even for some high courts the number of judges is wrongly reported. In such cases, can we rely on the number of judges reported in subordinate judiciary? Therefore, relying on data of one single day may not be the best practice. There are studies conducted by the Department of Justice as well\cite{dojeval}. While this study is very comprehensive, the results reported are different from ours and the parameters considered for evaluation are different as well. Various studies including \cite{vidhi1}, have conducted research on e-Court policies. The importance of data analysis of judicial data and role of computer science is also suggested in \cite{judmess}. The Department of Justice also encourages research conducted on judicial reforms by means of funding \cite{doj_projects}. Another rich source of information on pending cases are the annual reports published by the Supreme Court of India \cite{SCAR}. 

The most relevant related work in this area is the Daksh report on the state of the Indian Judiciary \cite{daksh_report}. Their approach, however, is very different from ours. They have conducted a ground level research by surveying and obtaining the first hand experience of the litigants and other stake holders. Our work on the other hand, relies completely on the data provided by the national judicial data grid for high courts (HC-NJDG). At the time Daksh report was being written, NJDG was still in its infancy. After more than two years, it is reasonable to expect that the data on the grid to be much more organized. 

Our results are however, very similar in some areas. For example, the Daksh report has also found that there is a lack of uniformity in the available data. There is no unanimous agreement on the number of judges/courts in the lower judiciary in the country. Our study claims that there are discrepancies even in the number of judges reported in the high courts, let alone the subordinate judiciary. 

The issue has been of utmost importance to all the Chief Justices of India including the current one \cite{ranjan_plan}. Hence, a lot of exciting research is being conducted in the area and our hope is to be able to contribute to that.


\section{High Court NJDG Data}
\label{sec:njdg}

High Court National Judicial Data Grid (HC-NJDG) was launched in July 2017 \cite{njdg_gamechg, njdg_hc}. We started collecting data from the portal on August 31, 2017. The last data used in this paper was collected on December 26, 2018. Thus, we have collected data for a period of around sixteen months sampled on 73 days. Table \ref{tab:dates} lists the dates on which the data was collected.

\begin{table}[h]
\centering
    \begin{tabular}{ |l|l|l| }
    \hline
    Year & Month & Day \\ \hline \hline
    \multirow{5}{*}{2017} & August & 31 \\
    & September & 2, 5, 6, 8, 11, 12, 14, 19, 21, 23, 28 \\
    & October & 1, 4, 5, 11, 30  \\
    & November & 7 \\
    & December & 13, 20, 30 \\ \hline
    \multirow{12}{*}{2018} & January & 6, 11, 17, 19, 23, 31 \\
    & February & 2, 10, 14, 22, 26\\
    & March & 5, 12, 22 \\
    & April & 3, 10, 17, 24 \\
    & May & 2, 11, 18, 24 \\
    & June & 1, 7, 15, 20, 28 \\
    & July & 4, 11, 18, 23, 29  \\    
    & August & 6, 13, 20, 24, 31 \\
    & September & 7, 14, 22, 27 \\
    & October & 3, 10, 17, 31 \\
    & November & 6, 12, 22, 29 \\
    & December & 9, 19, 26 \\ \hline
    \end{tabular}
    \caption{Dates on which data was collected from HC-NJDG portal corresponding to the individual high courts. }
    \label{tab:dates}
\end{table}

The methodology for downloading the data is described as follows.

\subsection{Methodology for data collection}
The dates on which we have collected data have been mentioned already in Table \ref{tab:dates}. The dates chosen are arbitrary with an only intention of collecting data every 7-10 days. On every date, the following procedure was followed:
\begin{enumerate}
\item Connect to the portal available at \cite{njdg_hc}.
\item Download the web frame containing the statistics on all the high courts by changing the options available under the drop-down menu.
\item Save the file corresponding to the high court.
\end{enumerate}

\paragraph{Example of collected data} To provide a glimpse of the data, some of the statistics, as collected on August 20, 2018, are provided in Table \ref{tab:njdg_pending} and Table \ref{tab:njdg_monitor}. The portal has more statistics available but we have chosen to present only some. In Table \ref{tab:njdg_pending}, data related to the number of pending cases in all the high courts in India is shown. The cases are divided into two different kinds. The rows in the table show the division of cases based on the age of the cases, i.e., how old are they. The columns show the division based on the types of the cases, i.e., whether the cases are civil, criminal or writs. Hence, on this date, more than 3.3 million cases were pending in the high courts in India.

\begin{table}[h]
\centering
\footnotesize
    \begin{tabular}{ | l | r | r | r | r |}
    \hline
    Cases Pending & Civil & Criminal & Writs & Total \\ \hline
    Over 10 years & 371332 & 138120 & 142642 & 652094 \\ \hline
    Between 5-10 years & 370192 & 176909 & 257987 & 805088 \\ \hline
    Between 2-5 years & 419455 & 214270 & 372992 & 1006717 \\ \hline
    Less than 2 years & 345801 & 240455 & 340827 & 927083 \\ \hline
    Total & 1506780 & 769754 & 1114448 & 3390982  \\ \hline
    \end{tabular}
    \caption{HC-NJDG data of pending cases in High Courts as on August 20, 2018.}
    \label{tab:njdg_pending}
\end{table}

Table \ref{tab:njdg_monitor} presents data on the number of monthly disposed and filed cases. It also shows the aggregate of the cases that were listed for hearing on that particular date in all the high courts. The columns, like the previous table, represent the type of cases. In this table, we have also included another field, the number of judges in the high courts. Note that there was an error in the number of high court judges reported on that day. According to HC-NJDG portal the number of high court judges in India was 810, whereas according to the vacancy positions document (dated August 01, 2018) available at the Department of Justice website \cite{doj_vacancy}, the number of working strength of high courts was 659 and the approved strength was 1079. The number on the portal is different from both these numbers and yet no reasons for the digression are mentioned.

\begin{table}[h]
\centering
\footnotesize
    \begin{tabular}{ | l | r | r | r | r |}
    \hline
    
        & Civil & Criminal & Writs & Total \\ \hline
    Cases Filed (last month)& 27663 & 42404 & 32009 & 102063 \\ \hline
    Cases listed (today) & 11656 & 11793 & 13013 & 36462 \\ \hline
    Cases Disposed (last month) & 28080 & 47368 & 36548 & 111996 \\ \hline
    Total Judges & & & &810 \\ \hline
    \end{tabular}
    \caption{Number of cases filed (monthly) and disposed (monthly) and the number of cases listed (daily) and the number of judges in the high courts in India as on August 20, 2018.}
    \label{tab:njdg_monitor}
\end{table}

\subsection{Scrutinizing the data so collected}

After the data collection, we had to make sure that the data being used does not suffer from errors made during the download or during saving the files. After a careful inspection, we deleted some of the files which had errors. Table \ref{tab:hcnjdg_details} lists the amount of data that we can actually use from the data collected during the said period. For each high court, we have retained information from as many dates as mentioned in the ``Appearance in data" column. This column represents the number of files after scrutinizing the data for each high court. Four high courts have joined HC-NJDG after we started collecting the data, namely, Allahabad High Court, Gauhati High Court, High Court of Jammu and Kashmir and High Court of Madhya Pradesh. Hence, they appear fewer number of times. However, majority of the high courts -- 20 to be precise -- had their presence on HC-NJDG when we started collecting the data, i.e., on August 31, 2017. This means that for 20 high courts the number of files should be 73 each. However, it is not so due to the errors incurred in the data collection process. After removal of the files with errors, we proceed to analyze the data. 

Note that the last day for data collection for this paper was December 26, 2019. The $25^{th}$ High Court for the state of Telangana was formed on January 01, 2019. Hence, in our analysis only 24 high courts appear.

\begin{table}
\centering
    \begin{tabular}{|@{\makebox[2em][r]{\rownumber\space}} | l | c| }
    \hline
    \gdef\rownumber{\stepcounter{magicrownumbers}\arabic{magicrownumbers}}
    High Court & Appearance in data  \\ \hline\hline
    Allahabad & 47\\ \hline
    Bombay & 73 \\ \hline
    Calcutta & 71 \\ \hline
    Chhattisgarh & 72 \\ \hline
    Delhi &  72 \\ \hline
    Gauhati & 55  \\ \hline
    Gujarat & 72  \\ \hline
    Himachal Pradesh & 73\\ \hline
    Jammu and Kashmir & 50 \\ \hline
    Jharkhand & 73 \\ \hline
    Karnataka & 73 \\ \hline
    Kerala & 62  \\ \hline
    Madhya Pradesh & 67  \\ \hline
    Madras & 73  \\ \hline
    Manipur & 73\\ \hline
    Meghalaya & 73 \\ \hline
    Orissa & 73 \\ \hline
    Patna & 73 \\ \hline
    Punjab and Haryana & 73 \\ \hline
    Rajasthan & 73 \\ \hline
    Sikkim & 73  \\ \hline
    Telangana \& Andhra Pradesh & 72  \\ \hline
    Tripura & 73  \\ \hline
    Uttarakhand & 73  \\ \hline\hline
    National level data of all HC & 72  \\ \hline 
    \end{tabular}
    \caption{The details of the data collected from NJDG for high courts between August 31, 2017 to December 26, 2018. Ideally, all should have 73 files but four high courts joined HC-NJDG late and for the remaining, there are less than 73 files due to error incurred in data collection. Hence, those files that had errors are not used for analysis.}
    \label{tab:hcnjdg_details}
\end{table}

\subsection{Understanding the graphs}

In this paper, we have shown hundreds of graphs. In order to maintain coherence and simplicity, and to have a reach to wider audience, we have restricted ourselves to only two kinds of graphs, as explained below. 
\paragraph{Temporal data graphs (Dates on horizontal axis)} The horizontal axis (also referred to as X-axis in the paper), consists of dates beginning August 31, 2017 to December 26, 2018 from left to right. All the dates, as in Table \ref{tab:dates}, are present on X-axis. To reduce cluttering on X-axis, we are printing every fifth date. Hence, there are four more points (dates) between two printed dates in such graphs. The format of date on X-axis is YYYY-MM-DD. For the parameter considered, if there is a corresponding value available for that particular date, then it is printed, which corresponds to the Y-axis or the vertical axis. The title of each graph is present on the top which states the name of the high court that plot corresponds to. The title also contains the number of data points in bracket. The number in bracket, or the data points, can lie anywhere between 0-73, depending on whether the data was available or not. If the data is available for all the 73 dates, all will have a corresponding Y-axis value. If some date does not have a valid data, then data is not shown against that date but the date is still present on X-axis in the all cases. \fref{fig:hcjudges} is an example of this kind of graph.

\paragraph{Spatial data graphs (High Courts on horizontal axis)} In these graphs, we fix one date and the data from all the high courts is plotted corresponding to that particular date only. The horizontal axis, or X-axis, in these graphs have either 24 or 25 points. Each point represents one of the 24 high courts, 25$^{th}$ point, if exists, represent the aggregate of all the high courts. This point is referred to as ``Total" in the graphs. Y-axis plots the value of the considered parameter on that date. If a high court does not have a valid data for that particular date, its name still appears in the X-axis but have no value on the Y-axis. The title of the graph is present at the top. 

Choice of semilog Y-axis for some curves is made to accommodate wide range of values in one graph. \fref{fig:pend_date_c} is an example of such use. The numbers 1 and 1000 and almost 10 lakh (= 1 Million) are clearly visible in the same graph which would be difficult to show if Y-axis were linear.

\subsection{Number of Judges in High Courts}

\begin{figure}[h]
\includegraphics[width=8.6cm]{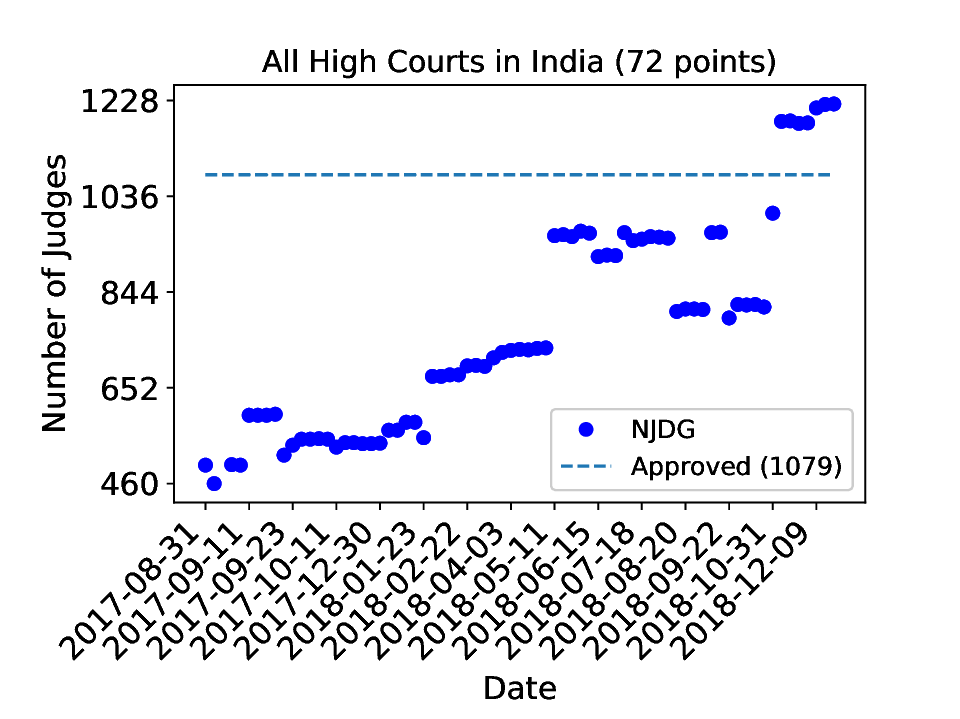}
\caption{The number of judges in the High Courts as provided on NJDG portal do not match with the approved or working strength. Data is plotted against the dates available from Table \ref{tab:dates}. However, to remove the cluttering of the data, every fifth date is printed.}
\label{fig:hcjudges}
\end{figure}

\begin{figure}[h]
\includegraphics[width=9.6cm]{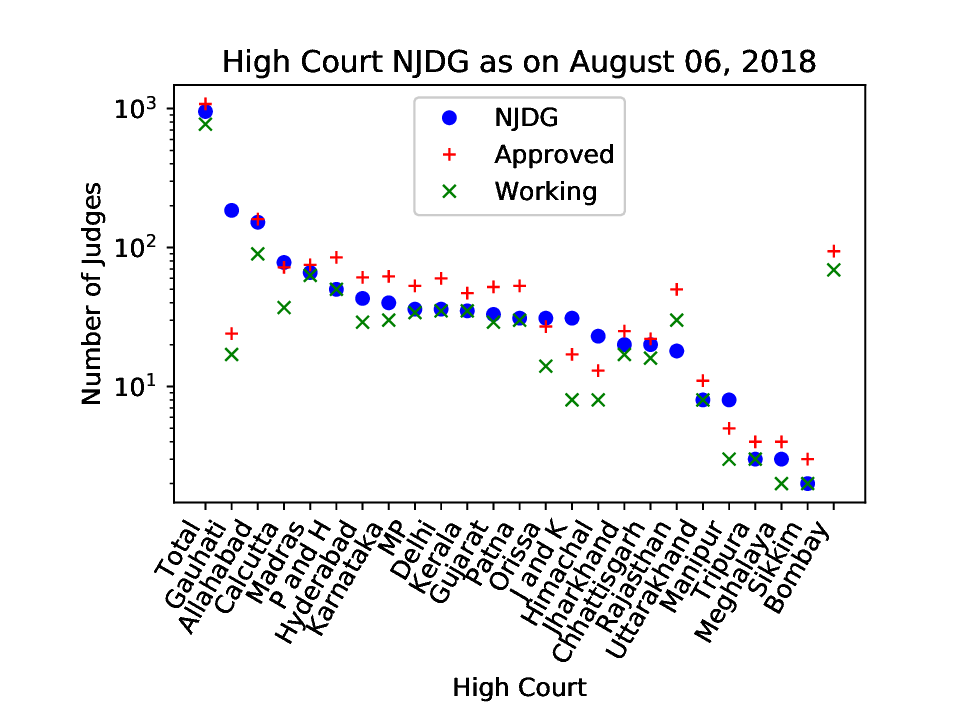}
\caption{Discrepancies found in data on the number of high courts on HC-NJDG. Nothing special about the date chosen, such discrepancies are throughout the data, refer to Section \ref{sec:judges_hc} for more details.} 
\label{fig:jud_date_j}
\end{figure}

\fref{fig:hcjudges} shows the aggregate number of judges in the high courts of India for each date on which the data has been collected. As shown in bracket in the title there are a total of 72 data points for this graph. This means that the national level data of high courts is present in 72 days out of 73 in our data. Hence, the point corresponding to that particular date will be missing. As it can be clearly seen, 73 dates are not shown in the figure but only 15 are shown so as to make the dates look tidy. In \fref{fig:hcjudges}, the date on which the data is missing is September 05, 2017. From the graph, the number of judges in high courts was around 460 in the first week of September 2017. After that we find another cluster around 550 during October 2017 to January 2018. There is yet another cluster at around 700 between
\begin{table}[h]
\centering
    \begin{tabular}{ | l | r | r | r |}
    \hline
    Date & NJDG & Working & Approved \\ \hline
    June 01, 2018 & 966 & 659 & 1079 \\ \hline
    July 04, 2018 & 963 & 668 & 1079 \\ \hline
    August 06, 2018 & 952 & 659 & 1079 \\ \hline
    September 07, 2018 & 963 & 652 & 1079 \\ \hline
    October 03, 2018 & 818 & 645 & 1079 \\ \hline
    December 09, 2018 & 1213 & 695 & 1079 \\ \hline    
    \end{tabular}
    \caption{HC-NJDG data as compared with the vacancy document available at \cite{doj_vacancy}. This document is made available by the Department of Justice on Day 1 of every month. Hence, we have chosen the closest date to 1st of respective months in our dataset.}
    \label{tab:njdg_doj}
\end{table}
February 2018 to May 2018 which again jumps to around 950 beginning June 2018. After a few frequent and drastic ups and downs in the number of judges during July to December 2018, it crosses 1200 mark. Analysis of this data compared with the vacancy document available on the Department of Justice website \cite{doj_vacancy} suggests that the HC-NJDG data on the number of judges in the high courts has some errors. In fact, according to HC-NJDG, the number of high court judges was more than the approved strength of 1079 during November and December 2018. In order to cross verify the data, we also started downloading the vacancy document as per the Department of Justice website starting June 2018. Table \ref{tab:njdg_doj} provides a comparison with the vacancy document. This clearly shows that the number of judges reported by HC-NJDG portal is approximately 300 more than the actual number. Hence, it is unclear what the number on HC-NJDG represents and why it keeps changing so frequently. The number of judges of high courts is a relatively stable number, it should not have doubled in one year and almost tripled in sixteen months!

In \fref{fig:jud_date_j}, we plot the spatial data. We have plotted the number of judges in each high court as available on August 06, 2018. Note that the Y-axis in this plot is on log scale. The data is plotted in the descending order of the number of judges in each high court from HC-NJDG data. The reasons for choosing this date is mainly because it is closest to August 01 which is the date when vacancy document was updated by the Department of Justice. It can be clearly seen that in some cases the number of high court judges reported by HC-NJDG is even higher than the approved strength provided in the vacancy document for that high court. To be precise, on August 06, 2018, the number of judges in High Court of Gauhati, Calcutta, Orissa, Jammu and Kashmir, Himachal Pradesh and Manipur was higher than the number of approved strength for the respective high courts.  

A more detailed analysis of the number of judges in each high court as mentioned on NJDG portal is done in Section \ref{sec:judges_hc}. 

\subsection{Pending Cases in High Courts}

\begin{figure}[t]
\includegraphics[width=8.6cm]{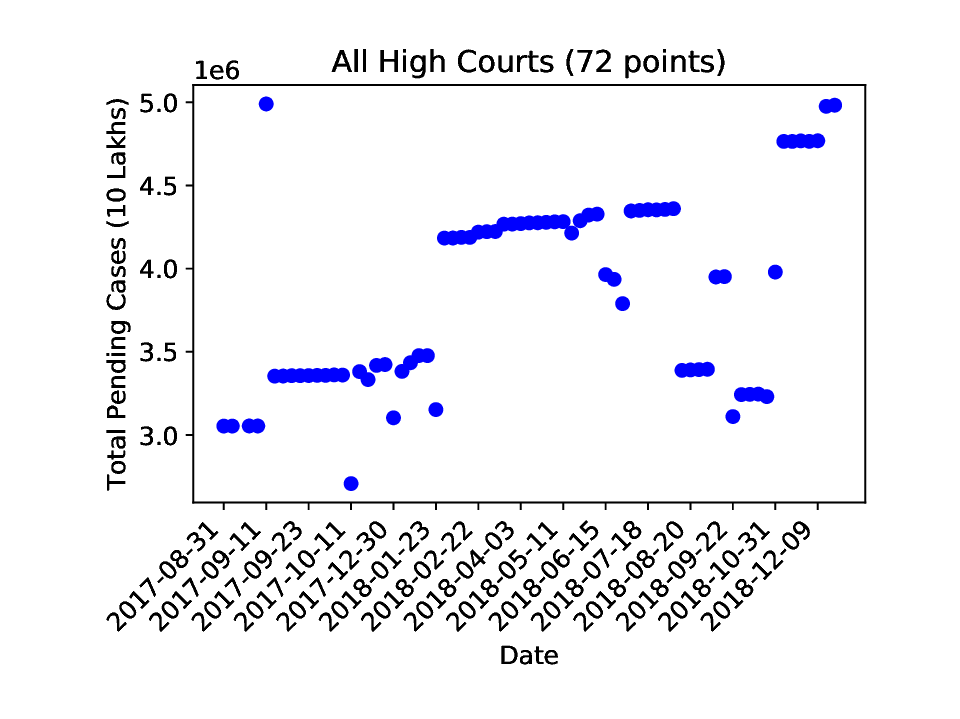}
\caption{Pending cases in all the High Courts of India.} 
\label{fig:hcpendency}
\end{figure}

\fref{fig:hcpendency} shows the number of pending cases in all the high courts of India. As in \fref{fig:hcjudges}, there are 72 points because the data set remains the same and only the parameter being plotted has changed. In this graph, we have plotted the total number of pending cases in the high courts in India as obtained from the NJDG portal in our data set. It can be clearly seen that the data has few continuous clusters and few sudden jumps. While initial sudden jumps can be explained by the fact that few high courts have joined NJDG late and they may be taking time to converge to report stable number, the overall graph does not represent a healthy update culture. Our data has some holes during November 2017, i.e., we do not have as much data as in other months. So a sudden jump may have been there due to lack of continuity in our data but frequent sudden jumps, isolated points and flat regions are indicatives of fewer updates. It also implies, in some cases that the updates have been erroneous. For example, on September 11, 2017, there is a completely isolated peak with pending number of cases greater than 49 lakh 90 thousands (around 5 million) whereas the adjacent dates on both sides have less than 35 lakhs. While it is true that the error was quickly ratified, there should be attempts towards not introducing such errors in the first place. In particular, after more than one year of inception, the graph should have started looking like a ``smooth continuous function", which as of now, it doesn't. For around six months between January 2018 to July 2018 the number of pending cases in high courts was reported to be more than 43 lakhs (4.3 million) which suddenly dropped down to 33 lakh (3.3 million) in August 2018. This essentially means that around 10 lakh cases were disposed in the week between August 06 to August 13, 2018! Such sudden jumps need explanations to be reliably used for any practical significance. The reasons for such jumps must be investigated so that such errors do not get repeated.

%



\begin{figure}[h]
\includegraphics[width=9.6cm]{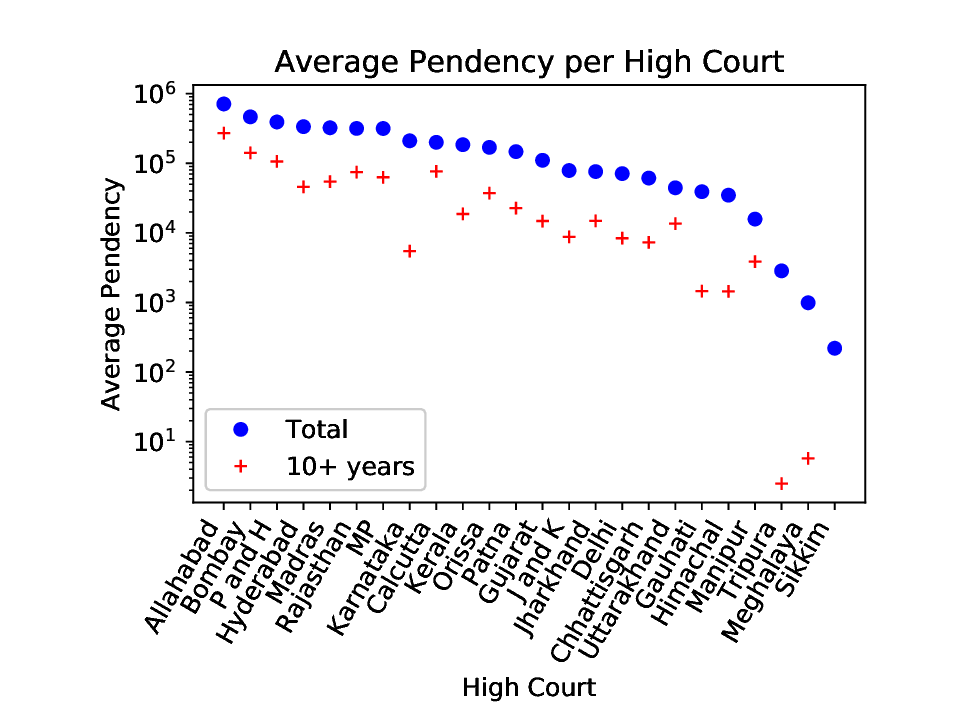}
\caption{Average of Pending cases in the High Courts during our data collection period.} 
\label{fig:pend_date_c}
\end{figure}

\fref{fig:pend_date_c} shows the average of pending cases during the data collection period for all the high courts . Note that the Y-axis is in log scale and the data is sorted in descending order of the total number of pending cases at a high court. Hence, the first place is occupied by Allahabad High Court that has more than 7 lakh cases pending. The figure also implies that most of the high courts have huge number of pending cases except Sikkim High Court and newly established high courts for the states of Meghalaya and Tripura. It is also worth noting from this figure that Sikkim High Court has no cases that are pending for more than 10 years. Meghalaya and Tripura High Court have less than twenty cases pending for more than ten years. The rest of the high courts are having the ten plus years pending cases as a substantial percentage of their total pendency.

\begin{figure}[h]
\includegraphics[width=9.6cm]{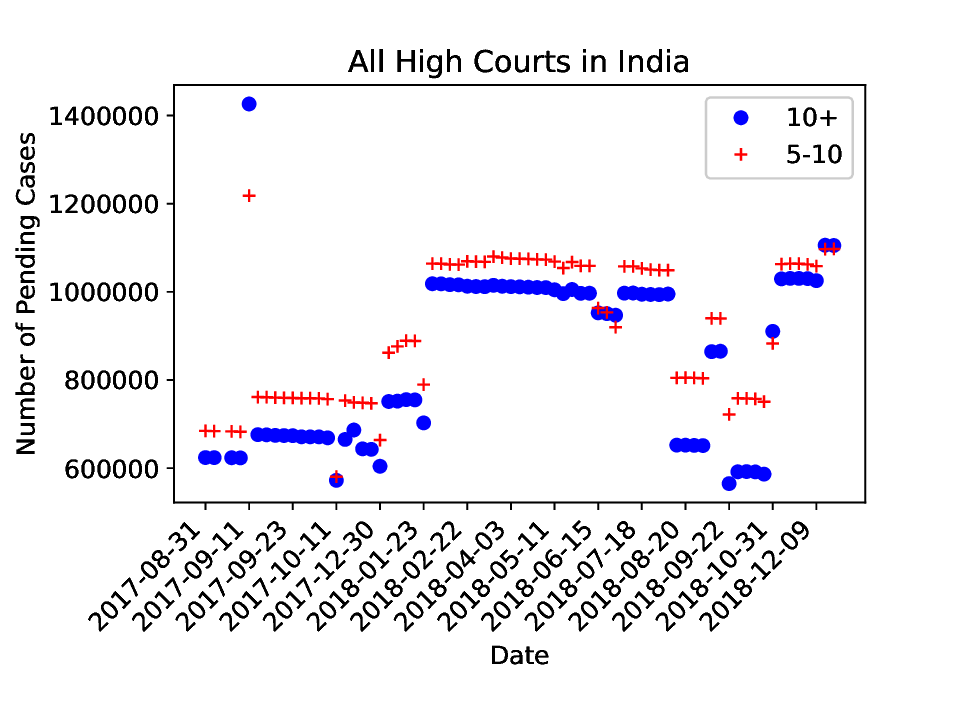}
\caption{Number of cases pending in the High Courts according to their age. We have plotted only two kinds of cases, those that have age more than ten years and those between five to ten years.} 
\label{fig:yearwise}
\end{figure}

\fref{fig:yearwise} shows the pendency in terms of the age of cases. From this data, we can see that the number of cases with age greater than five have mainly increased during last one year. However, the number of cases with 10+ years age, suddenly reduced from 9,95,031 on August 06, 2018 to 6,52,180 on August 13, 2018. A similar drop in number is seen for the cases pending for 5-10 years as well. The reasons for such an extraordinary decrease are unknown.

A detailed study of pending cases is done in Section \ref{sec:pending_hc}.

\subsection{Ratio of Pendency to Judges}

The total pendency, in itself, does not provide any information until the number of judges in the respective high court is also taken into account. This subsection considers the ratio of $\frac{\text{pending cases}}{\text{number of judges}}$ as a parameter for each high court.

\begin{figure}[h]
\includegraphics[width=9.6cm]{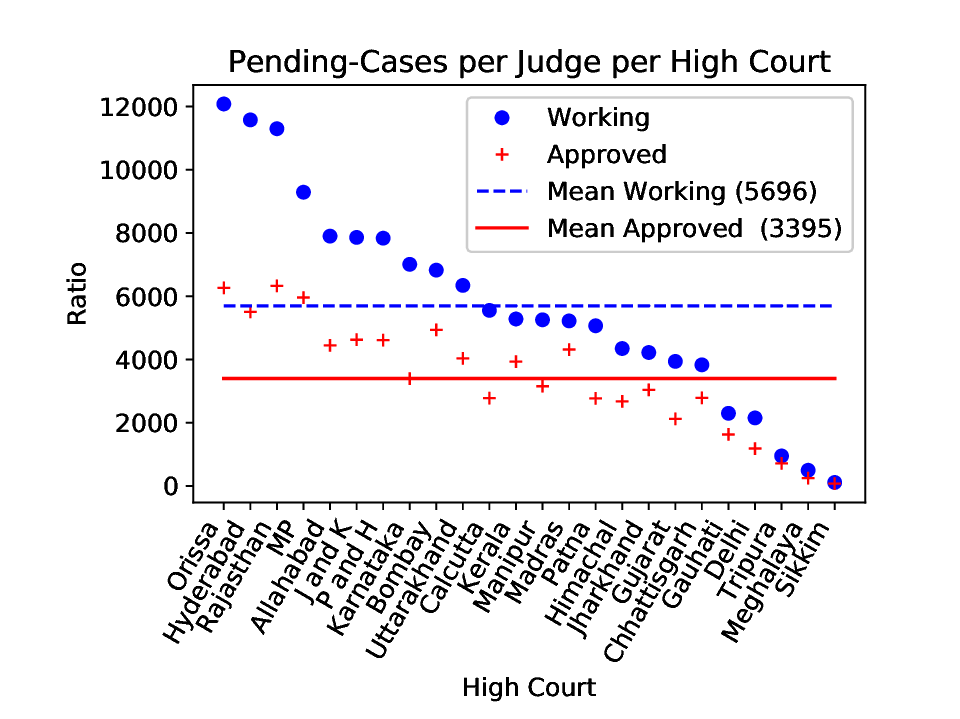}
\caption{Ratio of pending cases to the judges in High Courts. Instead of taking pending cases on one day, we have taken an average over the data we have collected. The number of judges as the working strength of a particular high court as available at the DoJ website on August 31, 2018. The average number of pending cases per judge, if the number of judges is equal to the approved strength of each high court, is also presented.} 
\label{fig:pend_date_r}
\end{figure}

\fref{fig:pend_date_r} plots the ratio $\frac{\text{pending cases}}{\text{number of judges}}$ for each high court. The number of pending cases is calculated by averaging over the data corresponding to each high court. The number of judges, however, are used as on August 31, 2018. The results are plotted in the descending order of the ratio so calculated. Note that the number of judges is taken from the vacancy document on the website of Department of Justice and not from NJDG. This graph provides the distribution of workload on each high court and judges thereof. The blue dots show the ratio $\frac{\text{pending cases}}{\text{working strength of judges}}$ for the working strength of each high court. For example, Orissa High Court has the maximum ratio of 12,080 cases for each judge whereas Sikkim High Court has the minimum ratio which is 110. Hence, statistically we can say that a judge of Orissa High Court has almost 110 times more load than a judge in Sikkim High Court. It can be seen that the situation is similar for most of the high courts. The mean of this ratio is 5696. It signifies that on an average each sitting judge of the high courts in India needs to dispose 5696 cases to get rid of pendency, provided no more cases are filed. On the other hand, the red `+' signs plot the ratio for each high court if we assume that the working strength of each high court is its approved strength. Then the mean of the ratio $\frac{\text{pending cases}}{\text{approved strength of judges}}$ is 3395. Hence, the number of pending cases per judge is huge and the numbers are so high that it would not be unfair to state that they are simply beyond the capacity of human beings, be those humans be the learned judges of high courts. Some input from technology is required to handle such huge numbers and ease the tasks of the judges without compromising on the quality of justice delivered.

\subsection{Disposed, filed and listed cases}

From the previous subsection, we understand that the average load on the judges of high courts is huge in terms of the number of pending cases. Now we see the rate of disposal, filing and listing of cases that can help us provide upper bounds on the time required to get rid of pendency. Unfortunately, NJDG as of now, updates data on monthly disposal and monthly filing of cases. The number of cases listed is provided on a daily basis. If the other two parameters could also be provided on a daily basis then our analysis and interpretations would be more accurate.

\begin{figure}[h]
\includegraphics[width=9.6cm]{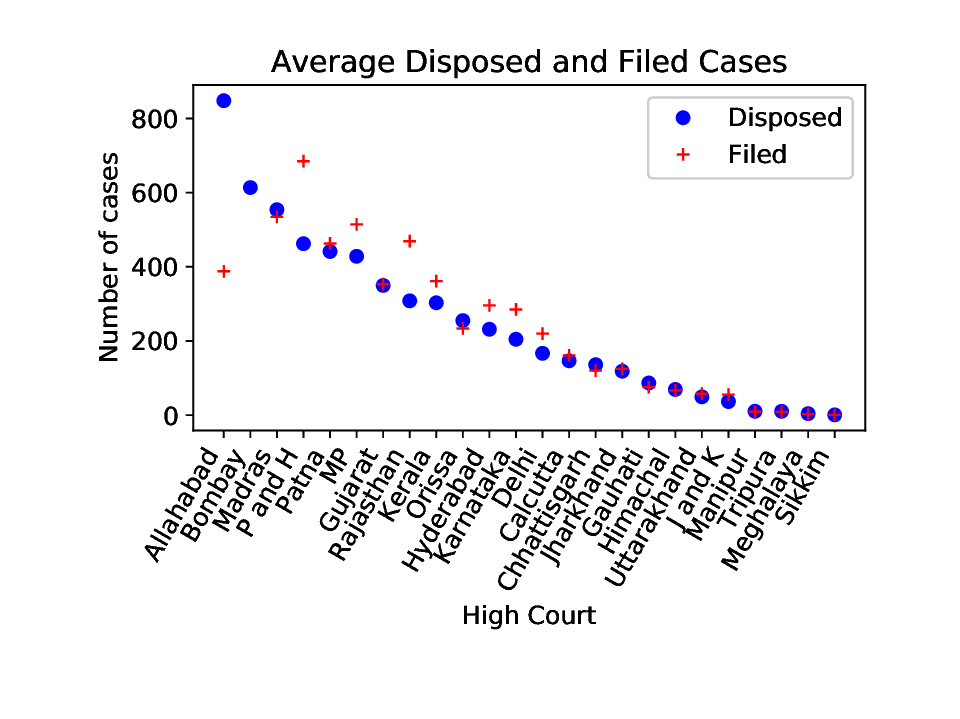}
\caption{Average number of cases disposed and filed monthly.} 
\label{fig:fdl_date}
\end{figure}

\fref{fig:fdl_date} shows the data corresponding to the number of disposed, filed and listed cases as available on NJDG on August 06, 2018. The number of cases disposed and filed are provided monthly. The plot is ordered in descending values of the cases disposed. Also note that the Y-axis is on linear scale and still there is a very little difference between the number of cases listed daily and the number of cases disposed monthly. This may mean that the number of listed cases in a day may be reduced, which will be for the benefit of all the stakeholders. The judges will have more time to hear a case unlike now \cite{daksh_report}. The litigants will be better off because the chances of hearing their case will increase. The advocates will save time in appearing for the cases and hence get more time to prepare for cases. The court staff will have lesser files to move and manage.

While we are not embracing the idea of using NJDG data of just one day, the above figure captures the gap between the three kinds of statistics that are closely related to piling up of pendency. The point here is that the causlists in the high courts should be prepared more scientifically. We back up our assertion by a more careful and rigorous analysis involving more data.

%

\begin{figure}[h]
\includegraphics[width=9.6cm]{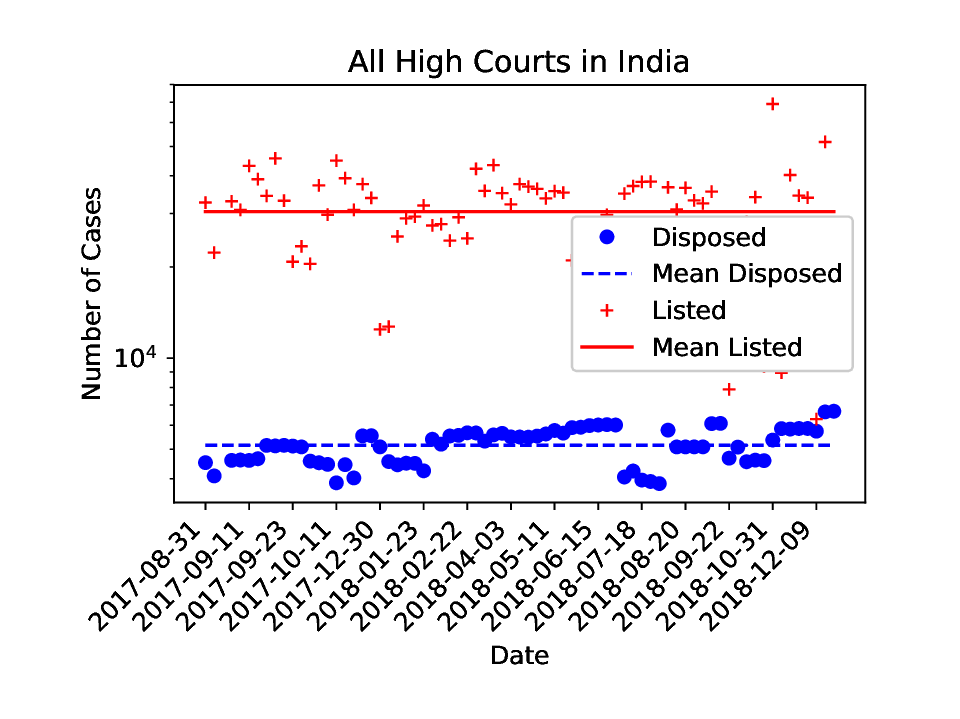}
\caption{The aggregate number of cases disposed (daily) and listed (daily) for high courts in India. We see that the number of cases listed is an order of magnitude more than the number of cases disposed.} 
\label{fig:fdl_hcall}
\end{figure}

It is shown in \fref{fig:fdl_hcall} that there is a huge gap between the number of cases listed and the number of cases disposed in a day. The number of cases disposed in a day is obtained by dividing the monthly statistics by 22 because we are assuming that there are 22 working days in a month. The goal should be to minimize the gap between the number of cases disposed and the number of cases listed on any given day. This would mean that most of the cases that are listed, should be disposed, rather than adjourned. This is the parameter that makes us claim that there is a room for more scientific preparation of causelists as there is a room for decreasing the gap between the number of cases listed and the number of cases disposed in a day. Similar is the case for individual high courts as presented in \fref{fig:fdl_hc}.

\begin{figure}[h]
\includegraphics[width=9.6cm]{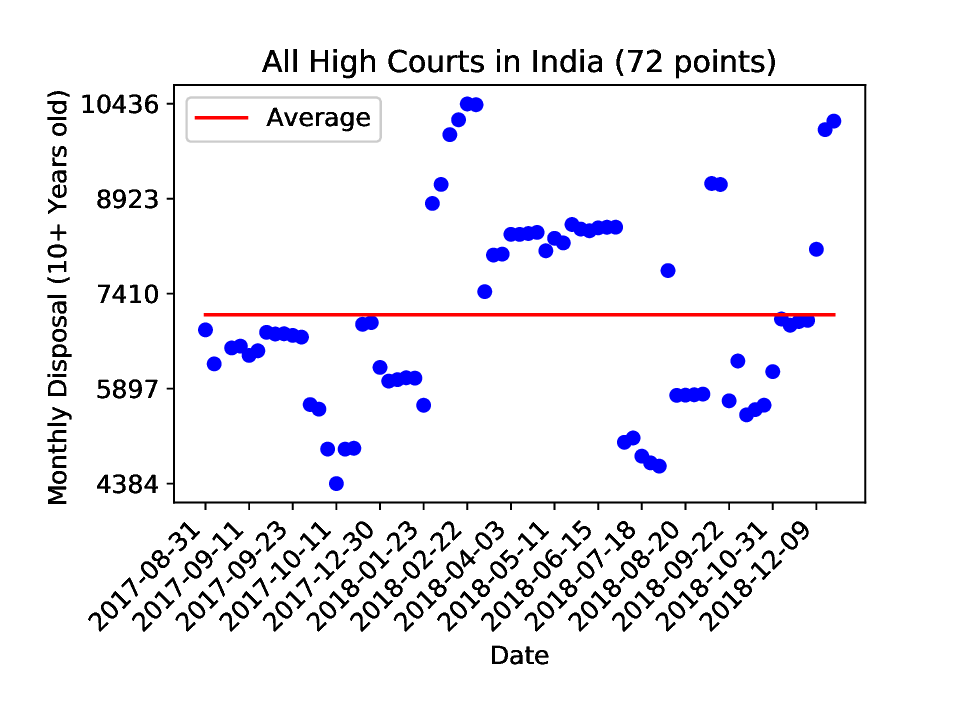}
\caption{The aggregate number of disposed cases that were pending for more than ten years for high courts in India.} 
\label{fig:fdl10_hcall}
\end{figure}

\fref{fig:fdl10_hcall} shows the number of disposed cases whose age was more than ten years. The graph is not expected to show any trend as the number of cases disposed depends on many circumstances. The only thing that can be reasonably concluded is that the number of disposal of more than ten year old cases was high during February 2018 to March 2018. It has decreased since then.

\section{Judges in High Courts}
\label{sec:judges_hc}

\begin{figure*}[h]
\includegraphics[width=4.4cm]{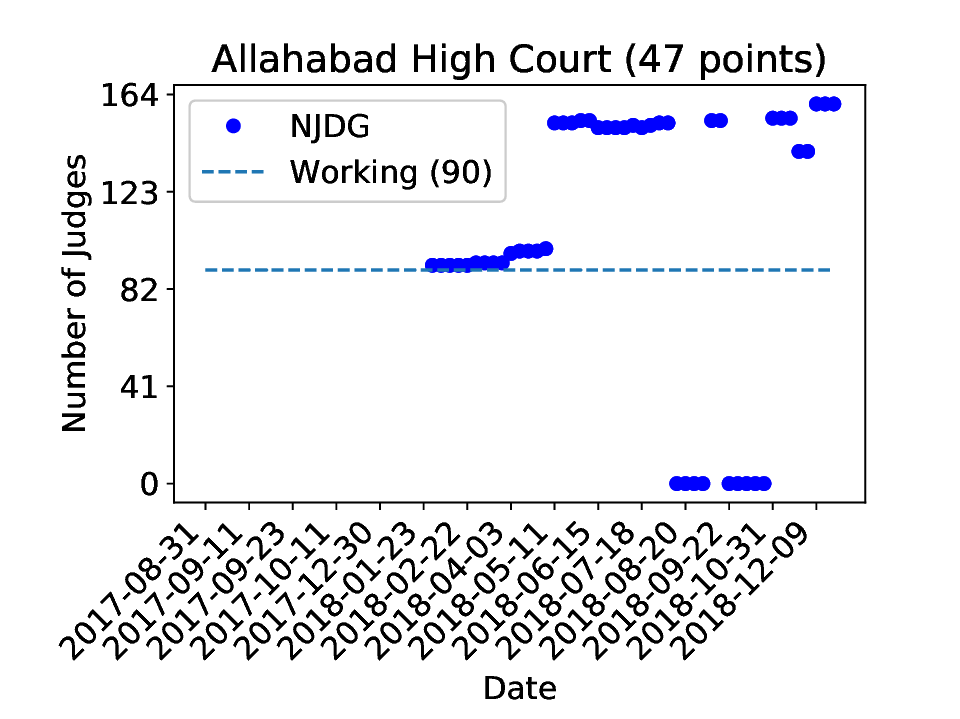}
\includegraphics[width=4.4cm]{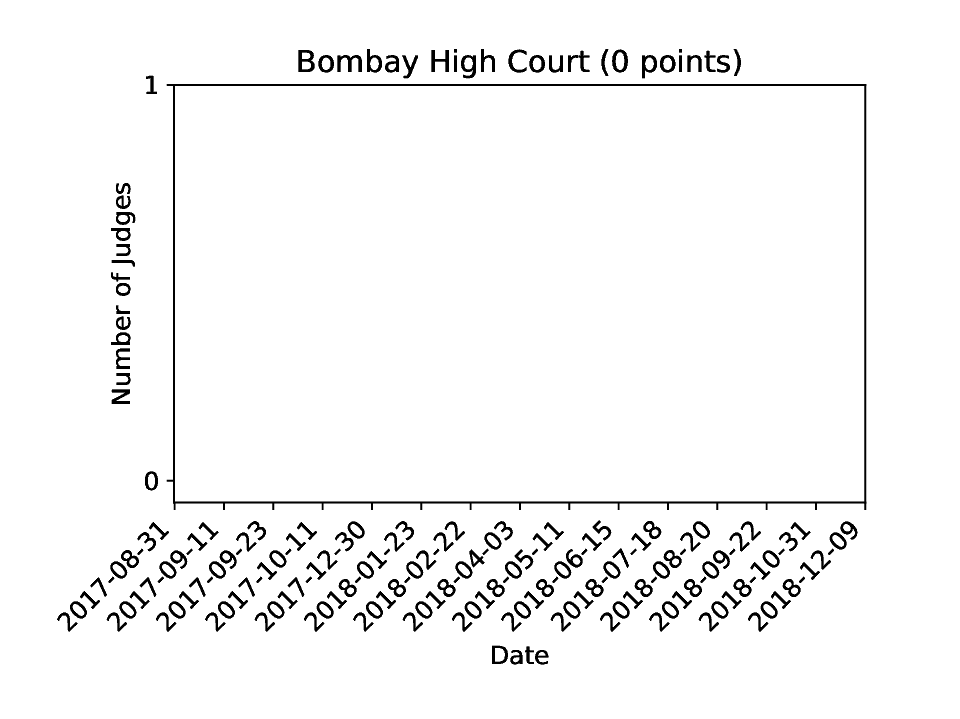}
\includegraphics[width=4.4cm]{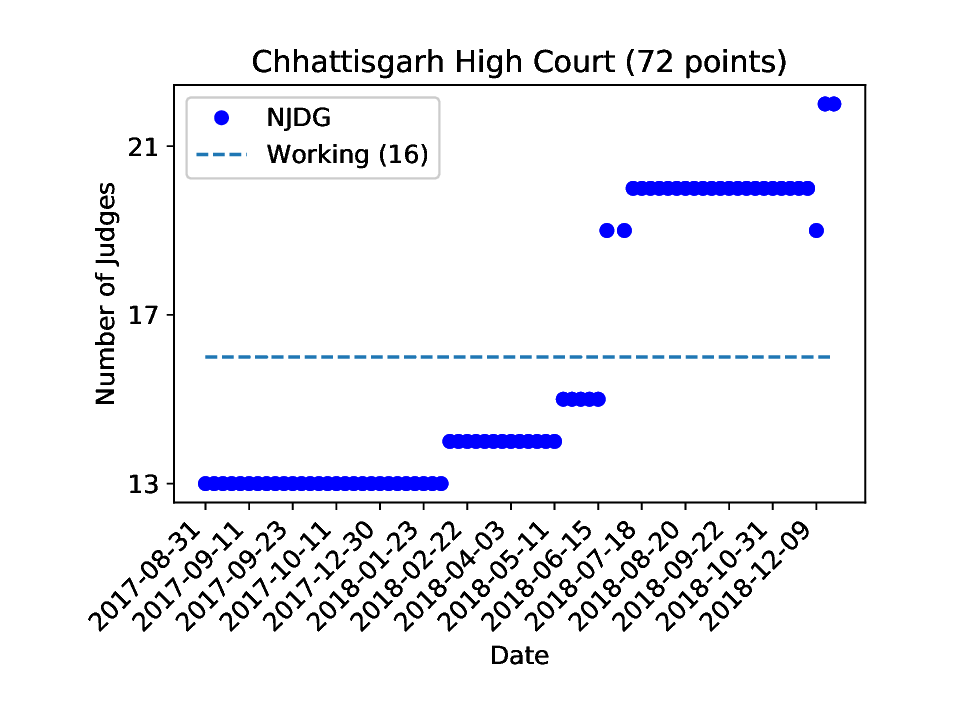}
\includegraphics[width=4.4cm]{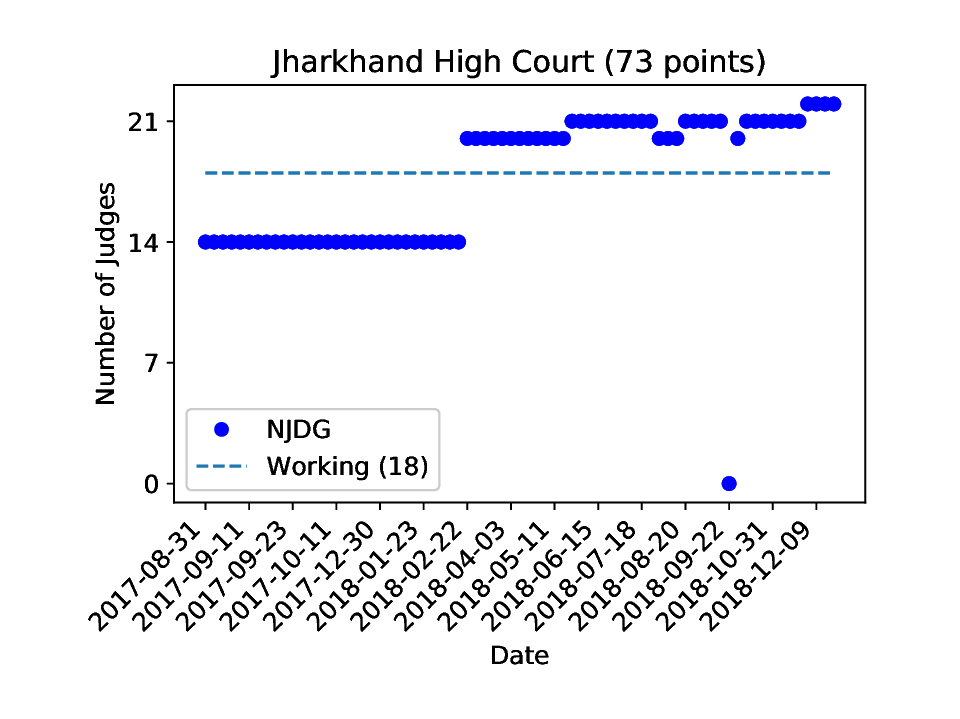}
\includegraphics[width=4.4cm]{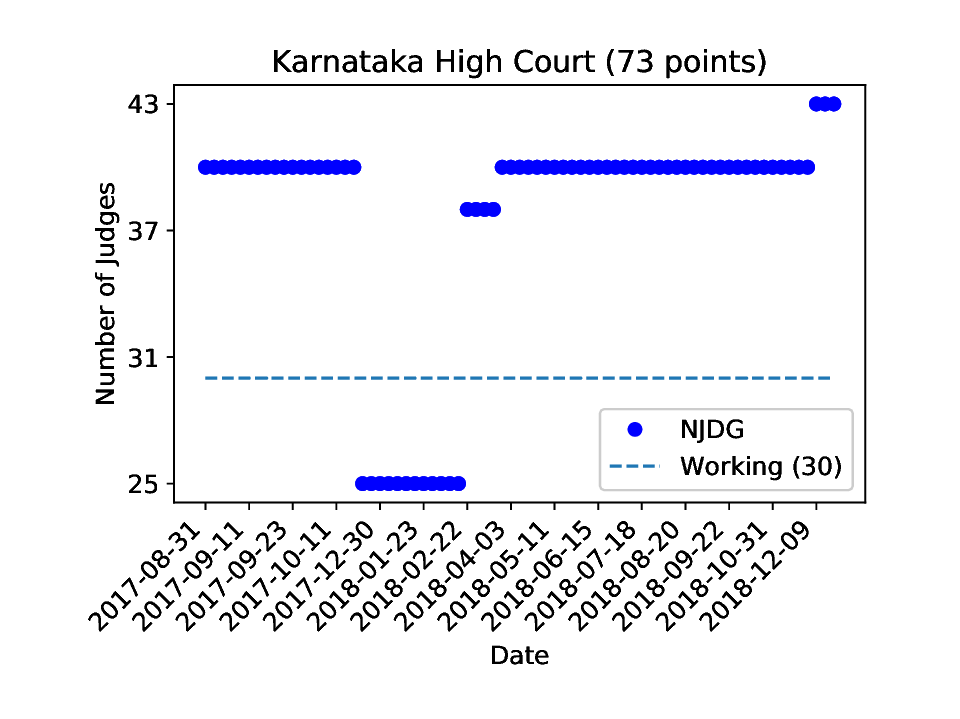}
\includegraphics[width=4.4cm]{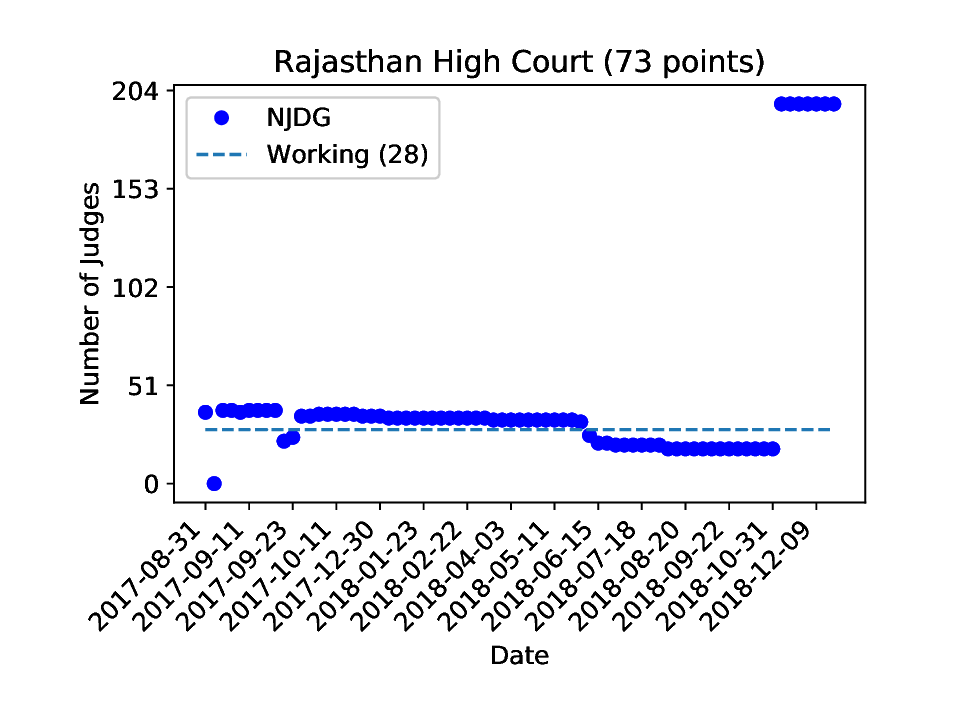}
\includegraphics[width=4.4cm]{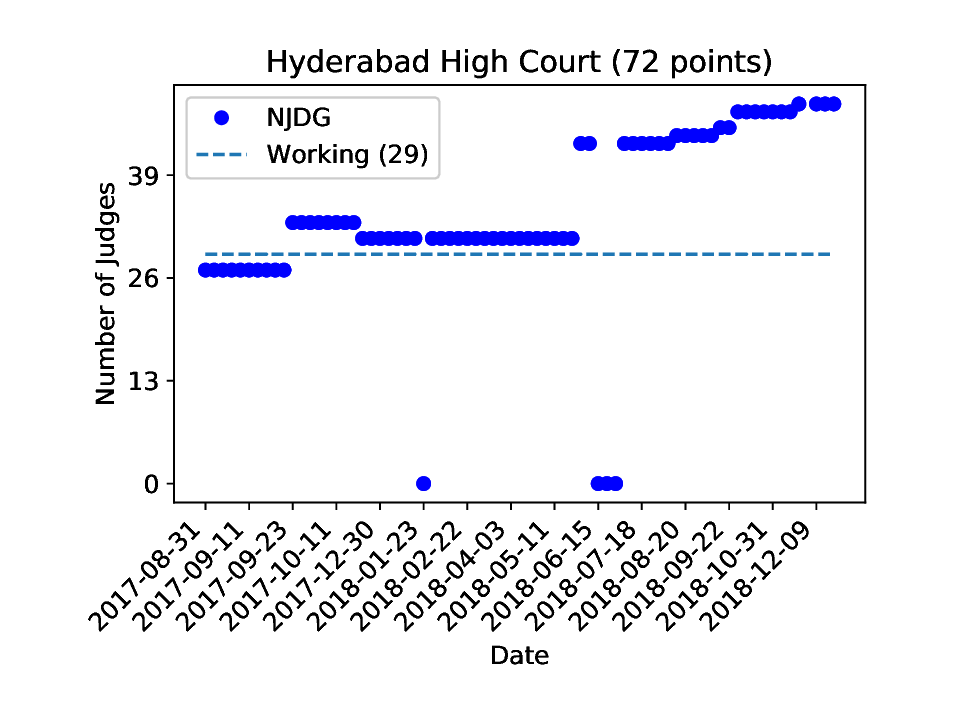}
\caption{High Courts that have unusual and unexpected data with respect to the number of judges in the respective high court. The number, though within the approved strength, is not close to the working strength as on August 31, 2018.} 
\label{fig:njudgeshc}
\end{figure*}

\begin{figure*}[h]
\includegraphics[width=5.9cm]{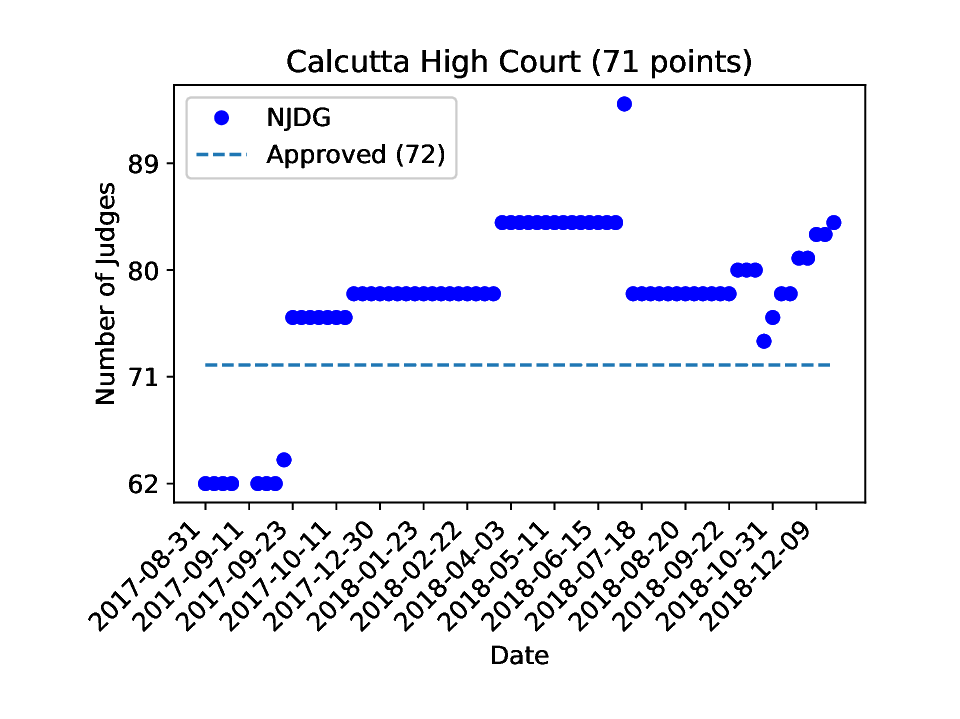}
\includegraphics[width=5.9cm]{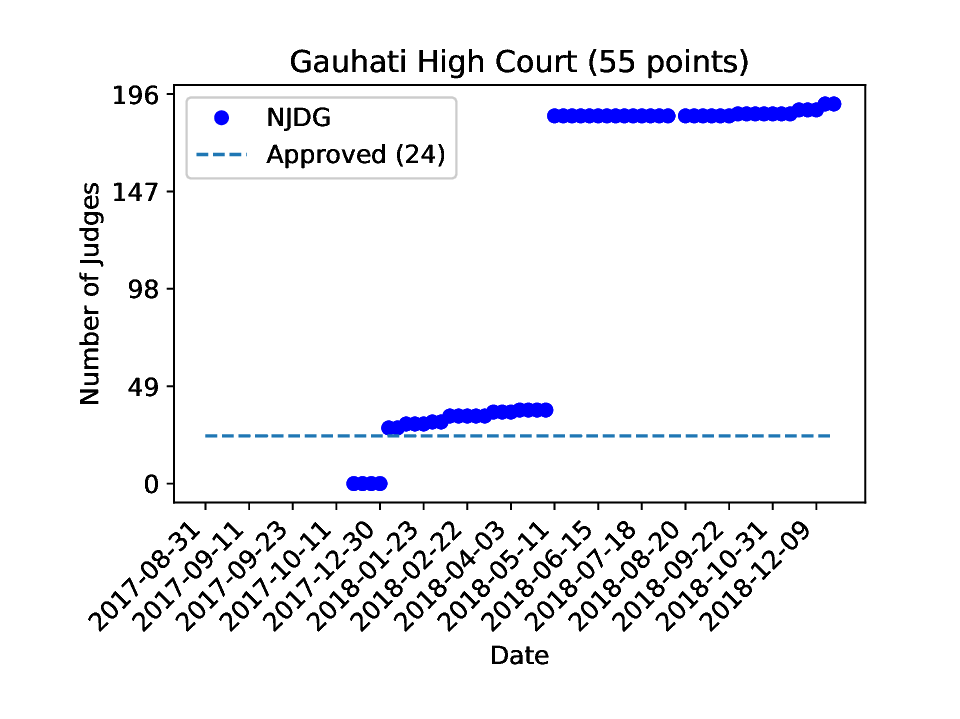}
\includegraphics[width=5.9cm]{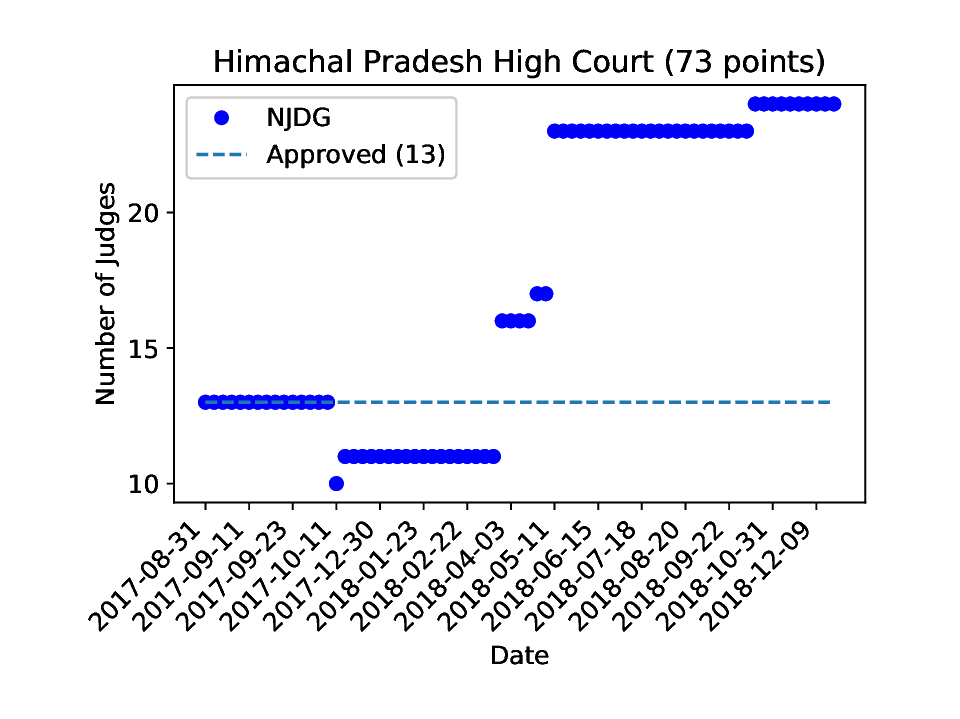}
\includegraphics[width=5.9cm]{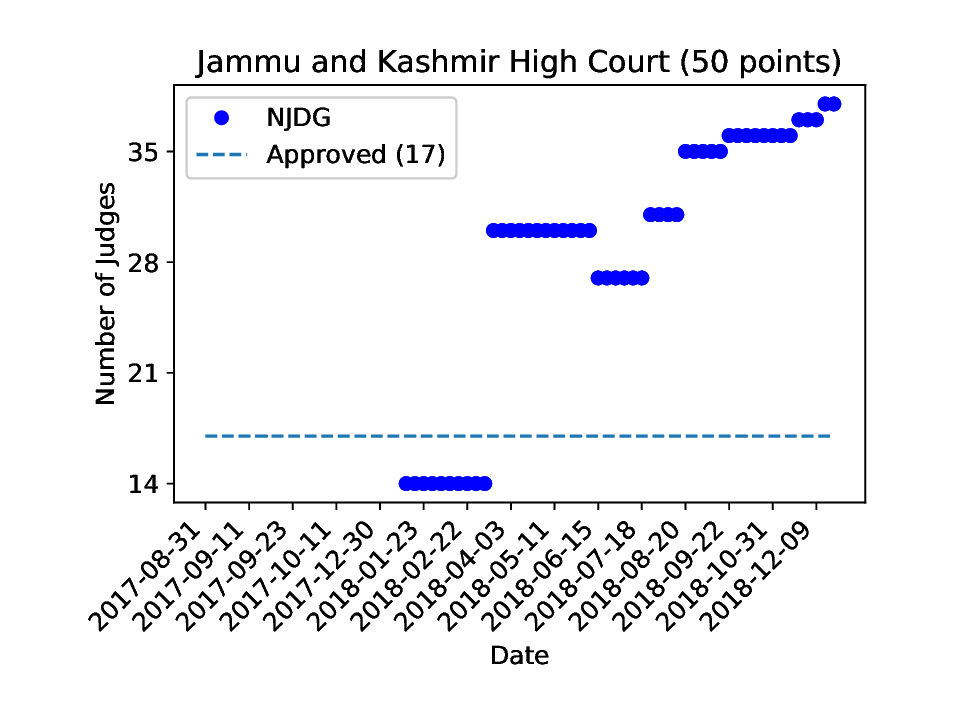}
\includegraphics[width=5.9cm]{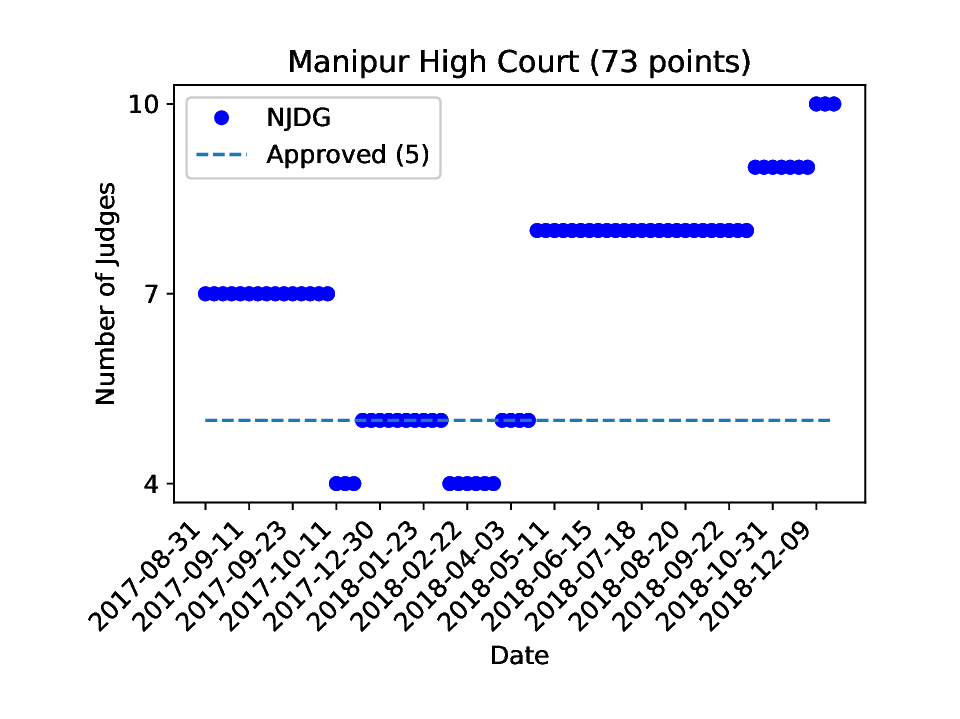}
\includegraphics[width=5.9cm]{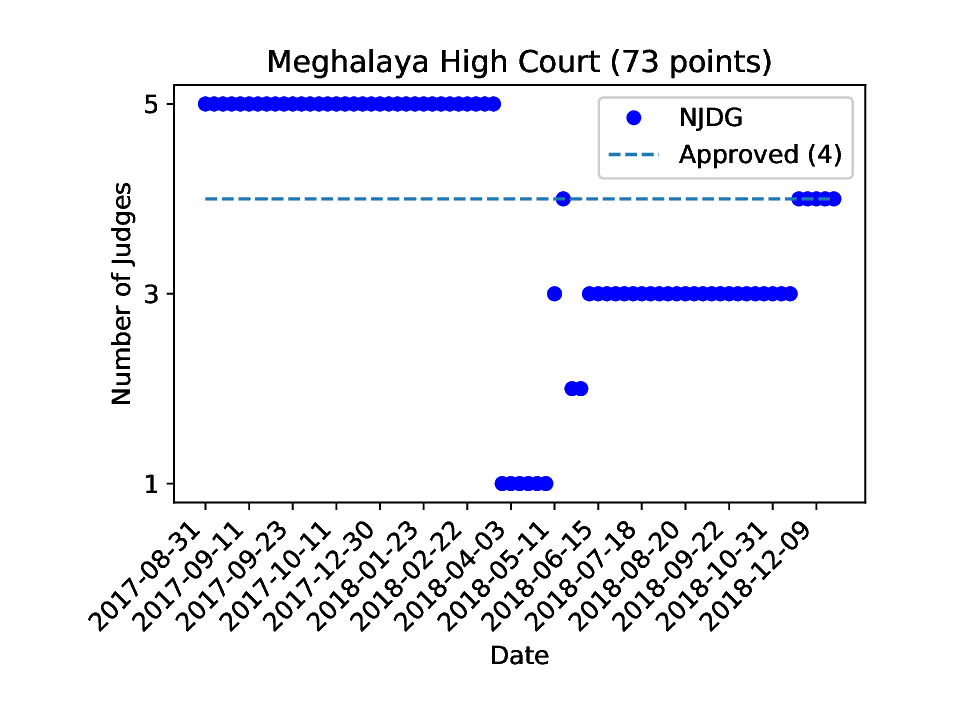}
\includegraphics[width=5.9cm]{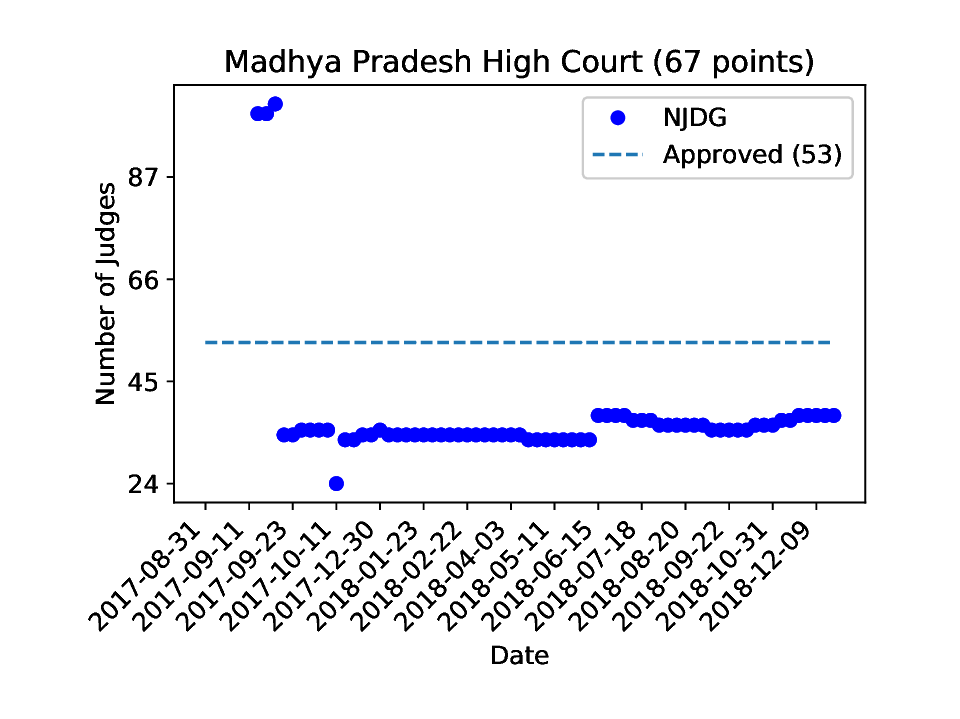}~~
\includegraphics[width=5.9cm]{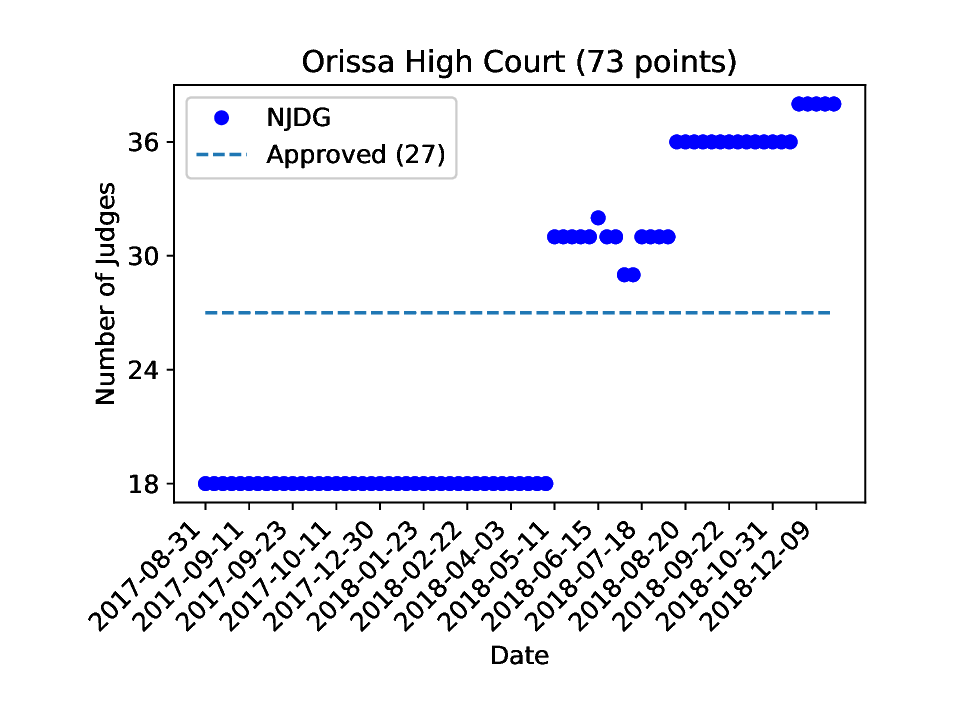}~~
\includegraphics[width=5.9cm]{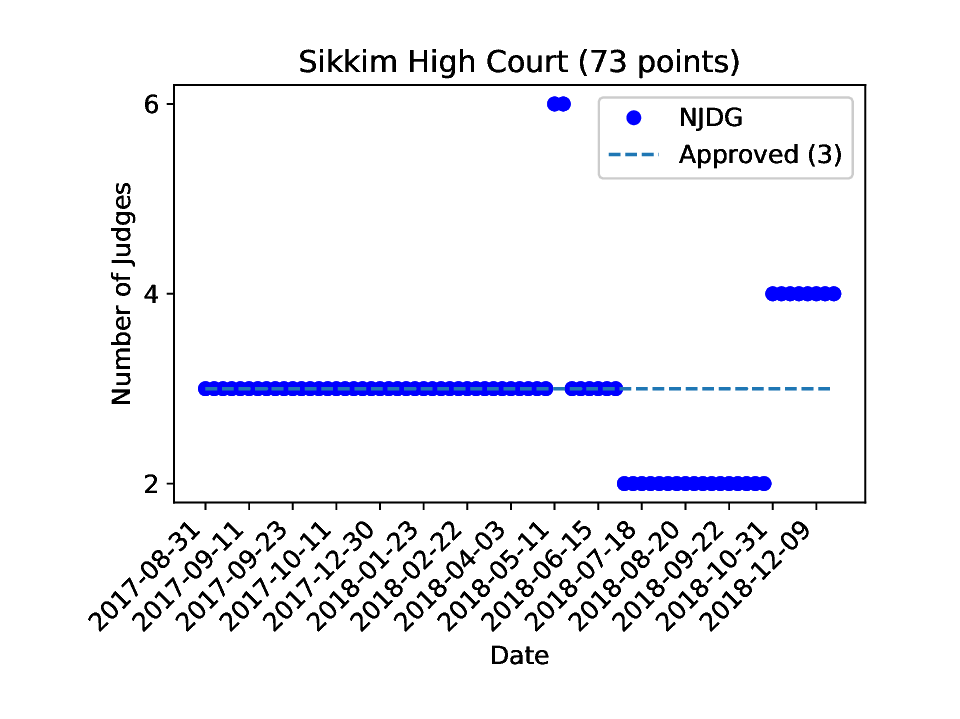}

\caption{High Courts that have reported, at least once, the number of judges more than the approved strength checked against the document on the Department of Justice website.} 
\label{fig:njudgeshc1}
\end{figure*}

\begin{figure*}[h]
\includegraphics[width=4.4cm]{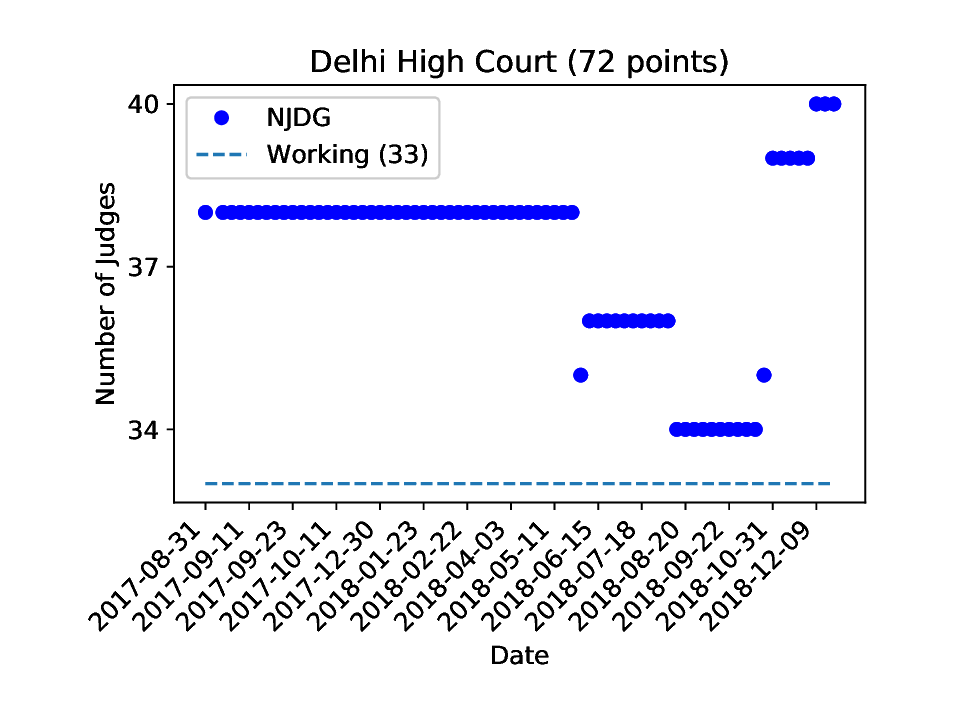}
\includegraphics[width=4.4cm]{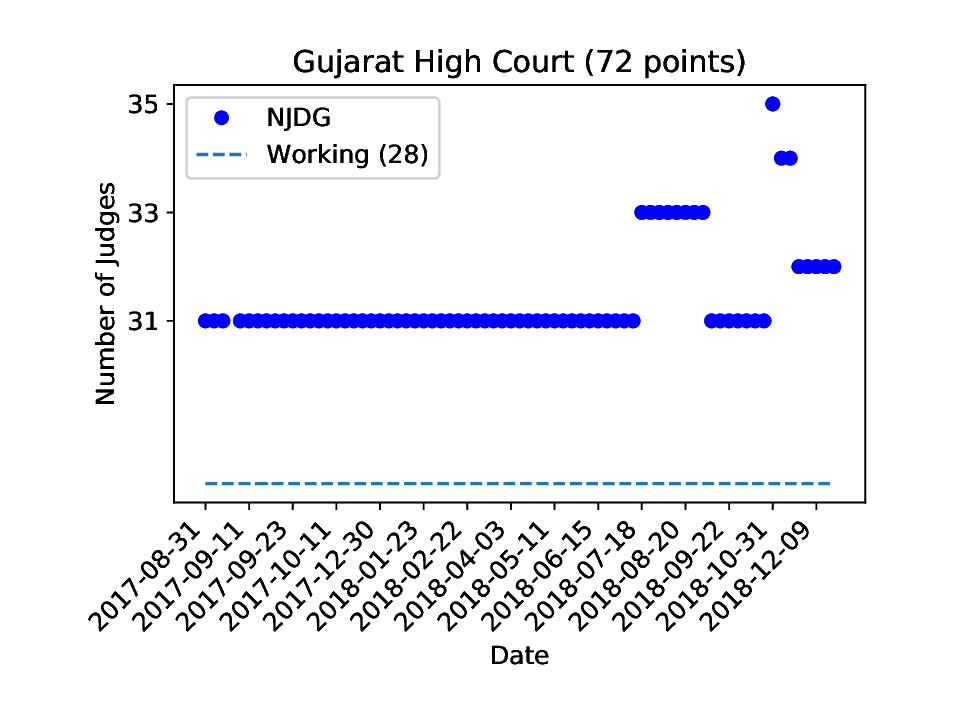}
\includegraphics[width=4.4cm]{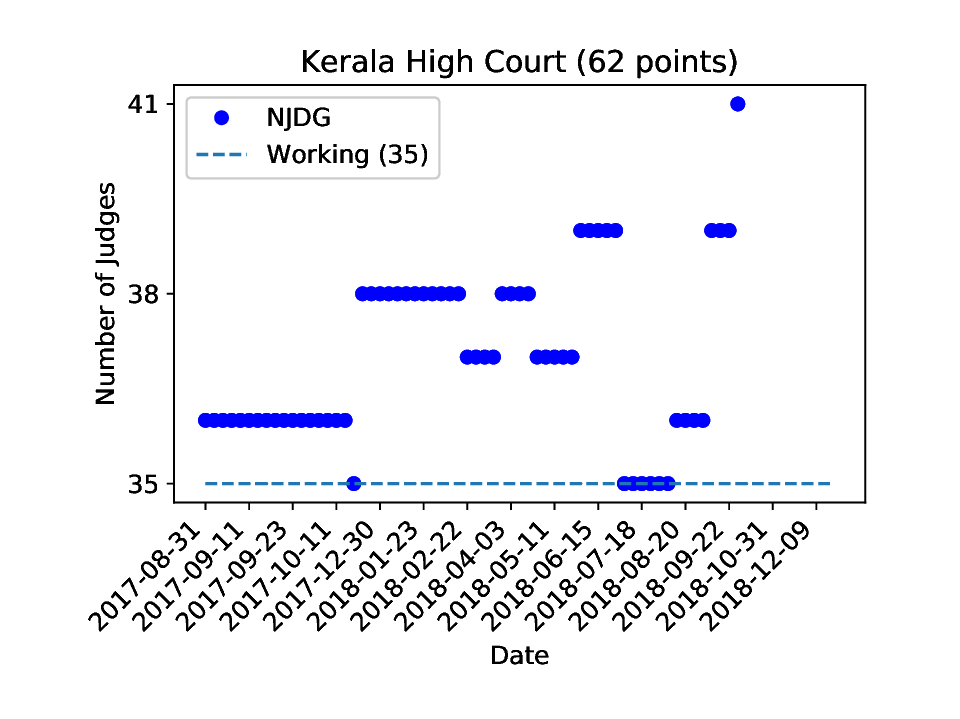}
\includegraphics[width=4.4cm]{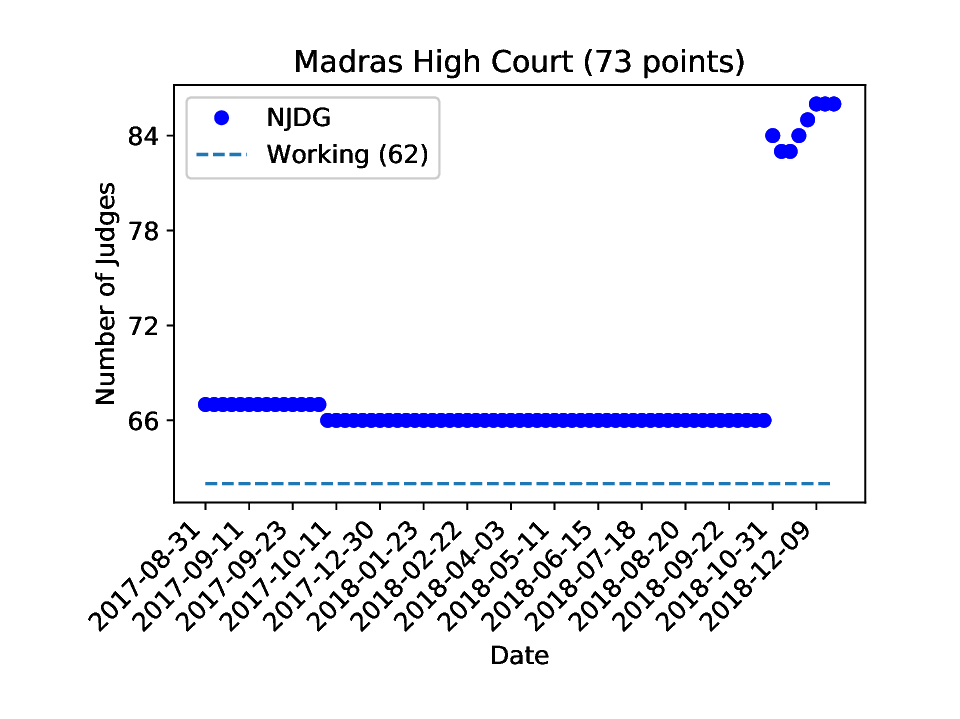}
\includegraphics[width=4.4cm]{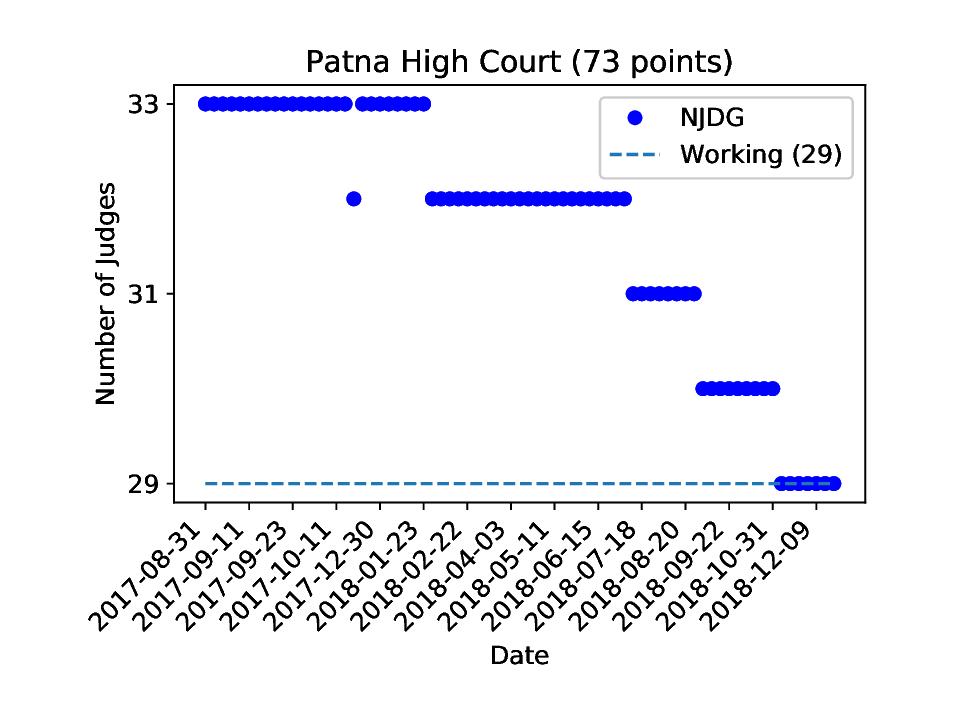}~
\includegraphics[width=4.4cm]{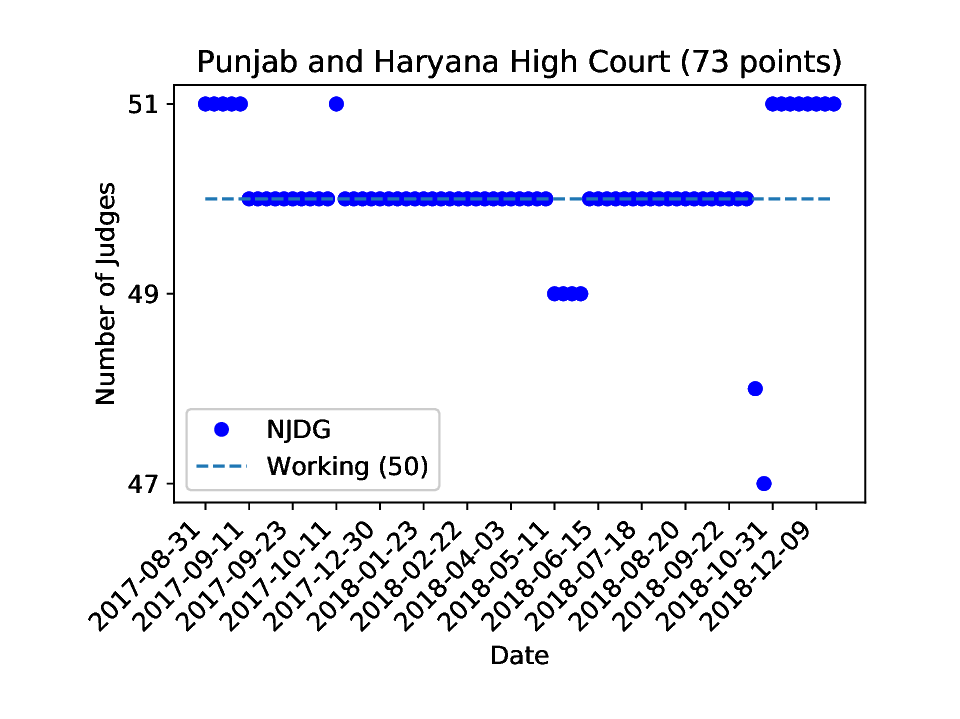}~
\includegraphics[width=4.4cm]{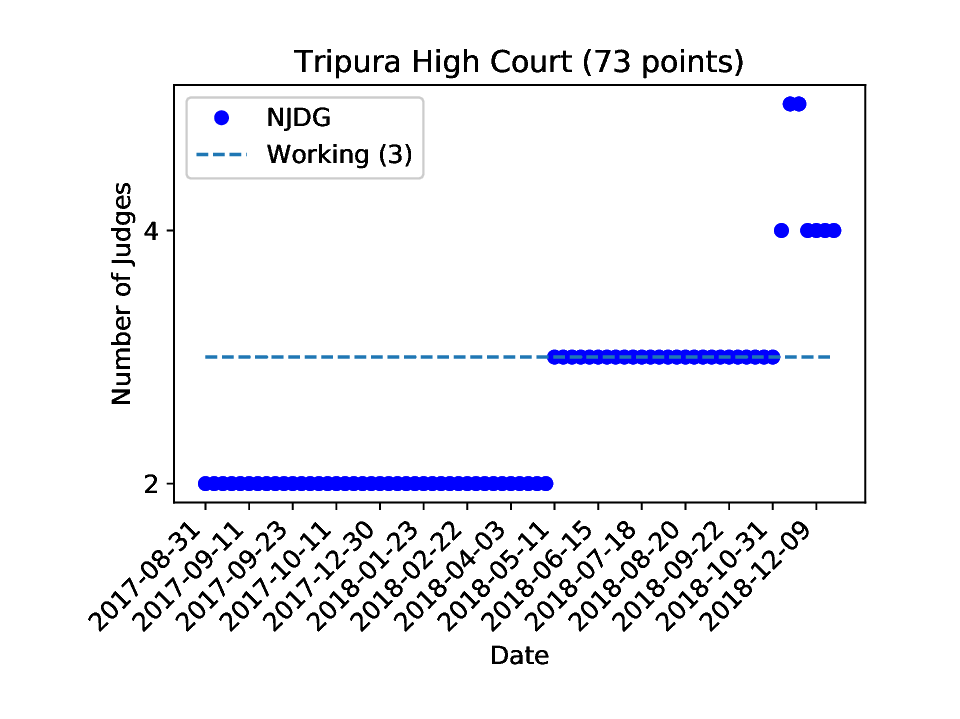}~
\includegraphics[width=4.4cm]{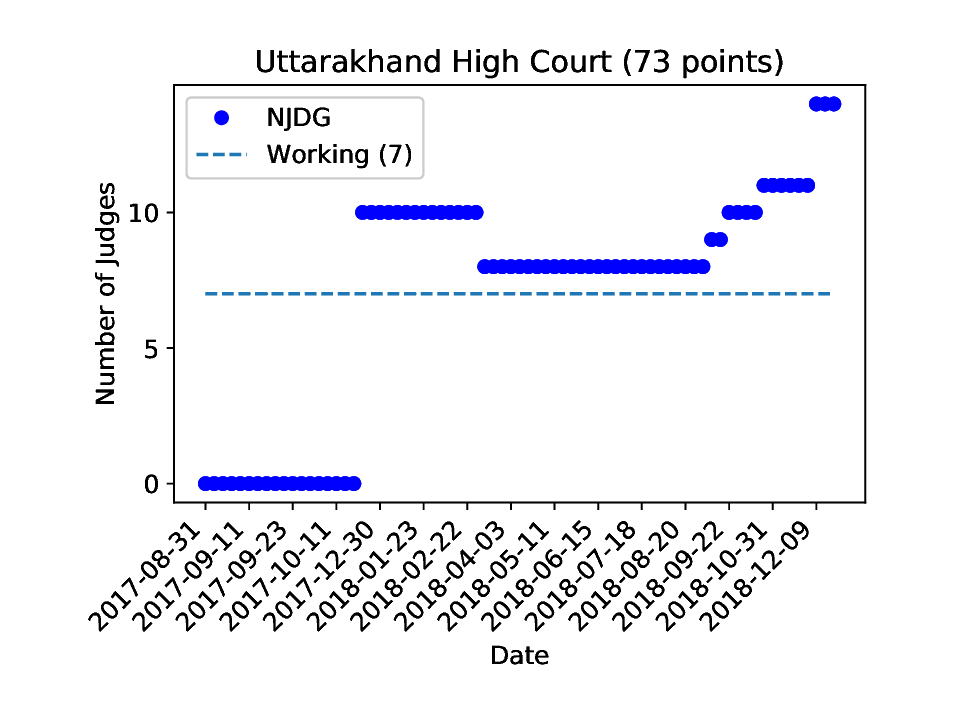}
\caption{High Courts with accurate or close to accurate working strength as on August 31, 2018 on HC-NJDG according to the vacancy document.}  
\label{fig:njudgeshc2}
\end{figure*}

This section presents the study on the number of judges in high courts. For the purpose of this section, we have taken the number of judges in High Courts as on August 31, 2018. As shown in \fref{fig:hcjudges}, the number of judges on the HC-NJDG portal is not in accordance with the vacancy document available at the Department of Justice website \cite{doj_vacancy}, which is also our main document for comparing correctness in this section. We provide more details on this parameter as available on NJDG. The number of judges, as available on NJDG, is divided into the following three broad categories:
\begin{enumerate}
\item High courts with the number of judges on NJDG very different from actual working strength (\fref{fig:njudgeshc}).
\item High courts with data on NJDG greater than the approved strength for that high court (\fref{fig:njudgeshc1}).
\item High courts with accurate or close to accurate NJDG data on the number of judges (\fref{fig:njudgeshc2}).
\end{enumerate}

\fref{fig:njudgeshc} presents the graphs of seven high courts that have unusual and unexpected ups and downs in the data on the number of judges in the high courts. Apart from the number of judges reported by NJDG, the working strength of each high court as on August 31, 2018 is also shown in parenthesis. The data is not close to reality. For example, corresponding to Allahabad high court, the number took a jump of 53 from 99 on May 02, 2018 to 152 on May 11, 2018. There can hardly be any justification for this. Hence, we would suggest that there should be some kind of checks built in the software itself that raises a warning whenever such unusual updates are being made. They are most likely to be clerical errors that have stood the test of time in absence of regular monitoring. During the complete data collection period, Bombay High Court has never provided the data on the number of judges. As it can be seen from our whole data analysis as well, Bombay High Court has seen the very few updates on HC-NJDG. Similar results can be seen for other high courts, namely, Chhattisgarh, Jharkhand, Karnataka and Hyderabad High Court as well. The case of Rajasthan High Court is a bit different in this regard. Sudden drop in statistics is due to incorrect update at the Principal Seat of High Court at Jodhpur. Since June 28, 2018, the number of judges, as shown on the NJDG establishment of Principal Seat at Jodhpur High Court is 0.

\fref{fig:njudgeshc1} presents all those high courts that have the number of judges as shown on HC-NJDG, greater than the approved strength as mentioned in the Department of Justice vacancy document \cite{doj_vacancy} dated August 31, 2018. The number of judges on HC-NJDG as well as the approved strength, as on the vacancy document, is also shown. The number in the bracket of legend \emph{Approved} is the number of the approved strength of that particular high court. We have included all those high courts whose NJDG data has such discrepancy at least once. Our emphasis here is that such data should have never made to the public portal. The software itself should be written such that it checks for such errors and raises warnings whenever such mistakes of fact occur.

Finally, \fref{fig:njudgeshc2} captures those high courts for which the data is more or less consistent with the vacancy document on the Department of Justice website. In these cases, the discrepancy is most likely because of the fact that we are using the working strength of the high courts as on August 31, 2018. The situation might have been different in the earlier months. We also see that for all these high courts, the number of judges have not varied much, apart from the last two months of data collection. However, the contribution of last two months in our data is only one-eighth which is not so significant. Moreover, the number of high court judges have increased substantially in last 4 months, which do not reflect the actual trend. It will take some time to see the impact of this increase.

\section{Pending Cases in High Courts}
\label{sec:pending_hc}

The problem of pending cases in India has taken an unimaginable form. In high courts alone, more than 20\% of the cases are pending for more than 10 years.  Hon'ble Supreme Court's decision of implementing e-Courts project have come timely, the things are improving, but there is a long way to go to reduce pendency. The energy and efforts put in e-Courts project must continue for few more years, if not decades, to see the real impact. 

Since there is no other benchmark for pending cases like the vacancy document of the Department of Justice \cite{doj_vacancy}, we shall assume the data obtained from NJDG to be correct. Then we will make use of some more fundamental arguments to see why not everything is right about the data and suggest scope for improvements.

\begin{figure*}[h]
\includegraphics[width=4.4cm]{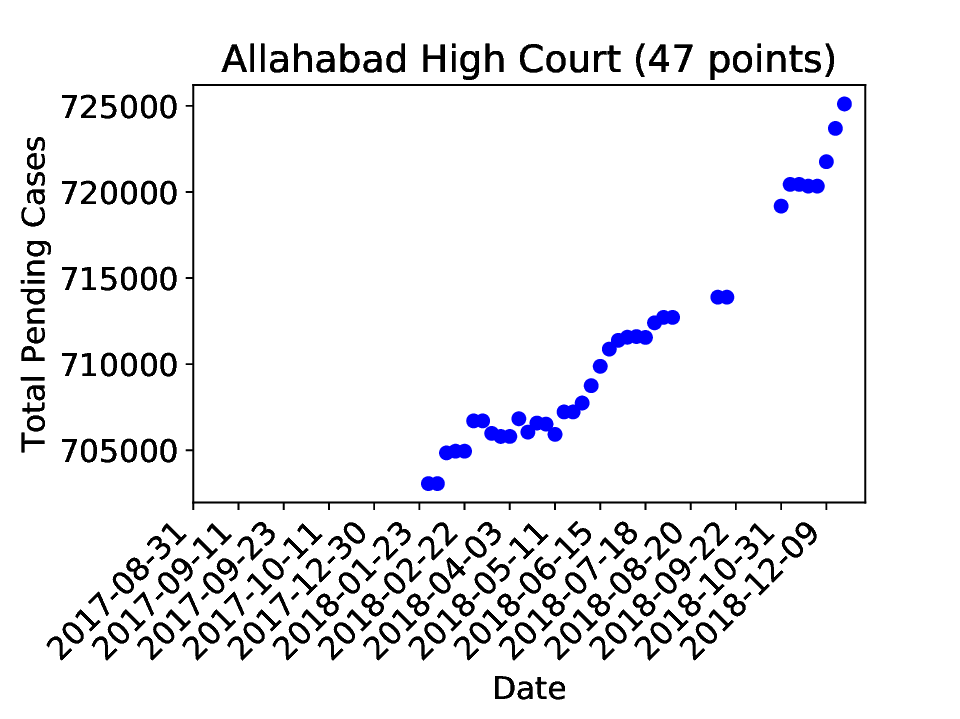}
\includegraphics[width=4.4cm]{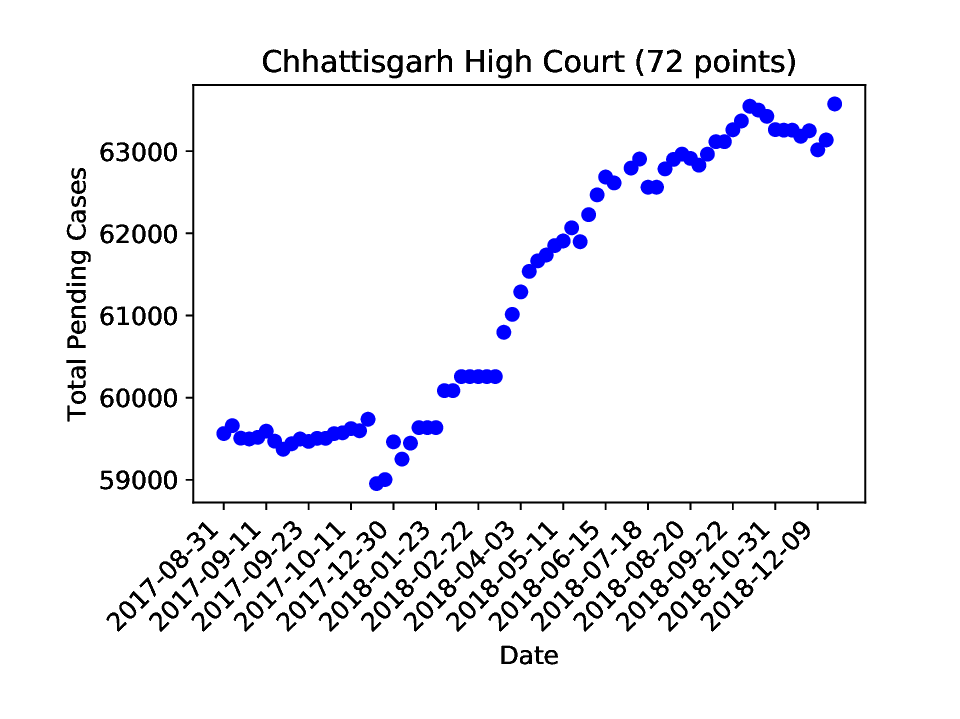}
\includegraphics[width=4.4cm]{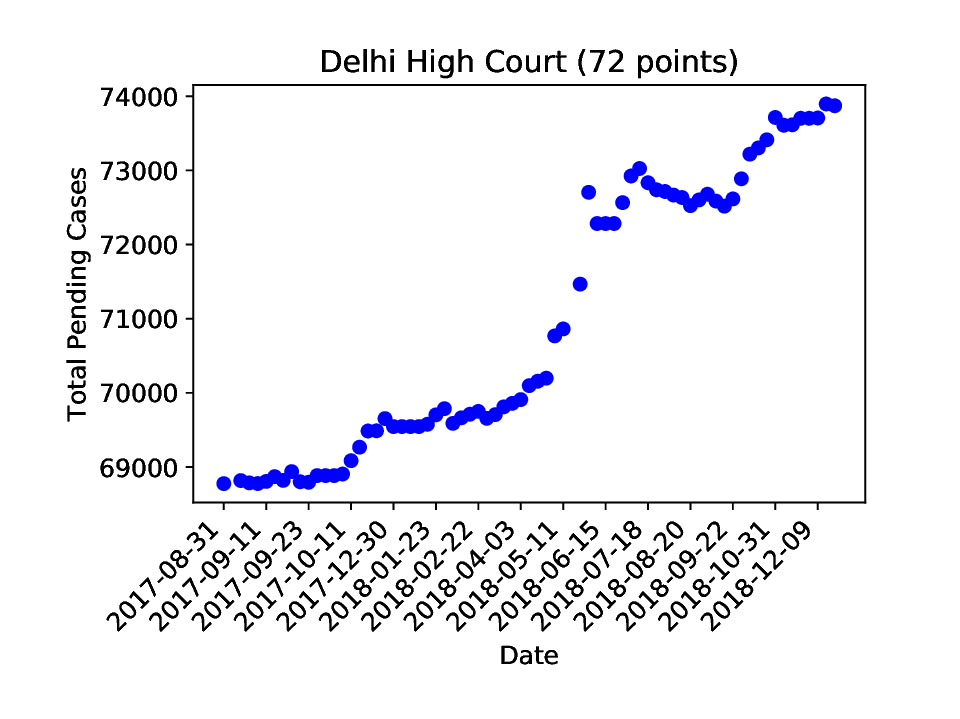}
\includegraphics[width=4.4cm]{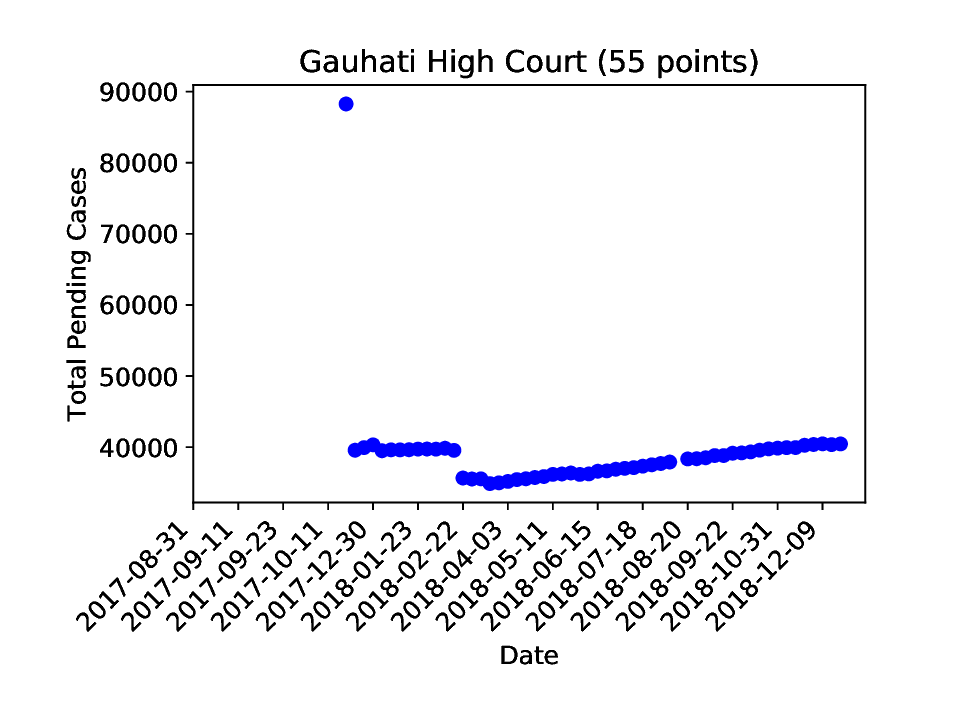}
\includegraphics[width=4.4cm]{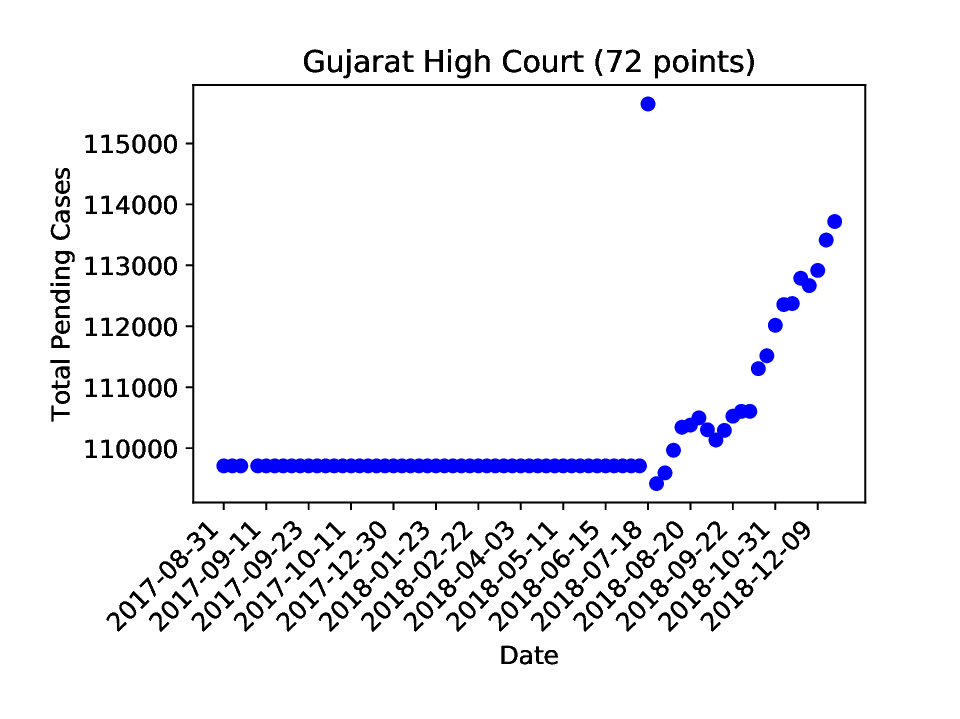}
\includegraphics[width=4.4cm]{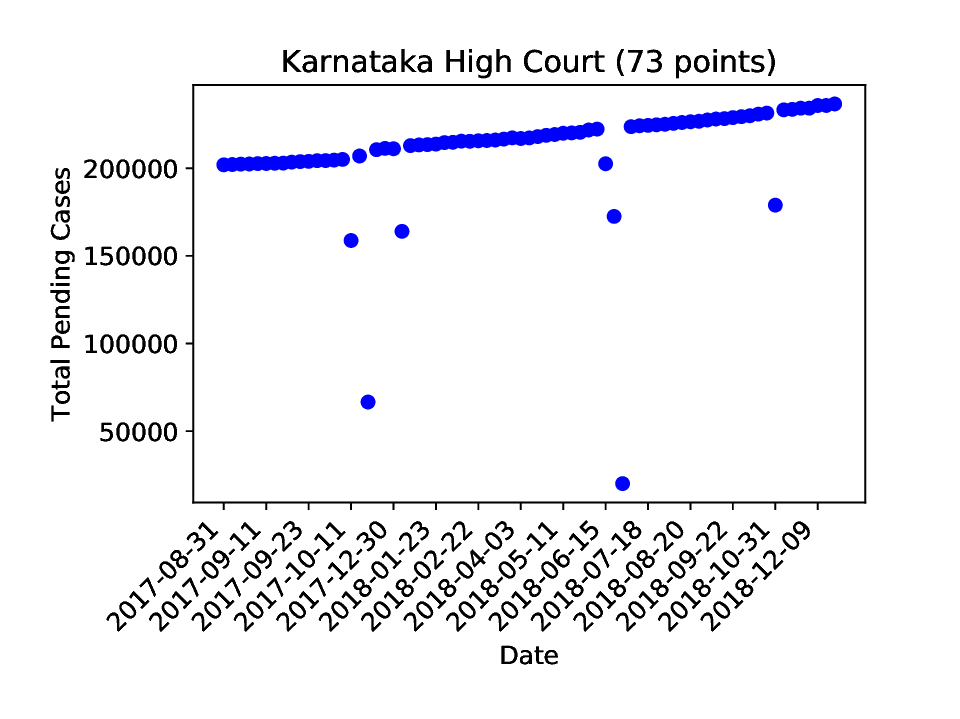}
\includegraphics[width=4.4cm]{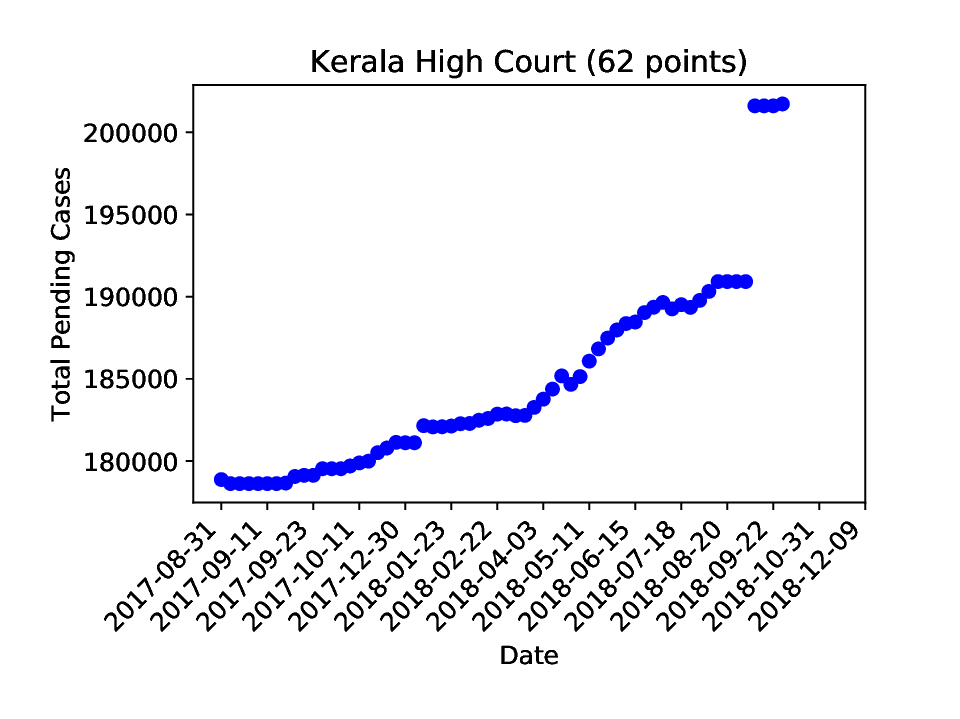}
\includegraphics[width=4.4cm]{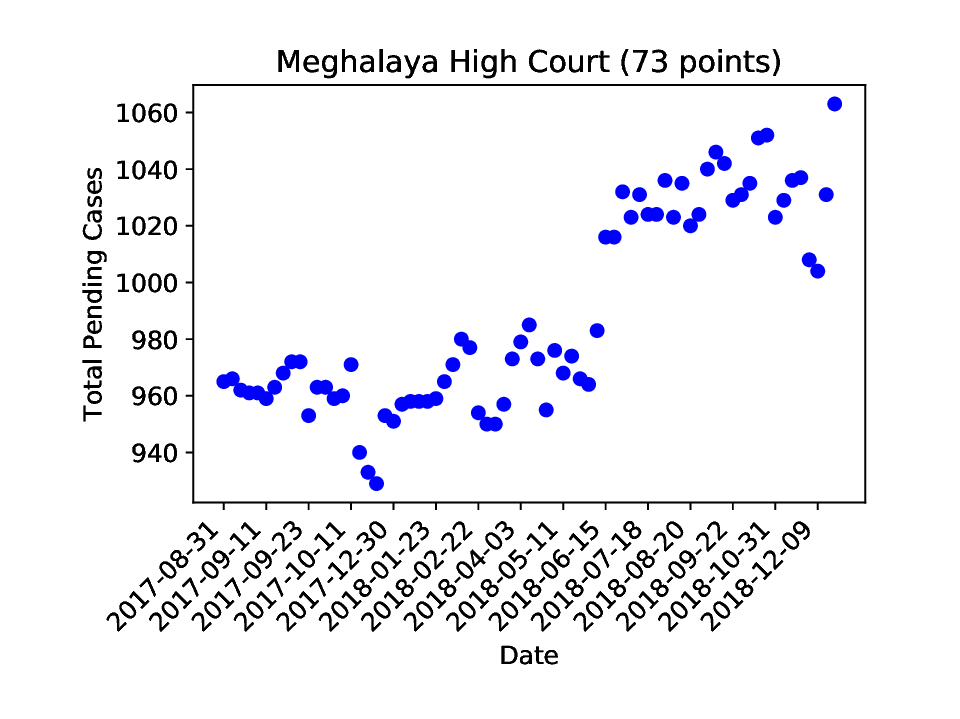}
\includegraphics[width=4.4cm]{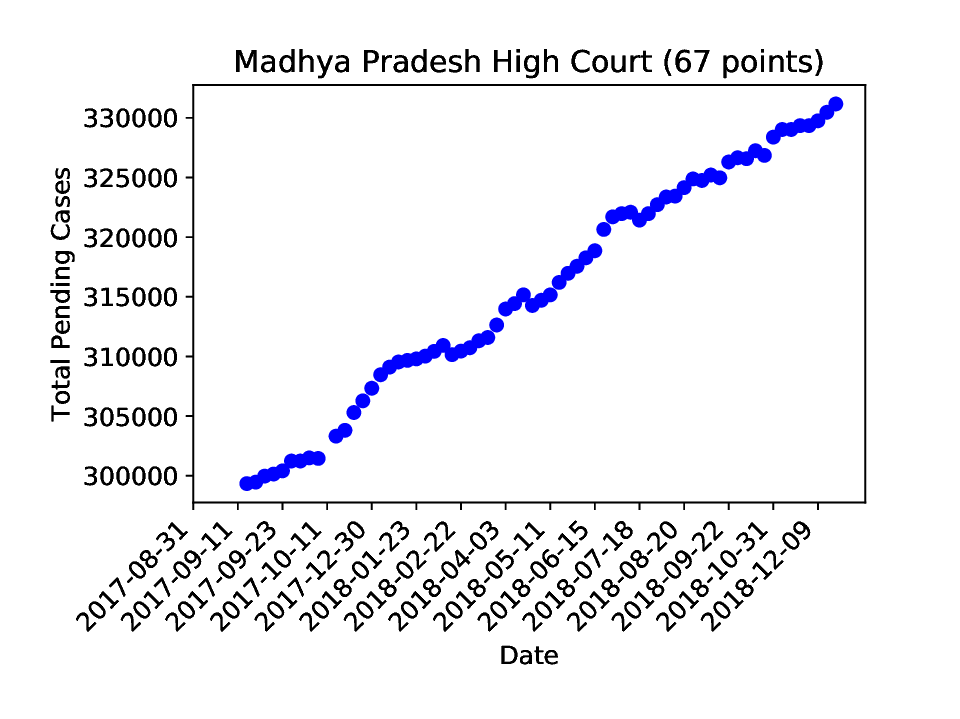}
\includegraphics[width=4.4cm]{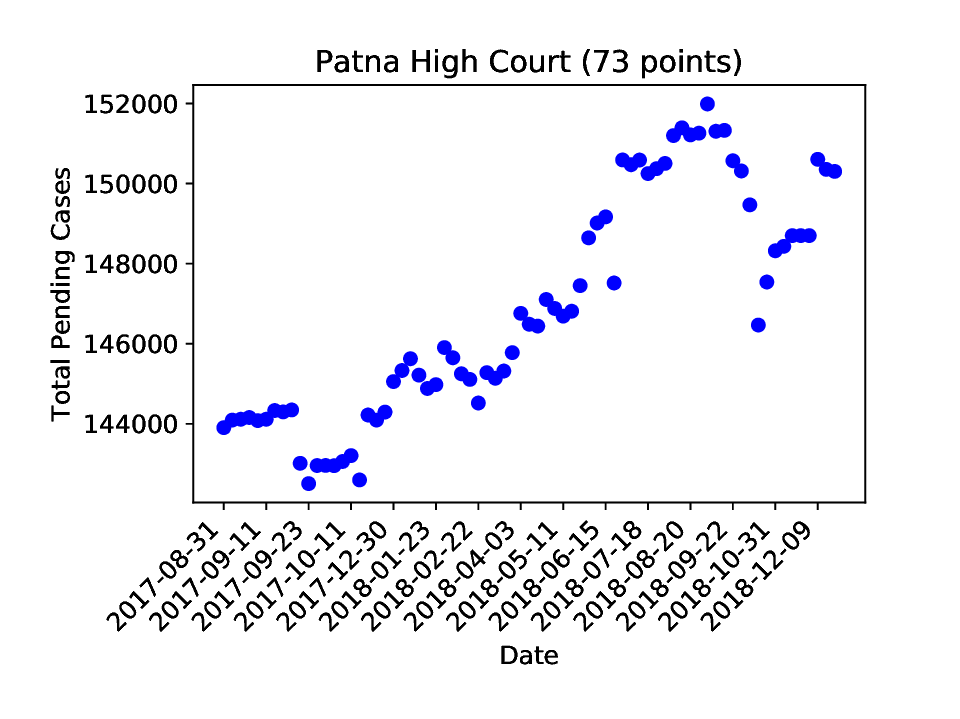}
\includegraphics[width=4.4cm]{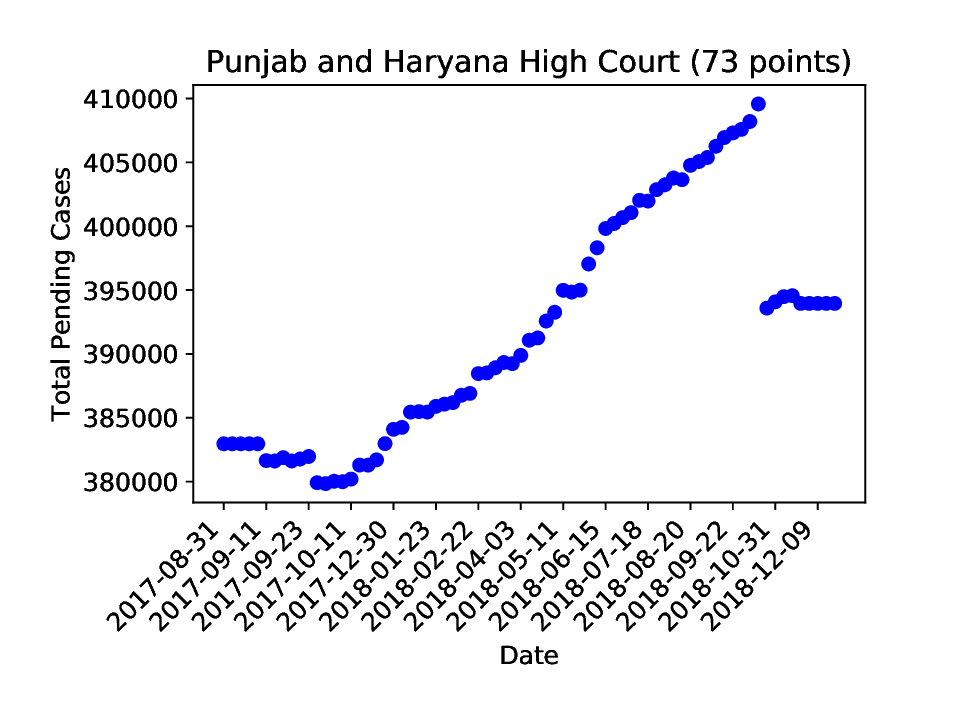}
\includegraphics[width=4.4cm]{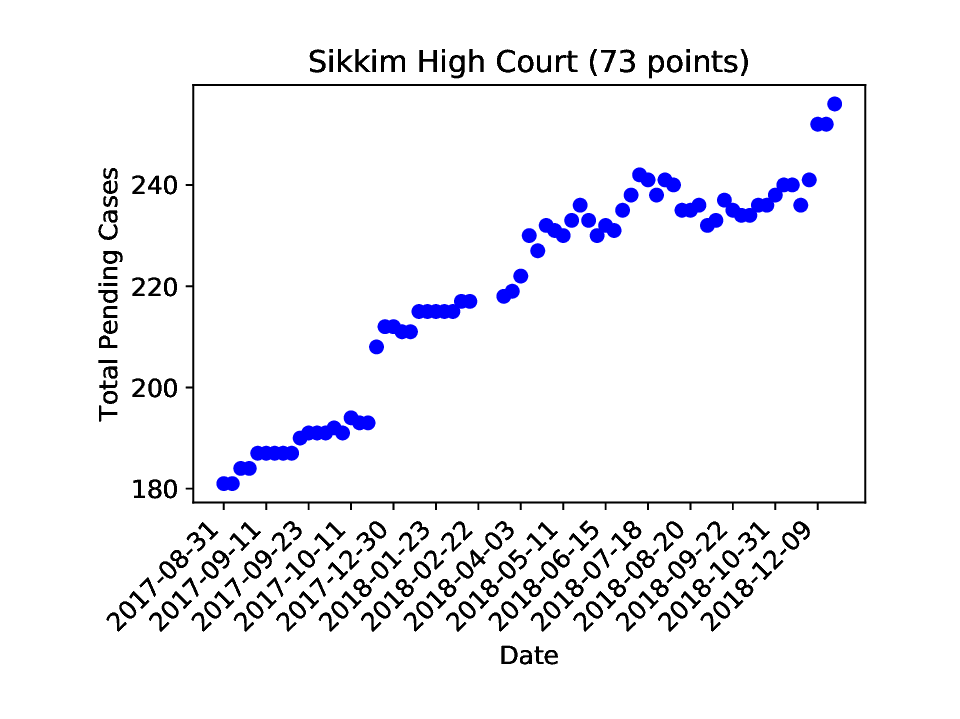}
\includegraphics[width=4.4cm]{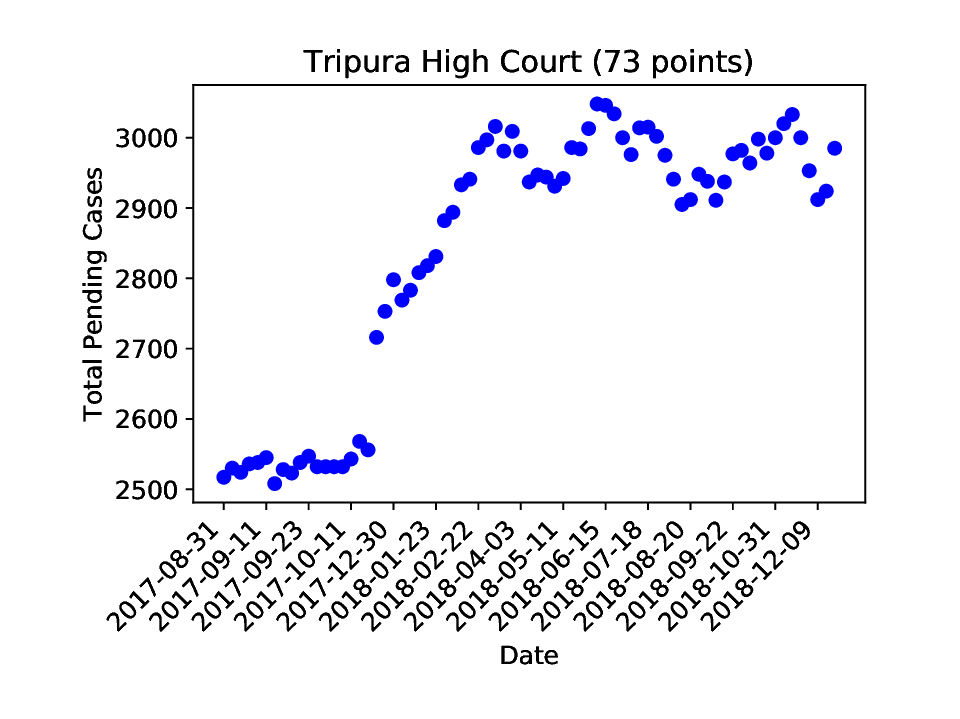}
\includegraphics[width=4.4cm]{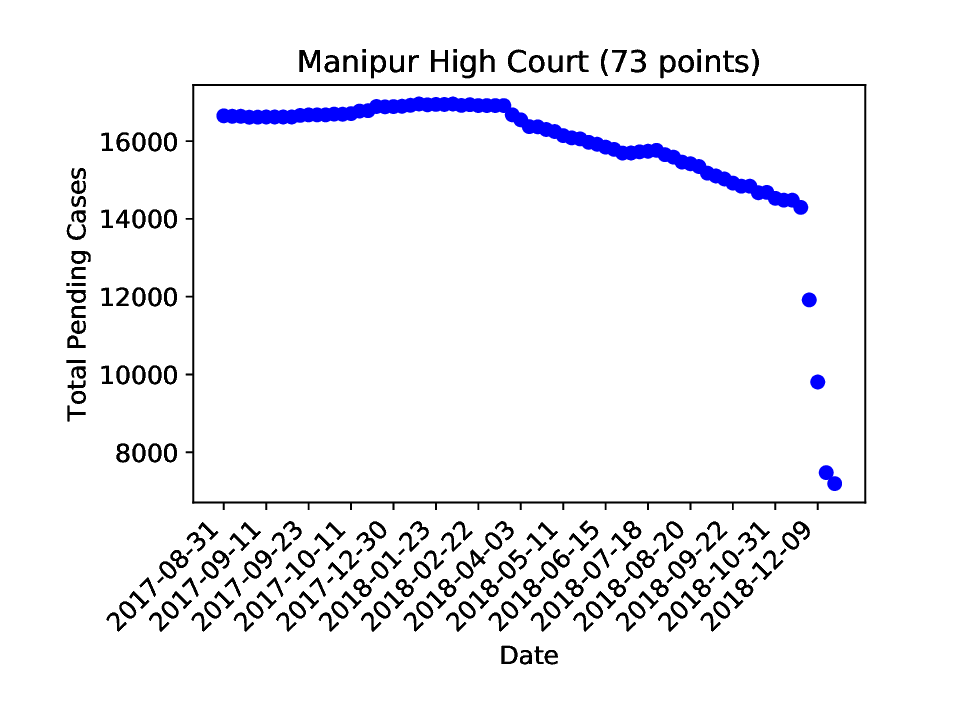}
\includegraphics[width=4.4cm]{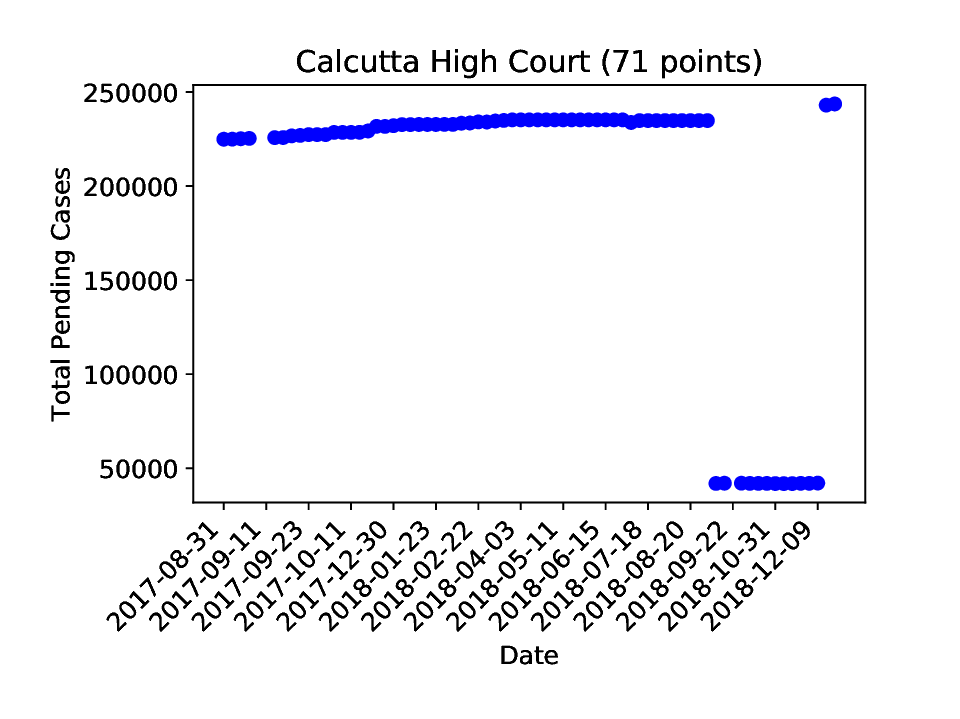}
\includegraphics[width=4.4cm]{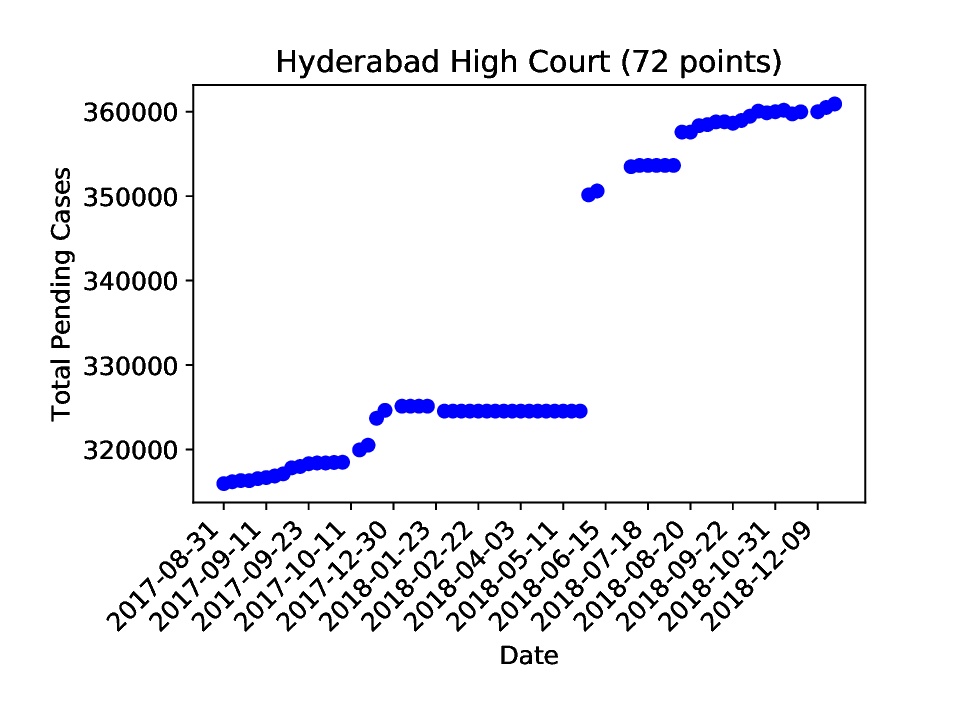}
\caption{Pending cases following a certain trend that implies that the update on HC-NJDG was regular. Apart from Manipur High Court, others have an increase in the number of pending cases.}
\label{fig:np_hc1}
\end{figure*}
\begin{figure*}[h]
\includegraphics[width=4.4cm]{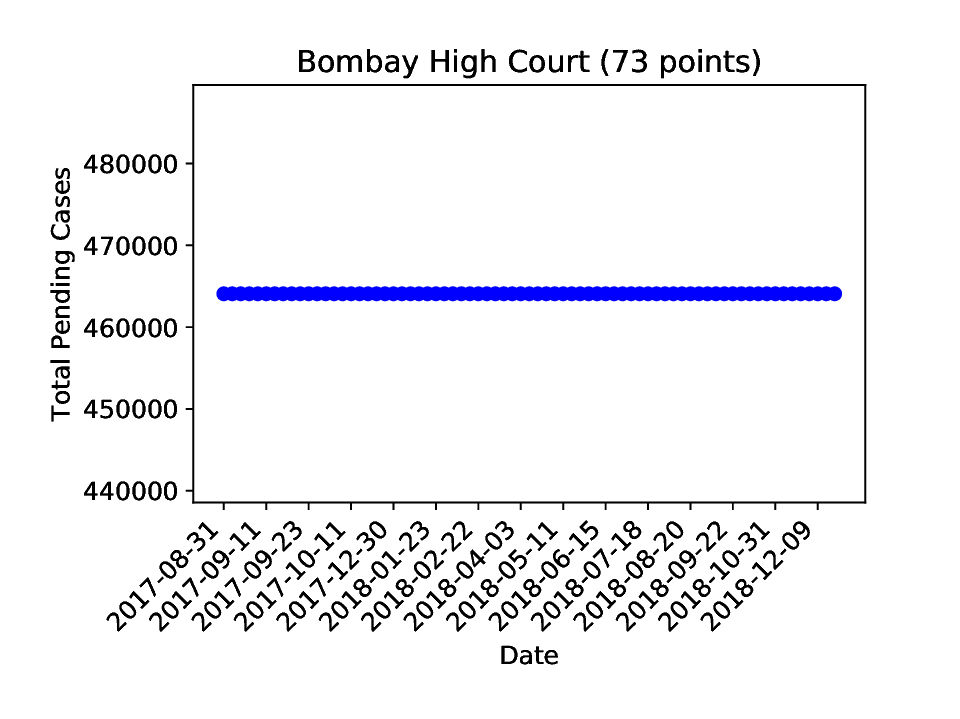}
\includegraphics[width=4.4cm]{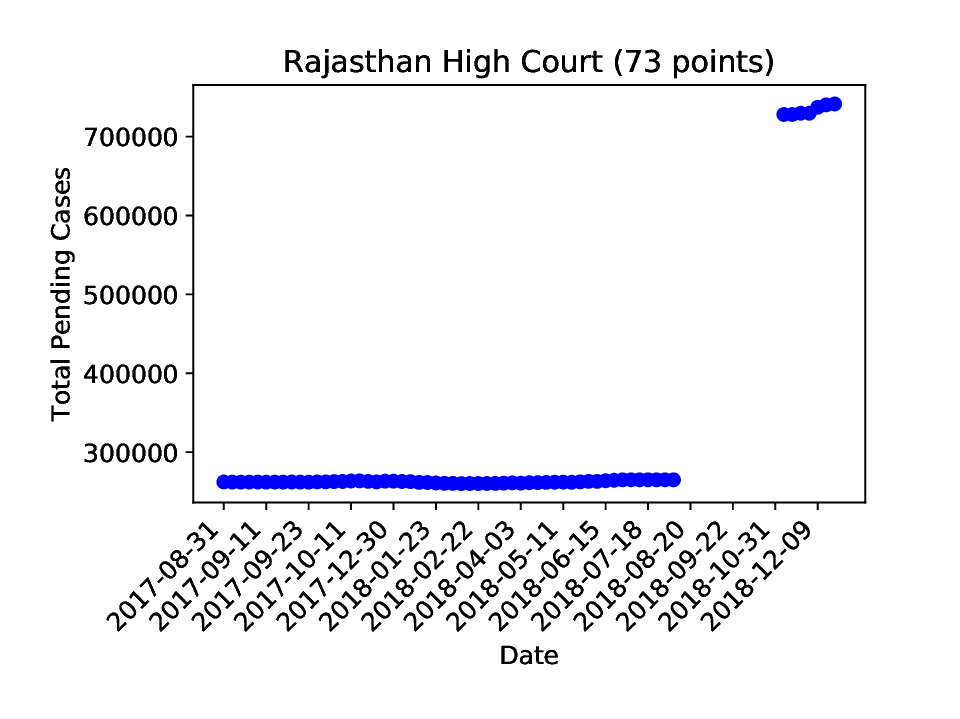}
\includegraphics[width=4.4cm]{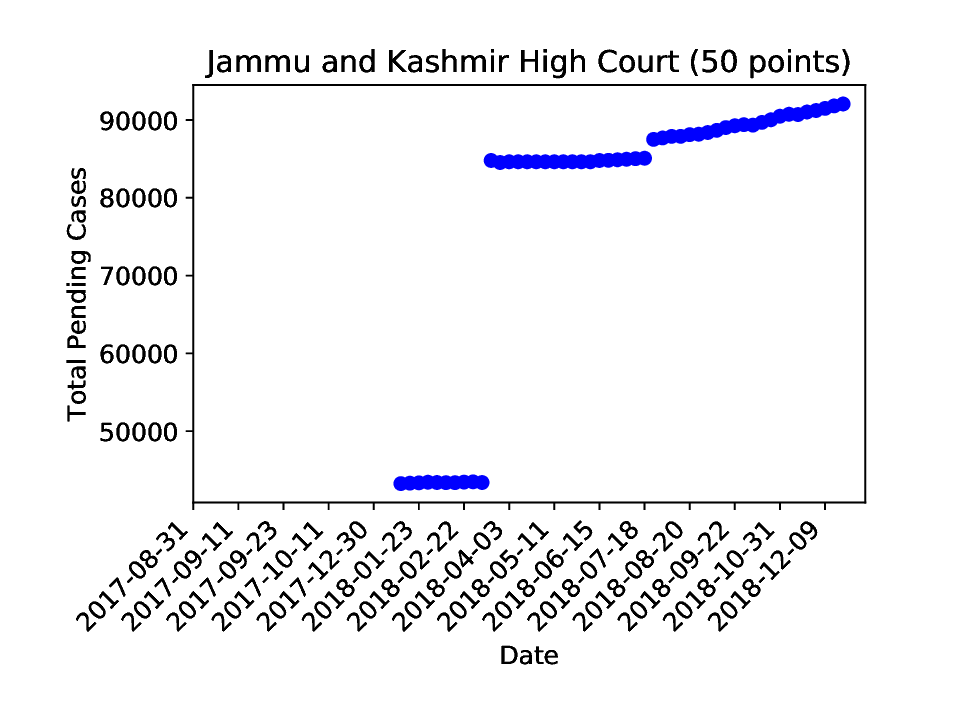}
\includegraphics[width=4.4cm]{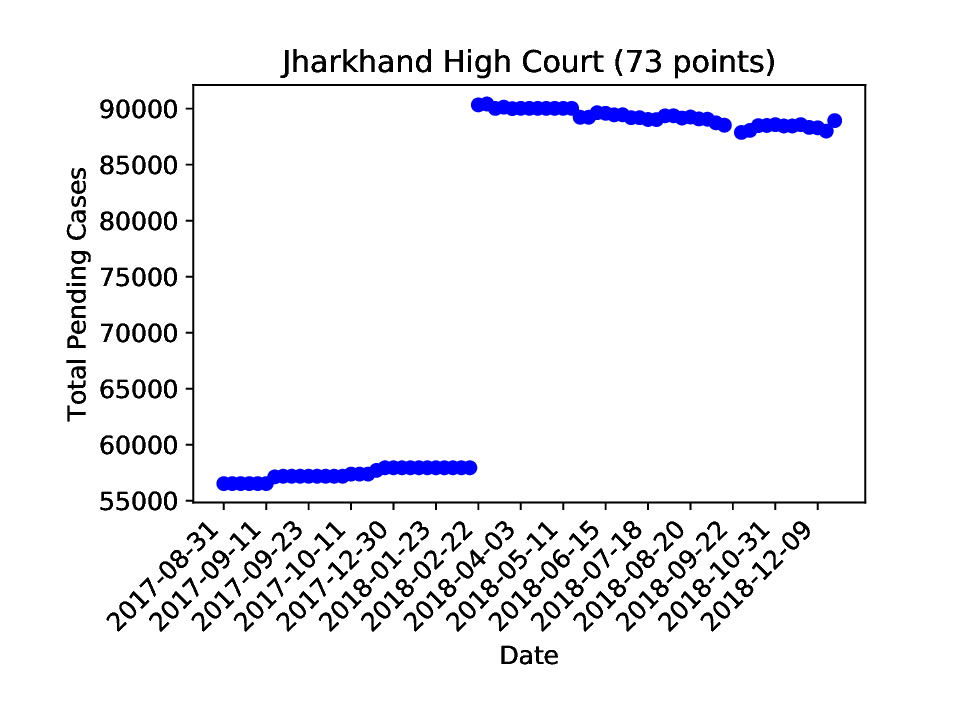}
\includegraphics[width=4.4cm]{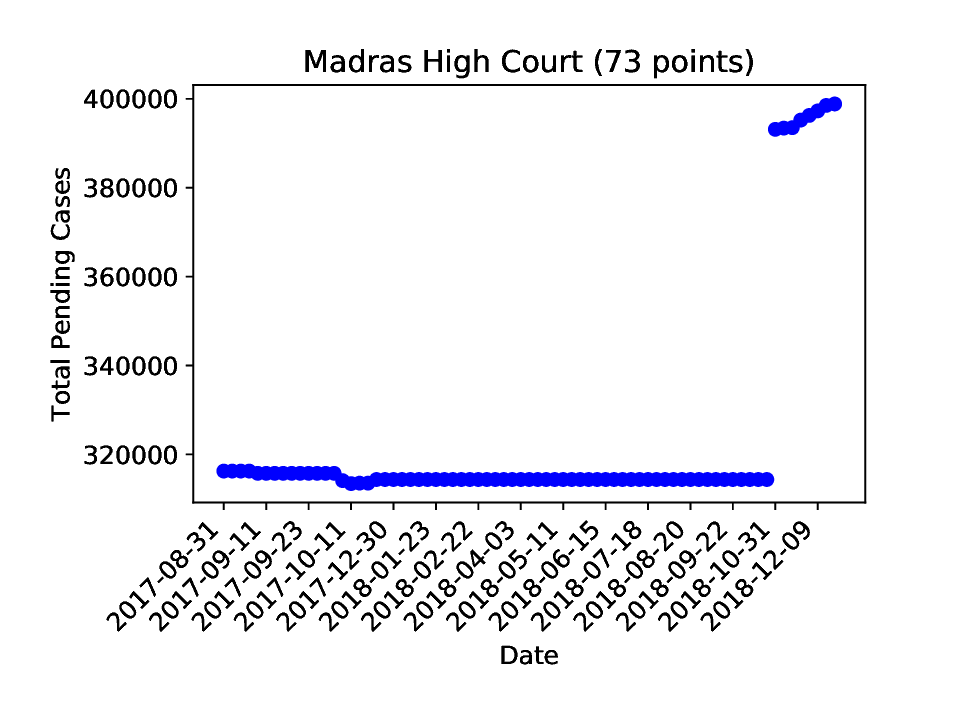}
\includegraphics[width=4.4cm]{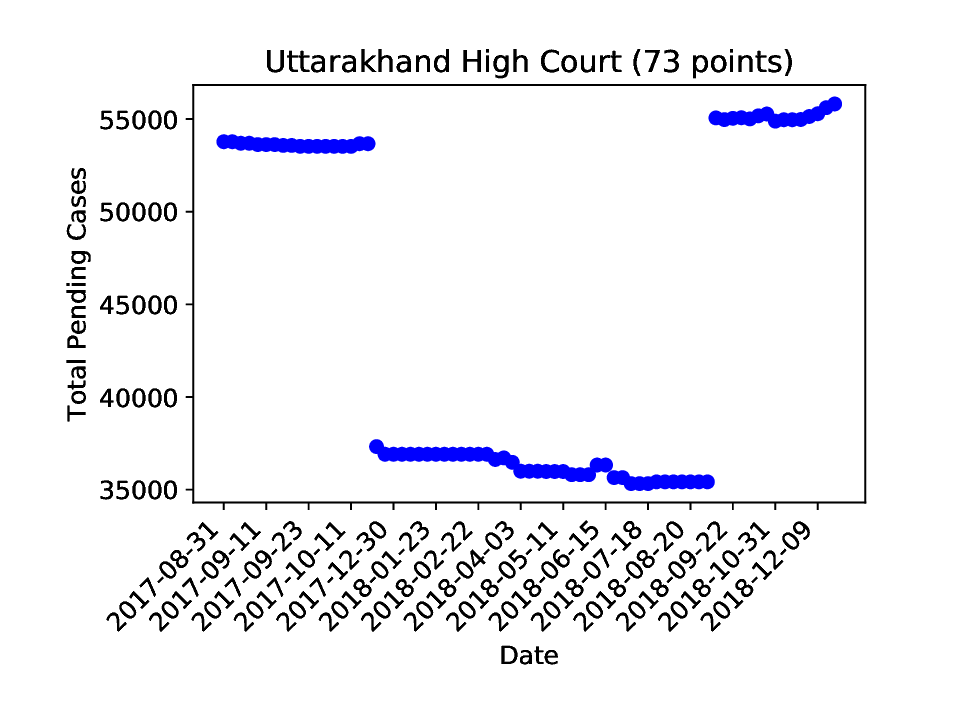}
\includegraphics[width=4.4cm]{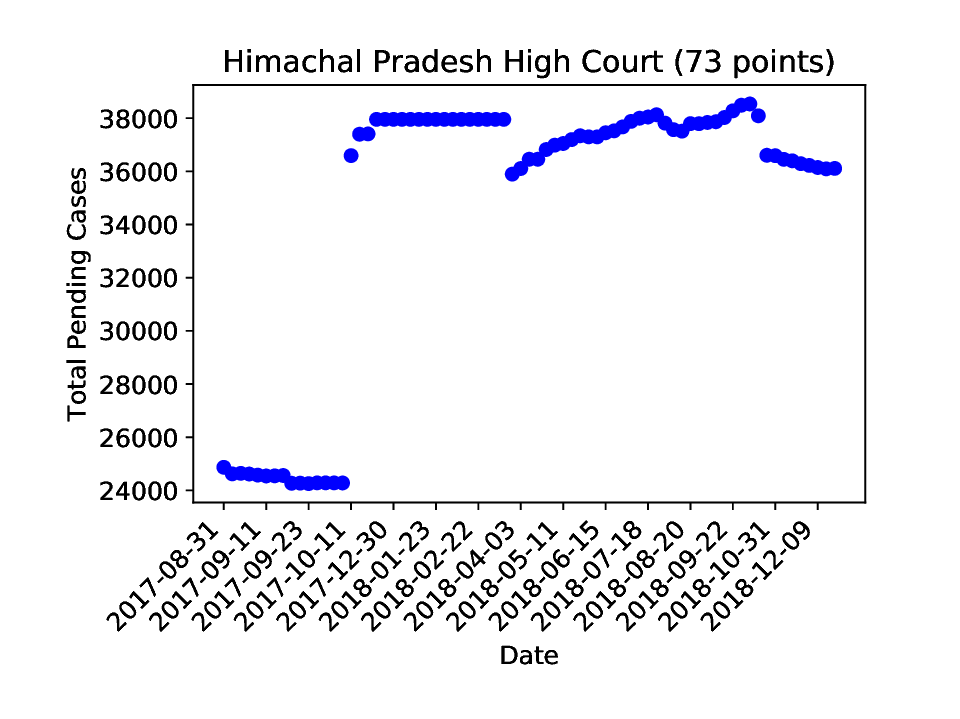}
\includegraphics[width=4.4cm]{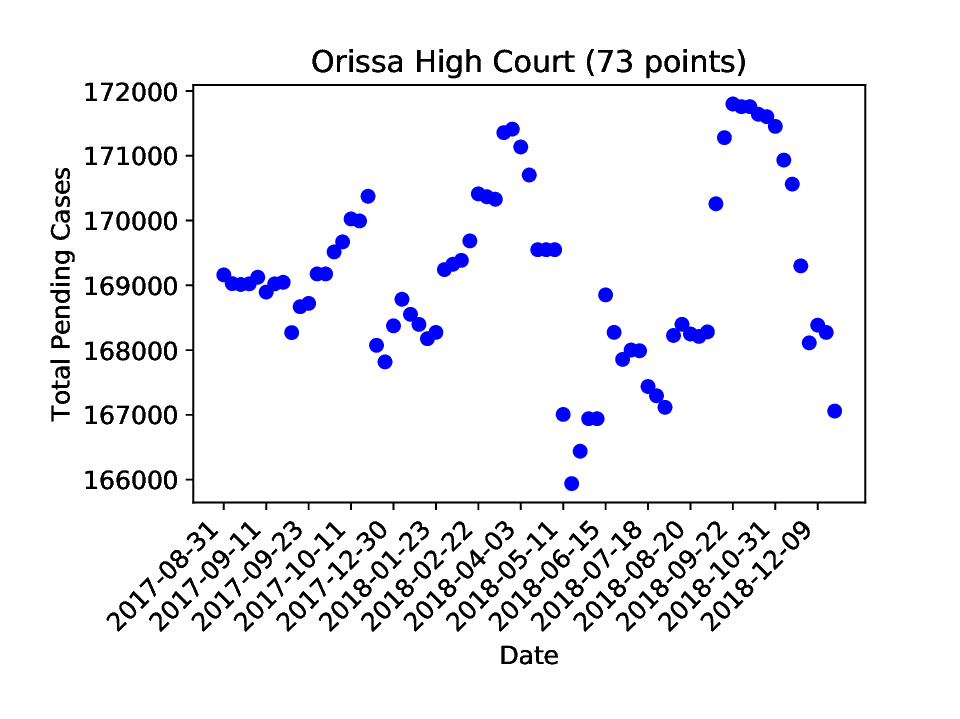}
\caption{High Courts that have not observed regular/stable updates on pending cases on NJDG during the data collection period.}
\label{fig:np_hc2}
\end{figure*}

We plot some graphs related to the total pending cases in the high courts in India. As for the statistics on the number of judges in high courts, we have divided the number of pending cases in high courts in following types. 
\begin{enumerate}
\item High courts with reasonable updates of pending cases during the data collection period (\fref{fig:np_hc1}). 
\item High courts having observed no or little updates with respect to the pending cases or whose data cannot be easily explained from the point of view of an ordinary observer (\fref{fig:np_hc2}). 
\end{enumerate}

\fref{fig:np_hc1} shows the high courts in which the data on pending cases have been updated regularly. It seems that apart from a few slips in the data here and there, most of the points in the graphs look reasonable or it is easy to extrapolate. In fact, the graphs of Madhya Pradesh, Gauhati, Allahabad, Chhattisgarh, Delhi, Sikkim, Meghalaya, Tripura and Manipur High Courts look very promising. Apart from Manipur High Court which has also seen almost linear decline in the number of cases with time, and one exception for Gauhati High Court, the rest of the above mentioned high courts have seen an anticipated and reasonable increase in the number of pending cases. There seems to be a uniform rate of increase in the pending cases in these high courts. Sikkim, Meghalaya and Tripura high courts can be seen as an exception to the above observation but it is mostly so because the number of pending cases at these high courts is hundred times smaller than the rest of the high courts and any small change will become visible in the graphs which may take away the smoothness of the graphs. Hence, from our analysis the updates by the above high courts have been very reasonable. Other high courts in the same class, namely, Gujarat, Karnataka, Kerala, Patna, Punjab and Haryana, Calcutta and Hyderabad, though the plots are not as uniform as the former ones, the updates still look reasonable. For all these high courts there is a period for which the updates have deviated from the usual trend of the respective high court. However, a reasonable trend can still be inferred. Gujarat High Court has started updating the data since July 2018 but it has been consistent since then. Before that it was not updating data at all. Karnataka High Court has been quite consistent in updating data apart from a few dates when its data falls very much apart from their normal curve. However, they have been good in fixing it quickly. Kerala High Court has followed a smooth increasing curve since the beginning, but it has witnessed an unexpected increase in the number of pending cases since September 2018. The updates from Patna High Court appear ad-hoc locally, i.e., if the time period is chosen to be month or so the updates do not follow any trend. However, over the period of an year, the rate still seems reasonable and steady. As opposed to Kerala High Court, Punjab and Haryana High Court has marked a sudden decrease in the number of pending cases sometime around October 2018. The trend of Calcutta and Hyderabad High Courts can be seen from our data because we have collected it for very long period. Otherwise there were some months during which the updates were very erroneous and stagnant. A careful observation of Tripura and Sikkim High Courts also witness the hole in our data. After November 07, 2017 the data was collected on December 13, 2018. This one month worth gap is clearly visible in the graphs of Sikkim and Tripura High Courts where the pendency has increased drastically compared to other dates in the graph.

Thus, we have seen that many high courts in \fref{fig:np_hc1} have been updating the data on HC-NJDG portal regularly, while others have been successful to bring the updates on track after falling off for a while. One of them, Manipur High Court, has also witnessed a decrease in the pendency. While increase in pendency is not a thing to cheer up, the fact that the updates on HC-NJDG has been very timely is laudable. Regular updates are the key to the success of NJDG and e-Courts project as a whole. NJDG can help decrease the pendency only if the updates are regular and accurate.

\fref{fig:np_hc2} shows the data of those High Courts whose update on HC-NJDG has not been regular or it is difficult to explain the trend of pendency. Flat graphs indicate that the number of pending cases have remained unchanged for a long duration. For example, the number of pending cases in Bombay High Court was 4,64,074 on August 31, 2017 the number of pending cases on December 26, 2018 was also the same. There has been no change in this number at any point during the period of data collection. Hence, from the point of view of observer, flat graphs mean no update is being done on the HC-NJDG portal. Rajasthan High Court has reported an unprecedented increase in its pendency after a gap of a couple of months for which it did not update the data at the NJDG portal. A sudden increase of more than 4 lakh cases only imply that there has been some error in the calculation or updating the data. However, the updates till August 20, 2018 look pretty promising. A similar case has happened for Jammu and Kashmir, Jharkhand, Madras and Uttarakhand High Courts. There have been long flat curves and then a sudden change in pendency, followed by another almost flat curve. In case of Jharkhand High Court, there was a sudden jump from 57,944 on February 14, 2018 to 90,335 on February 22, 2018. While it is true that there has been small updates related to pending cases in Jharkhand High Court, as can be seen in the graphs, there should have been more. It is true that the updates of small magnitude are not going to get reflected in these graphs, the counterargument is that high courts that have huge number of pending cases should have magnitude of updates that will become visible as we have already seen for high courts in \fref{fig:np_hc1}. Jammu and Kashmir High Court has the exact same explanation as Jharkhand High Court. Uttarakhand High Court too has a similar explanation. The only difference is that the number of pending cases have reduced from 53,669 on November 07, 2017 to 37,325 on December 13, 2017. HC-NJDG has seen only a few updates from Madras High Court except for a few updates towards the end of the year 2018.

Now we take up updates that are even more difficult to explain. The data from Himachal High Court does not follow one particular trend. It was difficult to understand what really is going on in these updates. There are many continuous curves setting different trends. Lastly, the most difficult data to interpret is from Orissa High Court. It is impossible for us to deduce anything apart from the fact that the updates are regular. We could not assign any trend to it. In fact, we are left with this question in mind, ``Are the updates correct?" The number of pending cases in the High Court is fairly large, so we do not expect the updates to be so irregular. Like all the other high courts we would expect these updates to follow some trend.

%

\section{Time Required to Combat Pendency}
\label{sec:combat}

Having discussed the number of judges and the numbers on pending cases in the high courts, we turn our attention to something that interests everyone on pending cases in India. Our attempt is the first -- to the best of our knowledge -- to be based on extremely rich statistical data obtained from NJDG to answer the question, ``How long will it take to combat pendency in the high courts?". Obviously, our analysis depends on the correctness of the data on NJDG. Also, we have seen earlier in the paper that the data on NJDG may not be correct. Hence, we do more careful analysis of the assumptions that we are going to make for the analysis of time required to nullify pendency in the high courts.

\subsection{Rate of Increase (RoI) of Pendency}
\fref{fig:hcpendency} presents increase in the number of total pending cases from August 31, 2017 to December 26, 2018. We make the following assumptions from this figure:
\begin{itemize}
\item The number of pending cases for the following two periods, ignoring any outliers, may be assumed to have increased linearly:
\begin{itemize}
\item September 12, 2017 to January 19, 2018.
\item January 31, 2018 to August 06, 2018. 
\end{itemize}

\item All the other points that are either isolated or do not belong to the above two linear increases may be ignored for the calculation of an average rate of increase of pending cases.
\end{itemize}

Hence, in \fref{fig:hcpendency}, we have identified two piece-wise linear rate of increase of pending cases. Assuming that the high courts function 210 days in an year, the rate of increase in pending cases from the first period turns out to be approximately 1660 cases per day. From the second period, the rate of increase in pending cases is 1632 per day.

For the sake of completeness, the rate of increase for the overall period is calculated to be 6984 cases per day. At this rate of increase, the pendency in high courts would increase by approximately 14.67 lakhs per annum, which clearly seems unreasonable and unrealistic. Hence, some kind of new updates are responsible for such hikes and not the actual increment in the cases. Whereas if we take the rate of increase from the two time periods mentioned before than the rate of increase turns out to be 17.43 lakhs in five years, which is much closer to the actual rate of increase. This justifies our choice of the above mentioned periods with linear rate of increase. These inferences are summarized in Table\ref{tab:rate}. 

\begin{table}[h]
\hspace{-0.5cm}
\footnotesize
    \begin{tabular}{ | l | c | c | c | c | c |}
    \hline
    Period & Start & End & Days & RoI/day & RoI/5 years\\ \hline
    First p/w linear & 3353808 & 3477056 & 129 & 1660 & 17.43 lakhs \\ \hline
    Second p/w linear & 4183883 & 4360437 & 188 & 1632 & 17.14 lakhs \\ \hline
    Total & 3053695 & 4982504 & 480 & 6984 & 73.33 lakhs\\ \hline
    \end{tabular}
    \caption{Rate of increase (RoI) of pending cases for the two piece-wise linear data that we have chosen in \fref{fig:hcpendency} as well as the total duration of data collection.}
    \label{tab:rate}
\end{table}

\begin{figure}[h]
\includegraphics[width=8.6cm]{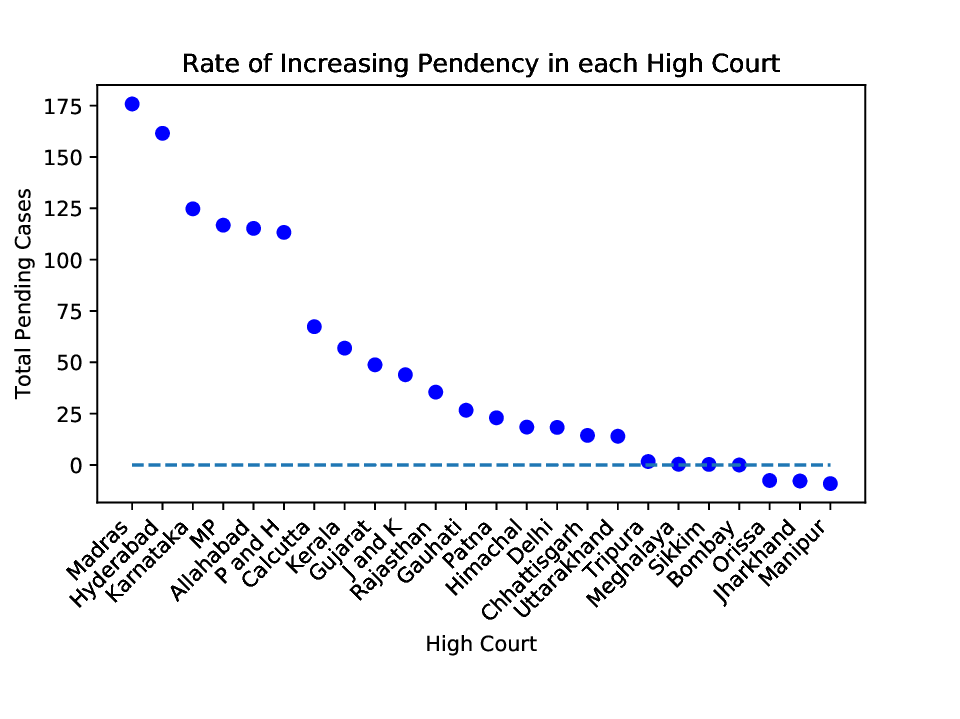}
\caption{Average rate of daily increase in the number of pending cases for each high court. The average is calculated by taking the difference of the cases on last and first day of the last piecewise "continuous" curve of that high court dividing it by the number of days. Hence, their sum may differ from the aggregate obtained above (which is 1660).} 
\label{fig:hcrate_all}
\end{figure}

Since the starting date for each high court to make data available on HC-NJDG may be different, a similar analysis is done for all the high courts. We plot the rate of increase of pending cases for the best period of each high court during the sixteen month period in \fref{fig:hcrate_all}. The best period is computed from the last continuous period for which the data was either increasing or decreasing in regular manner. This helps ignore the effect of the errors introduced by sudden jumps. 

The next level of possible scrutiny is for individual high courts. However, as we have seen before, not all the high courts have been updating data on a regular basis. Hence, there will always be some discrepancy corresponding to that. At the same time, from the above analysis, we can say with reasonable confidence that the cumulative daily increase in pendency for all the high courts should be somewhere between 1650 to 2000. This also accounts for the high courts who have not been publishing the data on NJDG regularly.

Thus, unless otherwise stated, we assume that the daily rate of increase of pending cases is 1650, which is also the aggregate of all the high courts as shown in \fref{fig:hcrate_all}. The major changes to this number may be introduced by receiving updated data from Bombay High Court, which has never updated the data during our data collection period. Hence, the rate of increase of pendency is zero as the data has never changed.



We observe that the number of pending cases in the high courts in India is increasing at a rate of approximately 1660 cases per day. In other words, it also means that the pendency will never get over rather increase with time. The graphs shown for various high courts that have been updating the data regularly very much confirm this observation (\fref{fig:np_hc1}).

\subsection{Comparison of Pendency with Filed/Disposed Data}
Another point of contention which is difficult for us to comprehend is that the number of filed and disposed cases are reported on a monthly basis whereas everything else is updated daily. If it were reported on a daily basis, then a straight forward formula to keep track of pendency, at the end of the day would be,
\[{P_c = P_o + (F - D)}\]
where,
\begin{itemize}
\item $P_c$ is the pendency at the closing time, i.e., in the evening
\item $P_o$ is the pendency at the time of opening, i.e., in the morning
\item $F$ is the number of cases filed on that day
\item $D$ is the number of cases disposed on that day. 
\end{itemize}

In this calculation, we are ignoring the pre-registration cases which are \emph{Cases-Under Objection} and \emph{Cases-Pending Registration}. Adding variables corresponding to these statistics may improve the above formula provided that the meaning of these statistics is clearly explained. 

If the above statistics, $F$ and $D$, are reported on a monthly basis and the number of pending cases is reported on a daily basis then some kind of rectification is required in one of them. From the above analysis, $(F-D)$ can be seen as an important metric which determines the growth of the pendency. \fref{fig:np_hc1} shows all the high courts with positive $(F-D)$, except Jharkhand and Manipur High Court, which have shown negative $(F-D)$ after March 2018. A similar case holds for the high courts shown in \fref{fig:np_hc2} but nothing can be said reliably about the High Courts of Bombay and Orissa High Court. These high courts represent the two extremes of the spectrum. One has never updated the data and the other has updated regularly but in a very unrealistic manner.  

\begin{figure}[h]
\includegraphics[width=9.6cm]{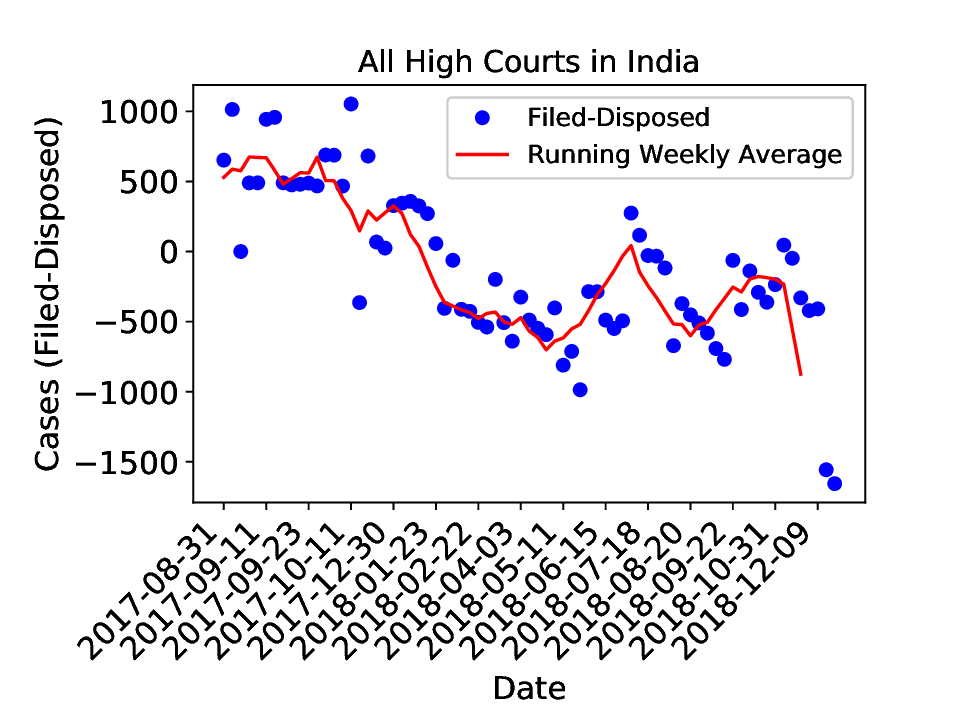}
\caption{Total ``Filed - Disposed" for all high courts. }
\label{fig:ratef-d}
\end{figure}

\fref{fig:ratef-d} plots $(F-D)$ for all the high courts in India. Before we proceed any further, we would like to make clear that the data plotted here is converted to daily statistics by simply dividing the monthly data by 22 (average working days in a month). It may introduce its own errors but we want to be consistent throughout the paper in the interpretation of numbers. In this figure, we have plotted the difference of $(F-D)$. Apart from plotting the exact values, we have also plotted the running weekly average of the data throughout the curve. Each point on the curve is plotted by taking average of the last five points. This shows that the gap between the number of cases filed and disposed has been decreasing. We see that until January 2018, $(F-D)$ was positive, whereas it has sharply decreased after that. Our claim is that if $(F-D)$ is negative in the second half of the data then the pendency should have seen a decline (in \fref{fig:hcpendency}) in last eleven months. However, this has not happened and thus more information is required to explain these contradicting phenomena. Ideally, the rate of increase (RoI) discussed before should be equal to $(F-D)$. Hence, these statistics are not in agreement with statistics on the number of pending cases. 


\subsection{Towards Computing Time to Combat Pendency}

\begin{figure}[h]
\includegraphics[width=9.6cm]{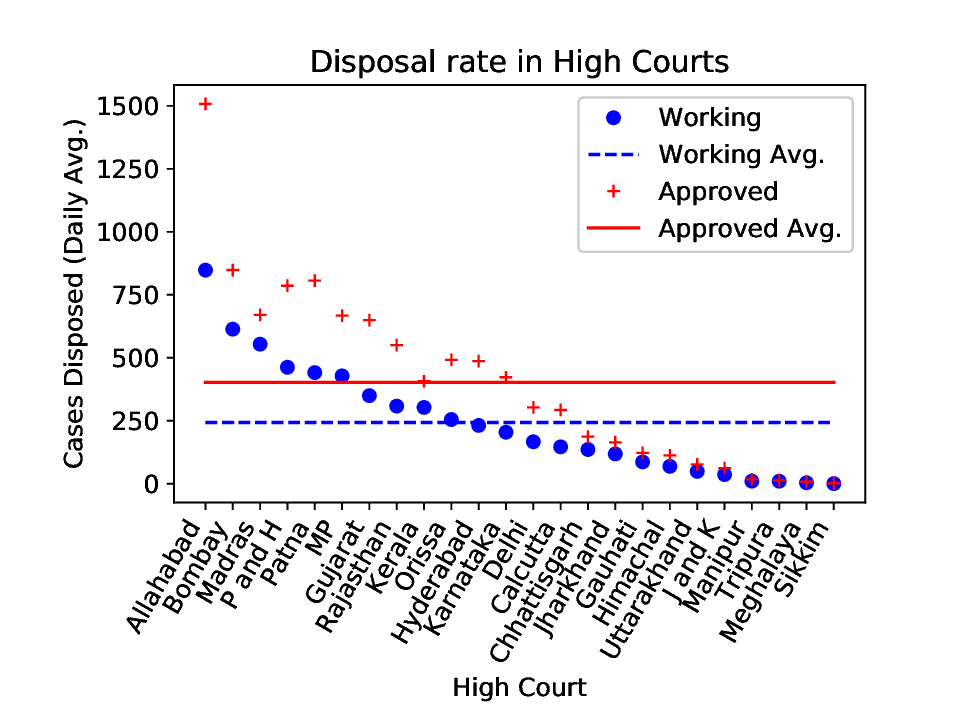}
\caption{Average disposal in High Courts.}
\label{fig:avg_disp}
\end{figure}

\begin{figure}[h]
\includegraphics[width=9.6cm]{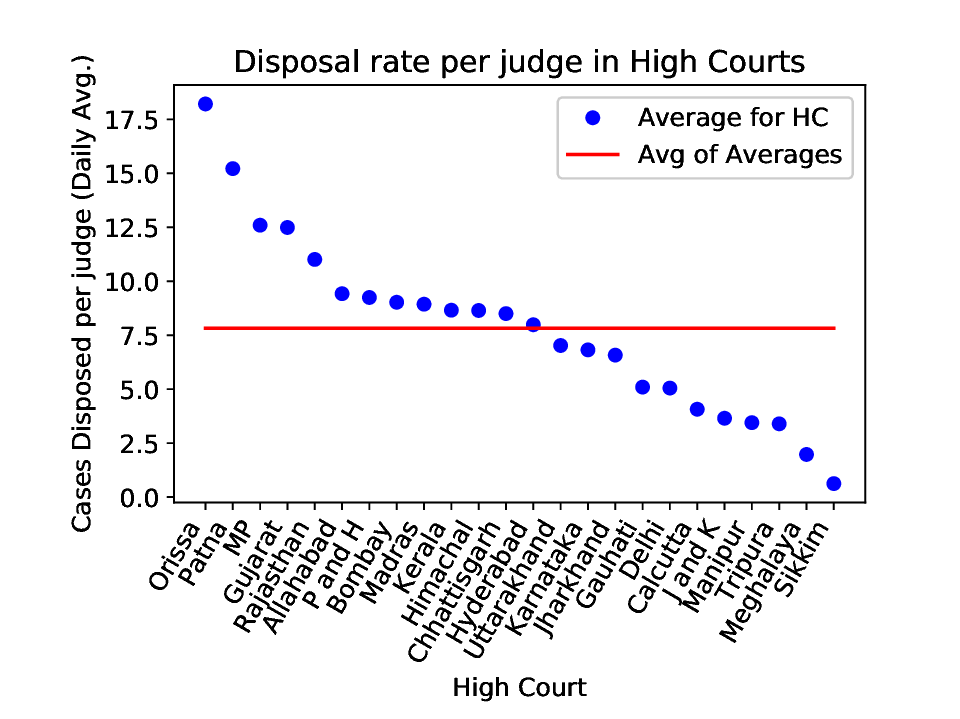}
\caption{Average case disposed per day per judge in High Courts.}
\label{fig:avg_disp_judge}
\end{figure}

\begin{figure}[h]
\includegraphics[width=9.6cm]{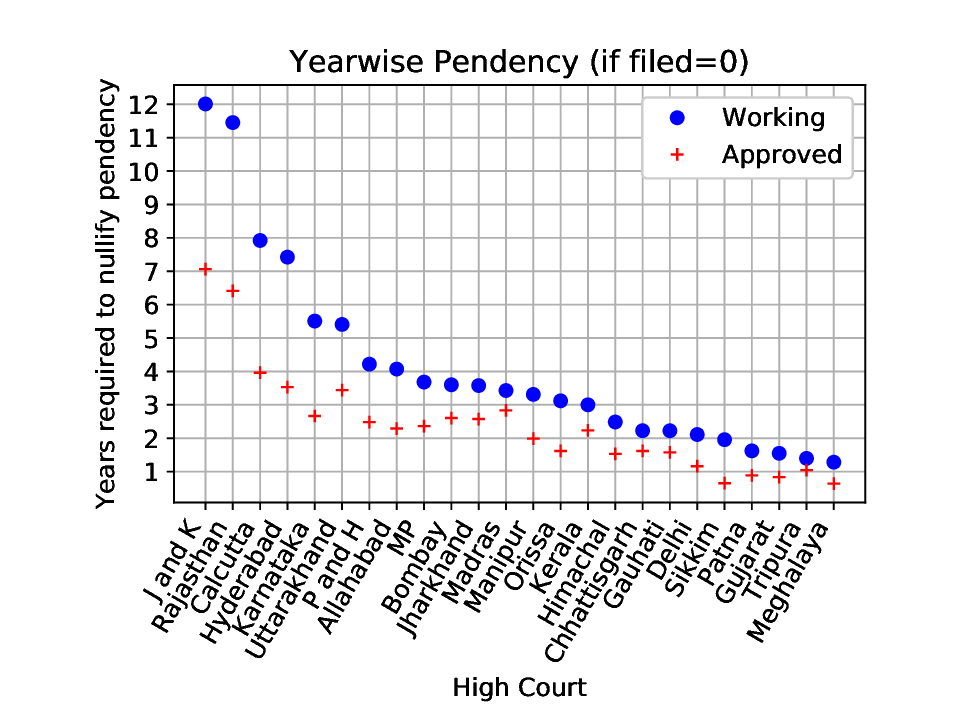}
\caption{Number of years to nullify pendency if no new cases are filed in High Courts.}
\label{fig:avg_time_filing_zero}
\end{figure}

\begin{figure}[h]
\includegraphics[width=9.6cm]{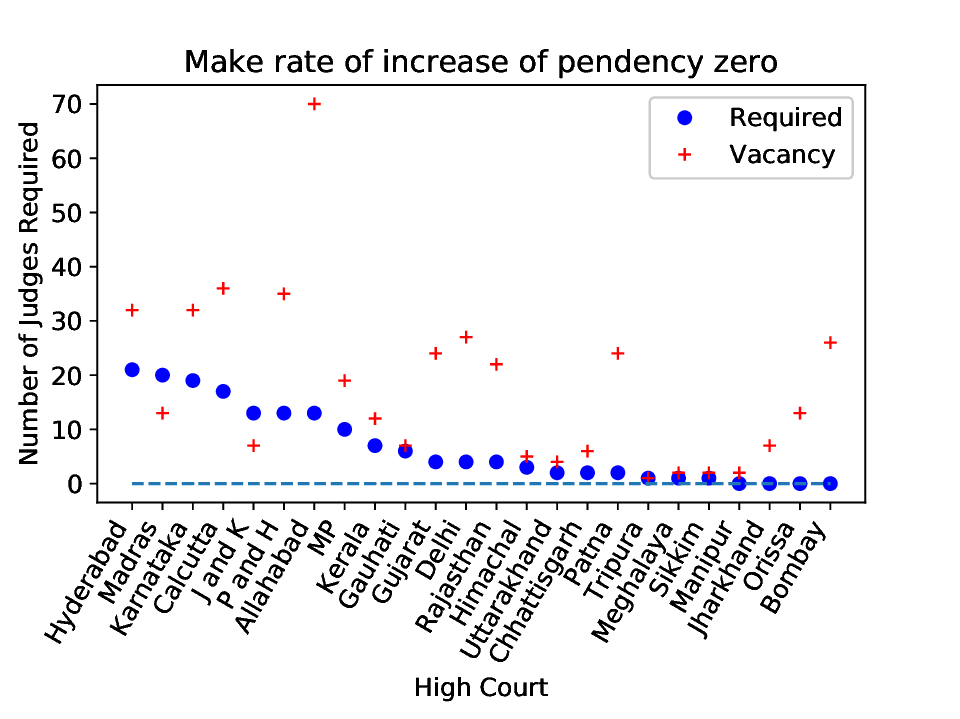}
\caption{Number of judges required to make the rate of increase of pendency to zero according to the analysis done on NJDG data.}
\label{fig:increase_judges}
\end{figure}

We use our analysis of NJDG data to find out answers to the following questions:
\begin{enumerate}
\item What is the rate of disposal of cases in high courts? (\fref{fig:avg_disp})
\item What is the rate of disposal of cases per day per judge in high courts? (\fref{fig:avg_disp_judge})
\item How long will it take to nullify pendency in the high courts if no new cases are filed? (\fref{fig:avg_time_filing_zero})
\item How many more judges in high courts are required so as to make the rate of increase of pendency of that high court to zero? (\fref{fig:increase_judges})
\item If the number of judges in high courts is made equal to the respective approved strength of each high court, and the average disposal rate used for a judge as provided in \fref{fig:avg_disp_judge}, then how many years required to nullify the pendency? (\fref{fig:years_combat_real})
\item If the number of judges in high courts is made equal to the respective approved strength of each high court, and the average disposal rate used for each judge as provided in \fref{fig:avg_disp_judge} but the minimum disposal rate used is the average, then how many years required to nullify the pendency? (\fref{fig:years_combat_national_avg_lowest})
\end{enumerate}

Now we discuss the questions and the figures referred above in detail. 

\fref{fig:avg_disp} plots the average daily disposal of cases for each high court. Disposal related statistics are provided on NJDG portal on a monthly basis. Thus, we have divided the number by 22 to get the daily figure. Our analysis would have been more accurate had these statistics were provided on a daily basis. Once we compute the daily rate of disposal of a high court for the working strength of that particular high court, we extrapolate it for the approved strength of the high court as well. Those high courts for which the working strength is close to the approved strength, there is not much difference between the average disposed cases. Apart from plotting the average for each high court, we also plot the average over all the high courts for both working strength as well as the extrapolated approved strength. The figure provides a clear hint that if all the vacancies in the high courts are filled then there will be a huge gain in the rate of disposal of cases. 

After computing the average disposal rate per high court, we can divide the disposal rate in  \fref{fig:avg_disp_judge} by the number of judges for their respective high courts to plot the number of cases disposed per judge per day for each high court. The average of all these averages is approximately 7.5 and the national average is 8.6. To provide a worst case analysis, we choose the average of averages for further analysis. This figure provides the average number of cases disposed by each high court judge in a day. We use these results to estimate the time required to nullify the pendency in different high courts in India.

\fref{fig:avg_time_filing_zero} provides an estimate of the years required to dispose all the cases if now new cases are filed. If no new cases are filed, then the courts will keep on disposing the cases at its current rate which will eventually lead to disposal of all the cases. We also plot the number of years required to dispose all the cases if no new cases are filed and the high courts are working at their respective approved strength. Obviously, in case of approved strength, the years required is lesser.

\fref{fig:increase_judges} does not assume that no new cases are filed. It instead provides an insight on the number of judges required if the rate of increase of pendency is to be made zero, i.e., the pending number of cases should neither increase nor decrease. This provides a good sign for most of the high courts as the number of judges required to make the rate of increase equal to zero is less than the vacancy in that particular high court. Note that only High Court of Jammu and Kashmir and Madras High Court have the required number greater than the vacancy in these high courts. This means that even if the number of judges is made equal to the approved strength, the rate of increase of pendency will still be non zero. This further implies that in these two high courts, the pendency may never decrease.

\subsection{Time Required to Combat Pendency}

We assume a linear decrease in the number of pending cases, i.e., if at time $t=0$ pendency is $p_0$, and the rate of decreasing pendency is $\alpha$ then at time $t$ pendency $p_t$ is given by 
\begin{equation} 
p_t = p_0 -\alpha\cdot t
\end{equation}

By putting $p_t=0$, and rearranging for $t$, we get, 
\begin{equation} 
t = \frac{p_0}{\alpha}
\end{equation}

We have enough information to compute the time required to nullify the pendency in high courts. From the analysis done till now, we know the following:
\begin{enumerate}
\item Disposal rate per judge per day (denoted as $d$) of a high court (from \fref{fig:avg_disp_judge}),
\item Pendency on any given day in a high court ($p_0$),
\item Working strength of a high court ($w$),
\item Approved strength of a high court ($s$),
\item Daily rate of increase ($r$) of pendency for a high court even though the working strength is $w$ (from \fref{fig:hcrate_all}),
\end{enumerate}

Provided the above information, we need to find the value of rate of decreasing pendency $(\alpha)$ in terms of the known parameters. For most of the high courts the rate of increase is positive if only the working strength of the high courts is considered. Hence, to make the rate of increase negative, or to make the rate of decrease positive, we consider that each high court is working at its approved strength. $r$ is the rate of increase of cases when working strength of high courts is used. Thus, the rate of decrease of cases can be given by 
\begin{equation} 
\alpha = d\cdot(s-w) - r 
\label{eq:alpha}
\end{equation}

If the rate $\alpha$ computed in Eq. \ref{eq:alpha} is positive then the pendency will become zero sooner or later. However, if $\alpha\leq 0$, then the pendency will never decrease, until either the disposal rate per judge per day increase or the approved strength is increased.

Putting all together, the following formula computes the number of working days required (denoted by $t$) to nullify the pendency in each high court:
\begin{equation}
t=\frac{p_0}{d\cdot(s-w)-r}
\label{eq:days}
\end{equation}

Since the above formula computes the number of working days, and each high court is supposed to function 210 days a year, the formula to compute the number of years (denoted by $y$) to nullify pendency is given by:
\begin{equation}
y=\frac{t}{210}
\label{eq:years}
\end{equation}

\begin{figure}[h]
\includegraphics[width=9.6cm]{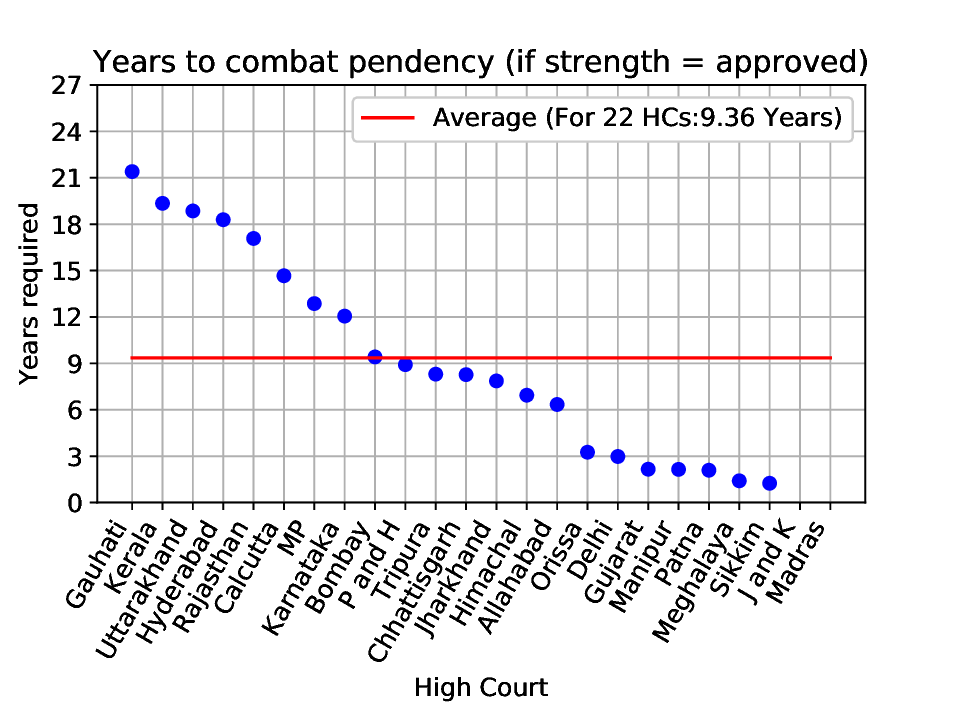}
\caption{Years required to nullify pendency if the working strength of each high court is assumed to be its approved strength and the rate of disposal of cases per day per judge is taken as reported in \fref{fig:avg_disp_judge}.}
\label{fig:years_combat_real}
\end{figure}

\fref{fig:years_combat_real} use formula in Eq. \ref{eq:years} to compute the number of years required to nullify the pendency for each high court. Recall that the working strength of each high court is assumed to be its approved strength. Note that there is no point corresponding to the high courts of Jammu and Kashmir and Madras High Court because as noted in \fref{fig:increase_judges}, the number of required judges is more than the vacancy in these high courts. Thus, under the current constraints, the rate of decreasing of cases cannot be made positive. Thus, the pendency in these two high courts will still keep on increasing. The maximum is for Gauhati High Court and the minimum is for Sikkim High Court.  The average for the 22 high courts turn out to be a bit over 9 years.

\begin{figure}[h]
\includegraphics[width=9.6cm]{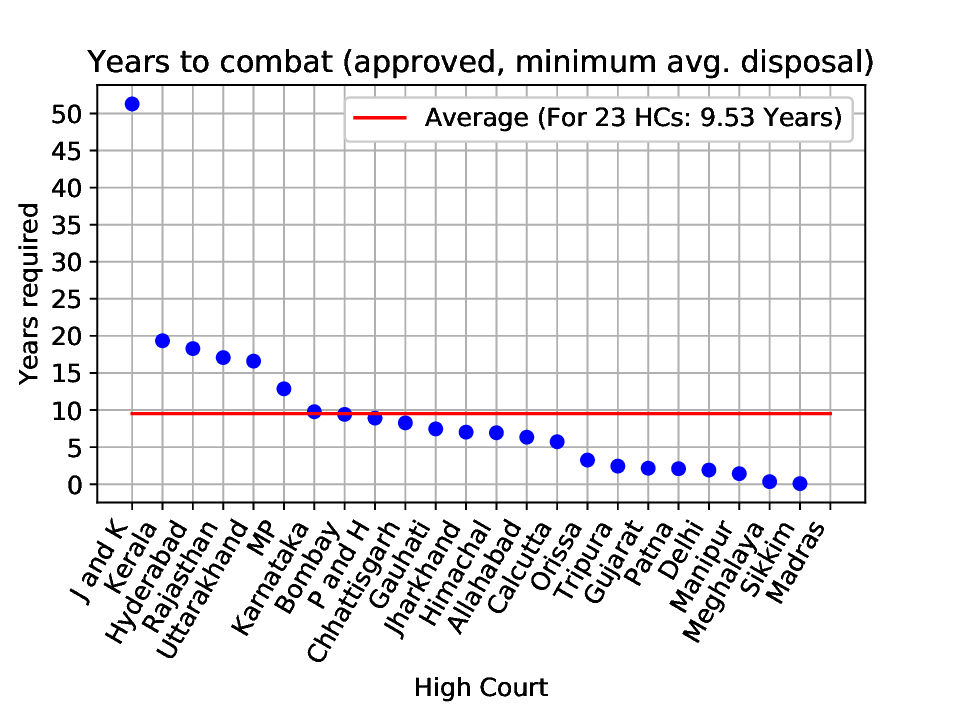}
\caption{Years required to nullify pendency if the working strength of each high court is assumed to be its approved strength and the rate of disposal of cases per day per judge is taken as reported in \fref{fig:avg_disp_judge} if the disposal rate is $>7.5$ and 7.5 otherwise, i.e., we use average of average disposal rates for high courts that have lesser  rate of disposal.}
\label{fig:years_combat_national_avg_lowest}
\end{figure}

\fref{fig:years_combat_national_avg_lowest} presents the number of years to nullify pendency if the disposal rate of those high courts is increased to 7.5 for which it is lesser than 7.5. In other words, we hypothetically increase the number of cases disposed per judge per day to 7.5, if it is lesser than that. This figure reports our final result. Note that due to increase in the disposal rate, High Court of Jammu and Kashmir now appears in the graph with a positive decreasing rate for pendency. Though according to this, the time taken to nullify pendency in High Court of Jammu and Kashmir is a bit more than 50 years but is possible nonetheless. In case of Madras High Court though, even this increase in the disposal rate does not result in the pendency decreasing to zero. 

\section{Miscellaneous Data}
\label{sec:misc}

HC-NJDG provides statistics about other metrics too. In this section, we study those other metrics in detail as well. The nature of the graphs remain same as before. Hence, understanding these graphs should be relatively easy. Explanation is provided in the corresponding caption of the figure.

\subsection{Cases Filed by Senior Citizens and Women}

\begin{figure}[h]
\includegraphics[width=9.6cm]{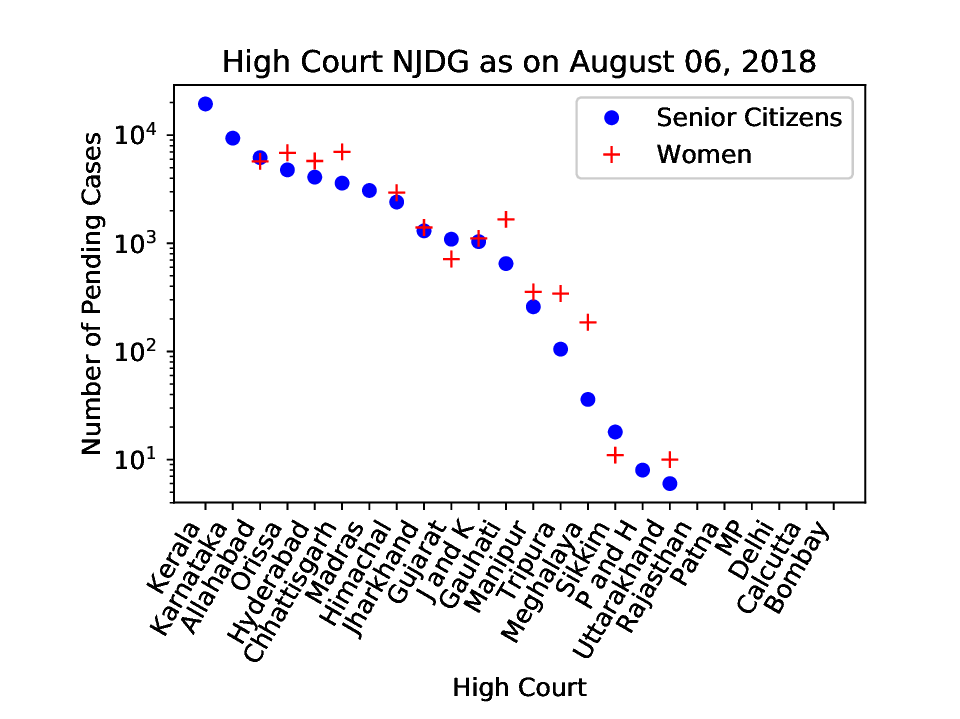}
\caption{The number of pending cases filed by senior citizens and women are much lesser than 5\% and 10\% respectively for most of the high courts as on August 06, 2018.} 
\label{fig:womsen_hc_date}
\end{figure}

\begin{figure}[h]
\includegraphics[width=9.6cm]{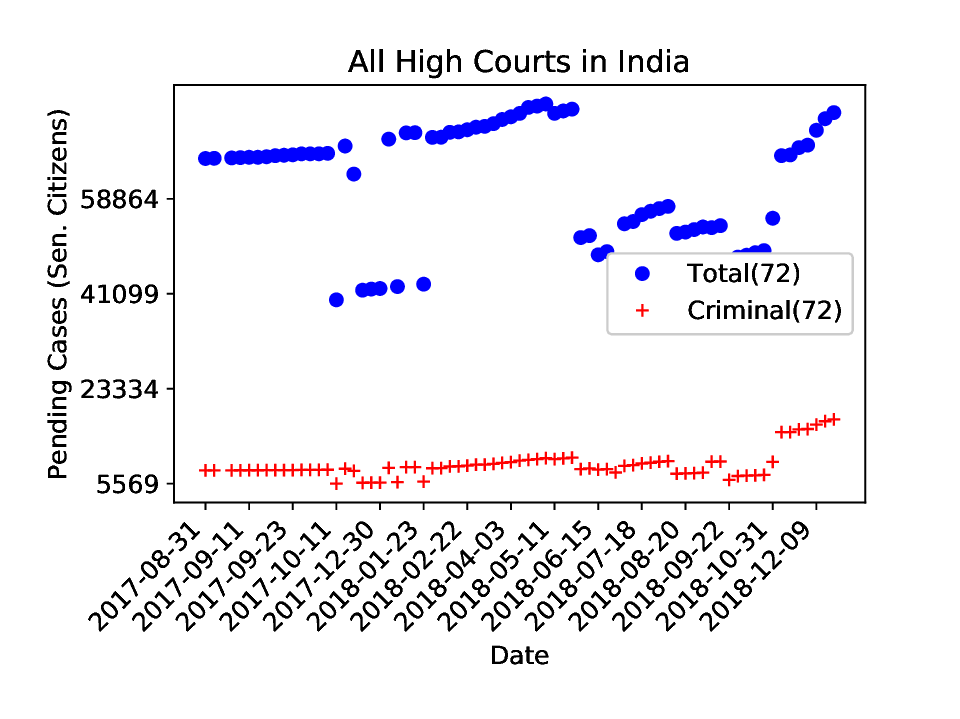}
\caption{The number of pending cases: filed by senior citizens.} 
\label{fig:sen_hc_all}
\end{figure}

\begin{figure}[h]
\includegraphics[width=9.6cm]{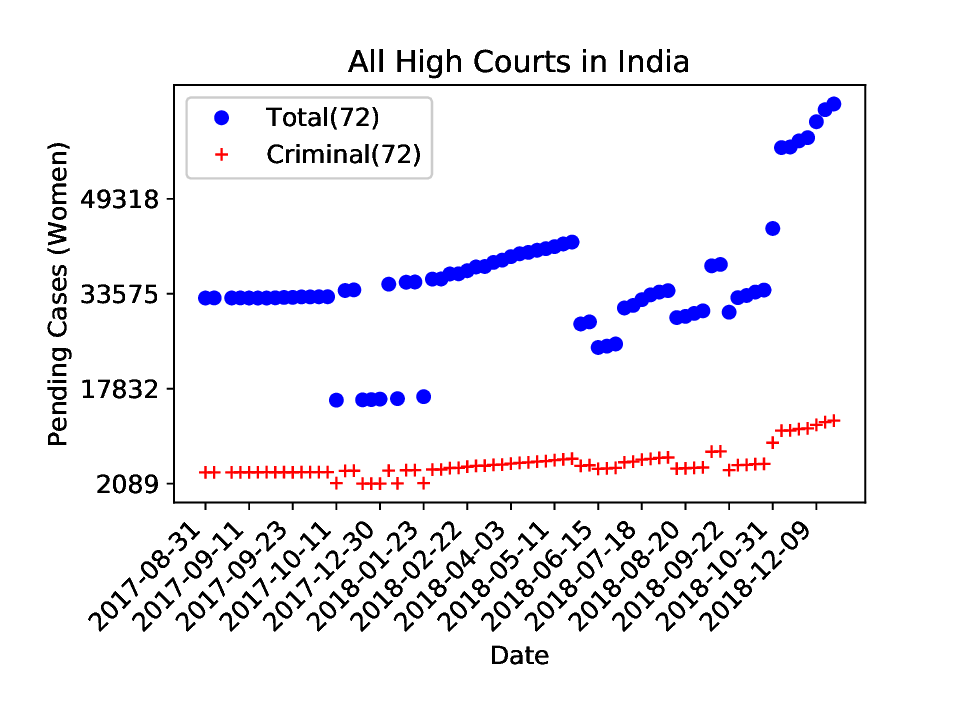}
\caption{The number of pending cases: filed by women.} 
\label{fig:wom_hc_all}
\end{figure}

\fref{fig:womsen_hc_date} shows the data related to the cases filed by senior citizens and women. It is worth noticing that in a country where the number of pending cases in high courts are roughly 3.3 million, very little percentage comes from the cases filed by senior citizens and women. For most of the high courts, together they constitute much less than 10\% of the total cases. We expect more data on HC-NJDG in this regard and once a complete picture is there, better conclusions can be made. It particularly impacts the aggregate data of all the high courts as presented in \fref{fig:sen_hc_all} and \fref{fig:wom_hc_all}.

In contrast with \fref{fig:womsen_hc_date}, when aggregate of the high courts is taken, \fref{fig:sen_hc_all} and \fref{fig:wom_hc_all} portrait the real scenario. We see that the number of cases filed by women are even less than 2\% by the end of December 2018. Similarly the number of cases filed by senior citizens is also less than 2\%. For the period from August 31, 2017 to December 26, 2018, the number of cases filed by the senior citizens and women has increased slightly. However, the criminal cases filed by them have not seen much increase. The criminal cases filed by women are less than 0.2\% of the total cases in High Courts in India. It has remained so throughout the data collection period. The situation is very similar for the criminal cases filed by senior citizens as well. This leads us to three hypotheses:
\begin{enumerate}
\item There are no crimes against women and senior citizens in India.
\item Women and senior citizens do not file cases to enforce their legal rights.
\item The data on NJDG with respect to these statistics is not updated.
\end{enumerate}

Our prima facie choice among these is that the statistics on NJDG need to be updated as the other two choices lead us to far reaching and non-trivial debates.

%


\subsubsection{Cases Filed by Senior Citizens}
\label{subsec:womsen_hc}

\begin{figure*}[h]
\includegraphics[width=4.4cm]{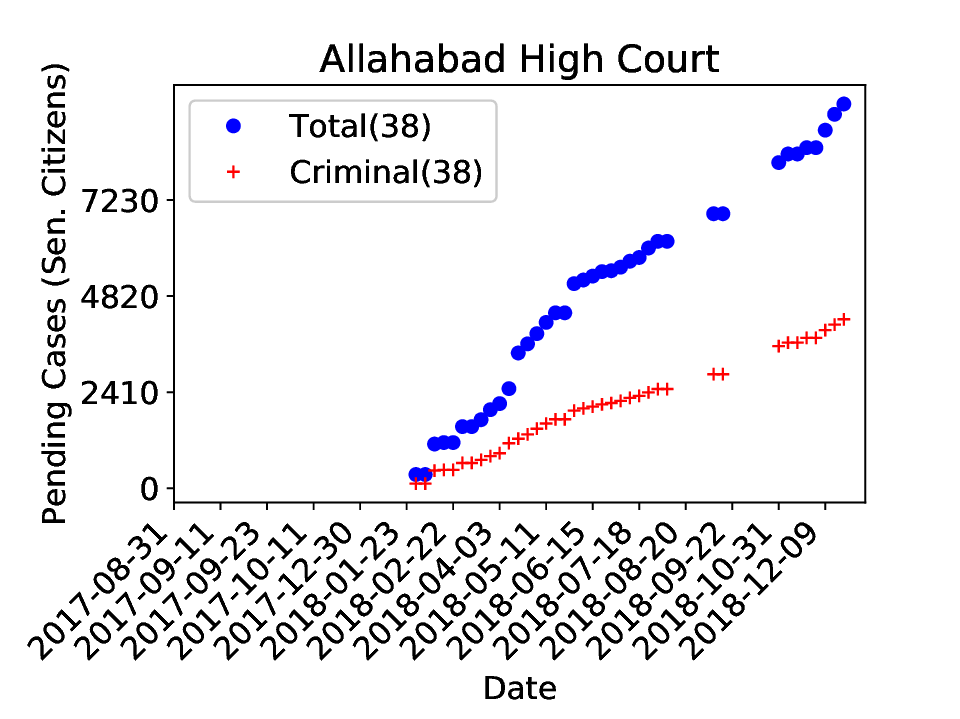}
\includegraphics[width=4.4cm]{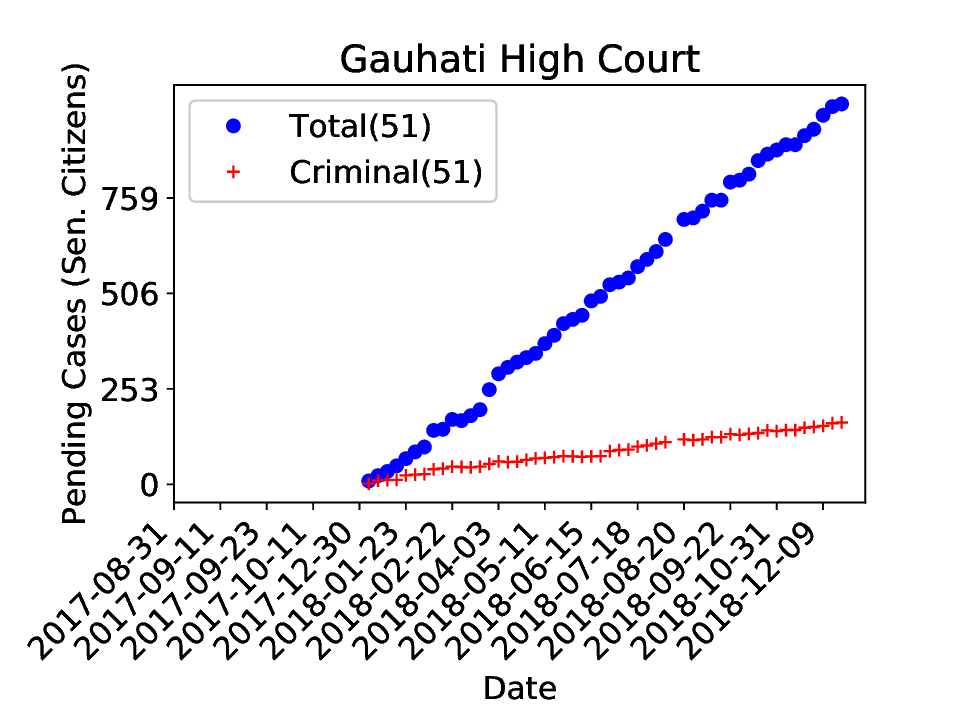}
\includegraphics[width=4.4cm]{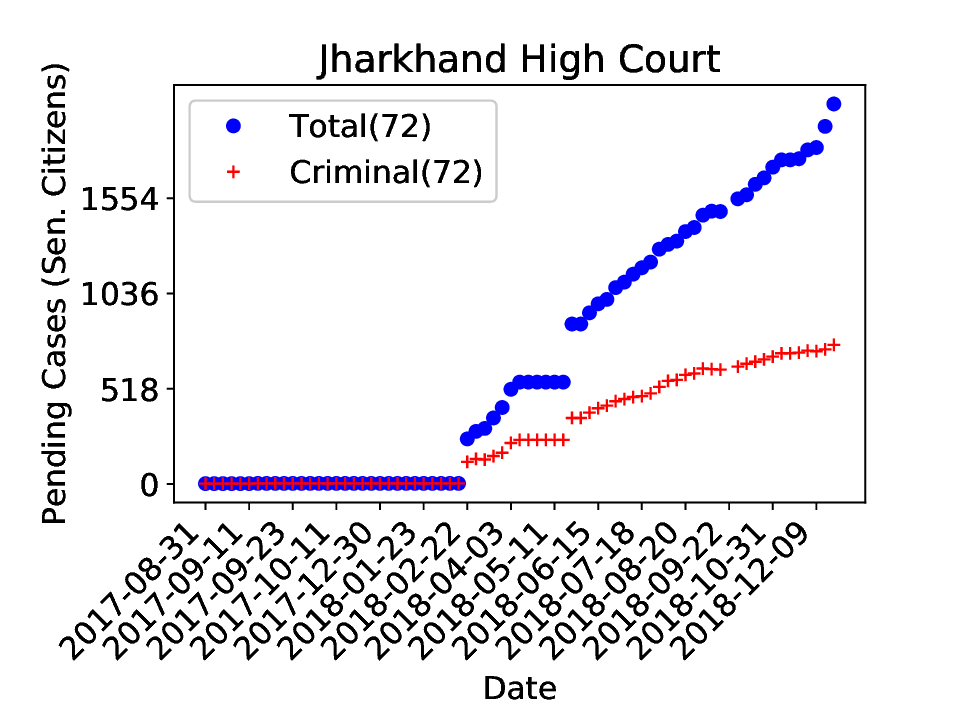}
\includegraphics[width=4.4cm]{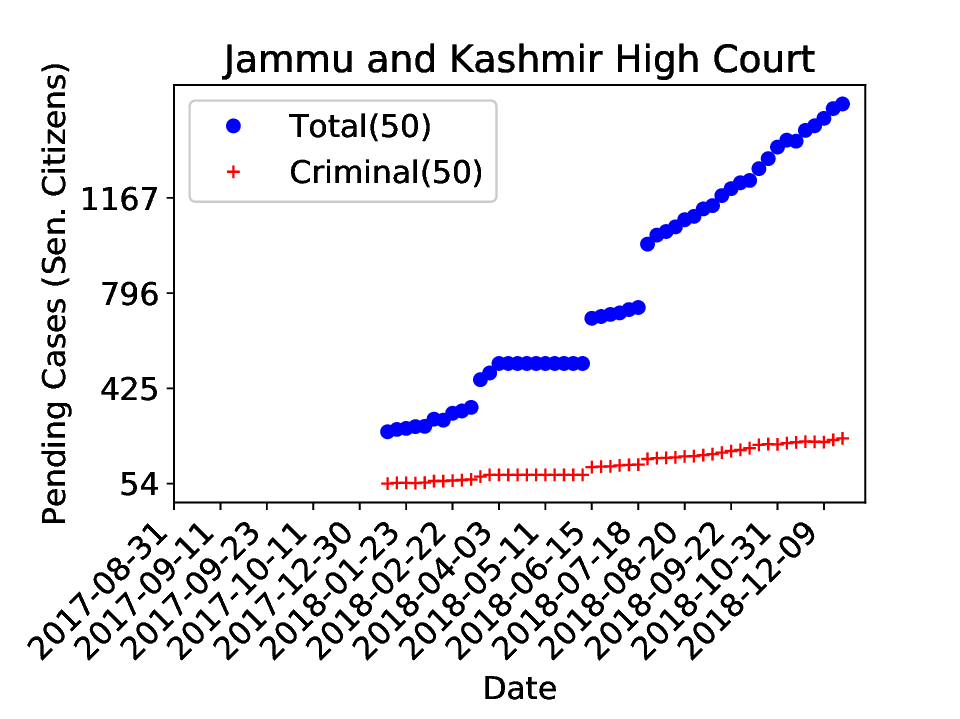}
\includegraphics[width=4.4cm]{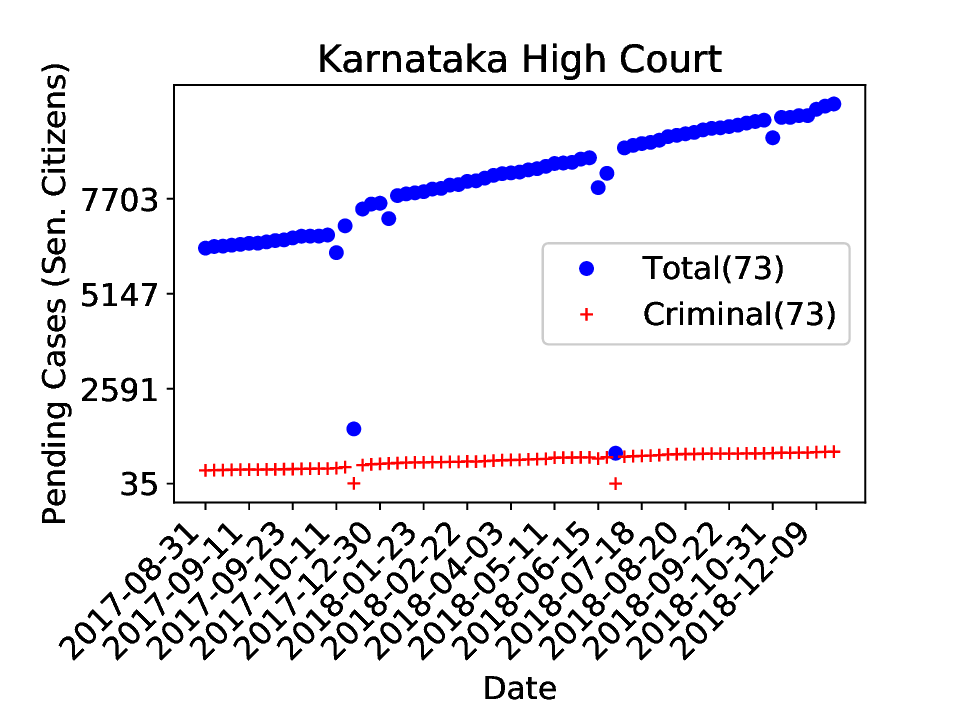}
\includegraphics[width=4.4cm]{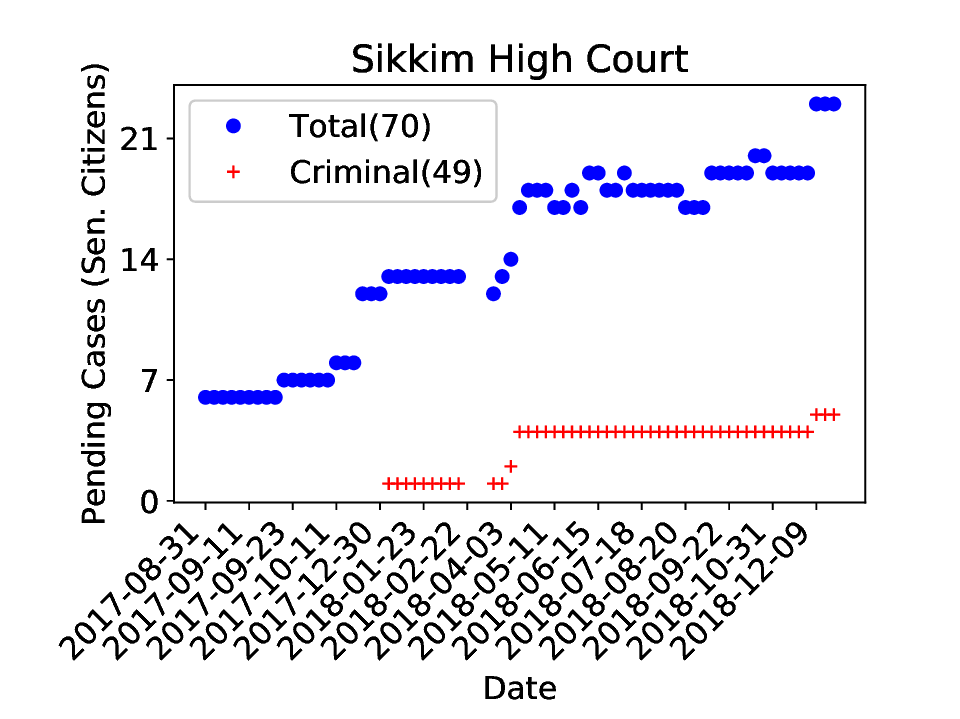}
\includegraphics[width=4.4cm]{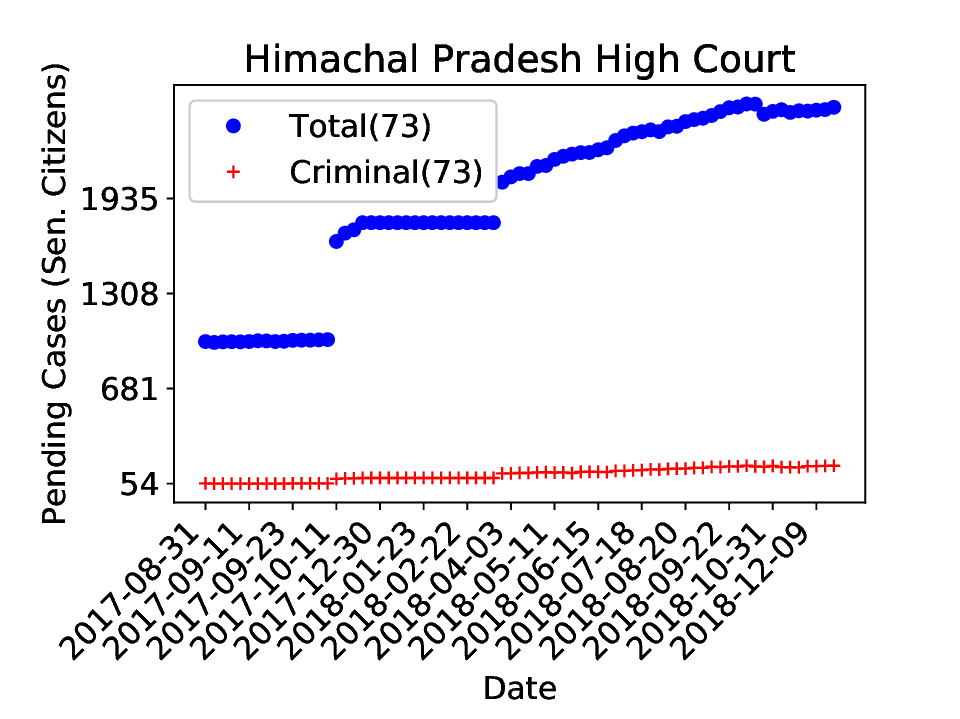}
\includegraphics[width=4.4cm]{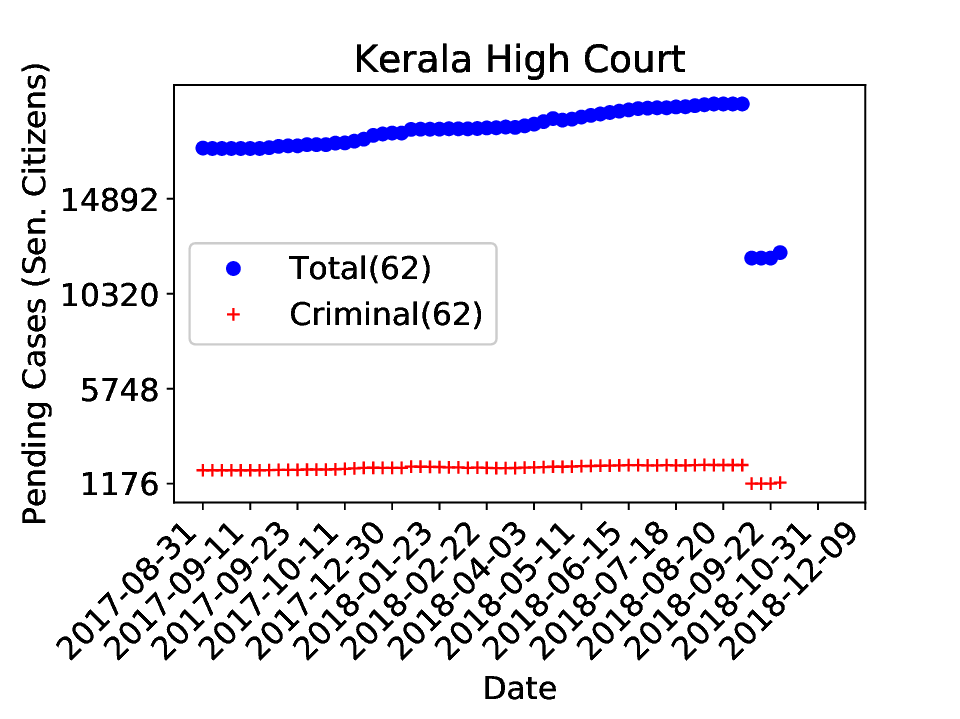}
\includegraphics[width=4.4cm]{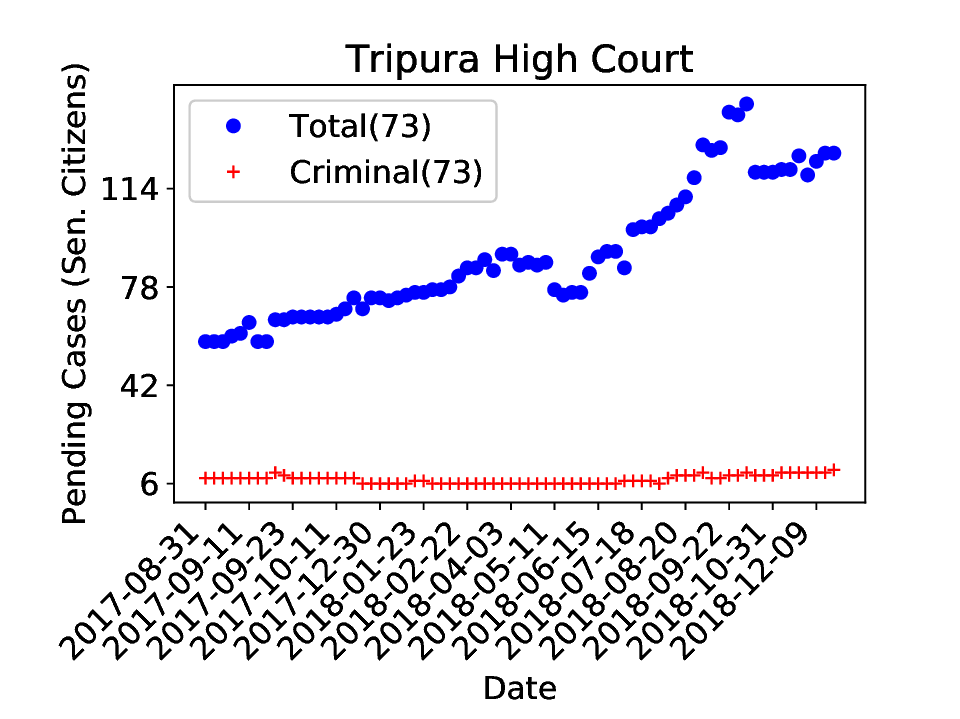}
\includegraphics[width=4.4cm]{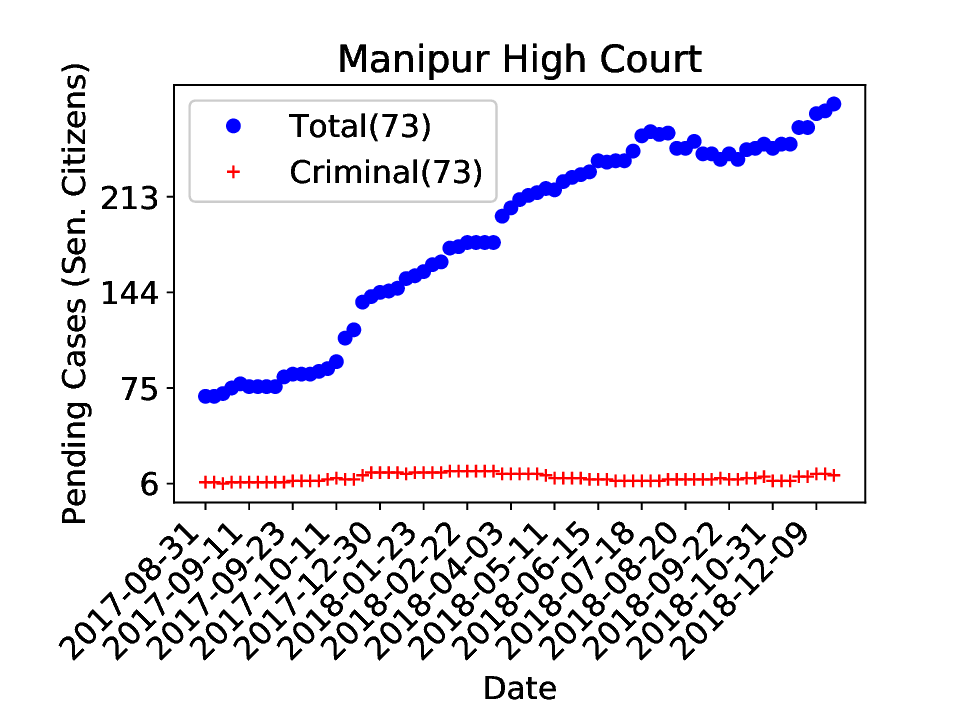}
\includegraphics[width=4.4cm]{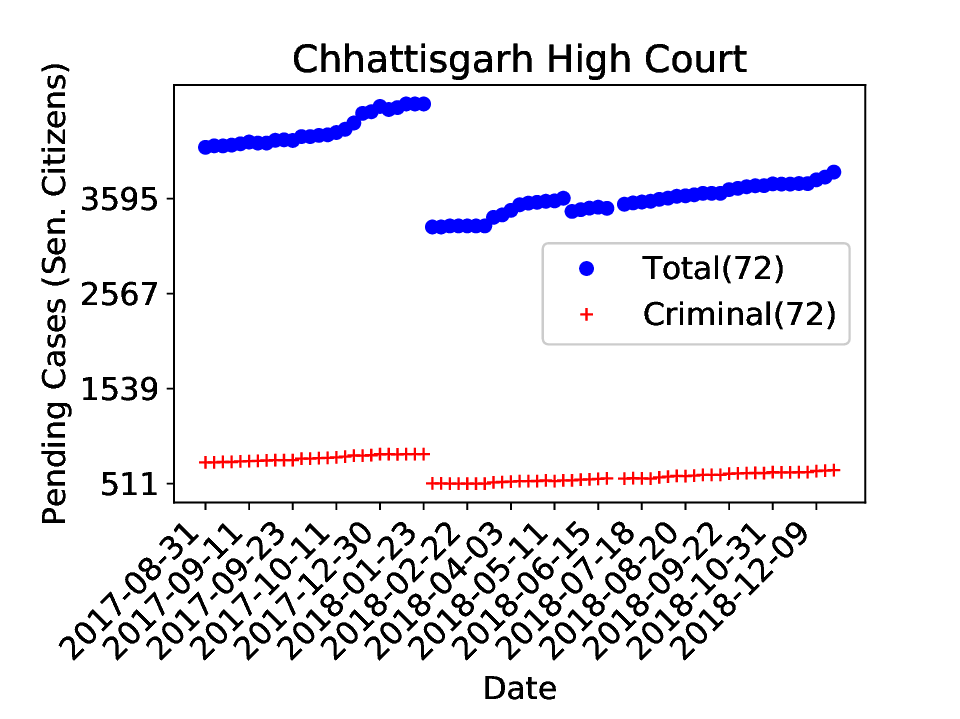}
\includegraphics[width=4.4cm]{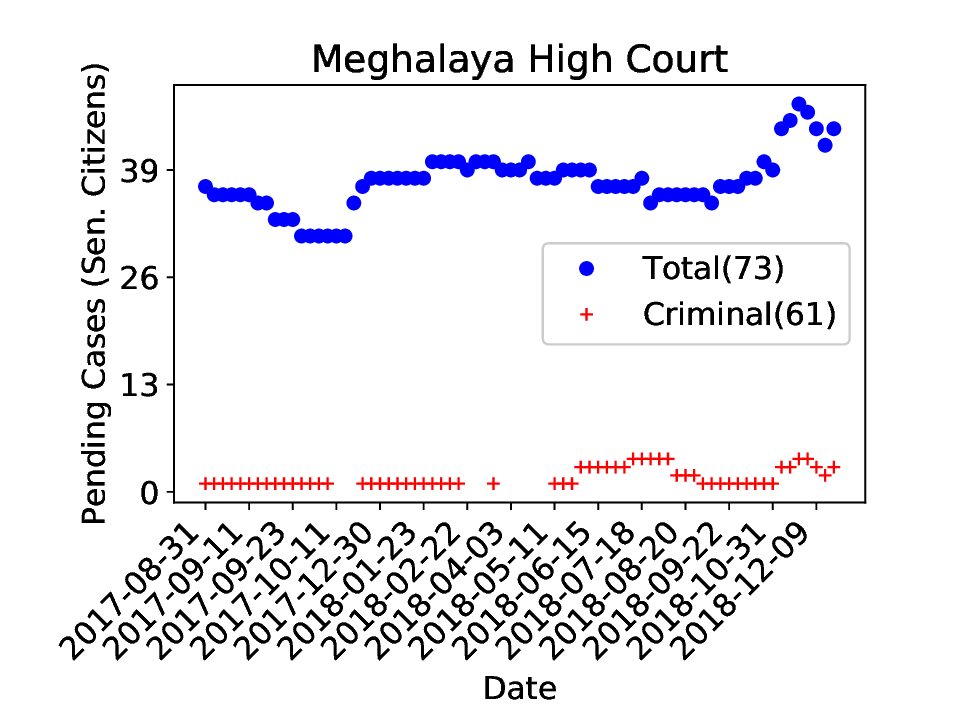}
\includegraphics[width=4.4cm]{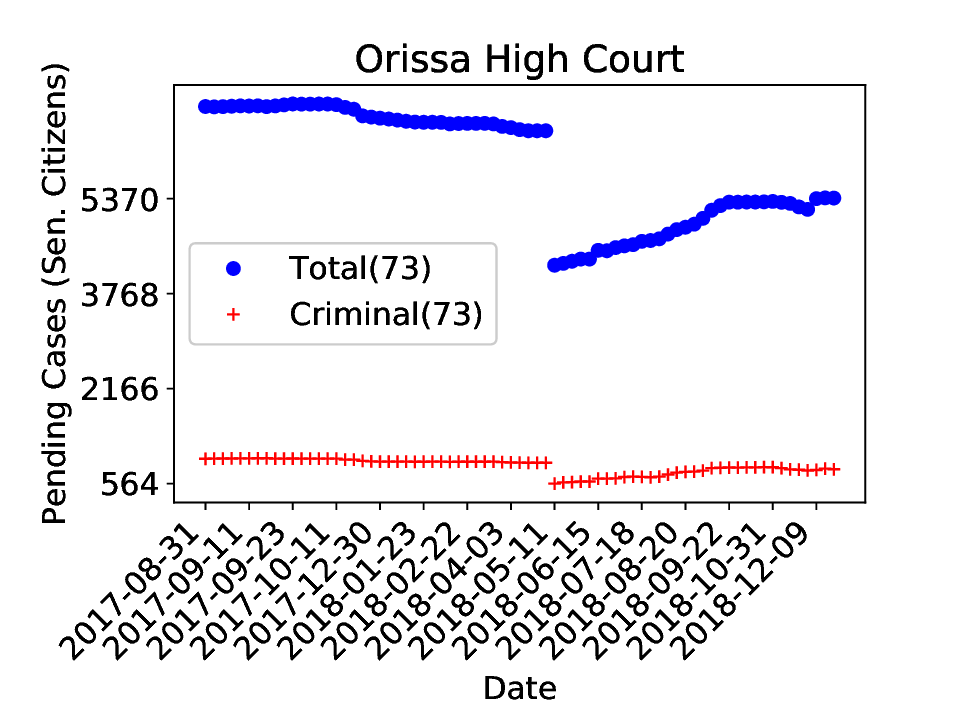}
\includegraphics[width=4.4cm]{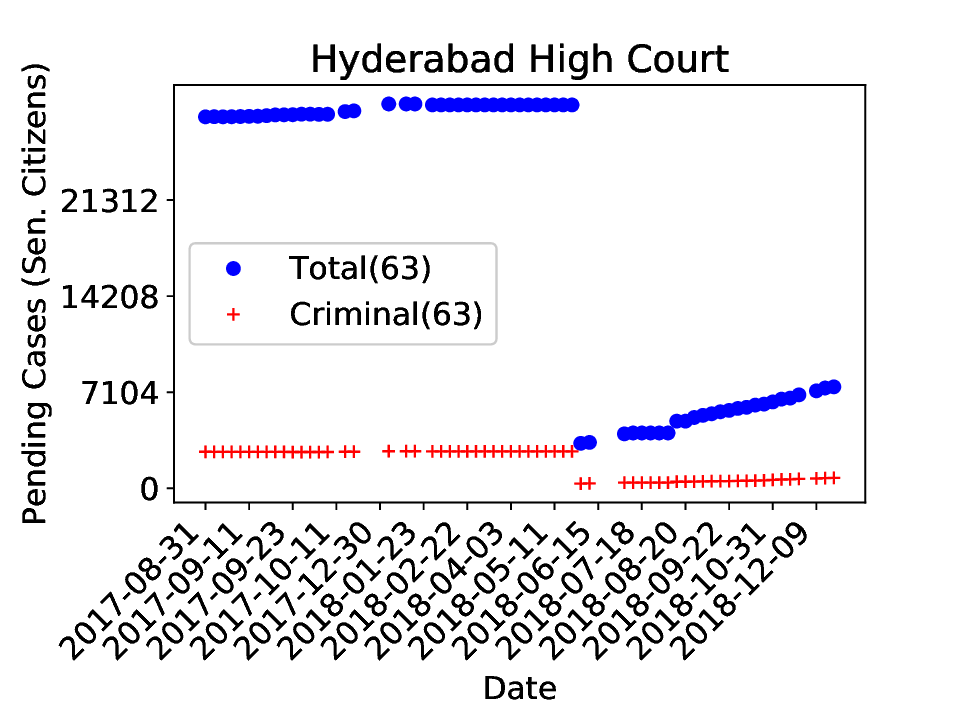}
\includegraphics[width=4.4cm]{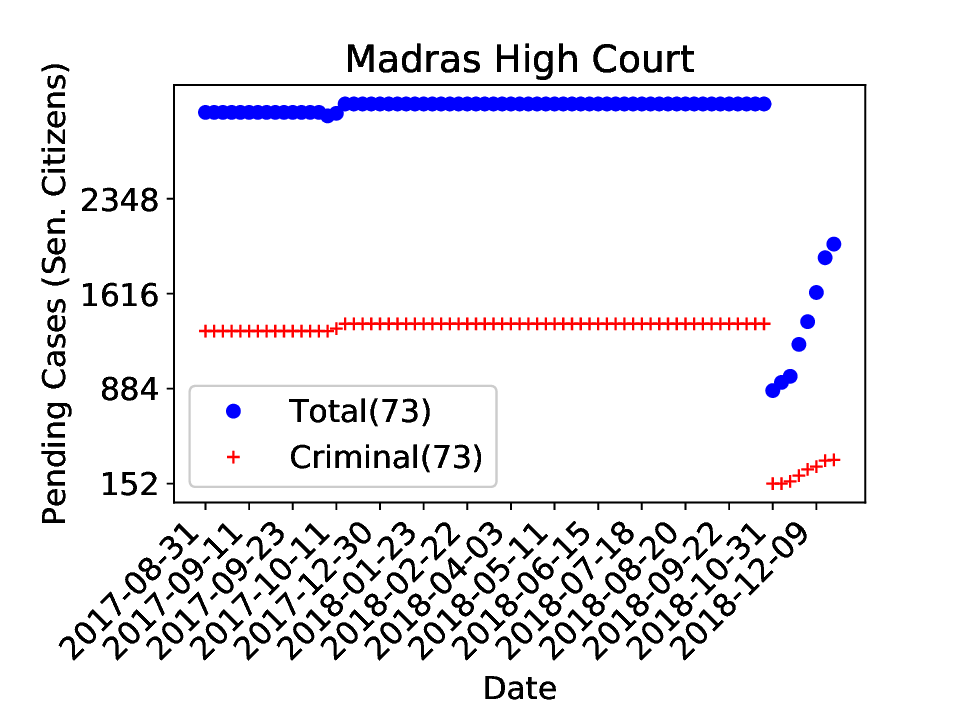}
\includegraphics[width=4.4cm]{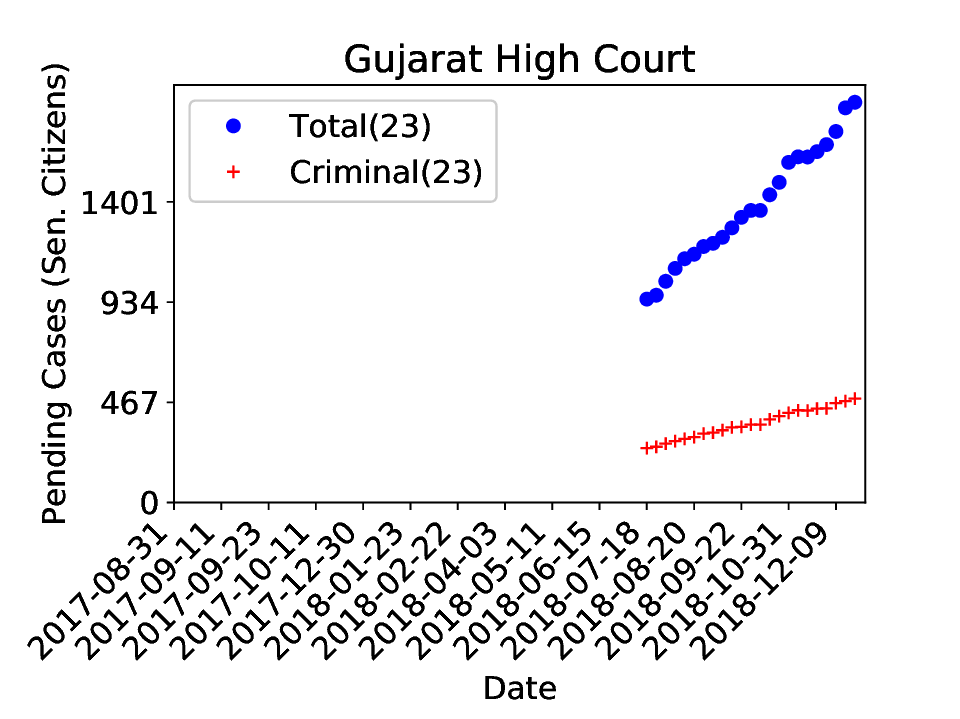}
\includegraphics[width=4.4cm]{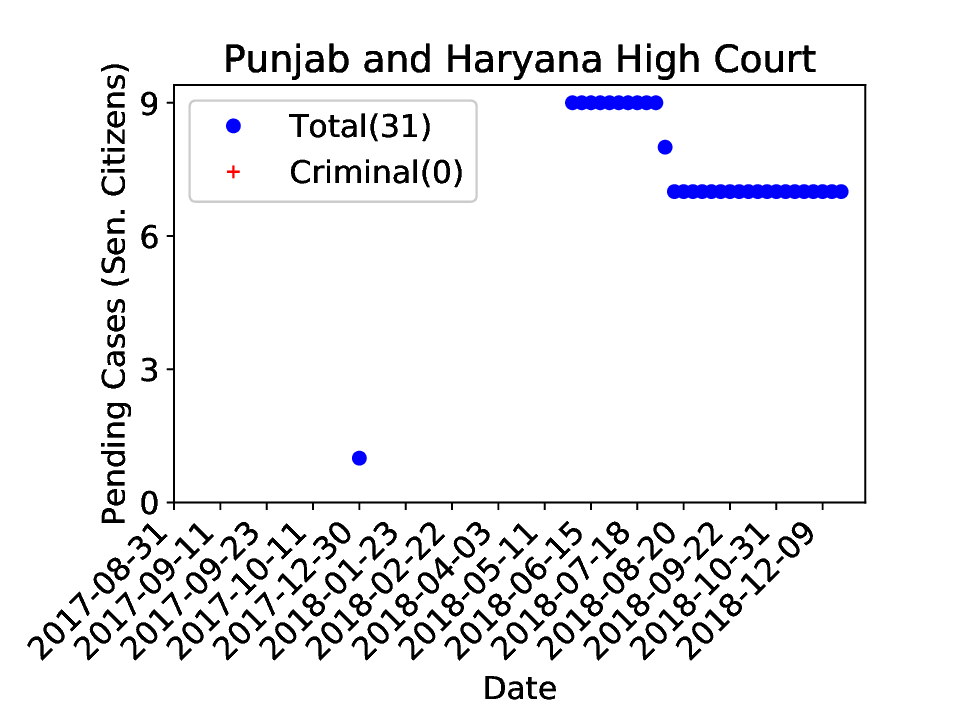}
\includegraphics[width=4.4cm]{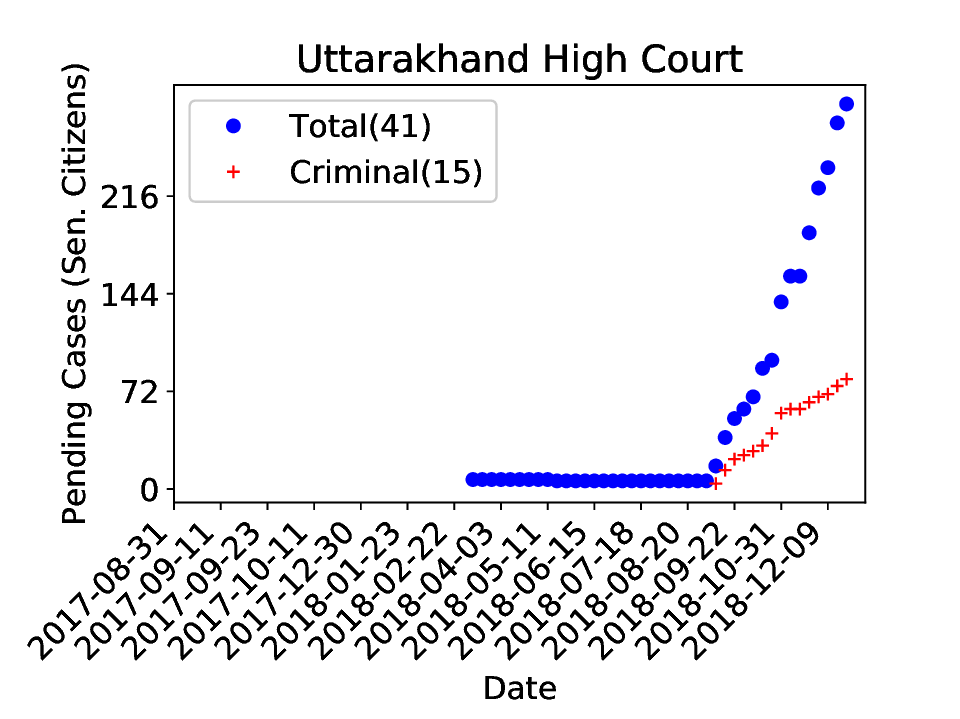}
\includegraphics[width=4.4cm]{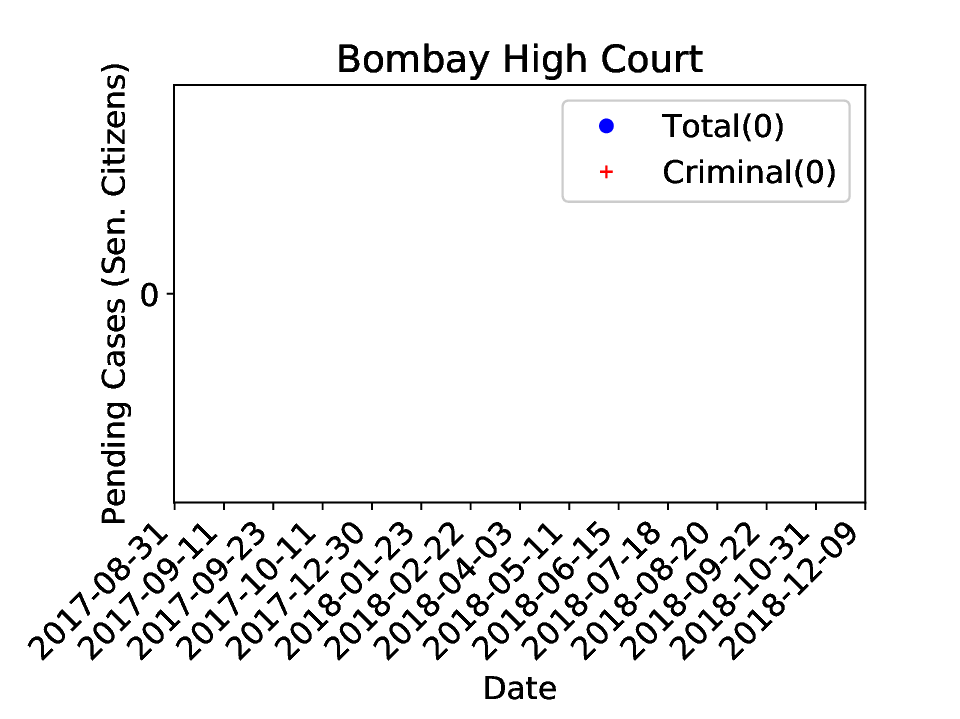}
\includegraphics[width=4.4cm]{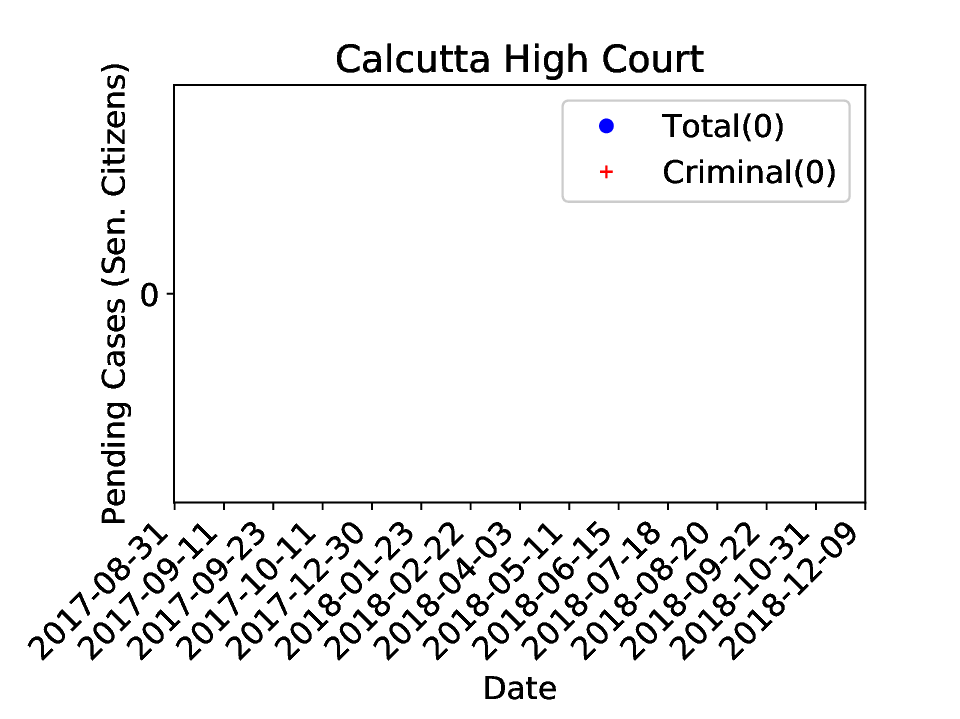}
\includegraphics[width=4.4cm]{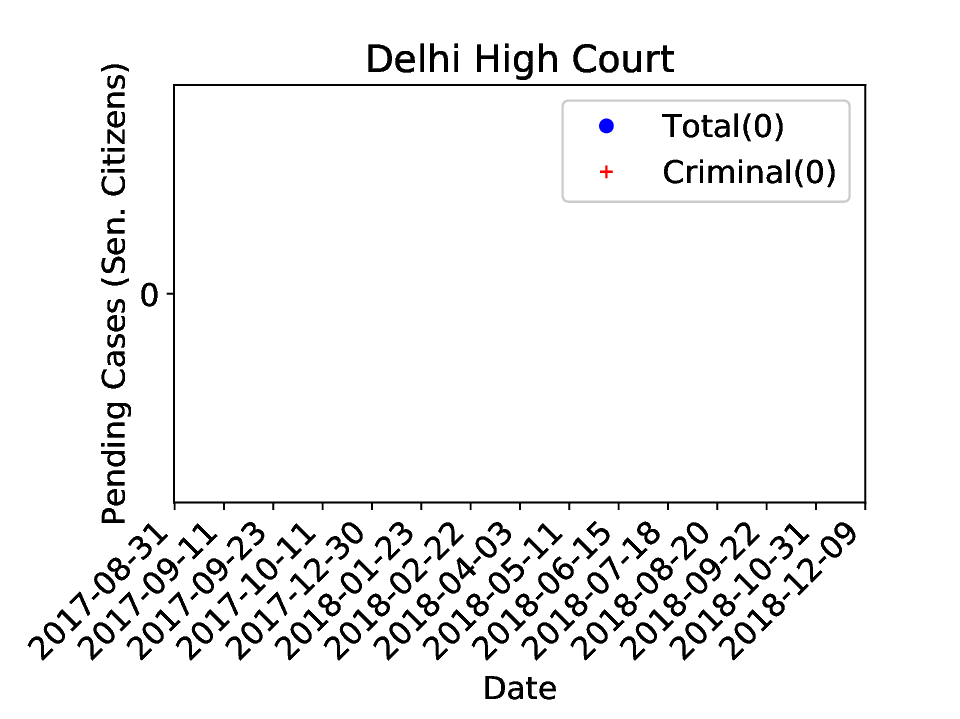}
\includegraphics[width=4.4cm]{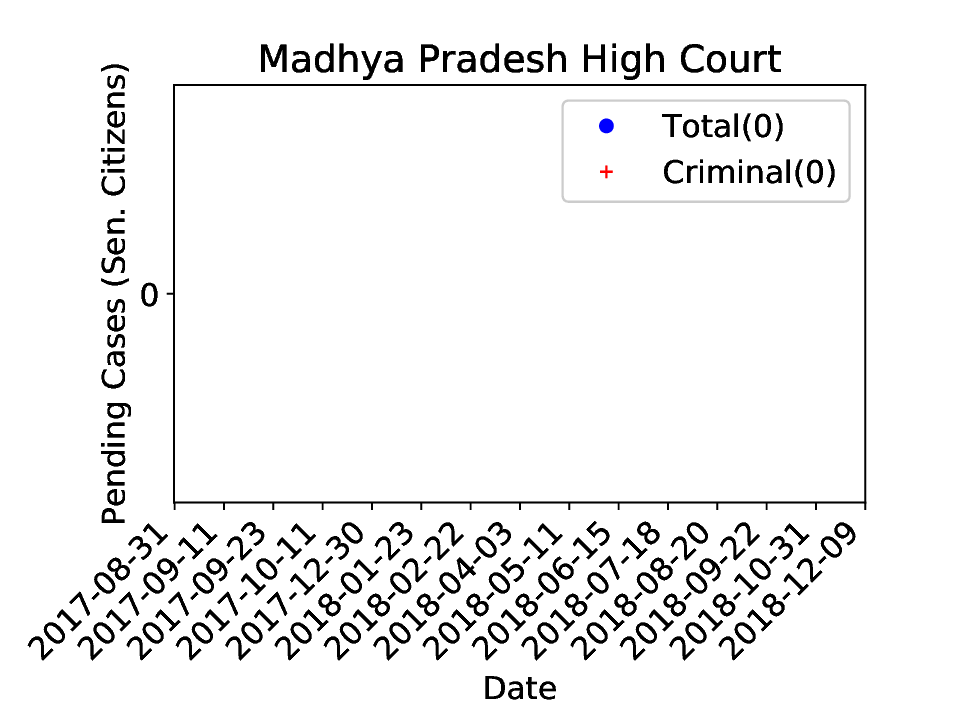}
\includegraphics[width=4.4cm]{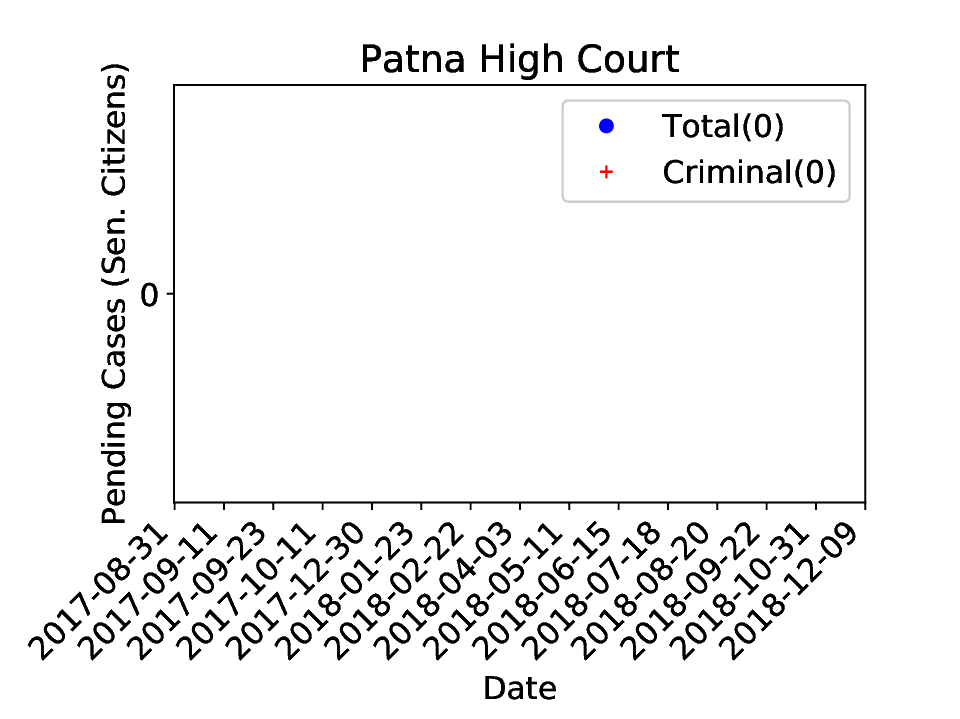}
\includegraphics[width=4.4cm]{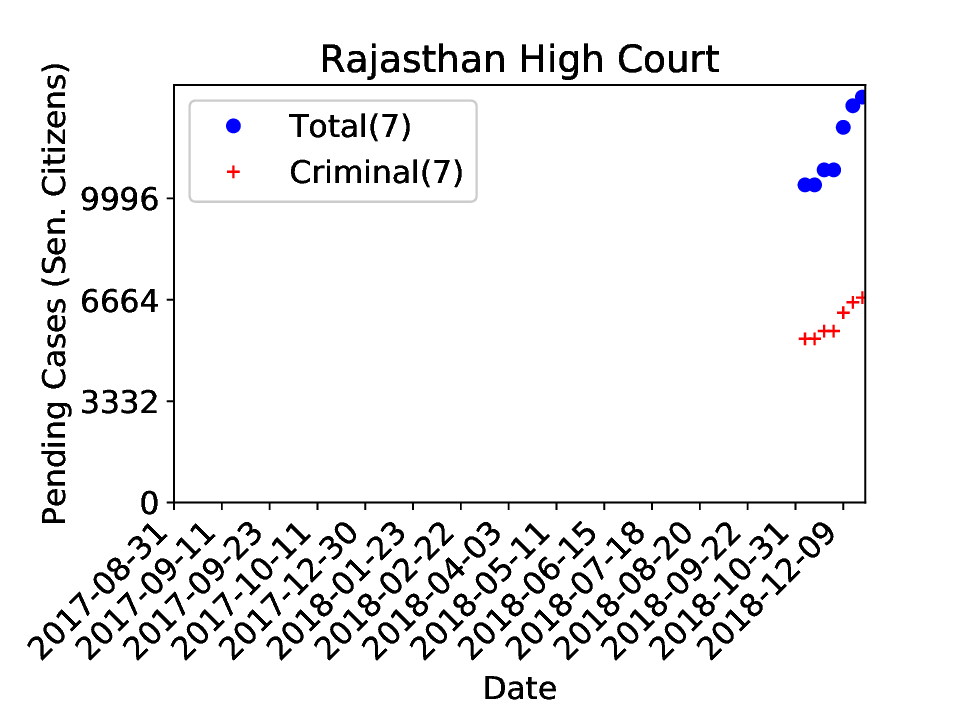}
\caption{The figure presents pending cases filed by senior citizens. We have chosen to draw only total cases and criminal cases filed by senior citizens. The number written in brackets is the number of data points. The data related to cases filed by senior citizens is either very little or none. This means that senior citizens usually do not file cases or HC-NJDG data is not complete. Bombay, Calcutta, Delhi, Madhya Pradesh, Patna and Rajasthan High Courts have no data on pending cases of senior citizens. While Madras High Court presents the data, it was almost never updated except for the last couple of months.}
\label{fig:sc_hc2}
\end{figure*}

\fref{fig:sc_hc2} shows the data of pending cases filed by senior citizens. Overall, we see that the total cases filed by senior citizens is increasing for almost all high courts. However, the number of pending criminal cases is either flat or increasing at a very little rate compared to the total cases filed by senior citizens. This implies that the senior citizens mainly file civil suits and writs in high courts. However, the overall numbers of cases filed by senior citizens is so small that they can be even given priority without practically affecting the other cases. 

\subsubsection{Cases Filed by Women}

\begin{figure*}[h]
\includegraphics[width=4.4cm]{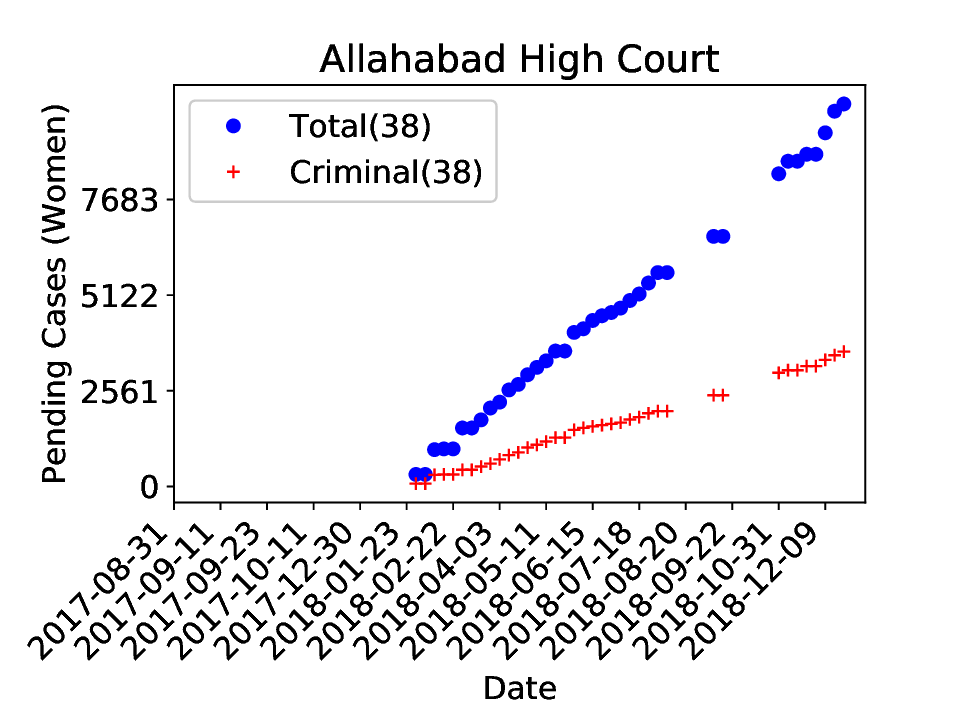}
\includegraphics[width=4.4cm]{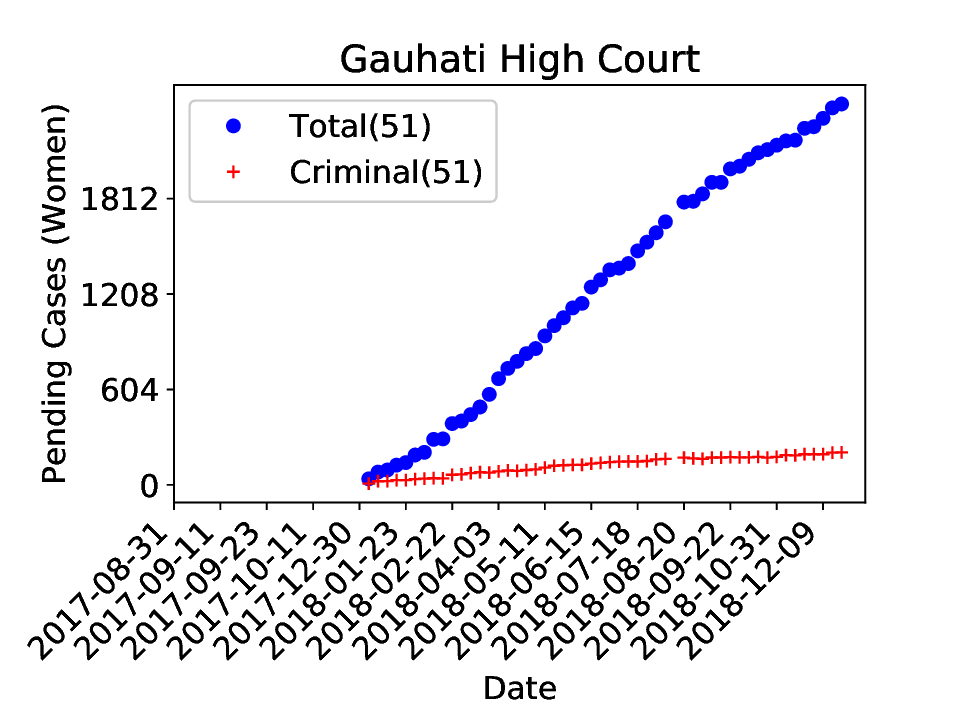}
\includegraphics[width=4.4cm]{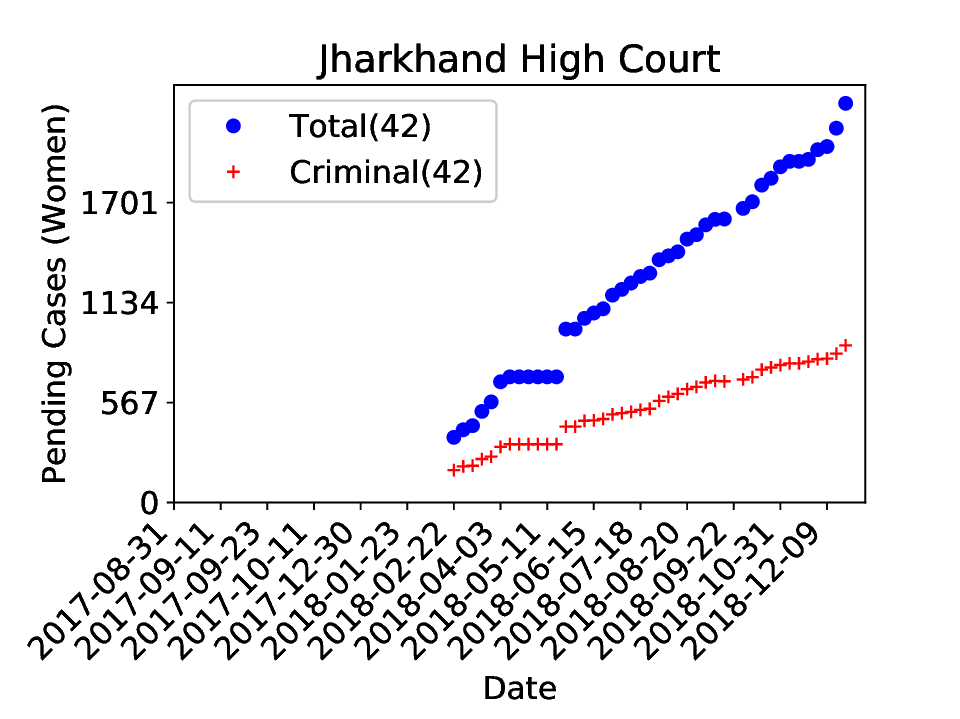}
\includegraphics[width=4.4cm]{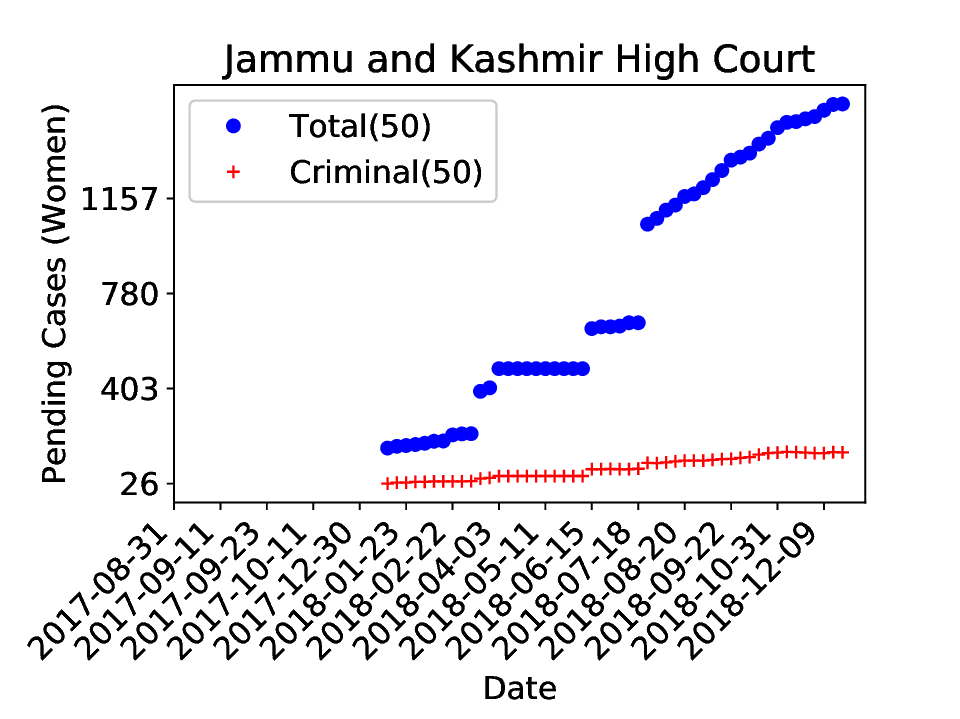}
\includegraphics[width=4.4cm]{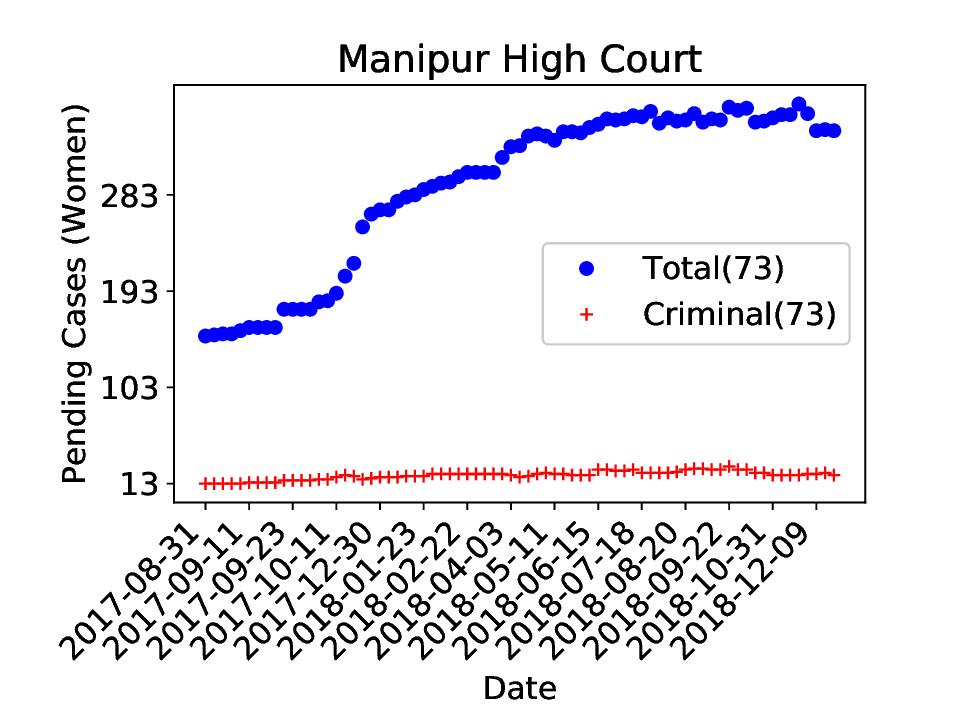}
\includegraphics[width=4.4cm]{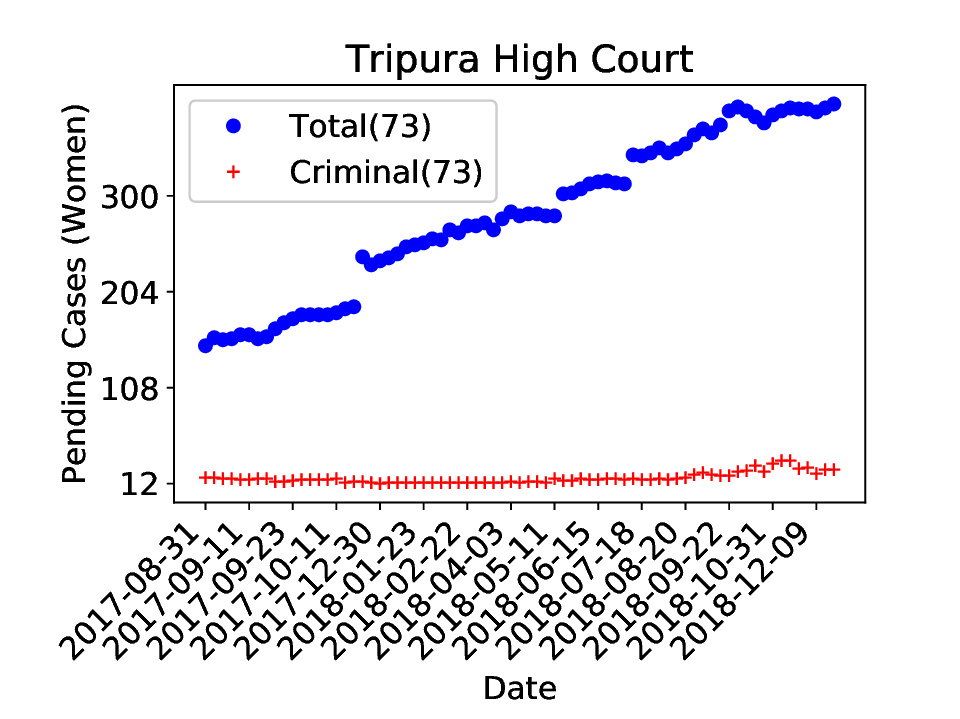}
\includegraphics[width=4.4cm]{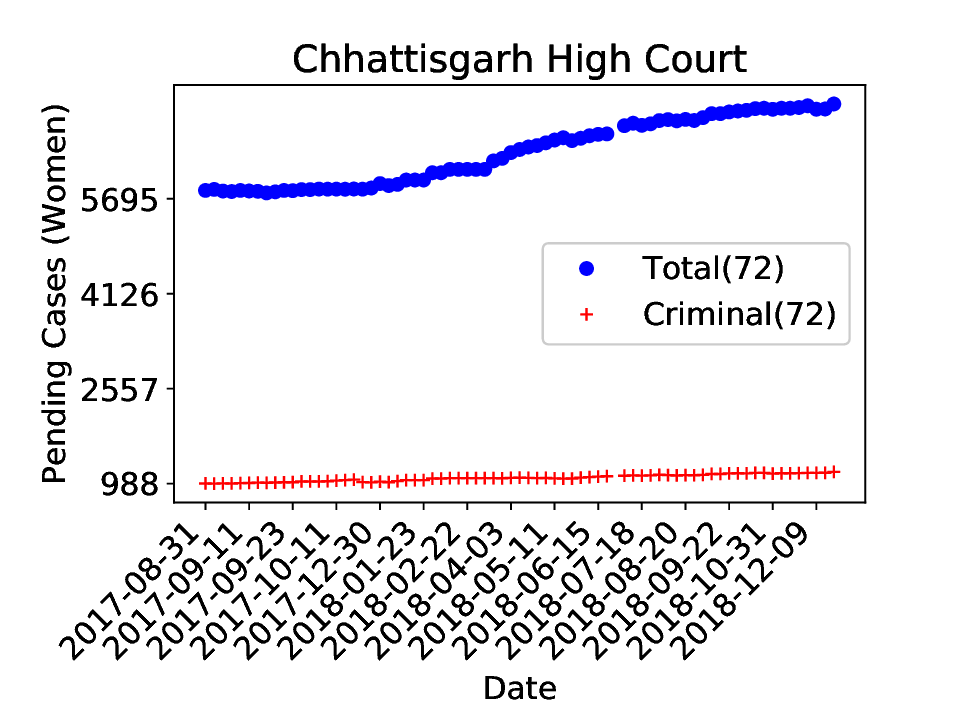}
\includegraphics[width=4.4cm]{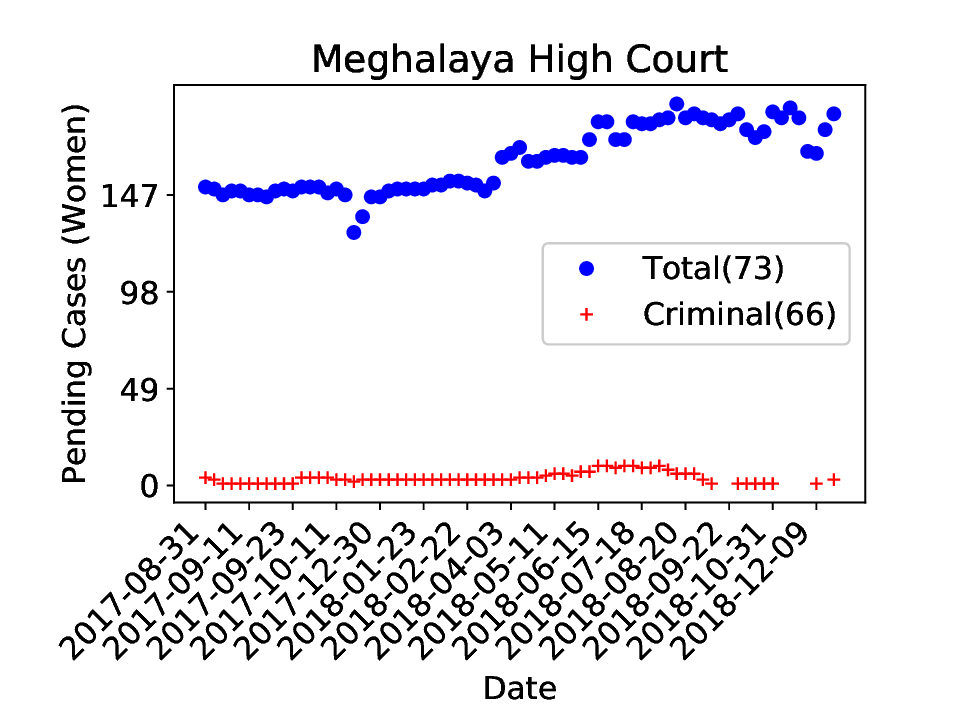}
\includegraphics[width=4.4cm]{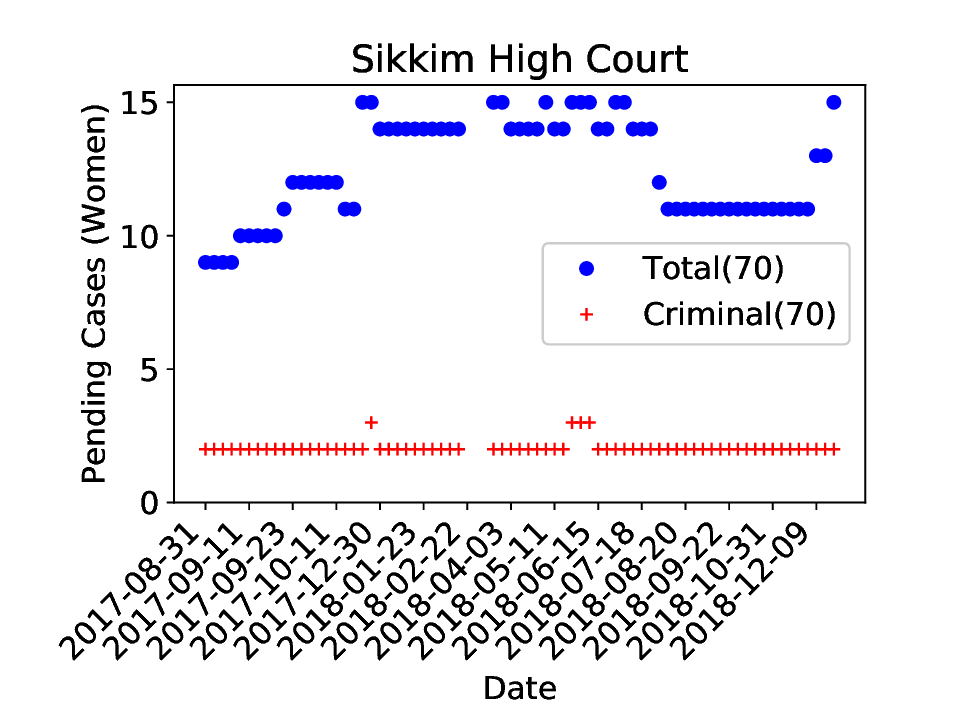}
\includegraphics[width=4.4cm]{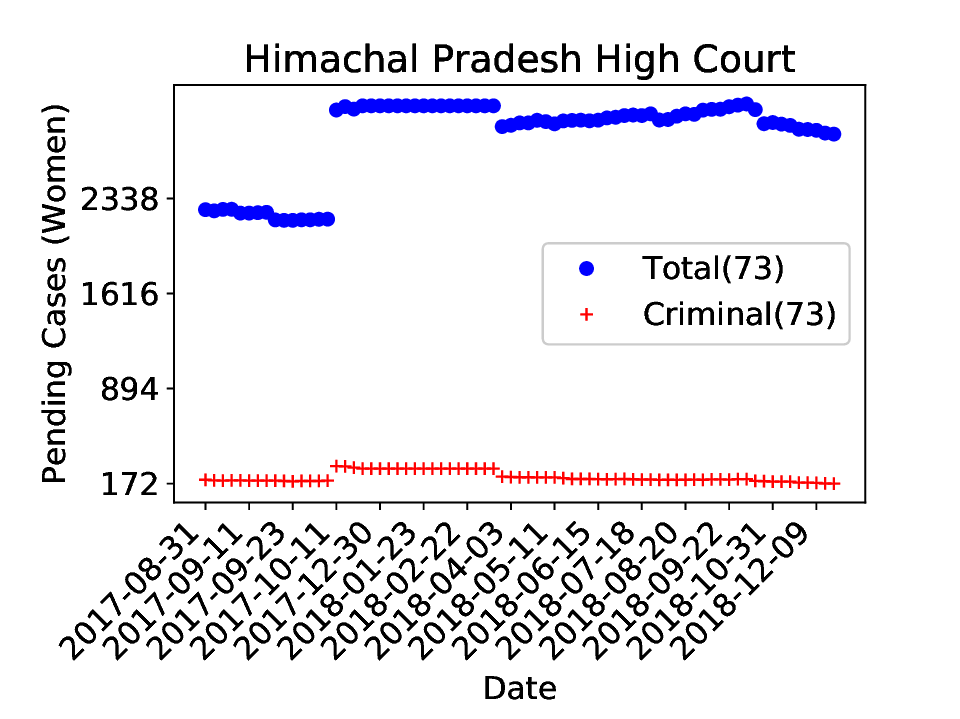}
\includegraphics[width=4.4cm]{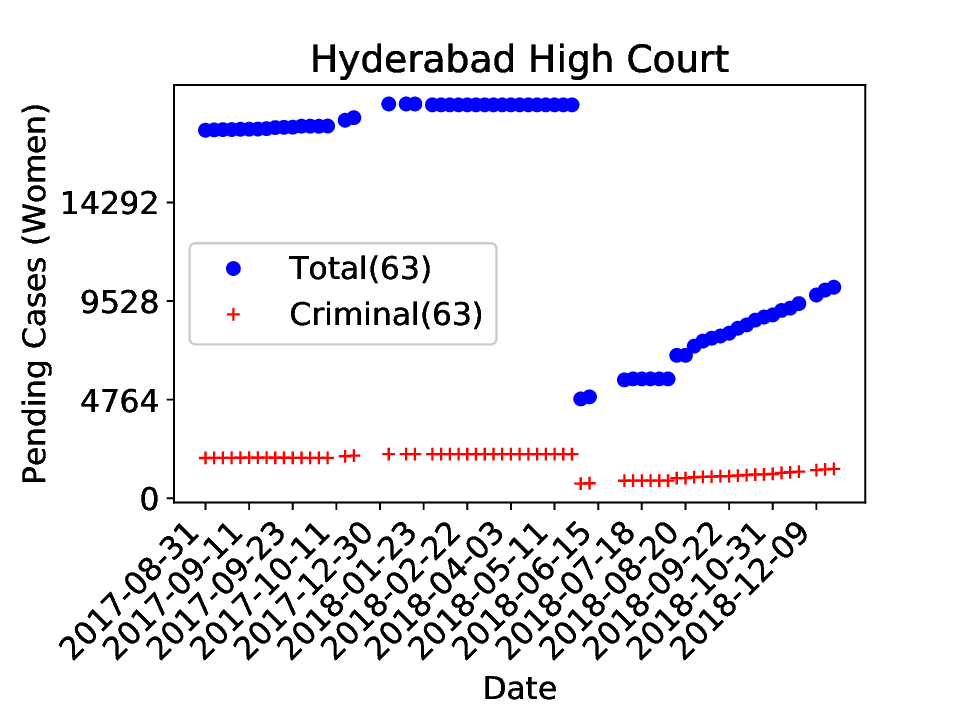}
\includegraphics[width=4.4cm]{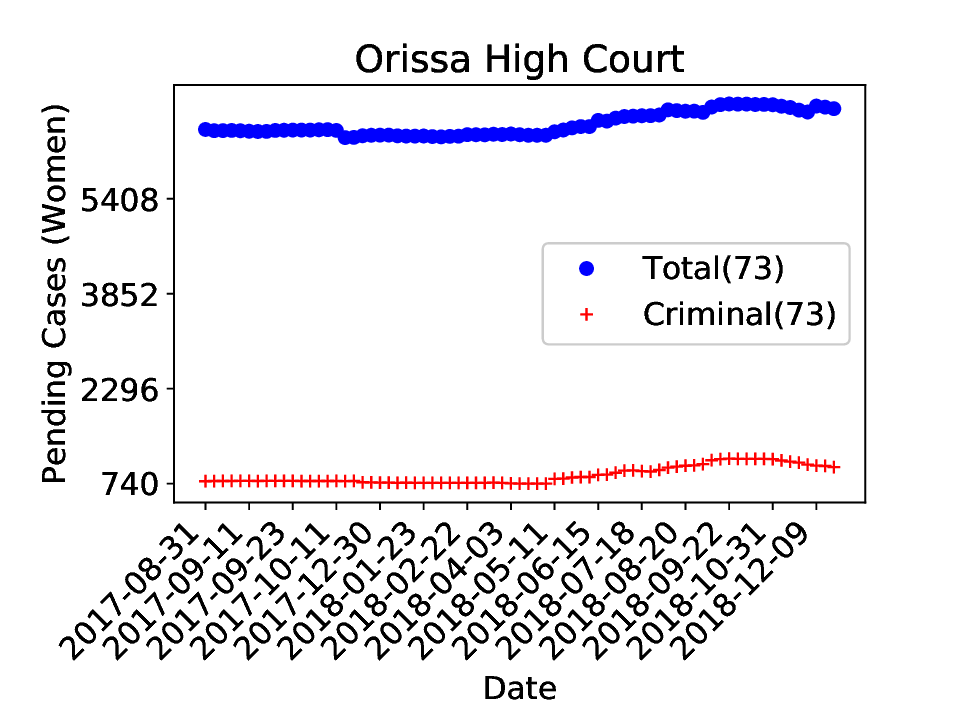}
\includegraphics[width=4.4cm]{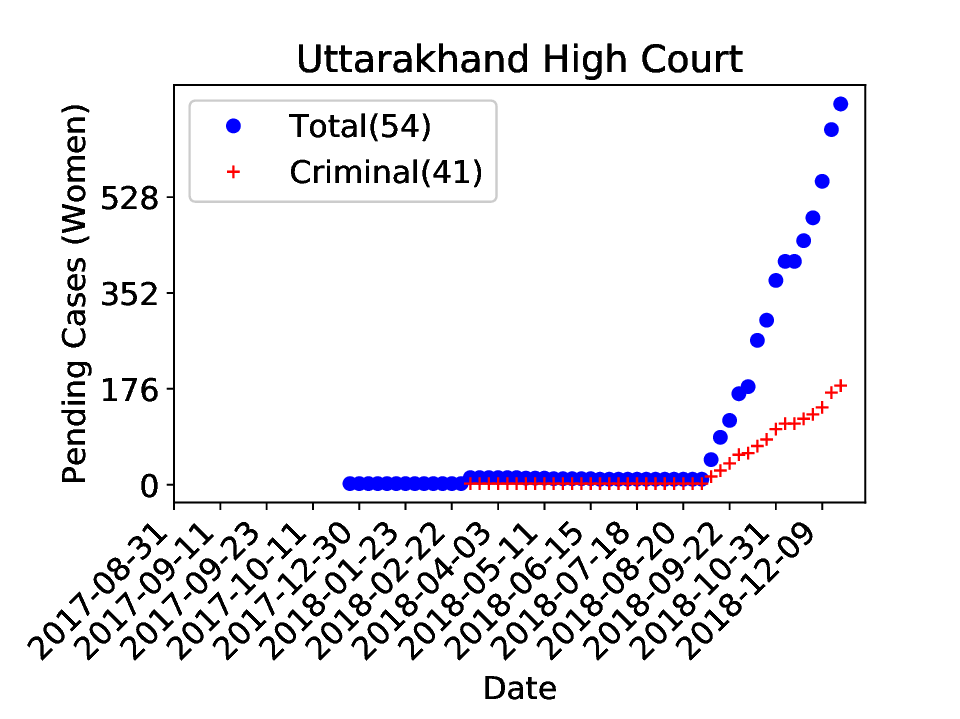}
\includegraphics[width=4.4cm]{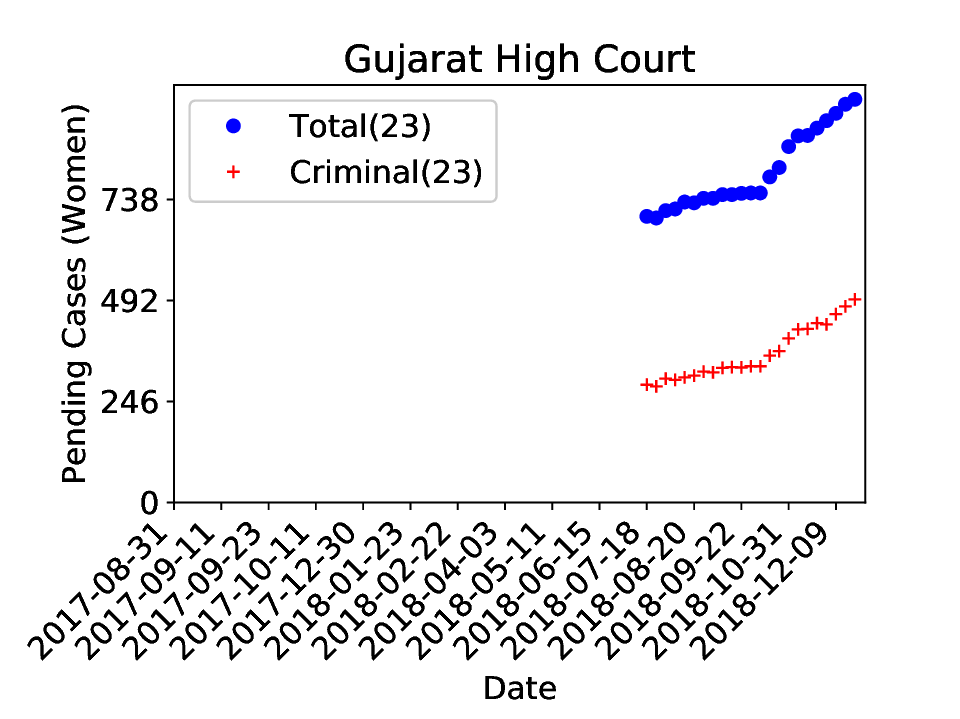}
\includegraphics[width=4.4cm]{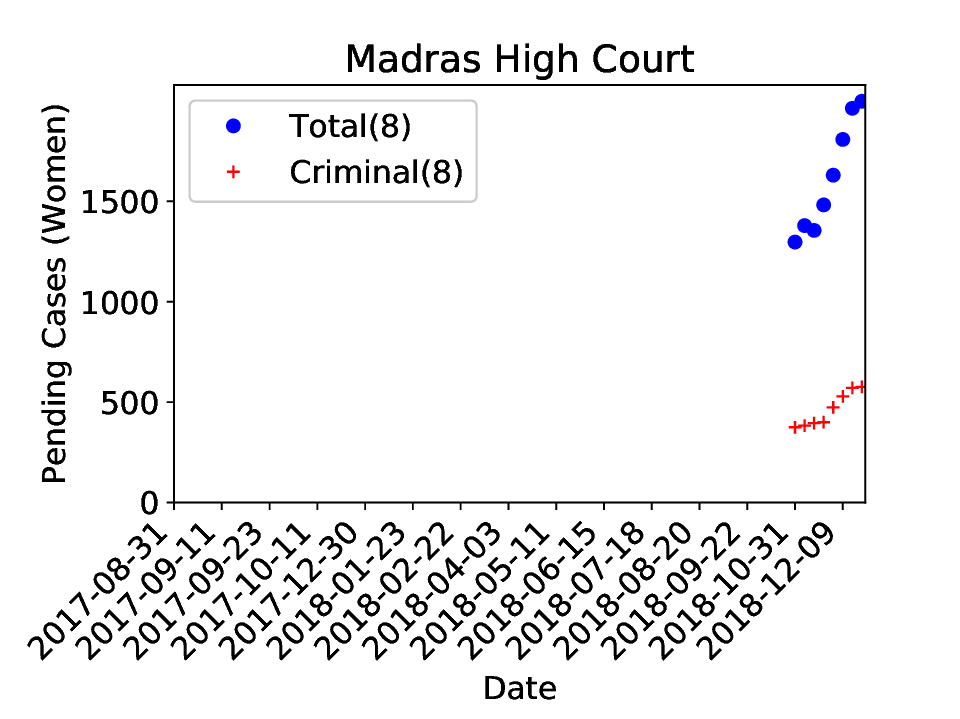}
\includegraphics[width=4.4cm]{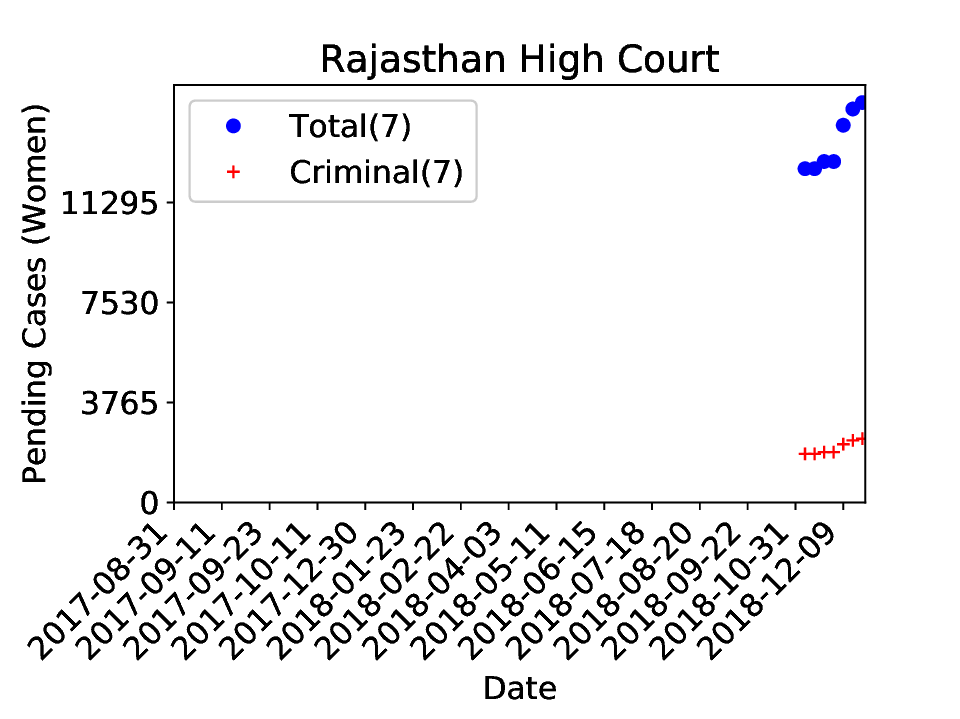}
\includegraphics[width=4.4cm]{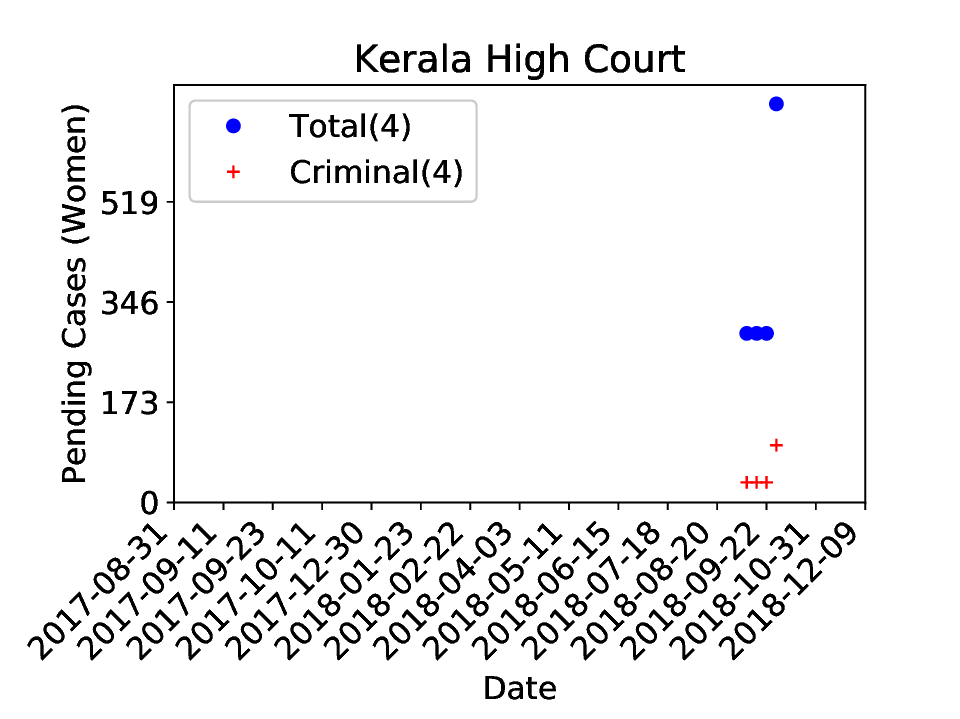}
\includegraphics[width=4.4cm]{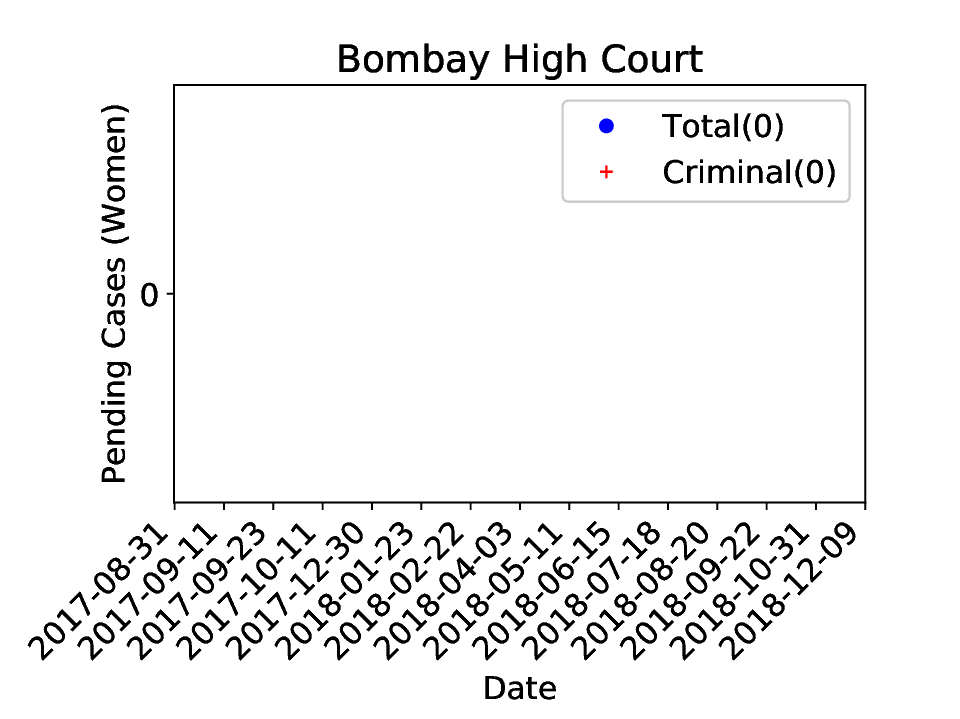}
\includegraphics[width=4.4cm]{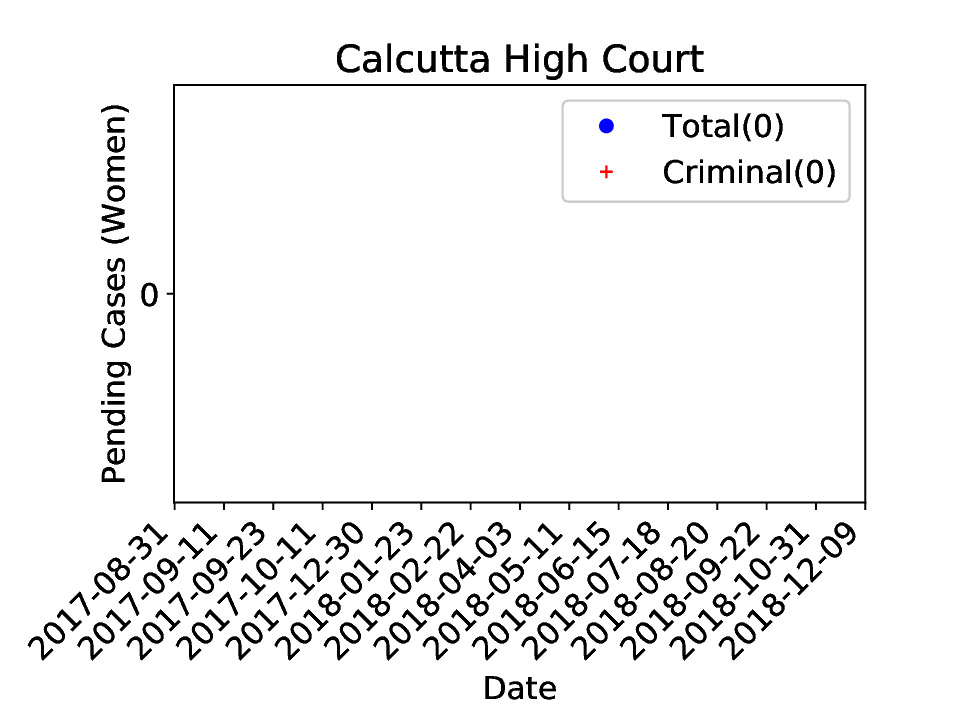}
\includegraphics[width=4.4cm]{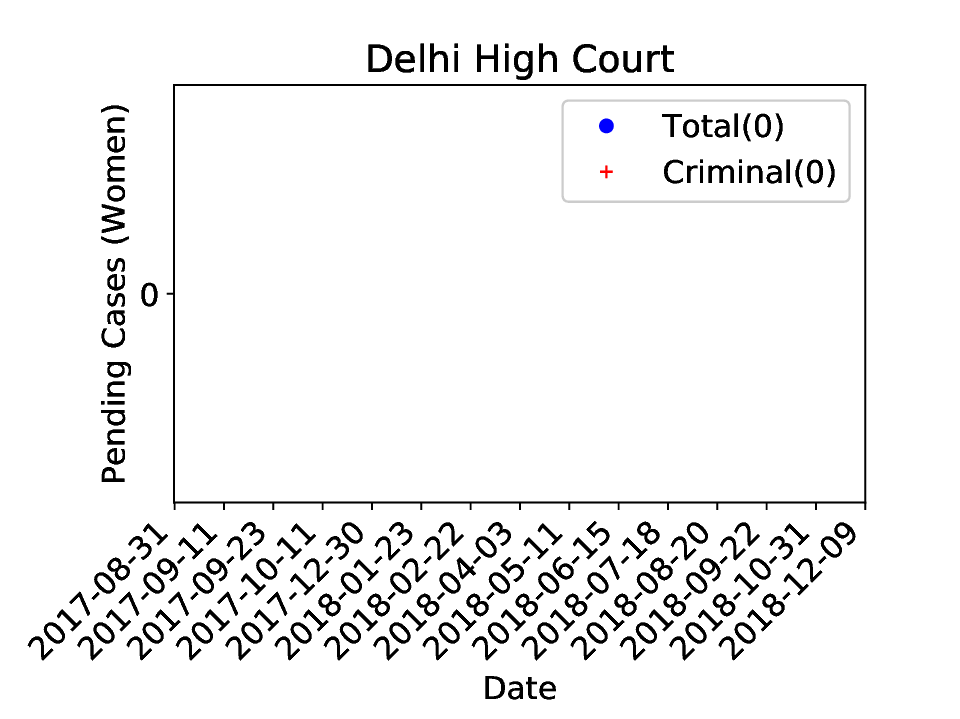}
\includegraphics[width=4.4cm]{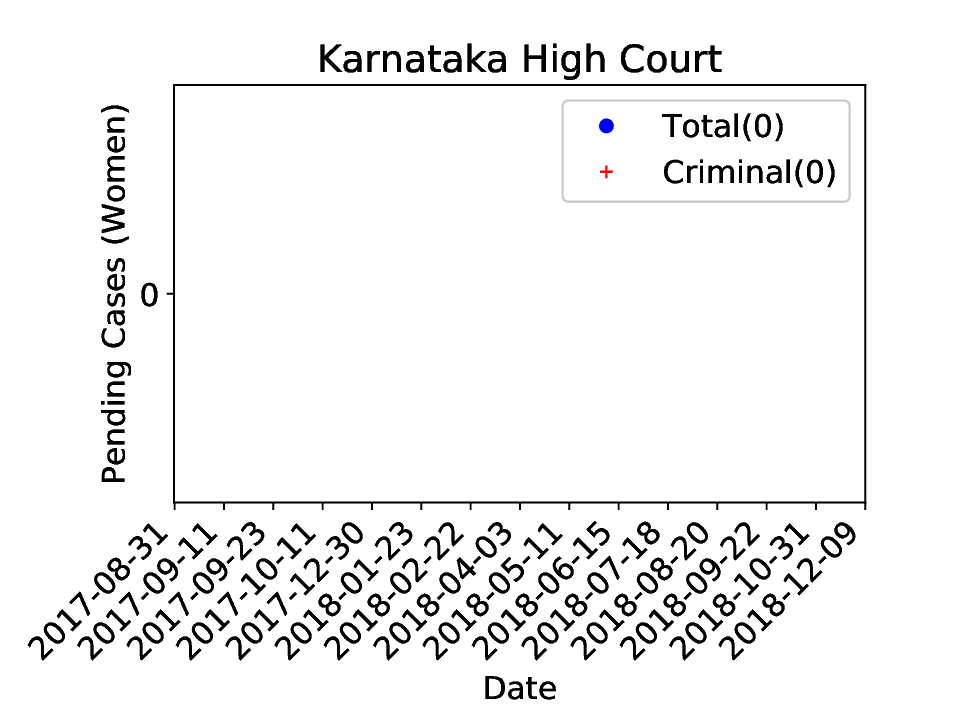}
\includegraphics[width=4.4cm]{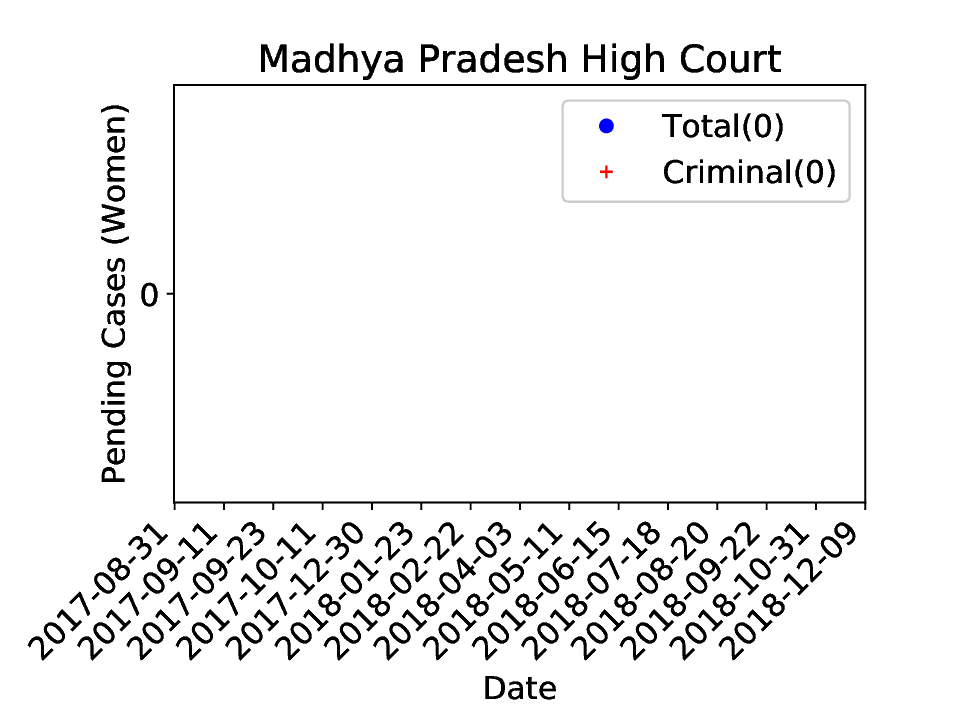}
\includegraphics[width=4.4cm]{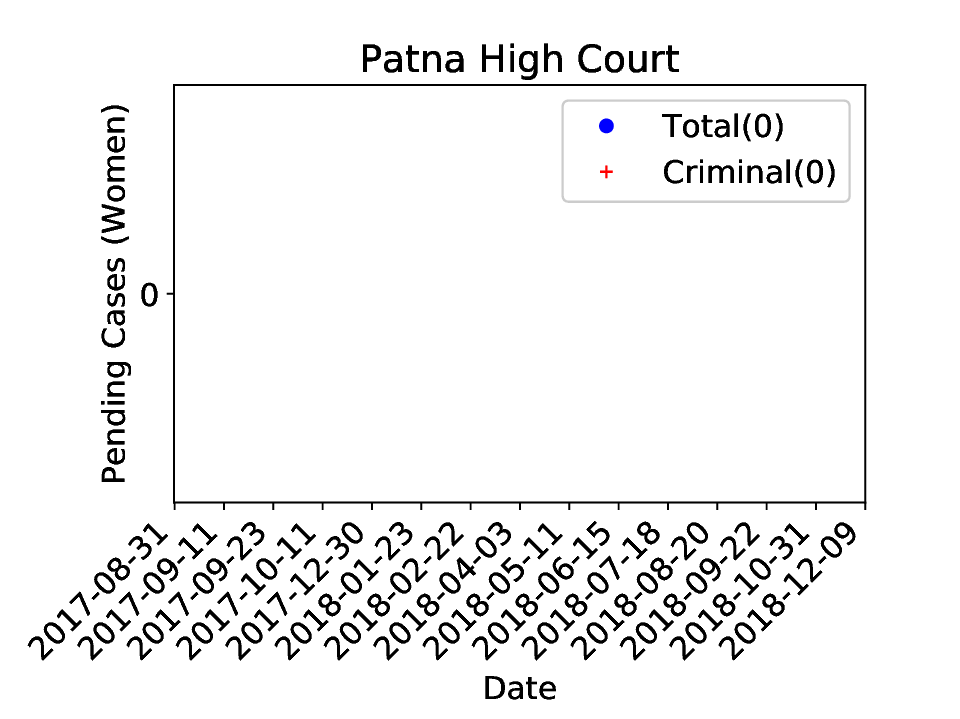}
\includegraphics[width=4.4cm]{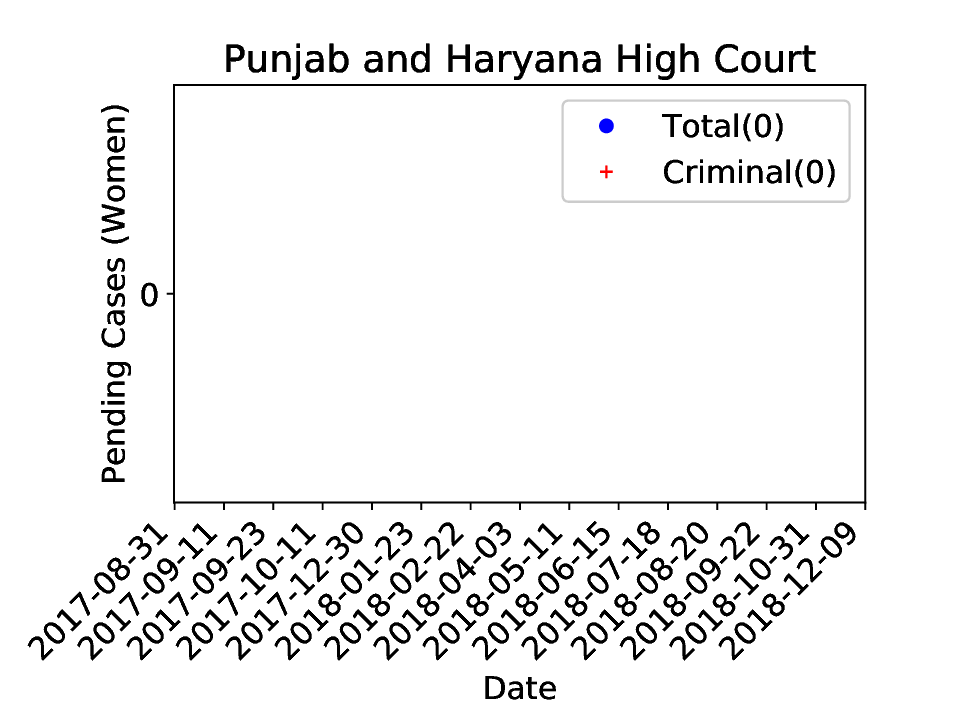}

\caption{The cases filed by women are very less in number. As for senior citizens, we plot total and criminal cases filed by women. Even though in some of the high courts the total cases filed by women are increasing, the number of criminal cases filed by women is almost constant. This usually would mean that women don't register cases against the crimes that they face or the data of HC-NJDG is not fully updated. Last ten high courts in this figure have no or very little data on cases filed by women.}
\label{fig:wom_hc}
\end{figure*}

As in the case of senior citizens, cases filed by women are also very less. The cases filed by women are almost two orders of magnitude smaller than the total pendency figures. For details refer to \fref{fig:wom_hc}.

\subsection{Year-wise pending cases}

\begin{figure*}[h]
\includegraphics[width=4.4cm]{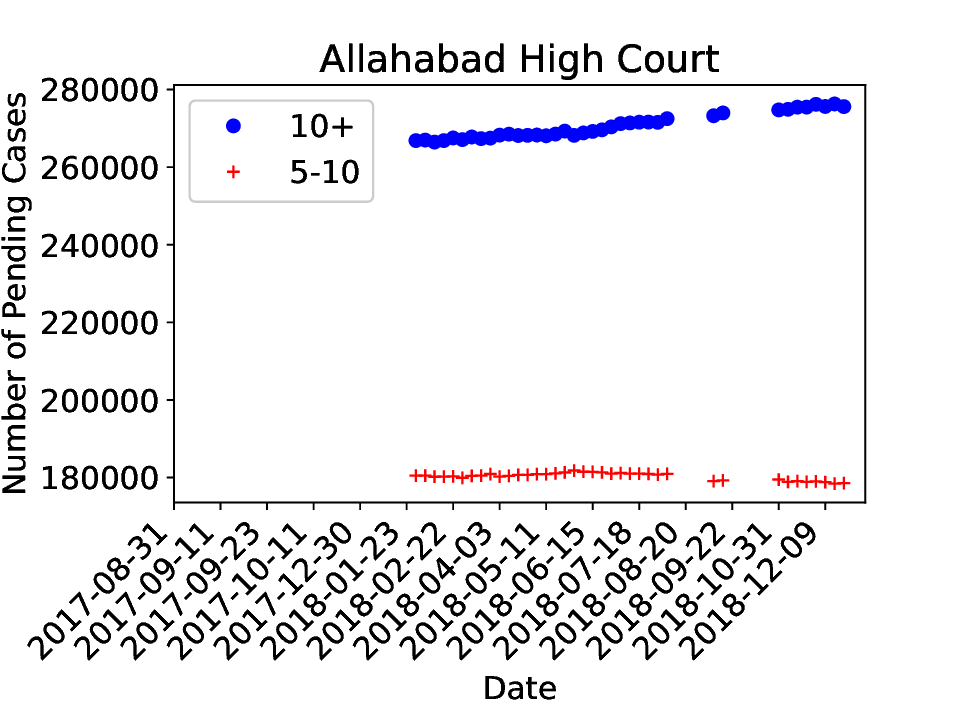}
\includegraphics[width=4.4cm]{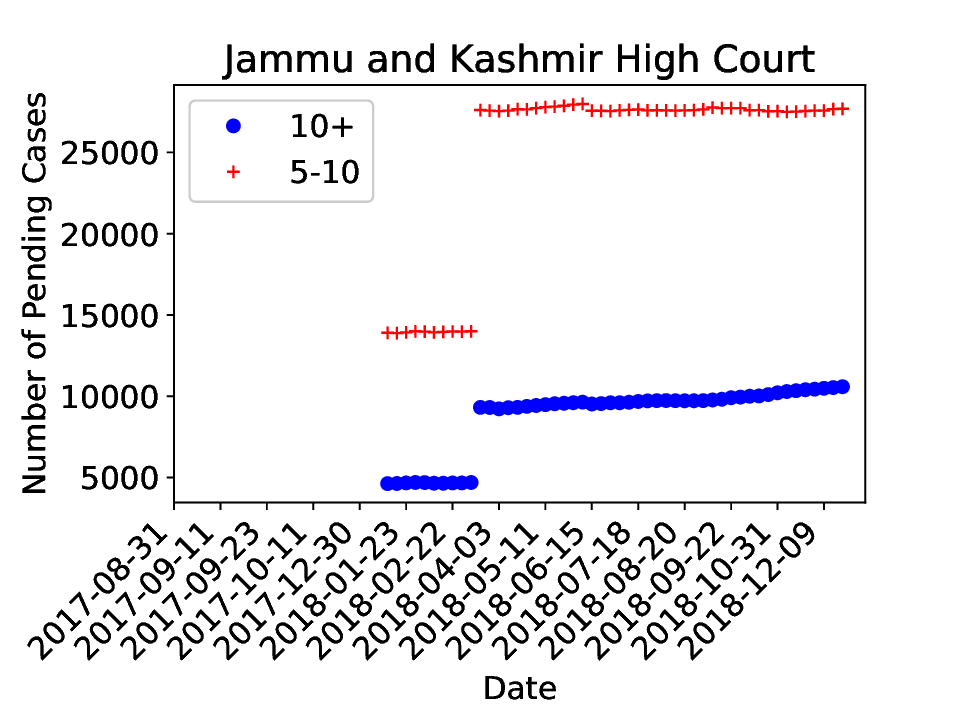}
\includegraphics[width=4.4cm]{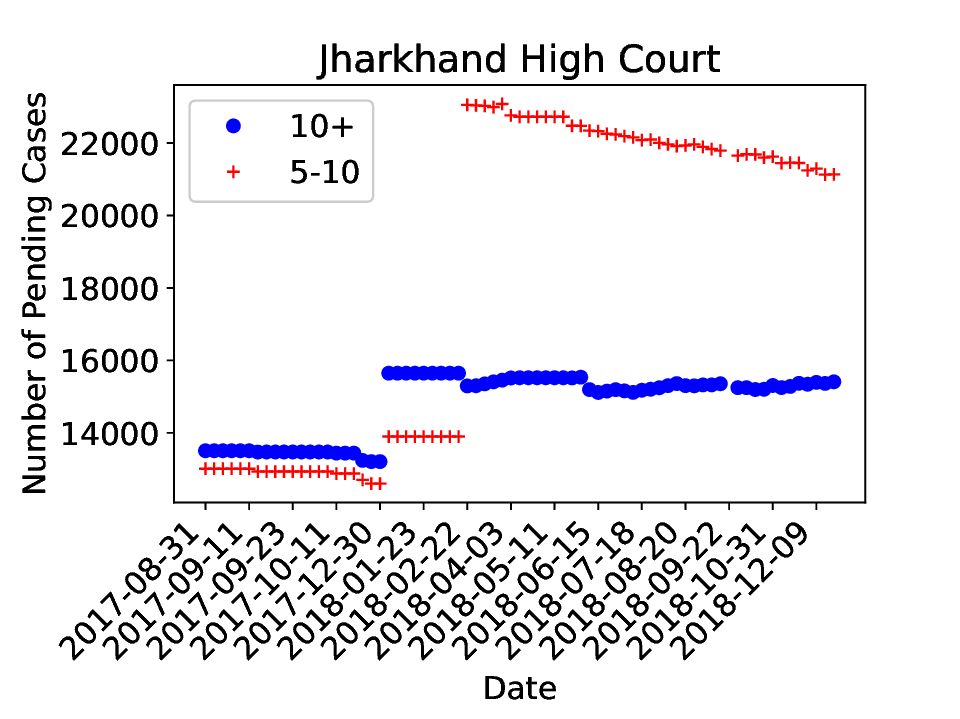}
\includegraphics[width=4.4cm]{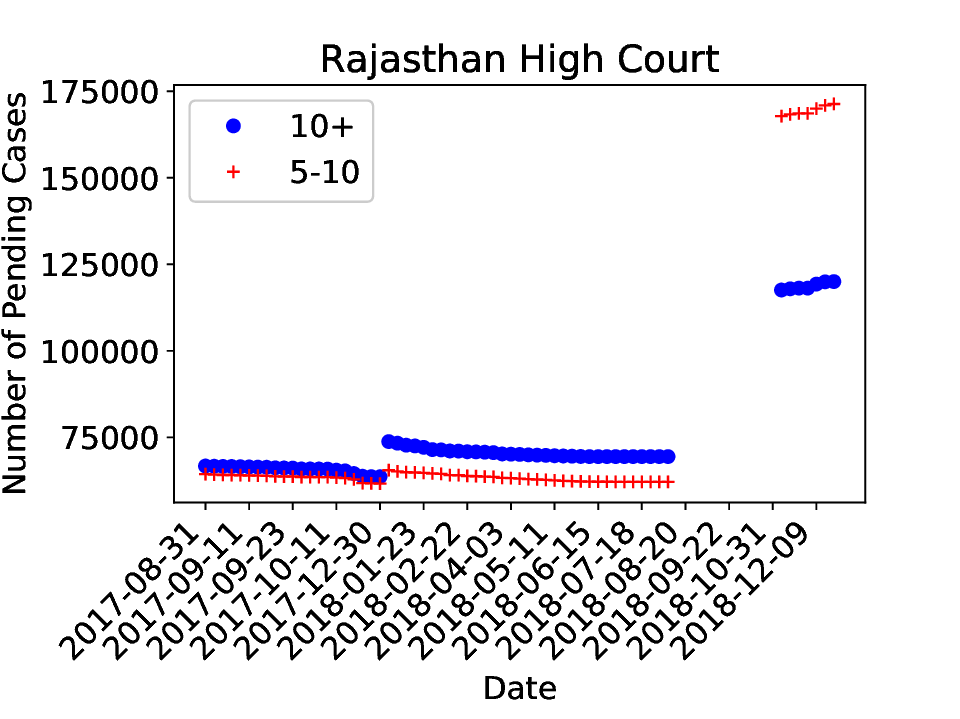}
\includegraphics[width=4.4cm]{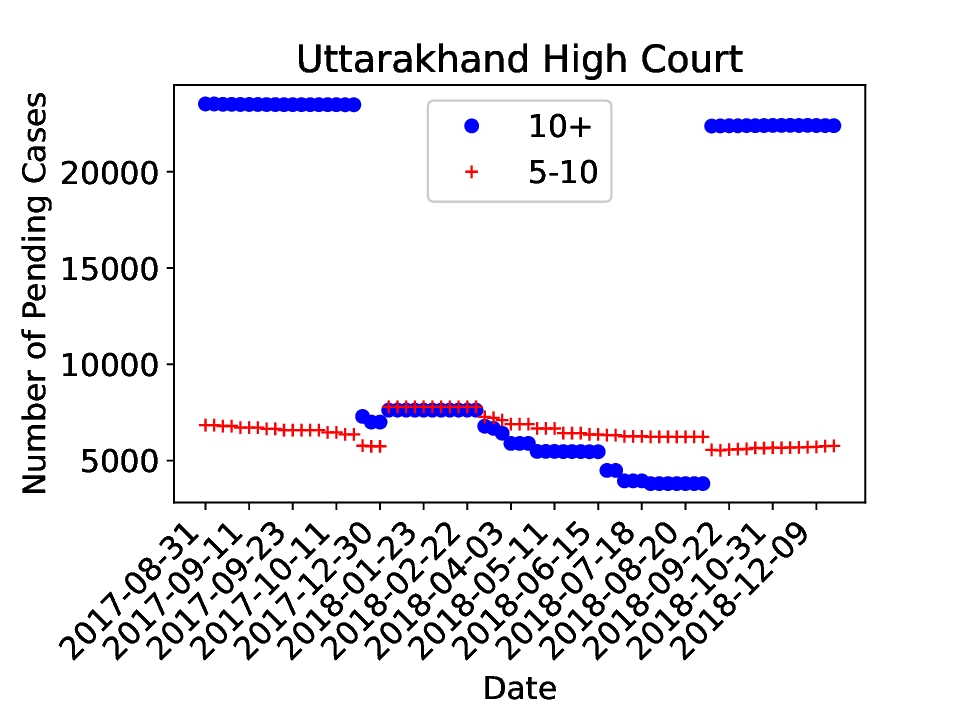}
\includegraphics[width=4.4cm]{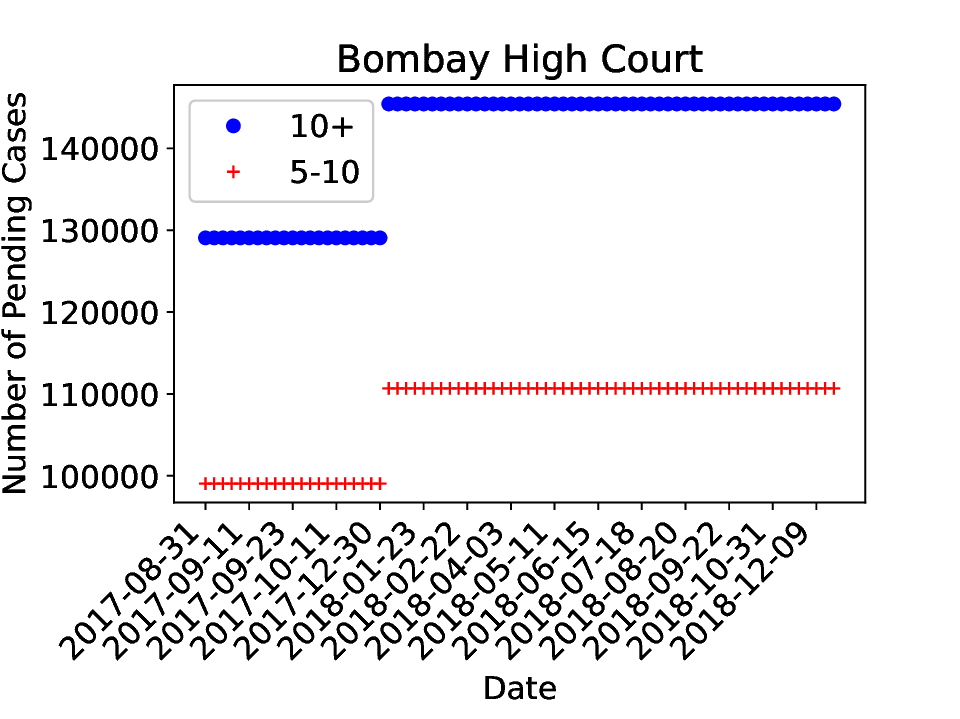}
\includegraphics[width=4.4cm]{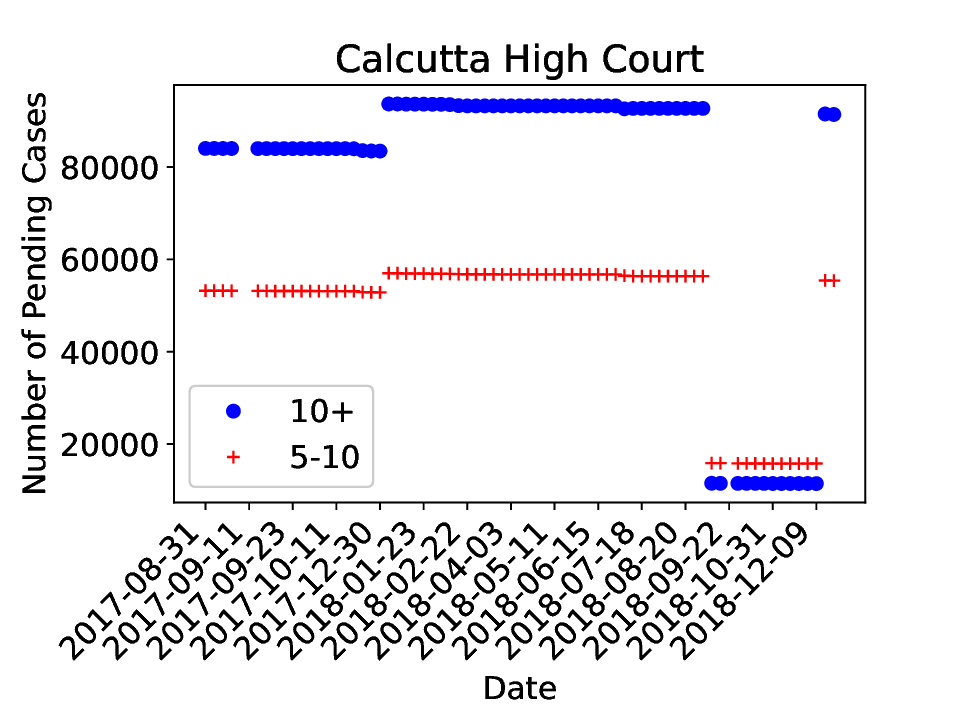}
\includegraphics[width=4.4cm]{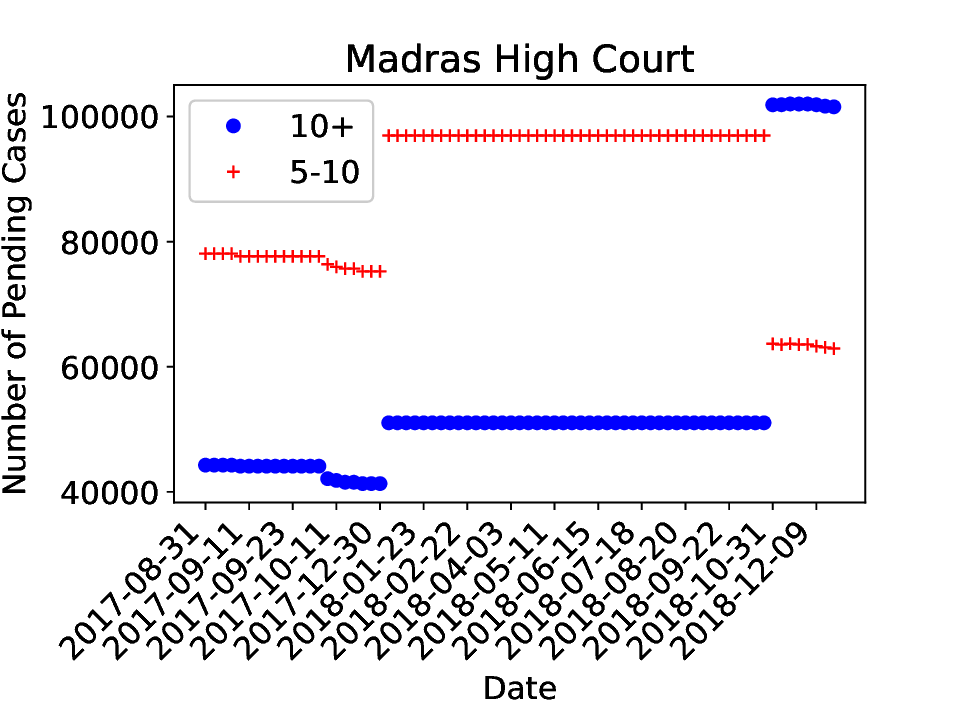}
\includegraphics[width=4.4cm]{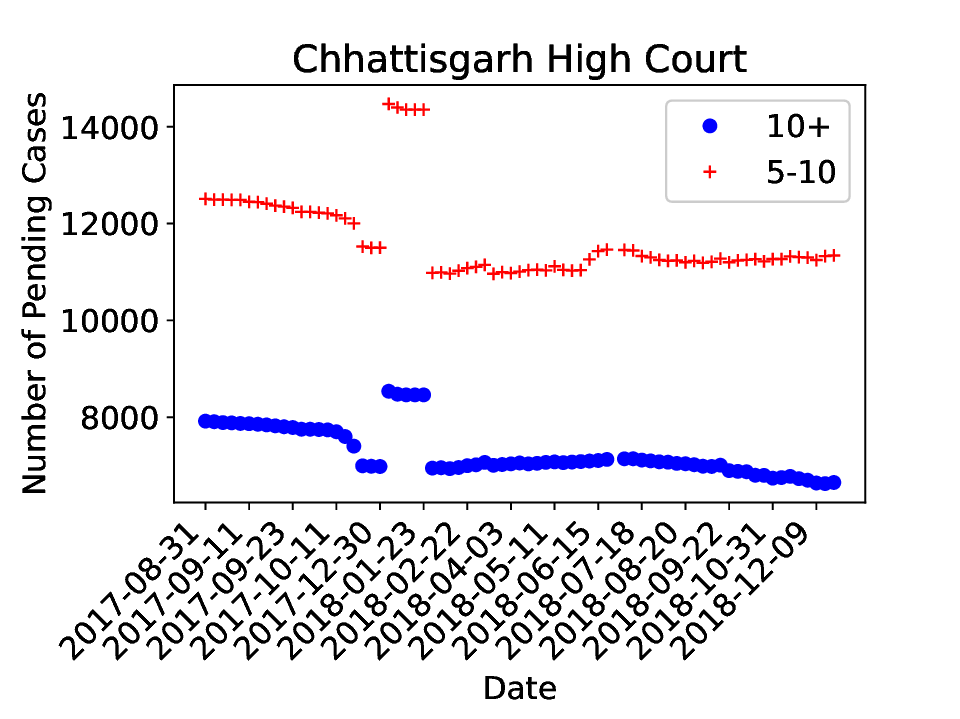}
\includegraphics[width=4.4cm]{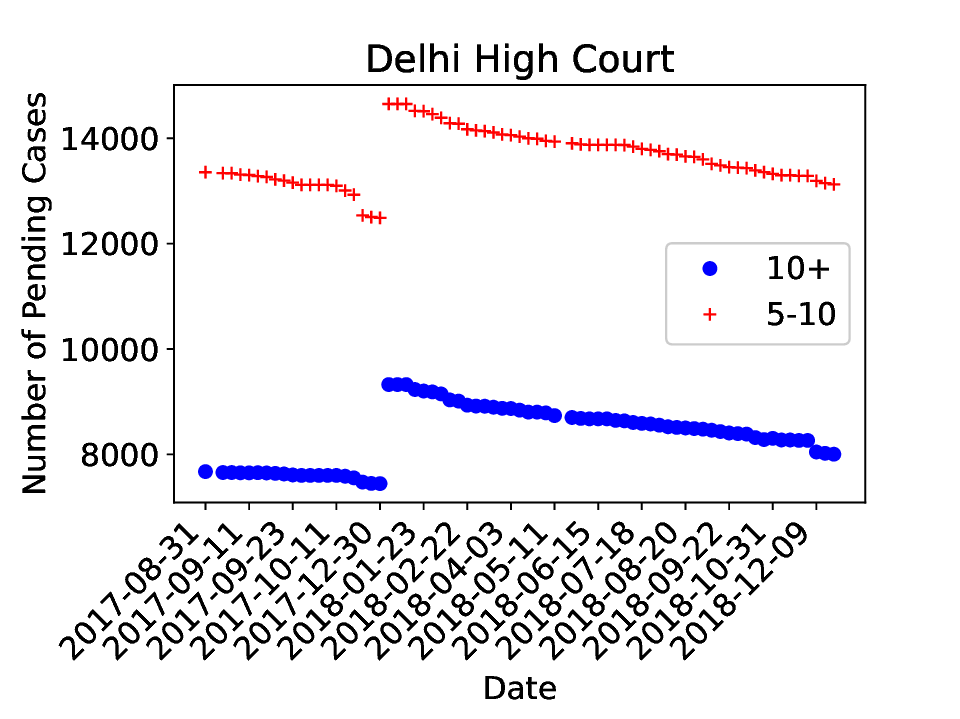}
\includegraphics[width=4.4cm]{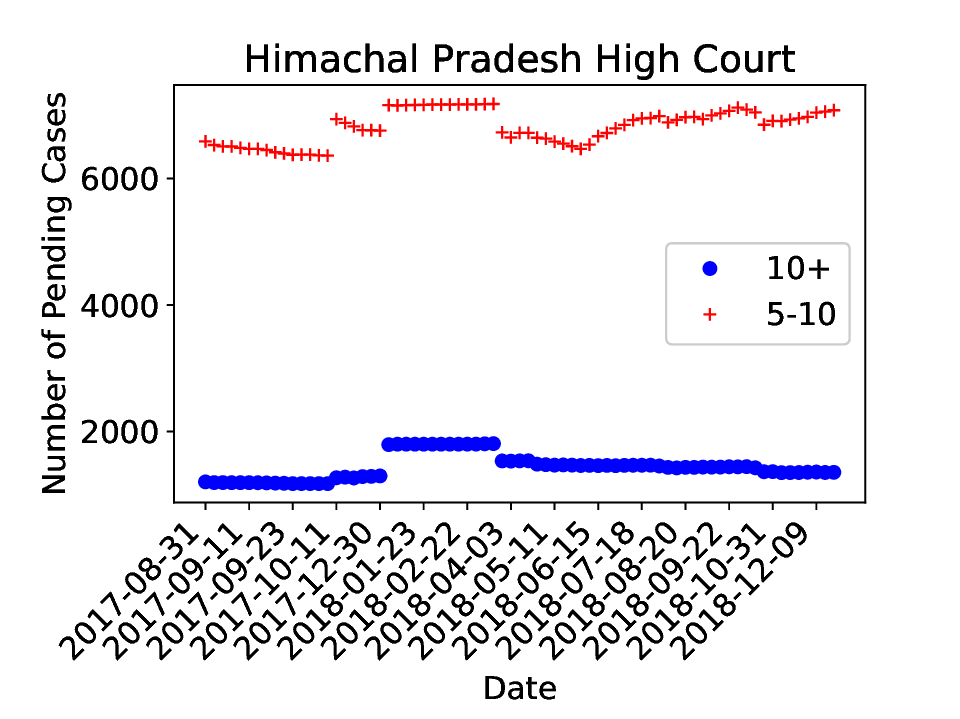}
\includegraphics[width=4.4cm]{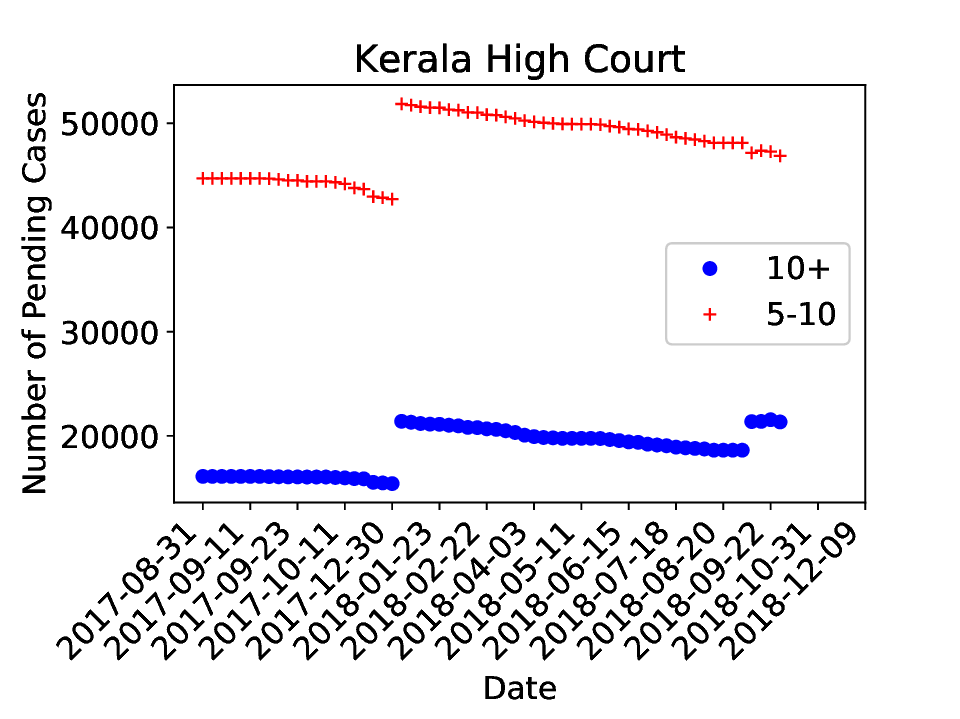}
\includegraphics[width=4.4cm]{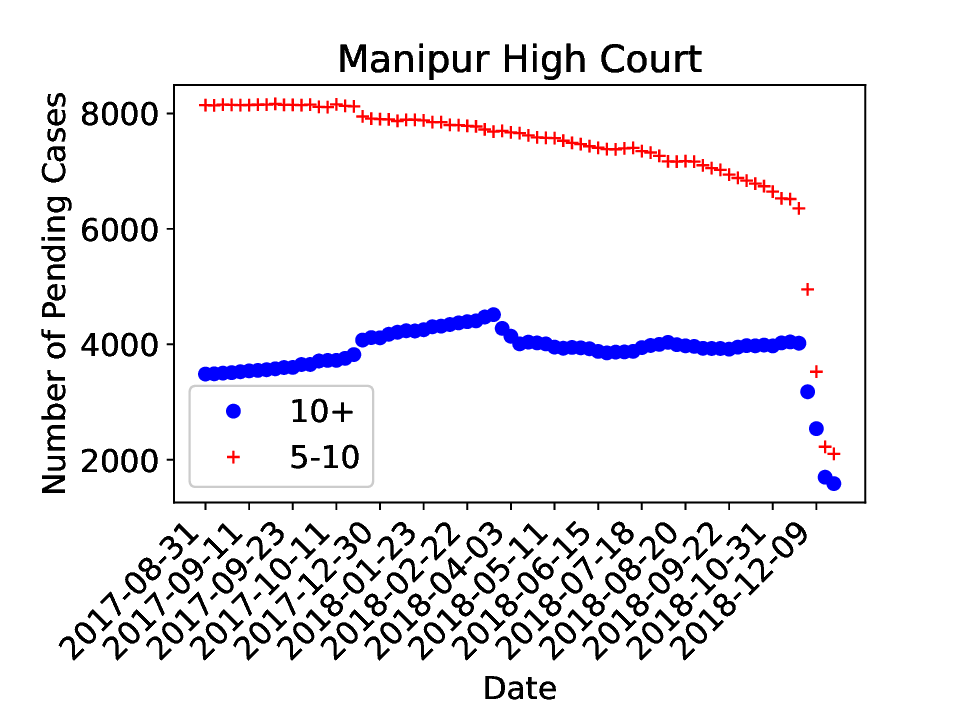}
\includegraphics[width=4.4cm]{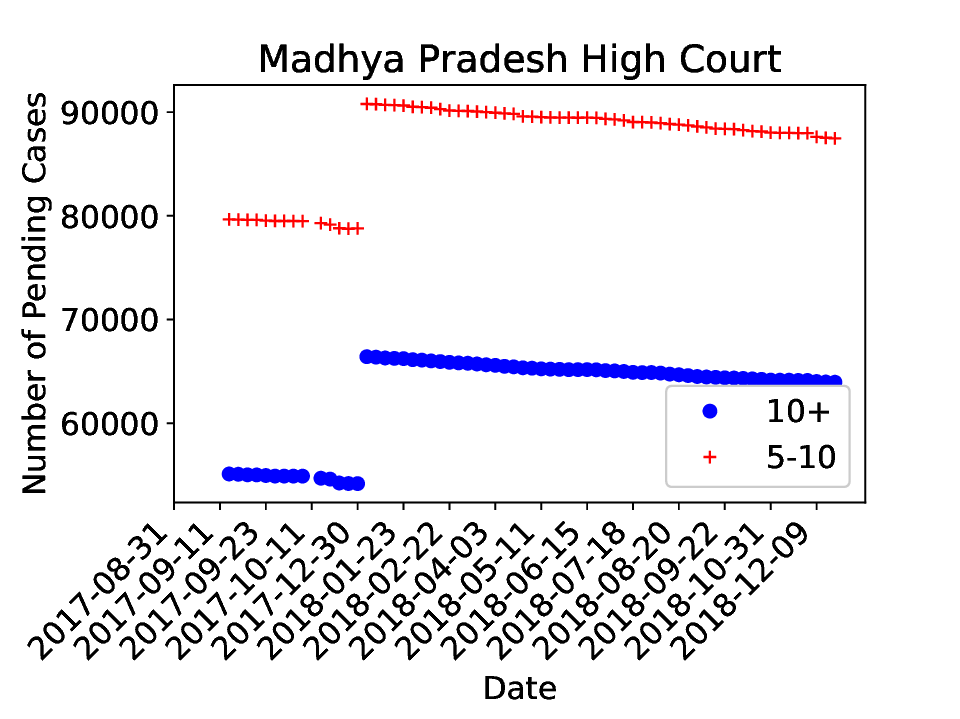}
\includegraphics[width=4.4cm]{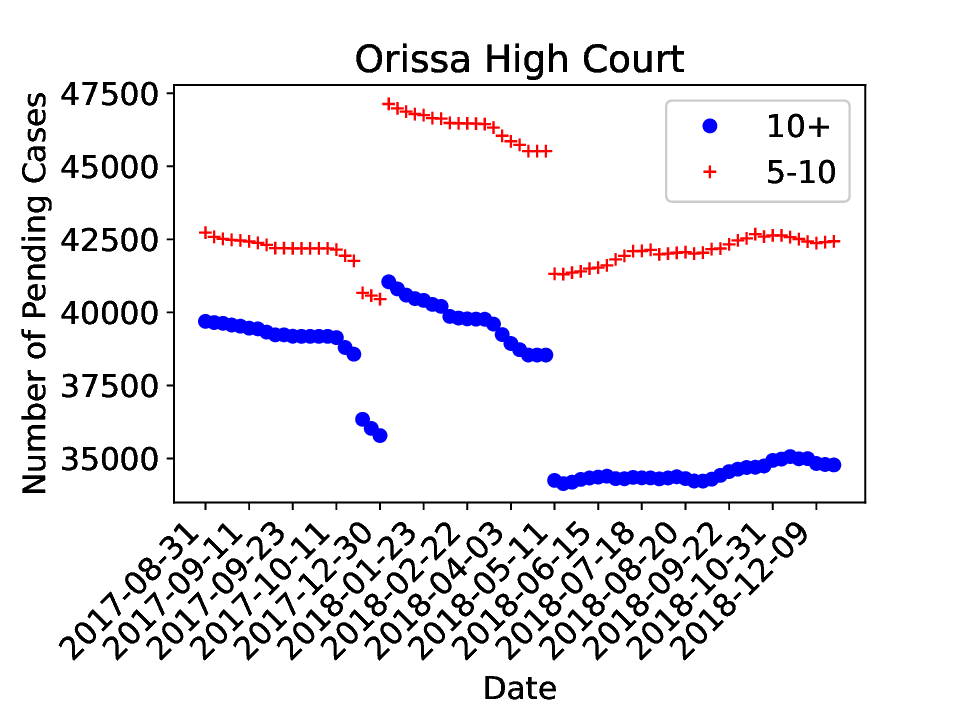}
\includegraphics[width=4.4cm]{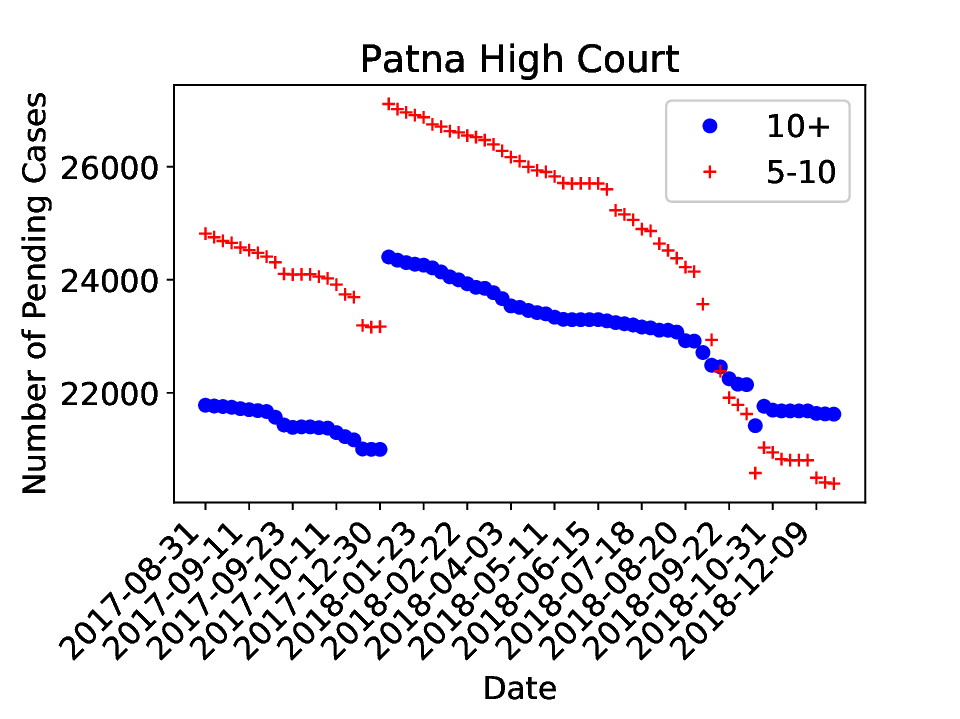}
\includegraphics[width=4.4cm]{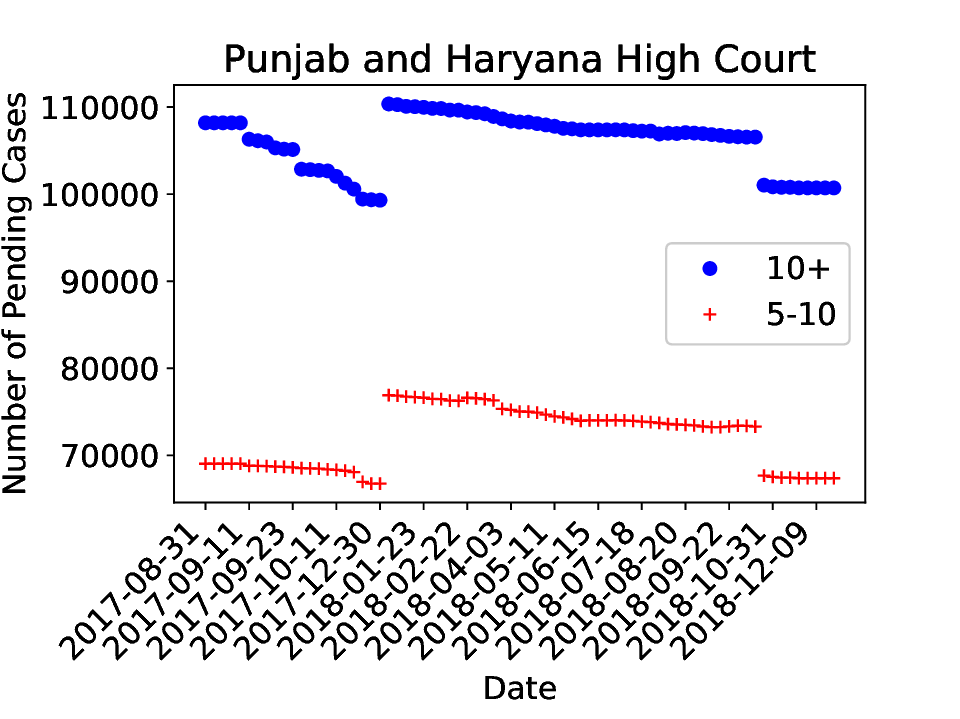}
\includegraphics[width=4.4cm]{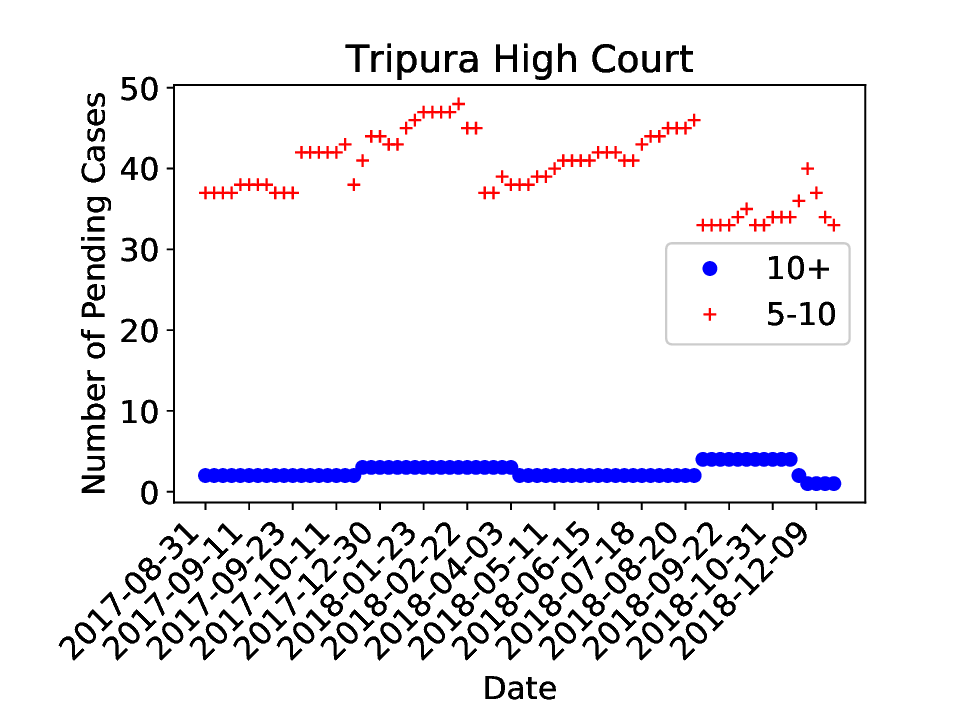}
\includegraphics[width=4.4cm]{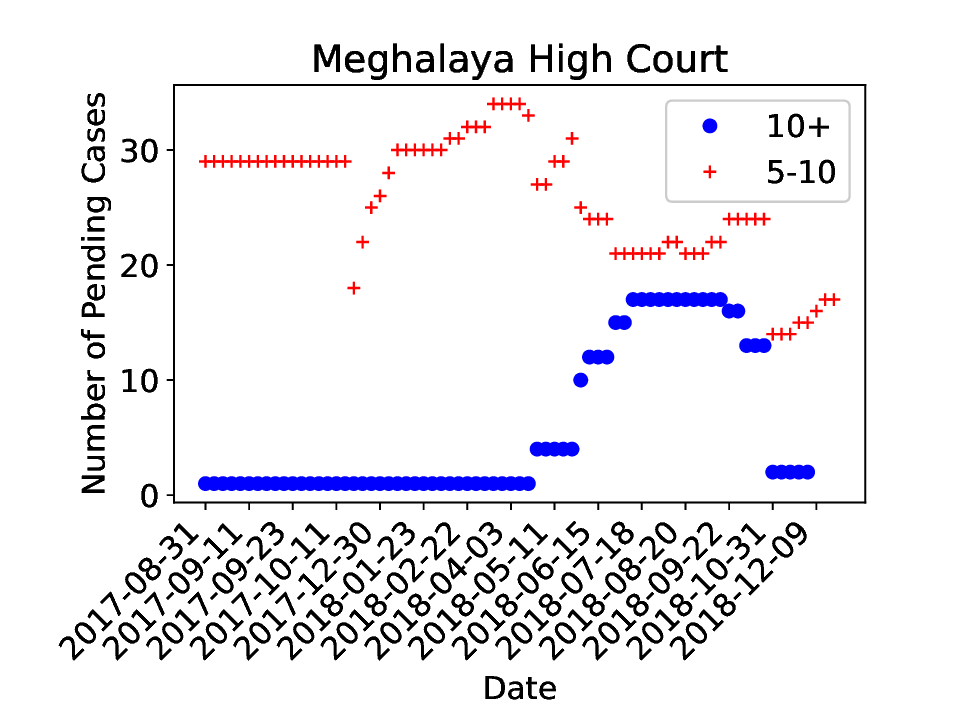}
\includegraphics[width=4.4cm]{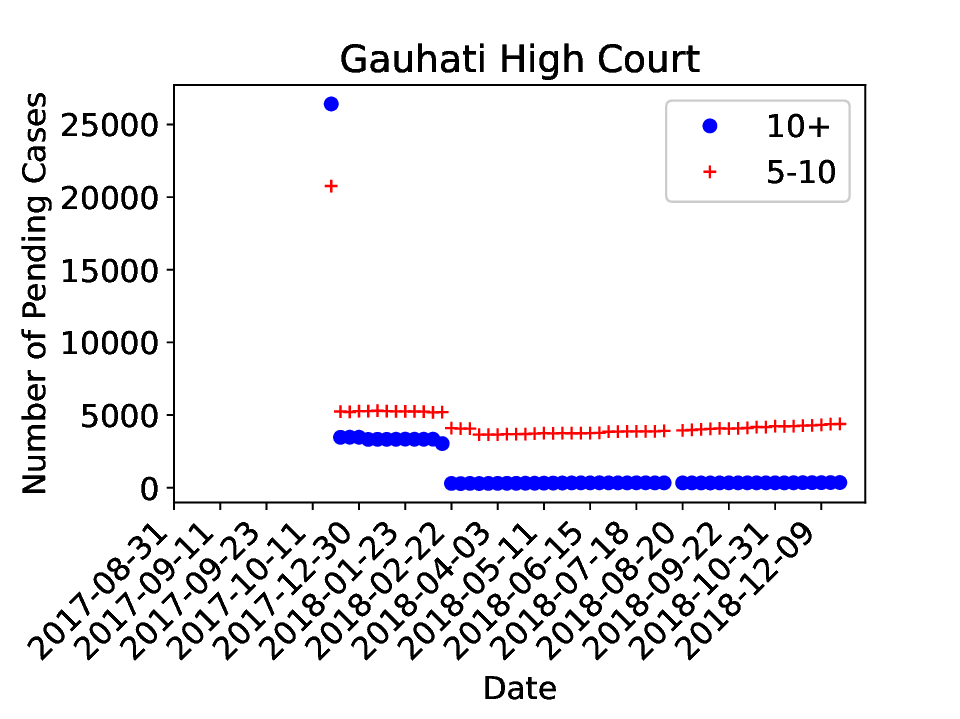}
\includegraphics[width=4.4cm]{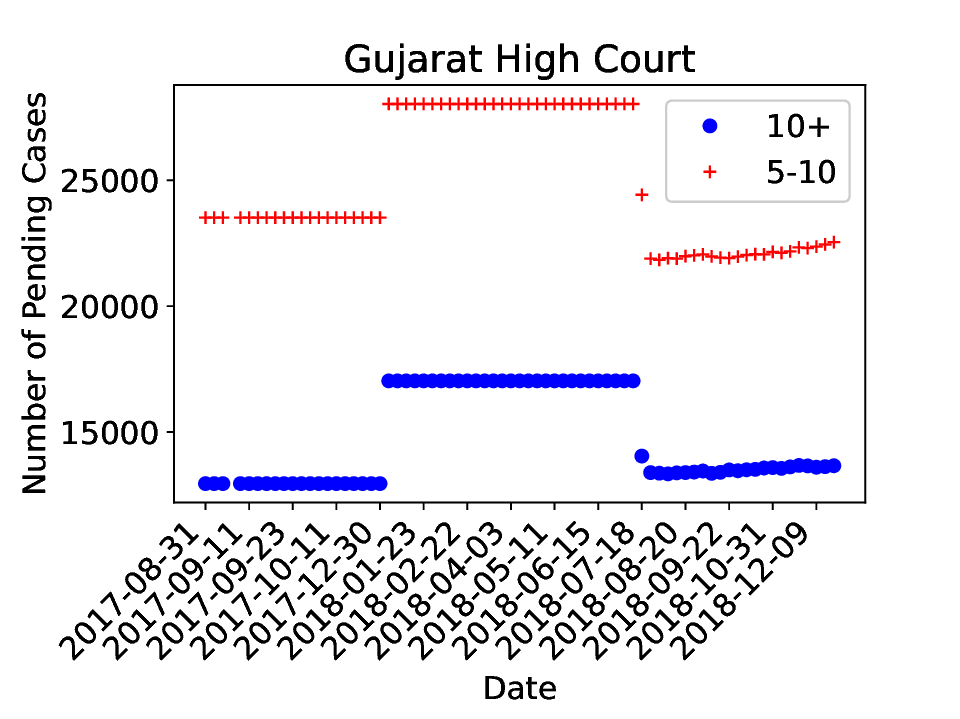}
\includegraphics[width=4.4cm]{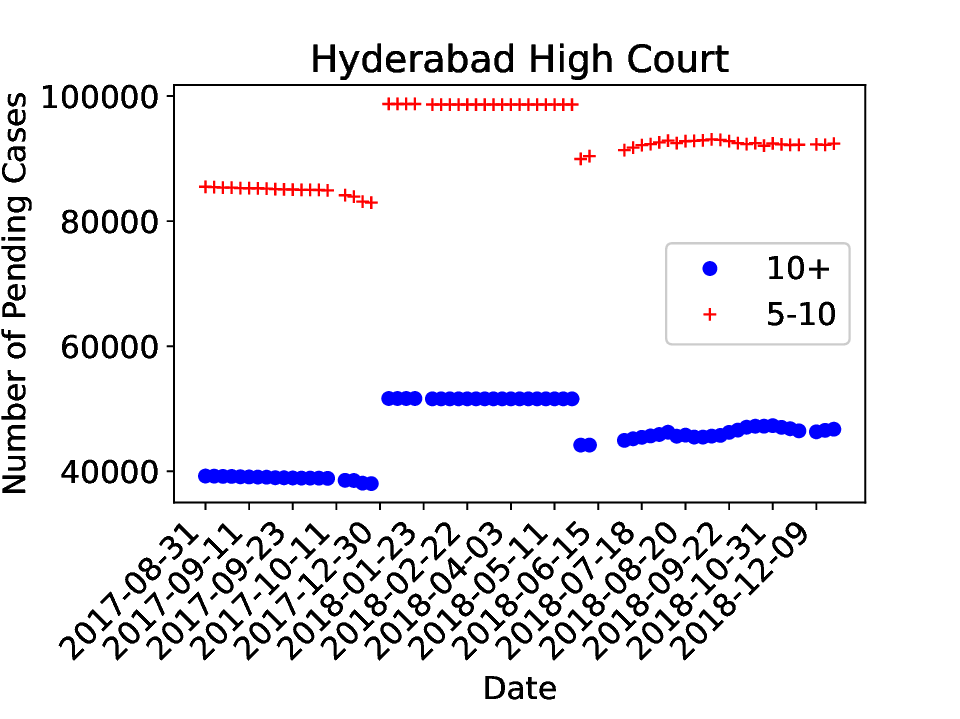}
\includegraphics[width=4.4cm]{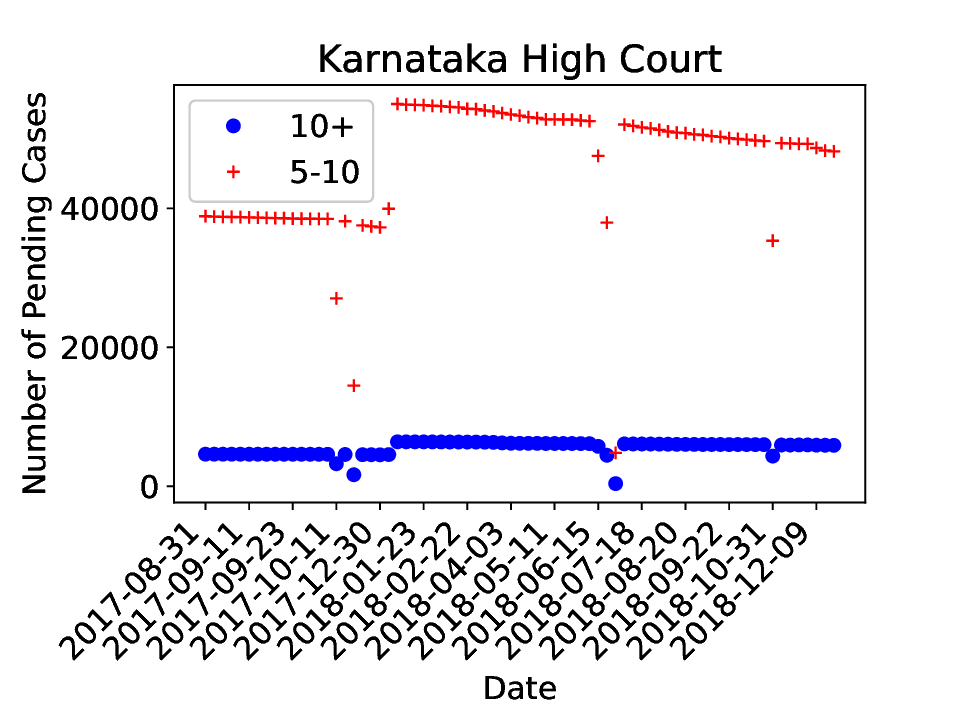}
\includegraphics[width=4.4cm]{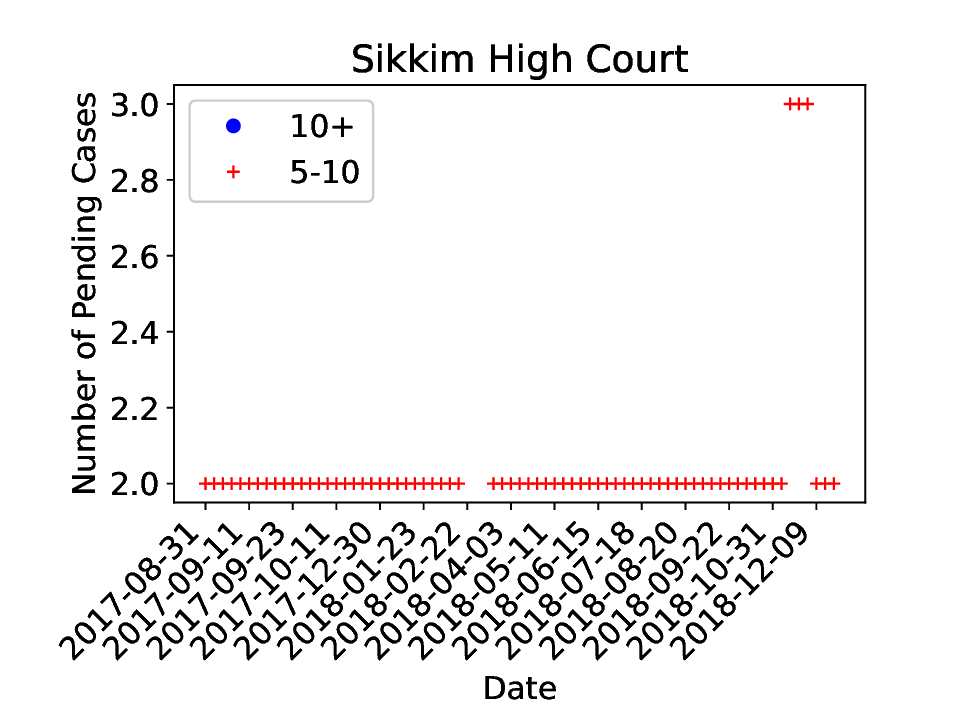}
\caption{Data related to pending cases between 5-10 years and 10+ years. Note that Sikkim High Court has no case pending for more than 10 years and the cases between 5-10 years is also very small. Overall the graphs of these statistics are not easily explainable. There are abrupt changes with most the continuous parts of the graph looking piece-wise constant. The reason for this may be that these statistics don't change that frequently and our data collection spans only 16 months, which is too small to capture changes that occur in a metric that considers half a decade or a decade as the least count.}
\label{fig:yw_hc2}
\end{figure*}

\fref{fig:yw_hc2} shows data related to cases pending for ten years or more and between five to ten years. Apart from Allahabad, Jammu and Kashmir, Jharkhand, Rajasthan, Uttarakhand, Bombay, Calcutta and Madras High Court all the other high courts have decreased the number of cases that are pending for ten years or more. However, the rate of decrease may still be increased so that the long pending cases are solved first. Something fundamental must have happened around the beginning of the year 2018 as most of the high courts have seen drastic increase in the number of pending cases lying for ten years or more. One reason for that may be that the new year has changed the status of many cases from 10- years to 10+ years. For almost all the high courts in this figure, the number of pending cases between five to ten years either decrease or remain same.

\subsection{Cases Filed, Disposed and Listed}
\label{sec:fdl_hc}

\begin{figure*}[h]
\includegraphics[width=4.4cm]{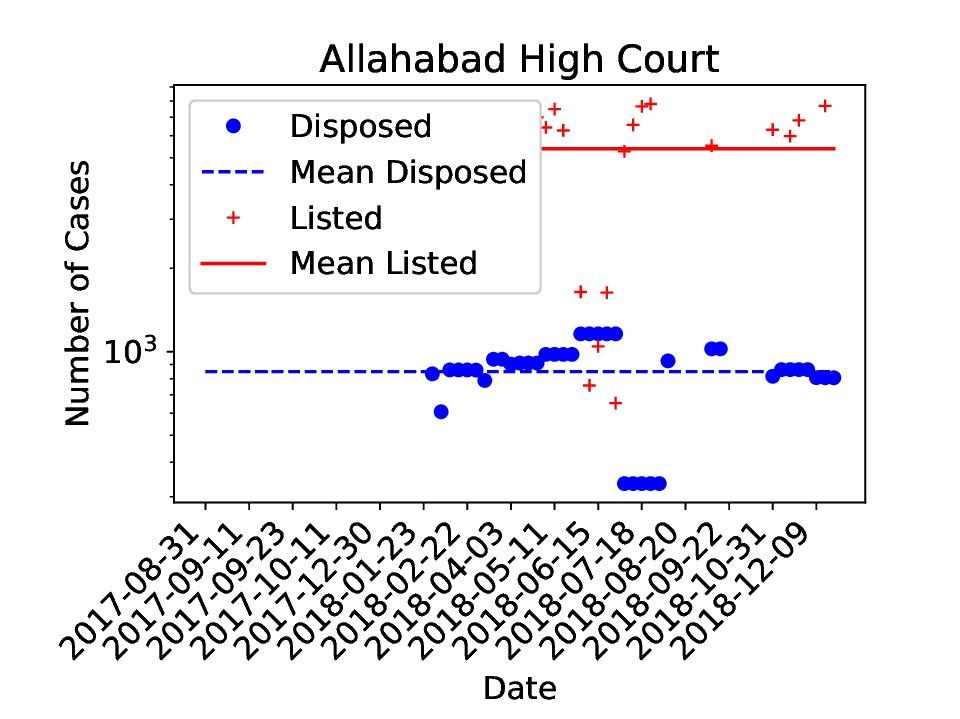}
\includegraphics[width=4.4cm]{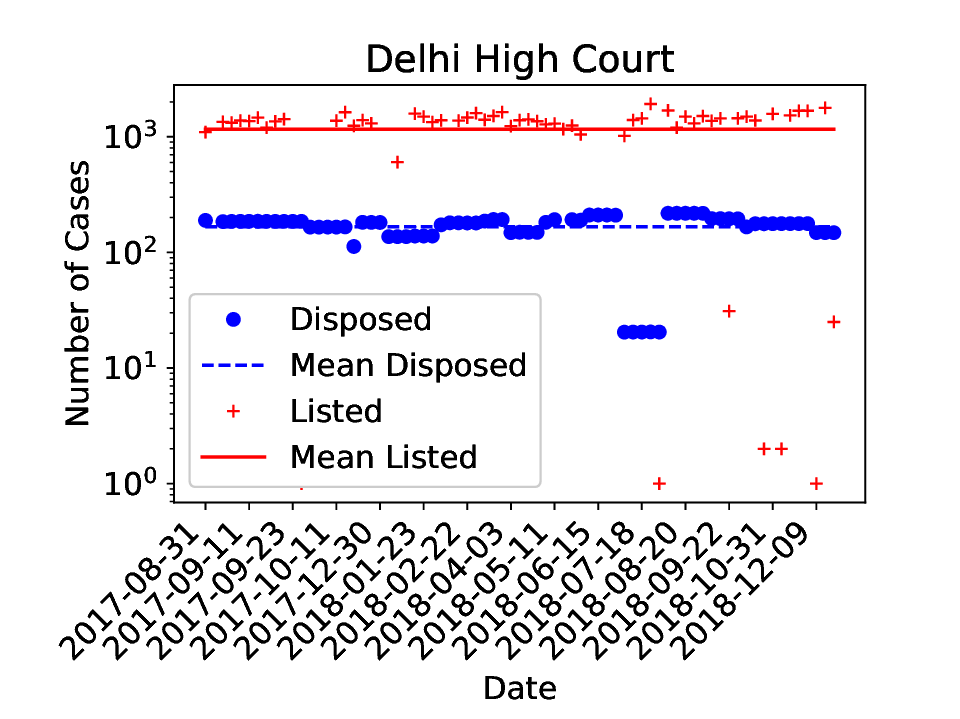}
\includegraphics[width=4.4cm]{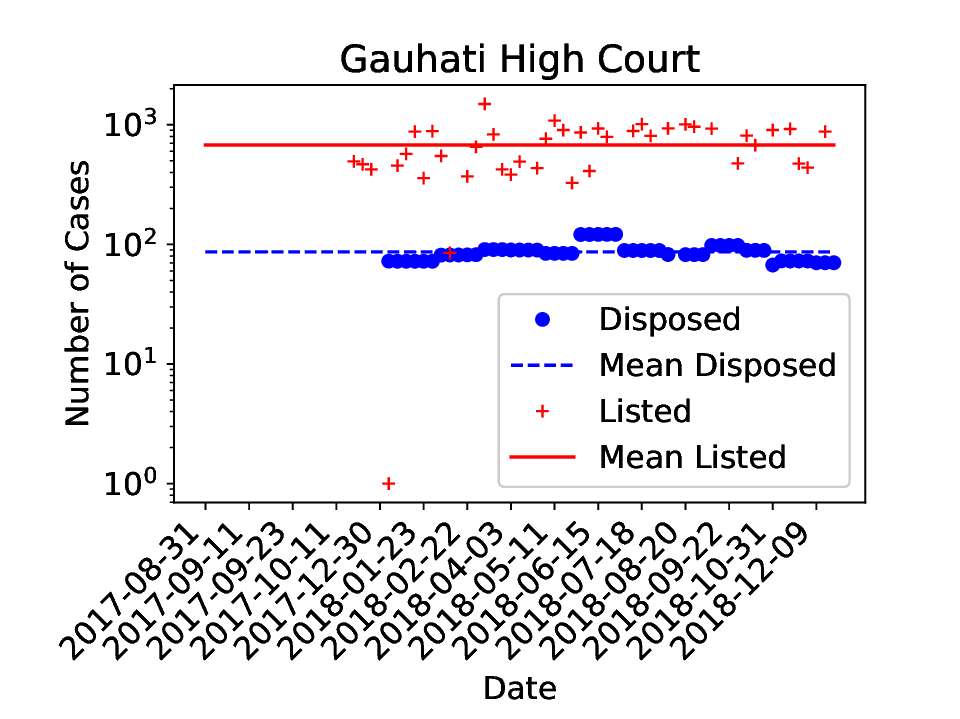}
\includegraphics[width=4.4cm]{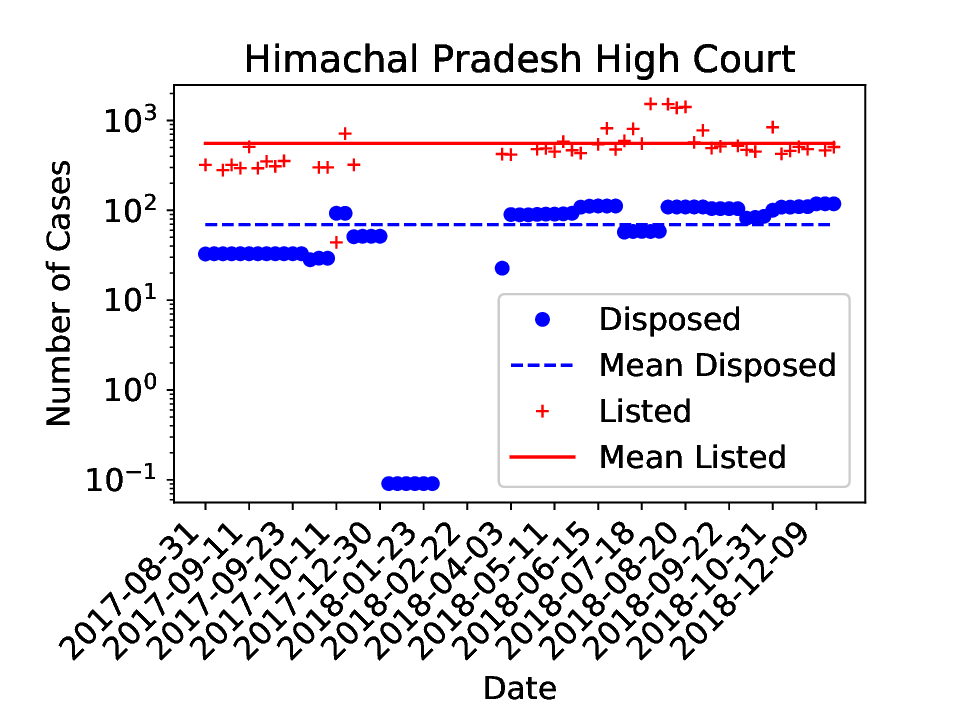}
\includegraphics[width=4.4cm]{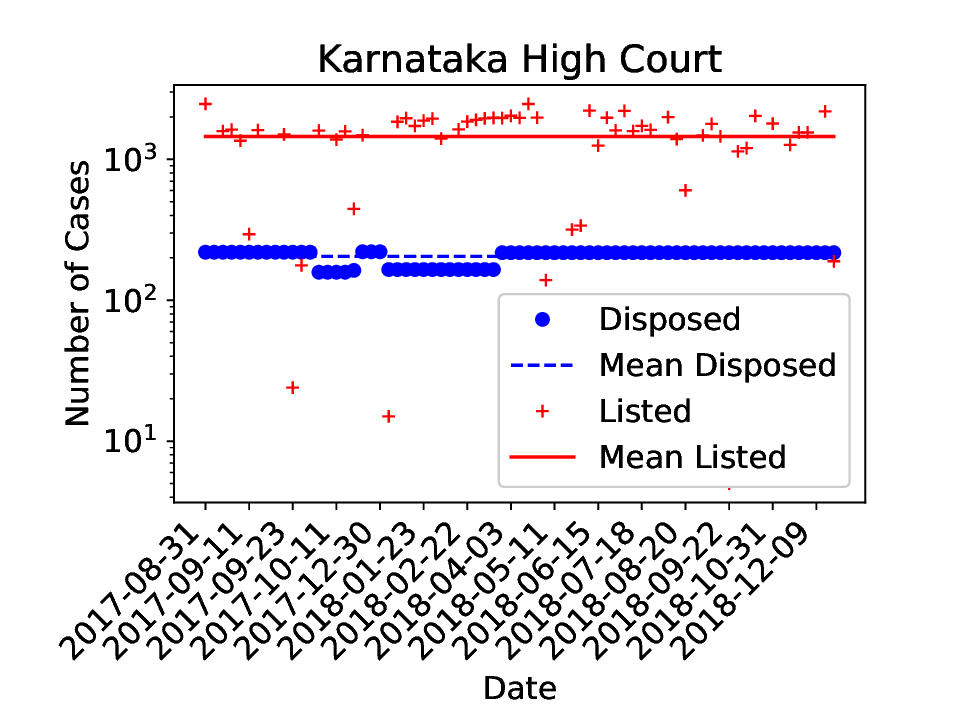}
\includegraphics[width=4.4cm]{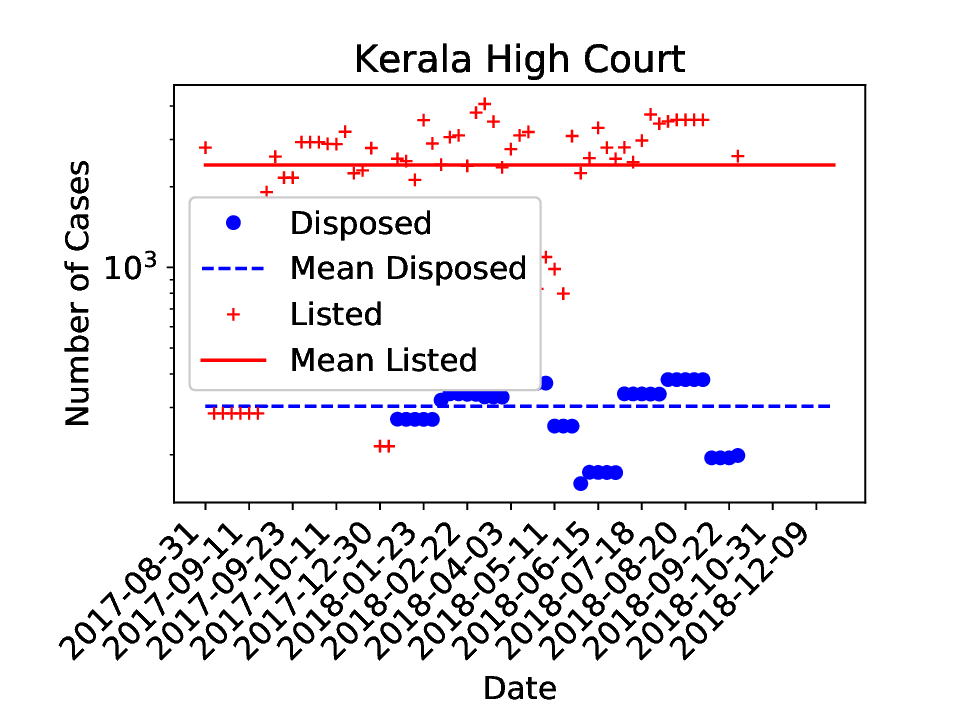}
\includegraphics[width=4.4cm]{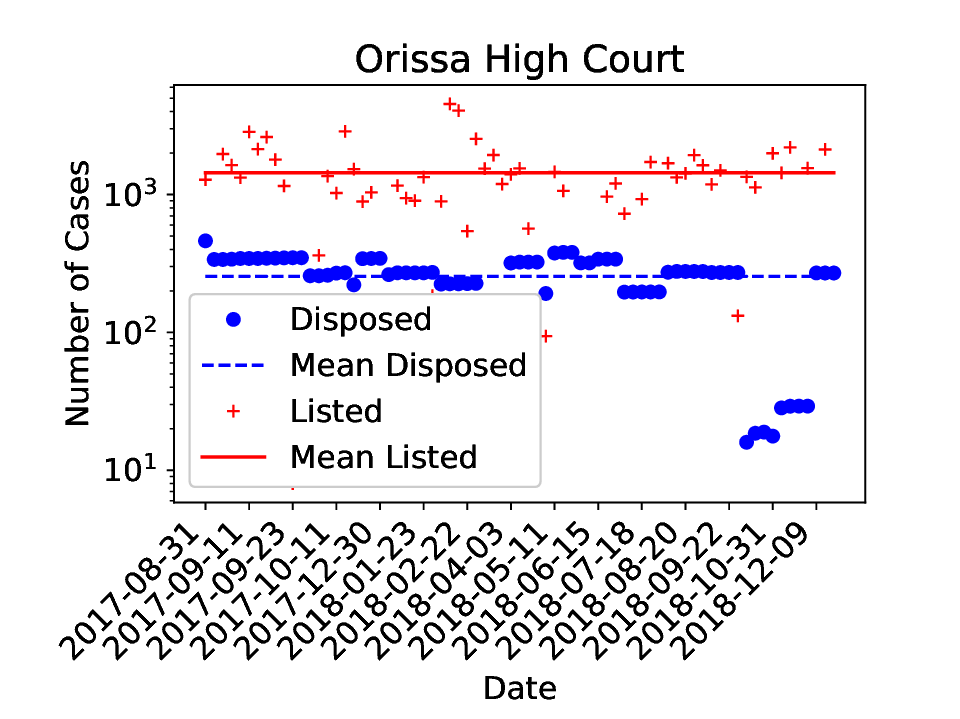}
\includegraphics[width=4.4cm]{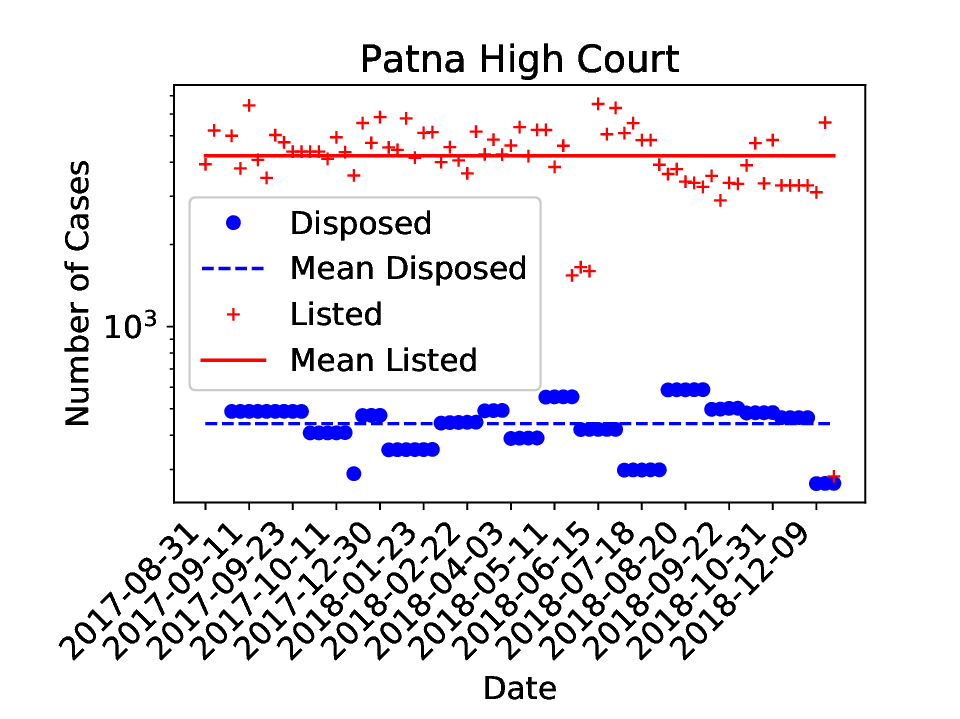}
\includegraphics[width=4.4cm]{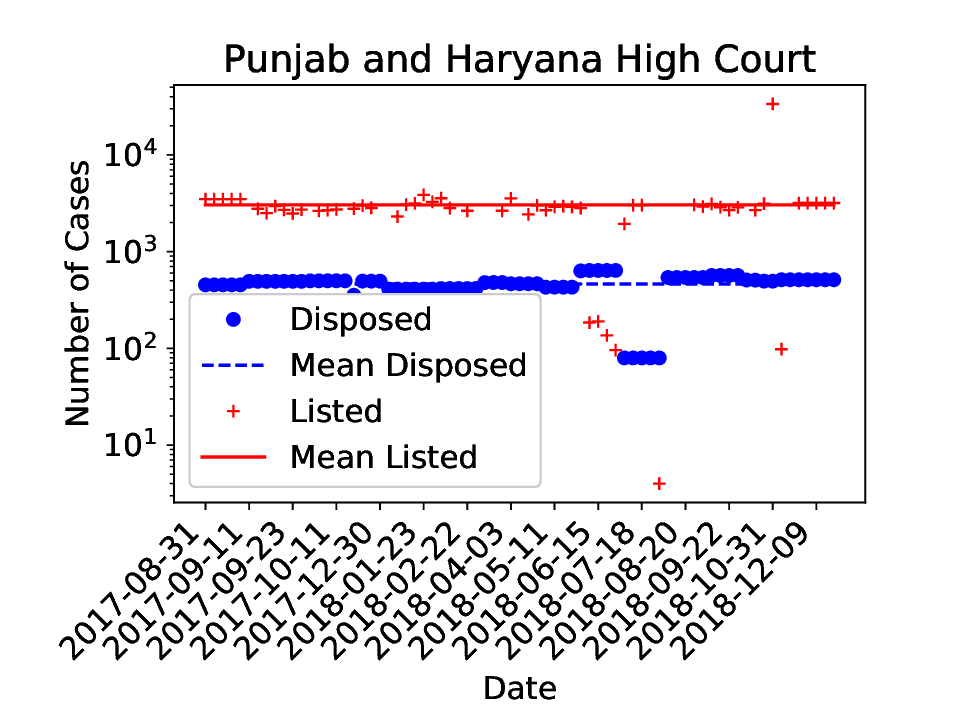}
\includegraphics[width=4.4cm]{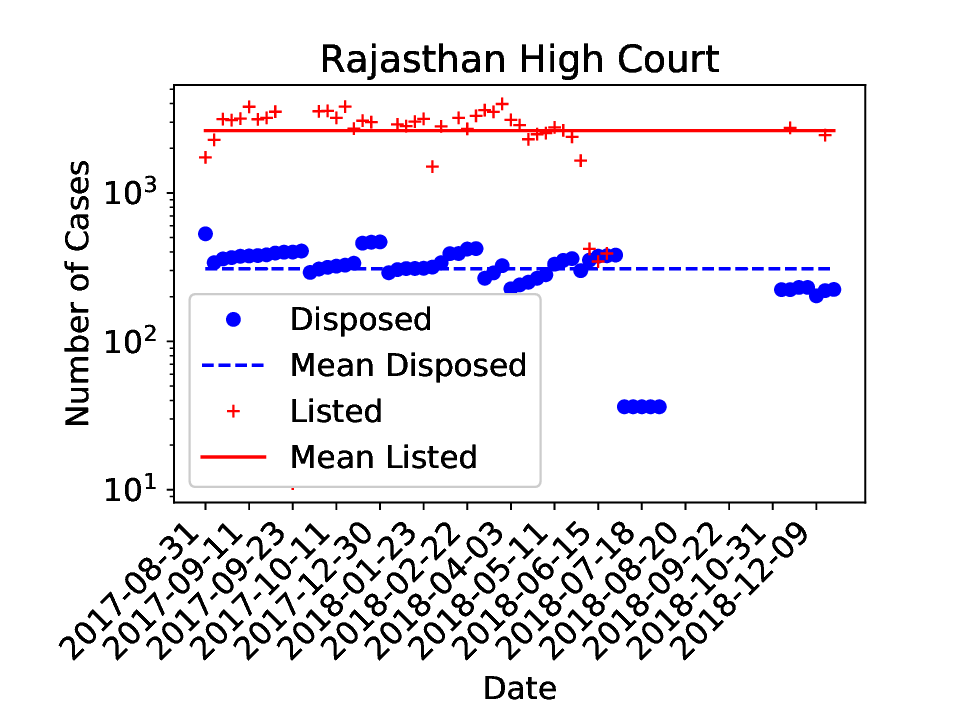}
\includegraphics[width=4.4cm]{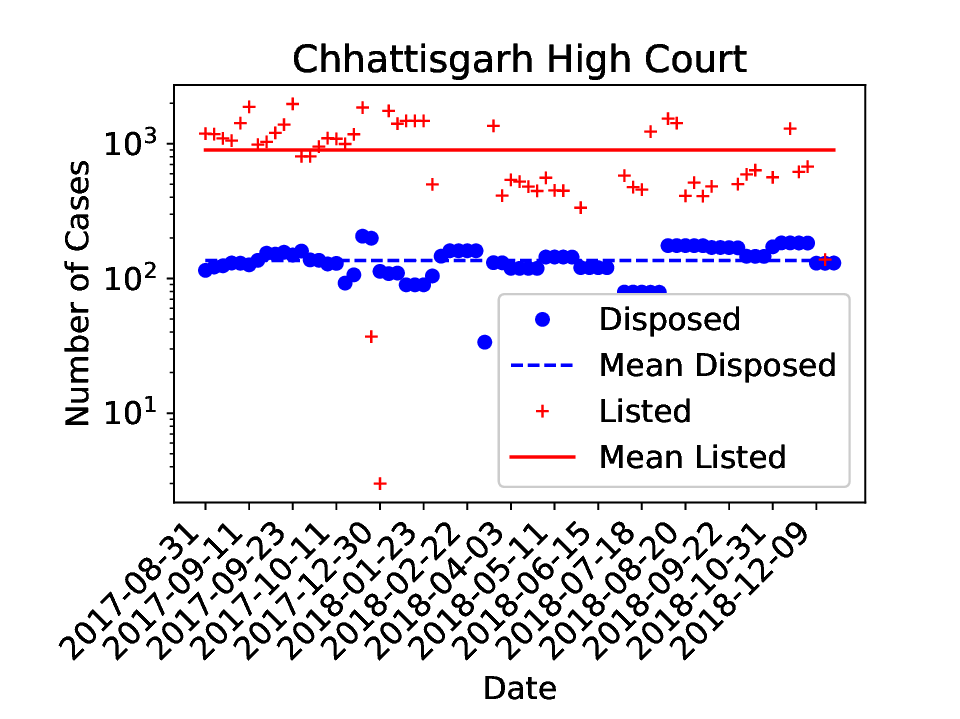}
\includegraphics[width=4.4cm]{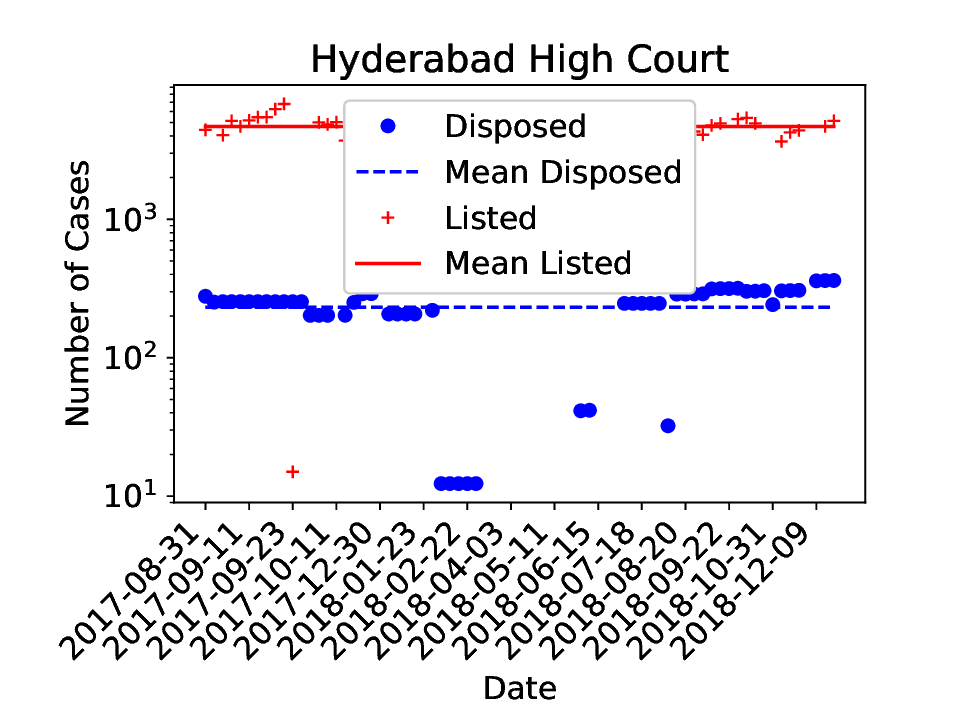}
\includegraphics[width=4.4cm]{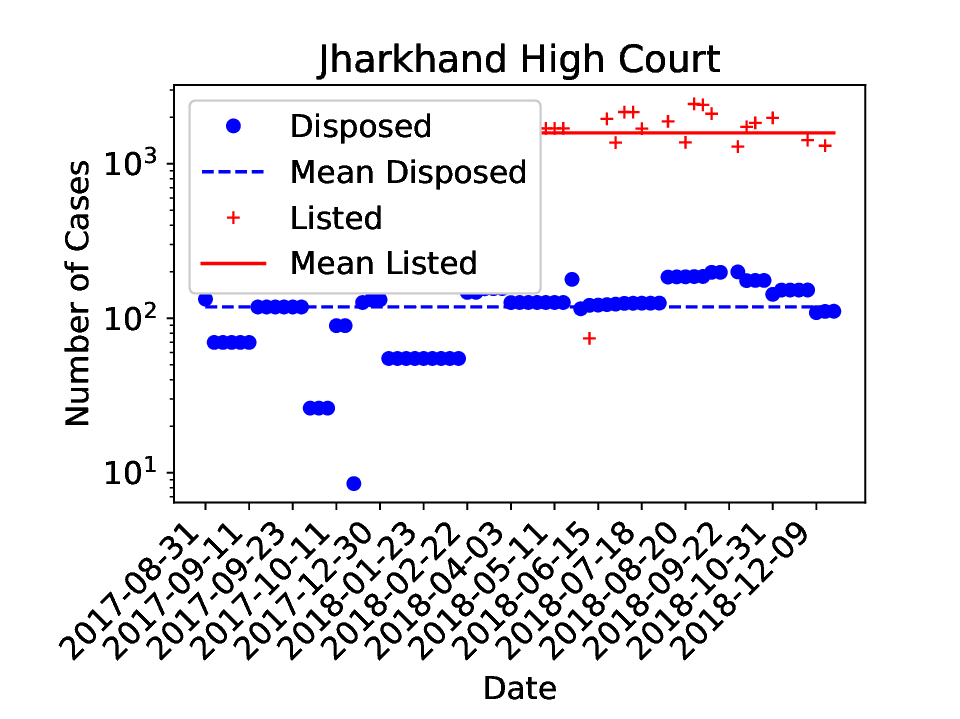}
\includegraphics[width=4.4cm]{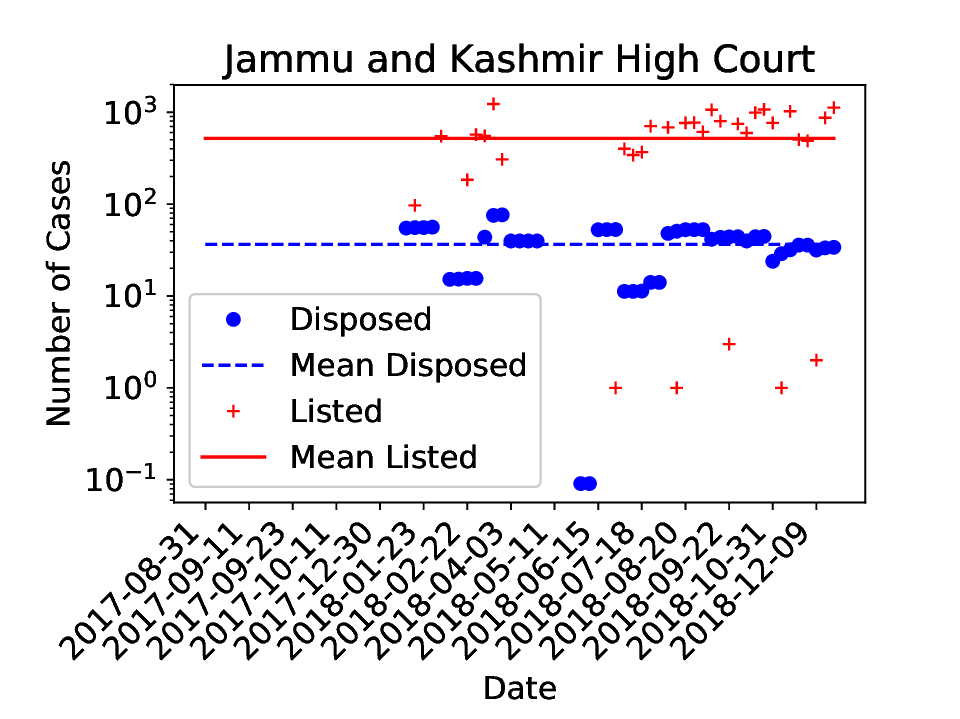}
\includegraphics[width=4.4cm]{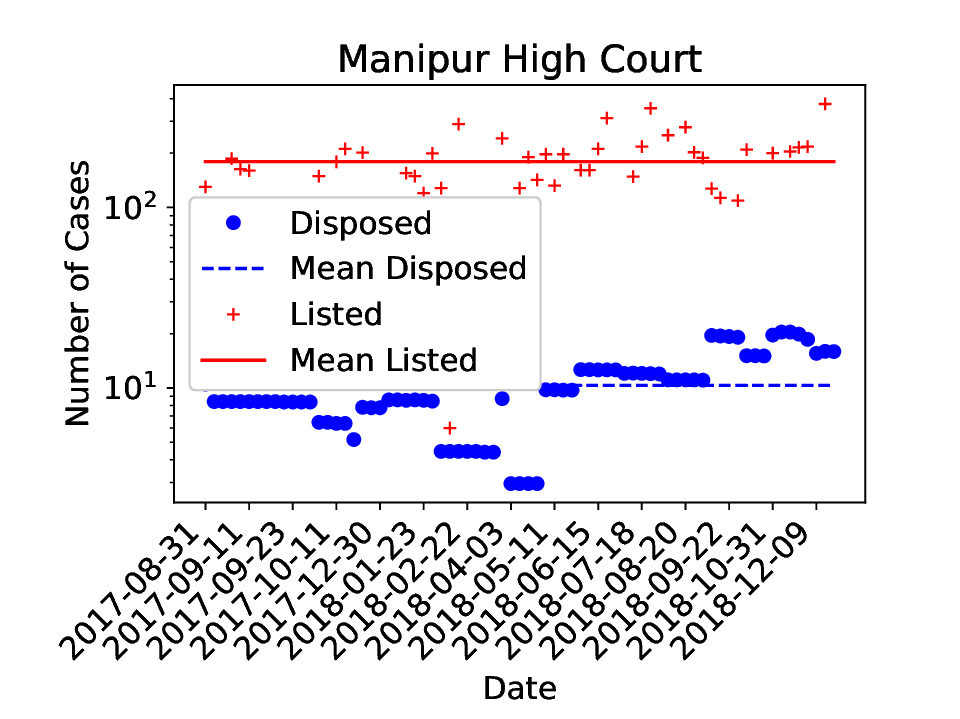}
\includegraphics[width=4.4cm]{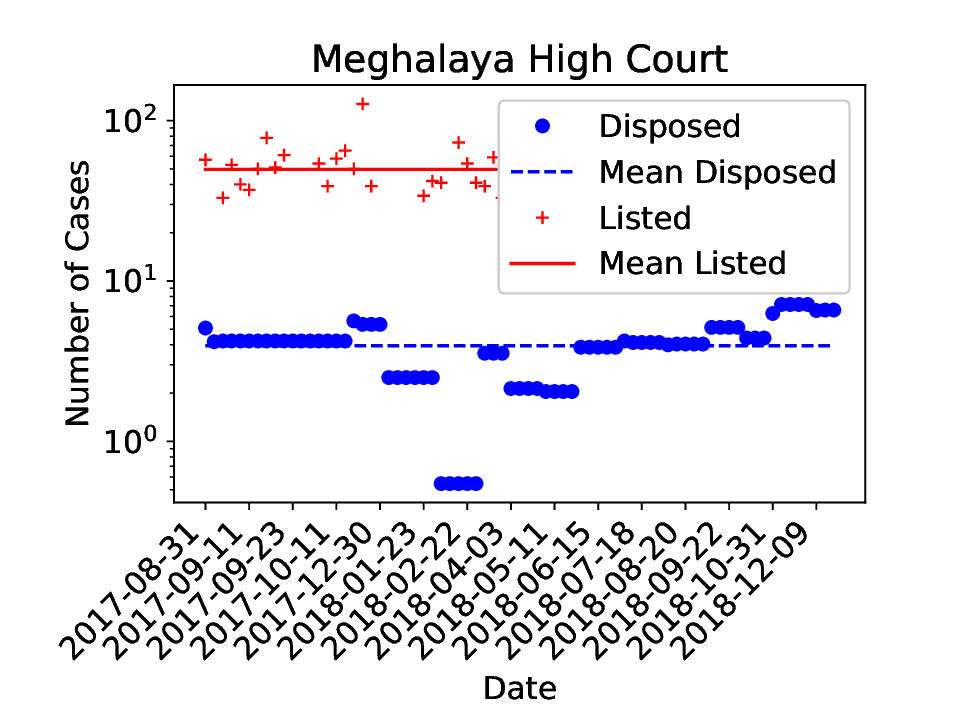}
\includegraphics[width=4.4cm]{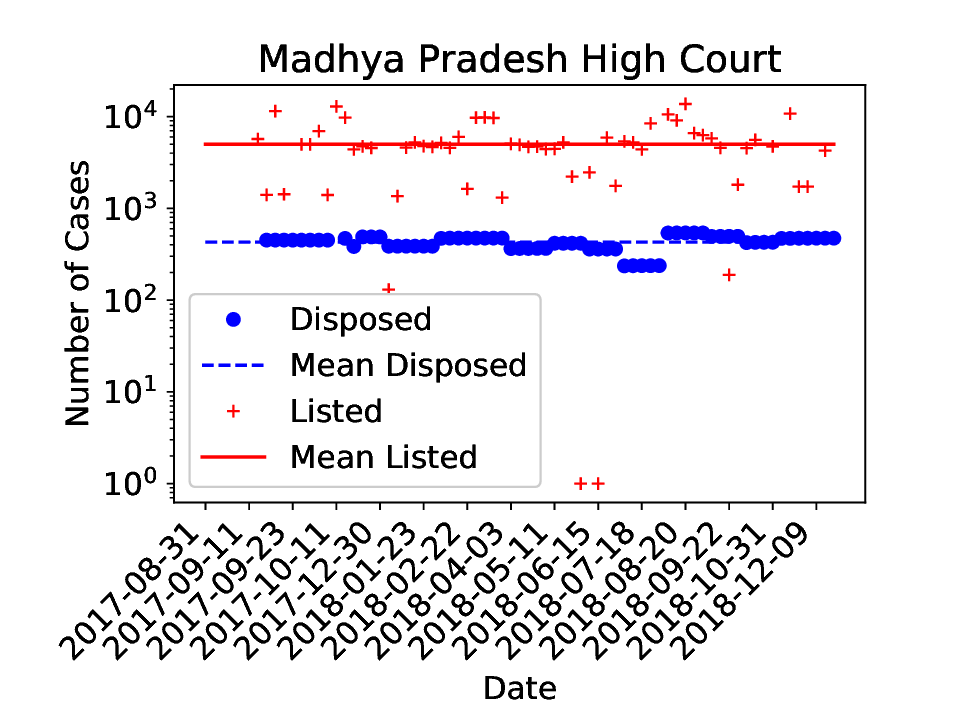}
\includegraphics[width=4.4cm]{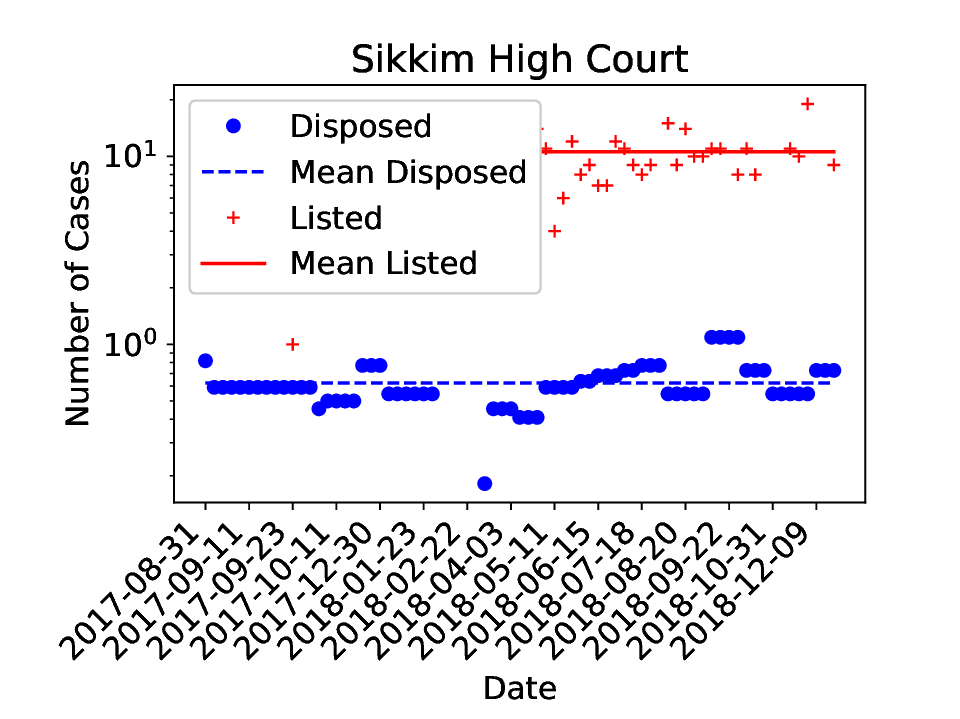}
\includegraphics[width=4.4cm]{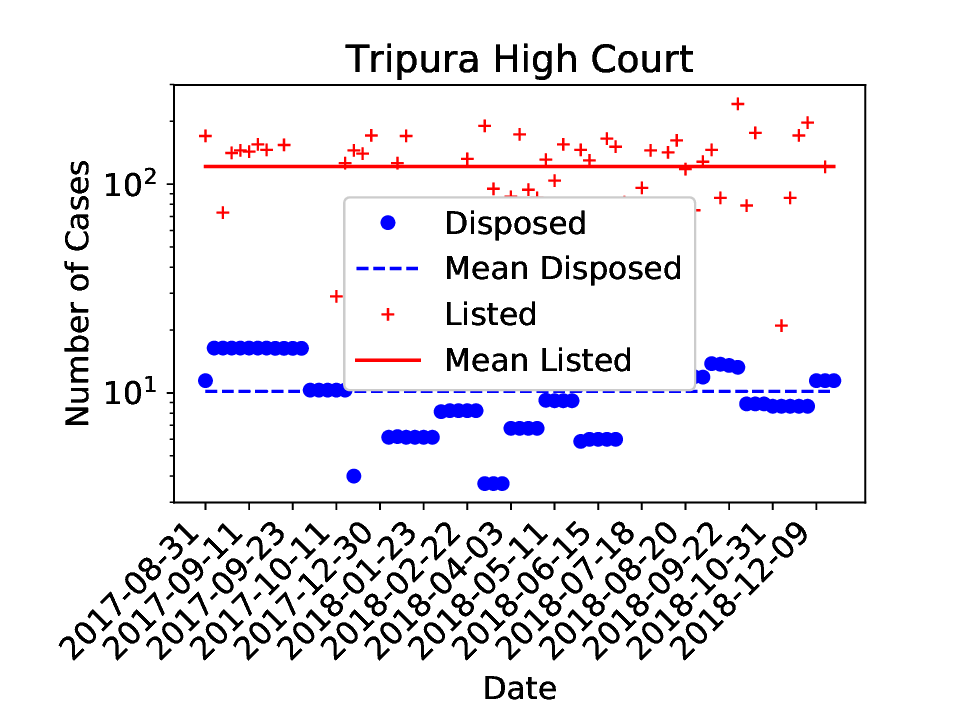}
\includegraphics[width=4.4cm]{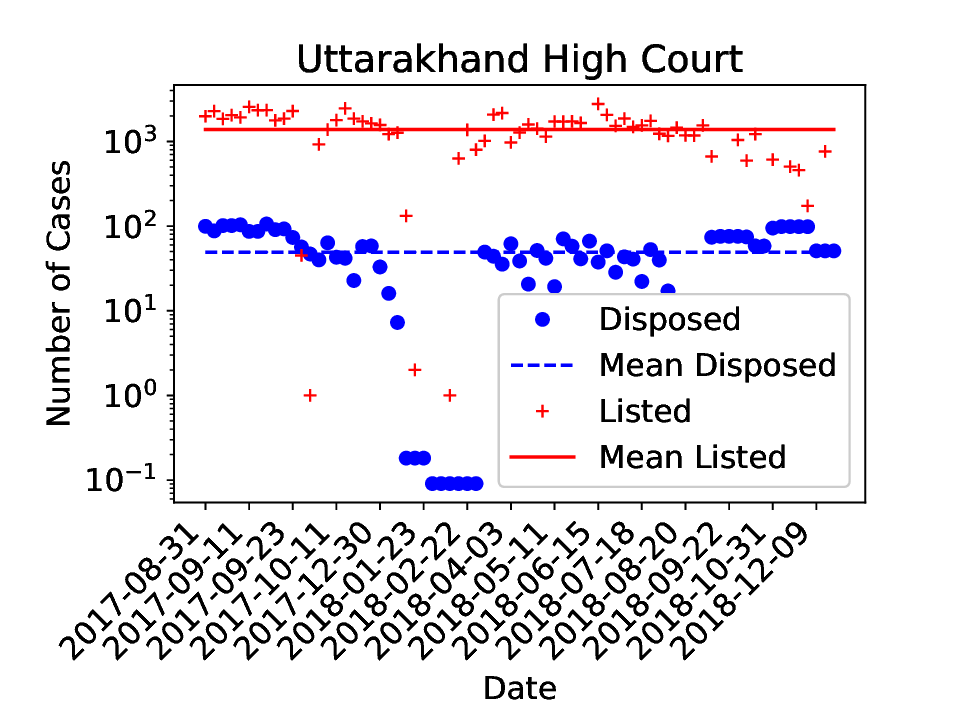}
\includegraphics[width=4.4cm]{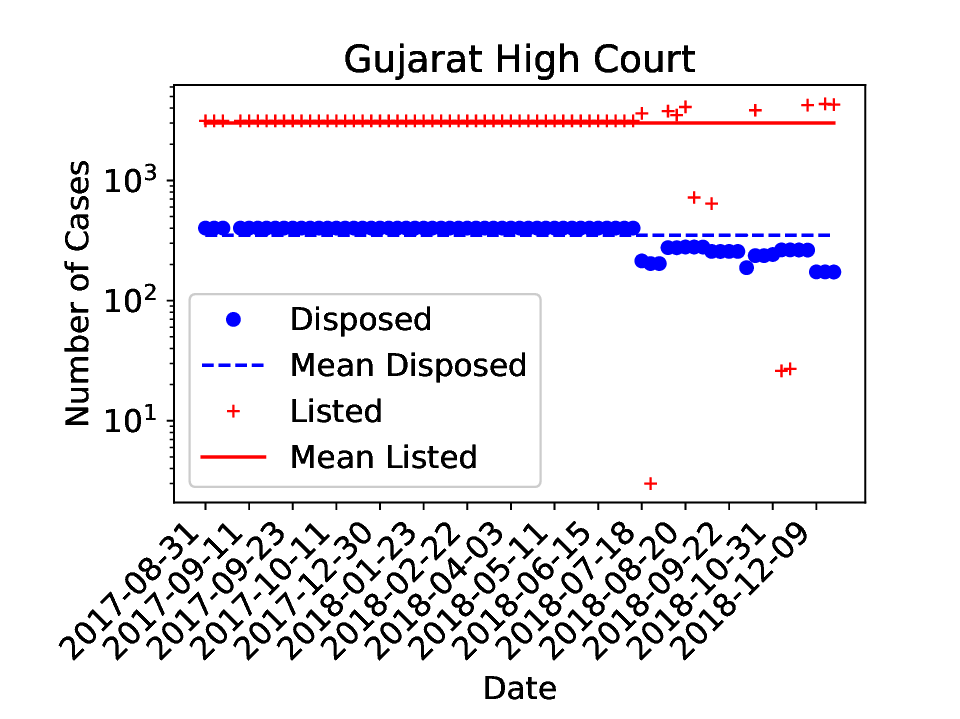}
\includegraphics[width=4.4cm]{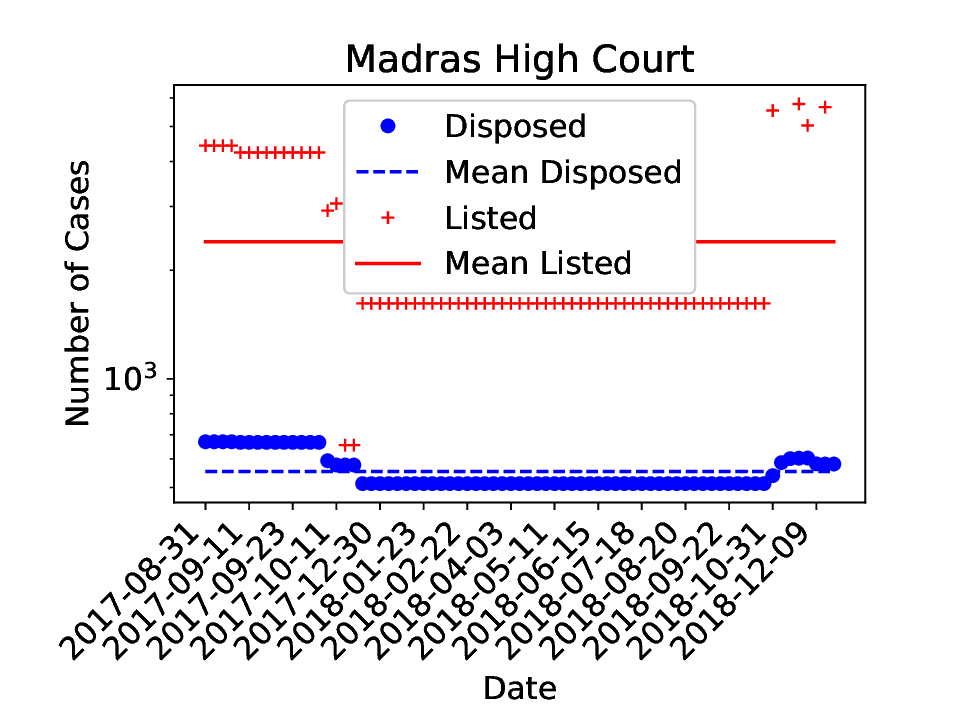}
\includegraphics[width=4.4cm]{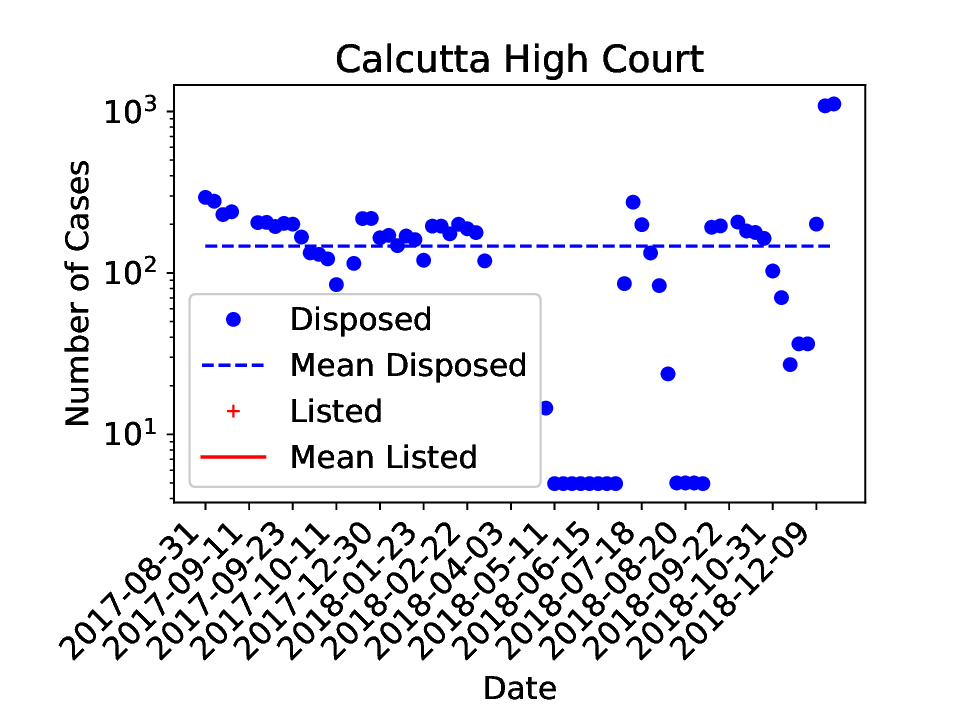}
\includegraphics[width=4.4cm]{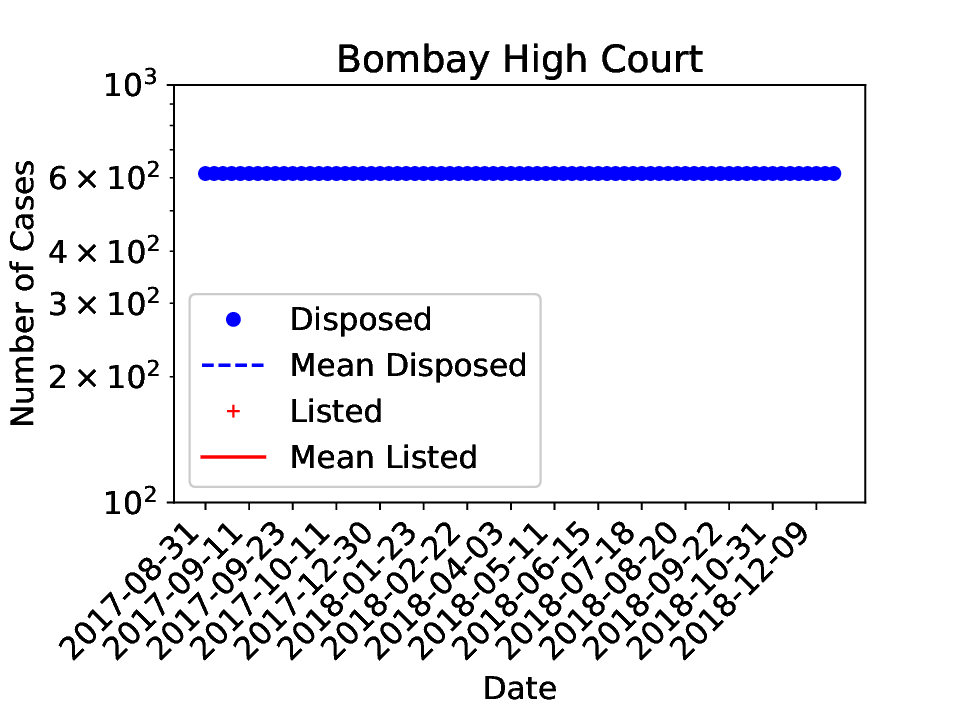}
\caption{The data on disposed cases on HC-NJDG is provided monthly, whereas the listed cases are daily. Hence, in order to bring them on the same platform, we divided the monthly disposal data by 22, assuming the number of working days in each month to be 22 only. Hence, division by 22 has rendered fractional values for some high courts. Daily listed cases are almost an order of magnitude more than the disposed (daily average) cases. So lesser number of cases can be scheduled in a particular day giving more time to spend on one case. This is one metric which looks like similar for all the high courts meaning that for all the high courts, the number of listed cases per day is much more than the number of disposed cases per day. If this gap is narrowed, it can help all the stakeholders. We can infer from this figure that scientific and more efficient ways of preparing causelists should be deployed to make the system work more efficiently.}
\label{fig:fdl_hc}
\end{figure*}

\fref{fig:fdl_hc} presents the data of listed and disposed cases in a day. The ratio of cases listed to disposed is very high. This means that more efficient ways of preparing causelists should be found. The goal should be to minimize the gap between the average number of cases listed on a day and the average number of cases disposed in a day. If statistics on the number of cases heard in a day are also reported by NJDG then this can give rise to a very effective metric to access the efficiency of judicial process in India.

\begin{figure*}[h]
\includegraphics[width=4.4cm]{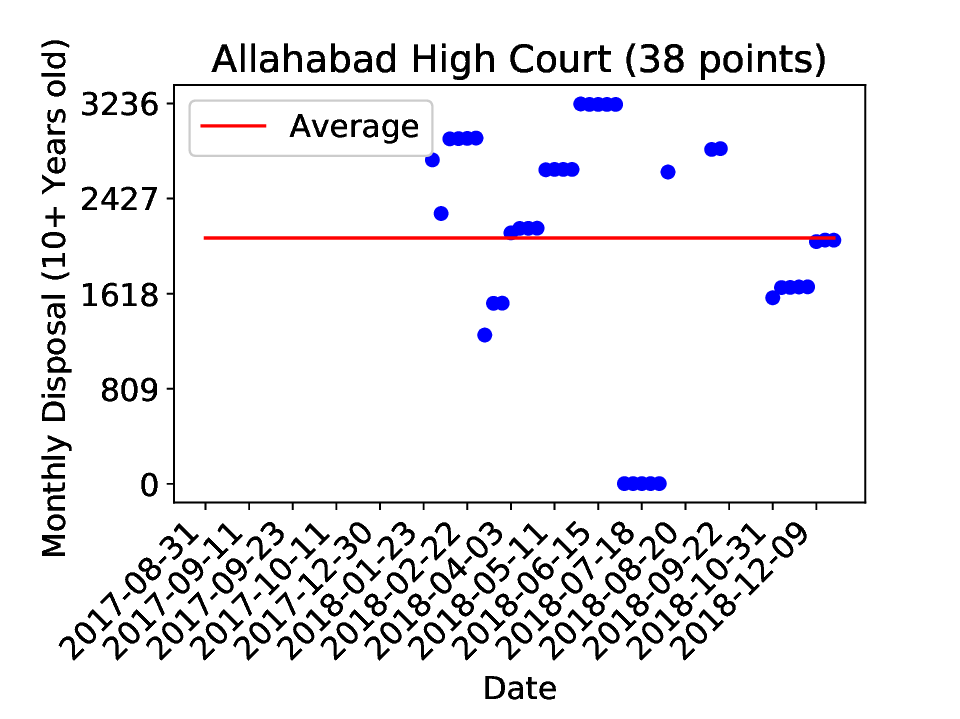}
\includegraphics[width=4.4cm]{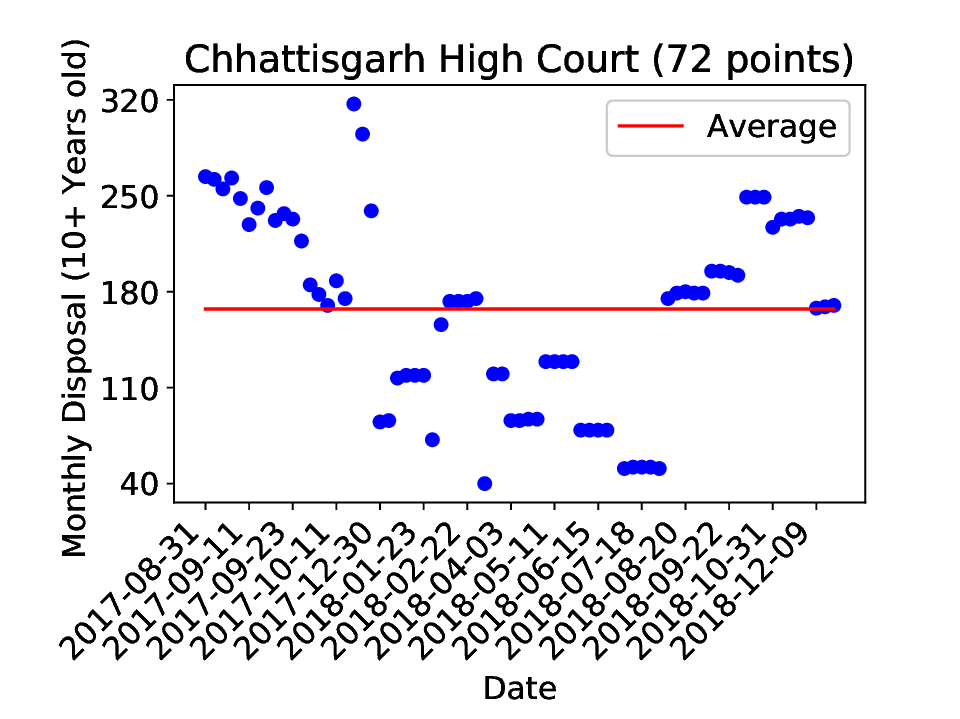}
\includegraphics[width=4.4cm]{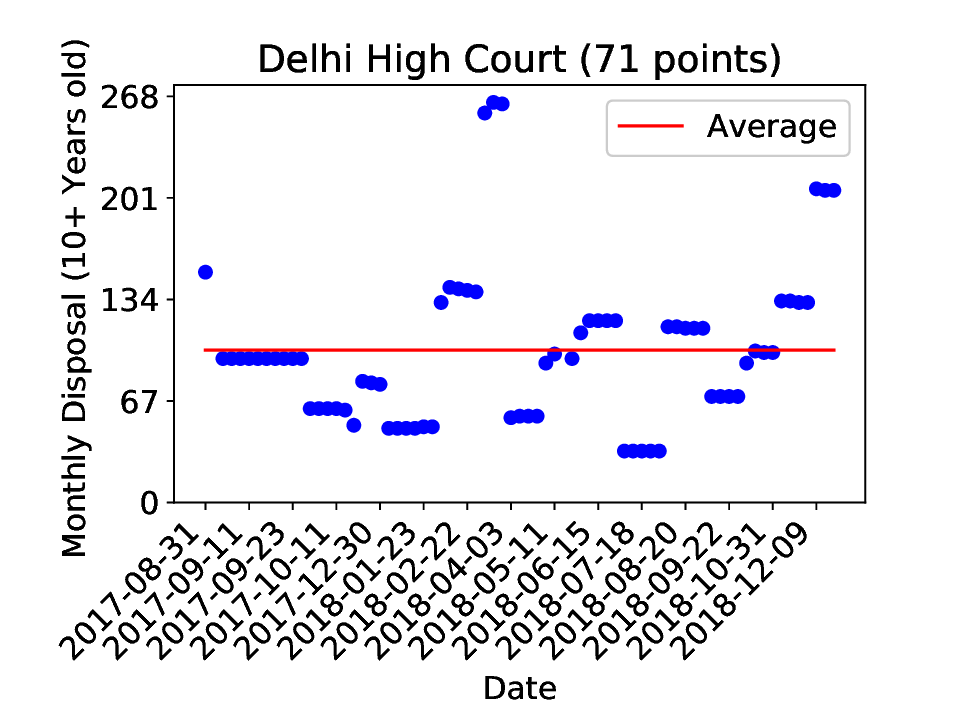}
\includegraphics[width=4.4cm]{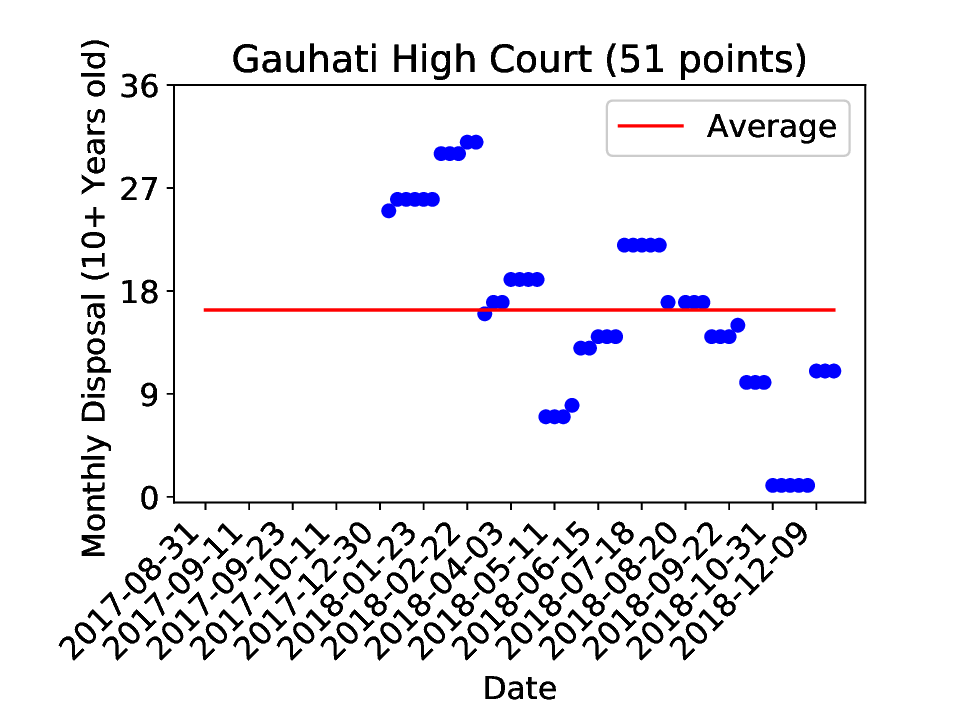}
\includegraphics[width=4.4cm]{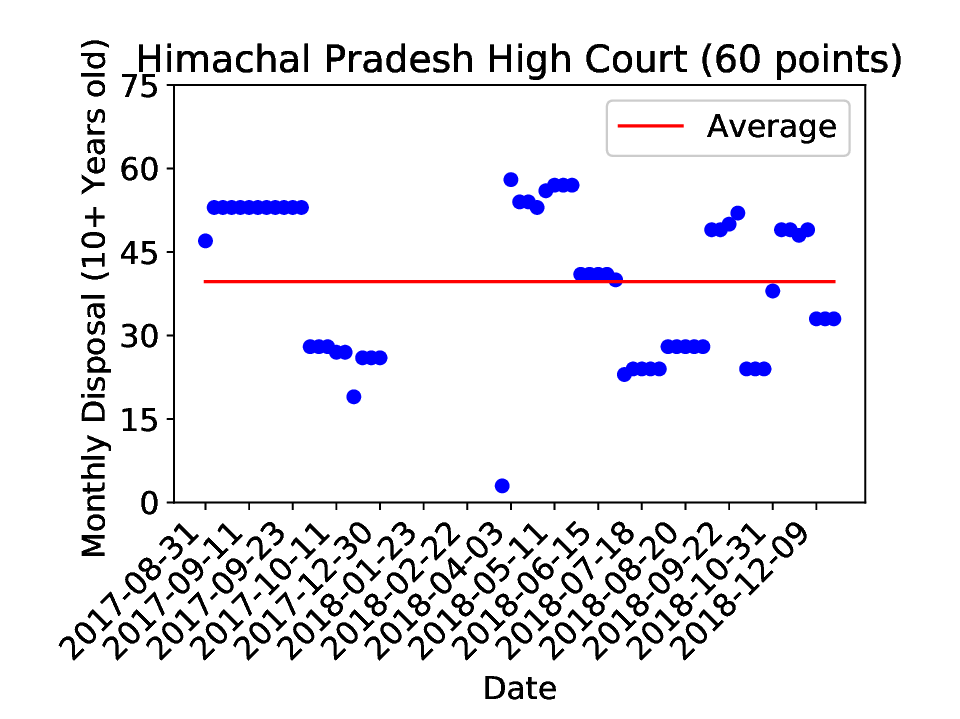}
\includegraphics[width=4.4cm]{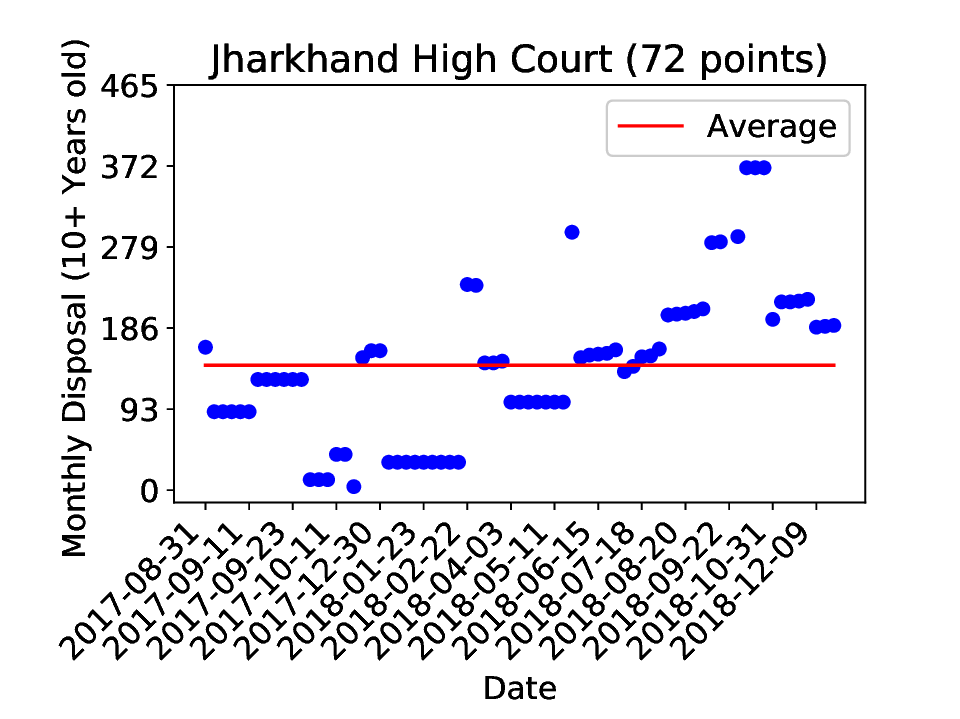}
\includegraphics[width=4.4cm]{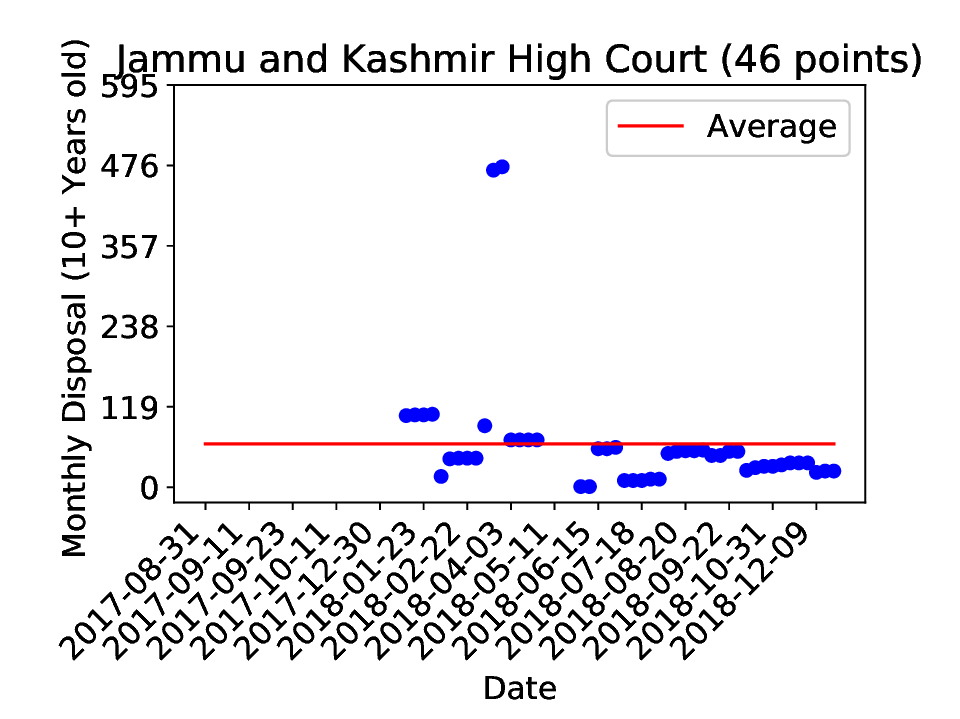}
\includegraphics[width=4.4cm]{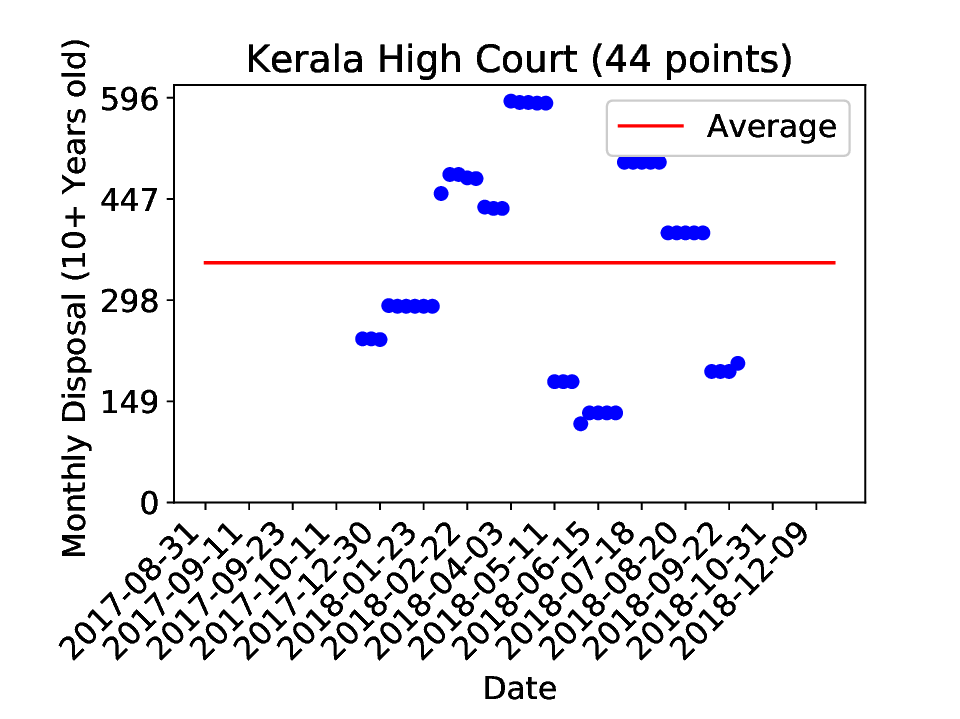}
\includegraphics[width=4.4cm]{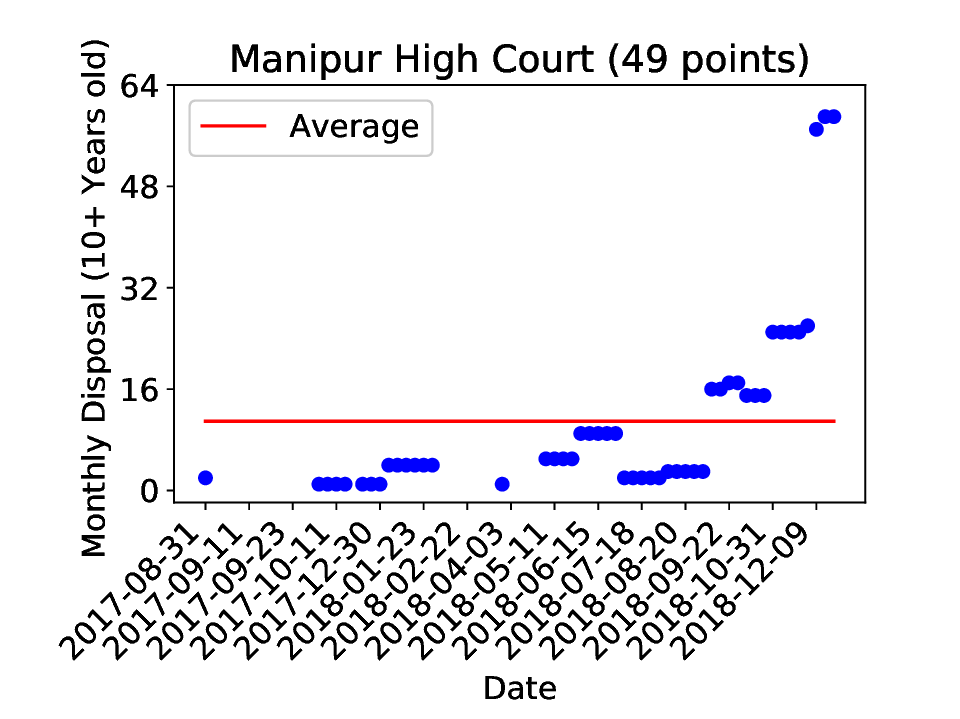}
\includegraphics[width=4.4cm]{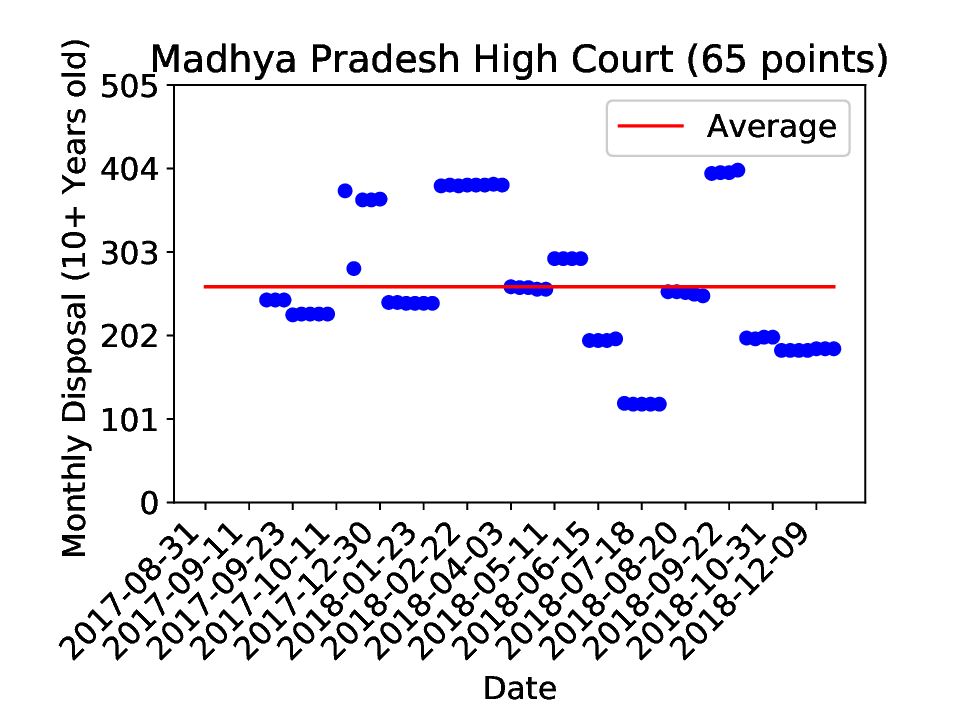}
\includegraphics[width=4.4cm]{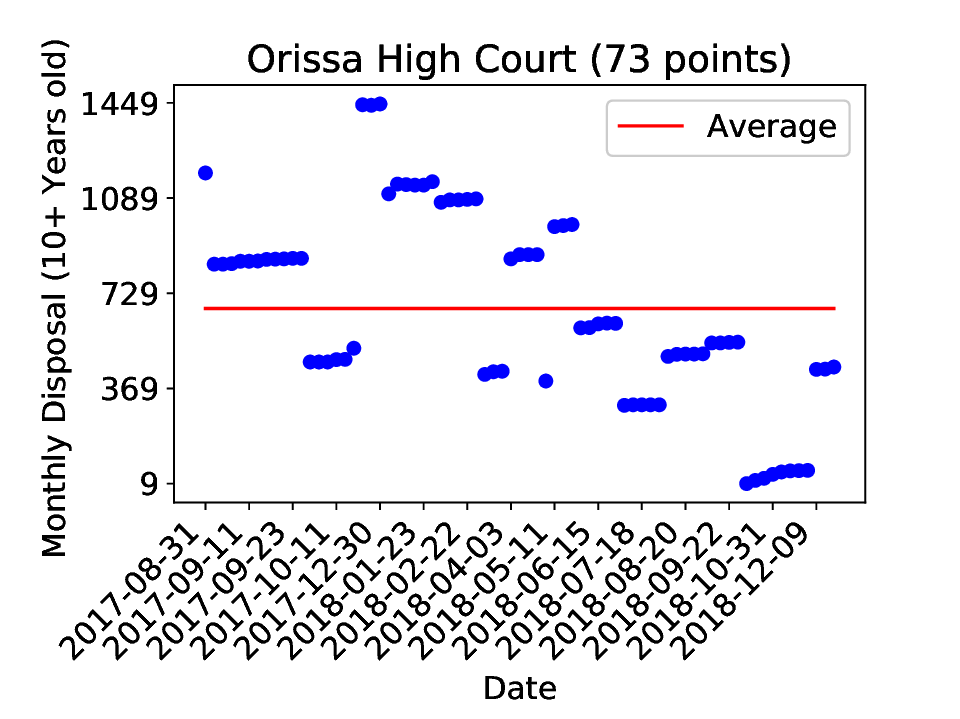}
\includegraphics[width=4.4cm]{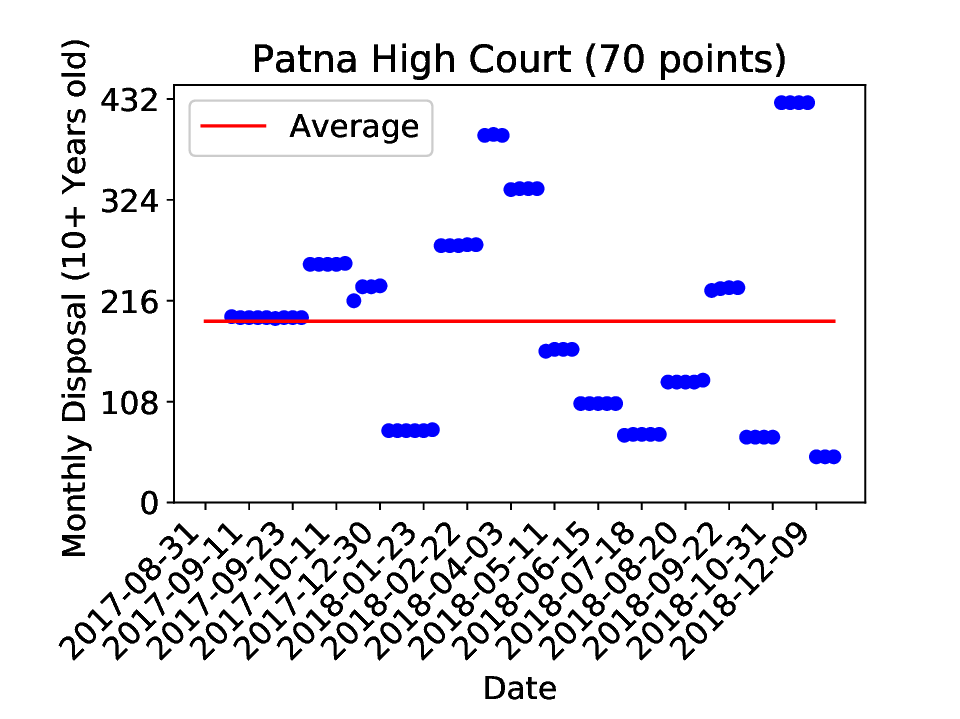}
\includegraphics[width=4.4cm]{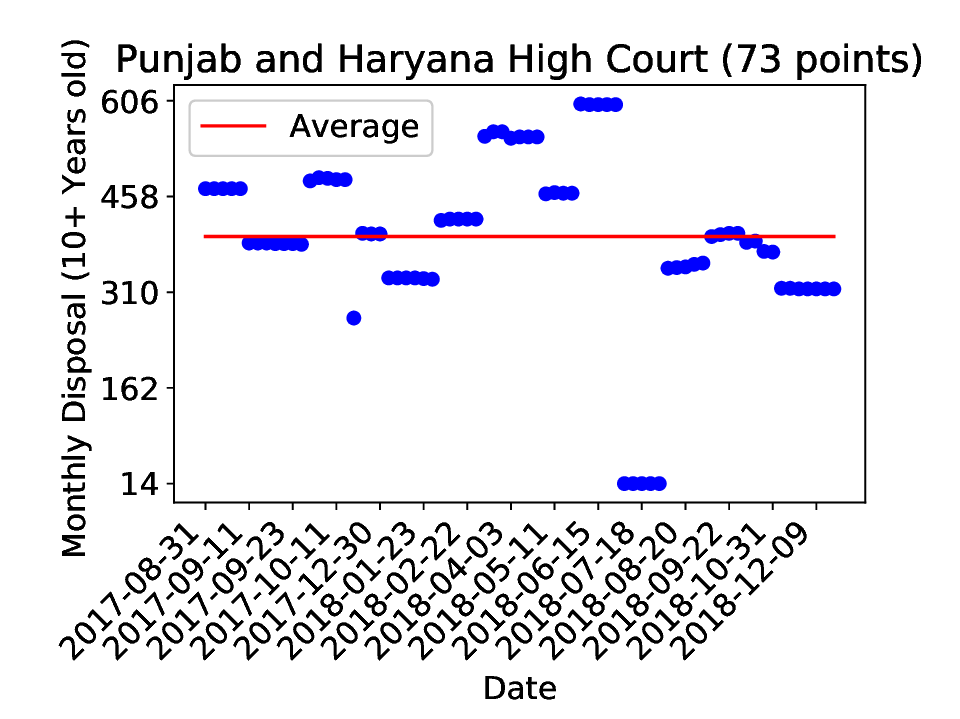}
\includegraphics[width=4.4cm]{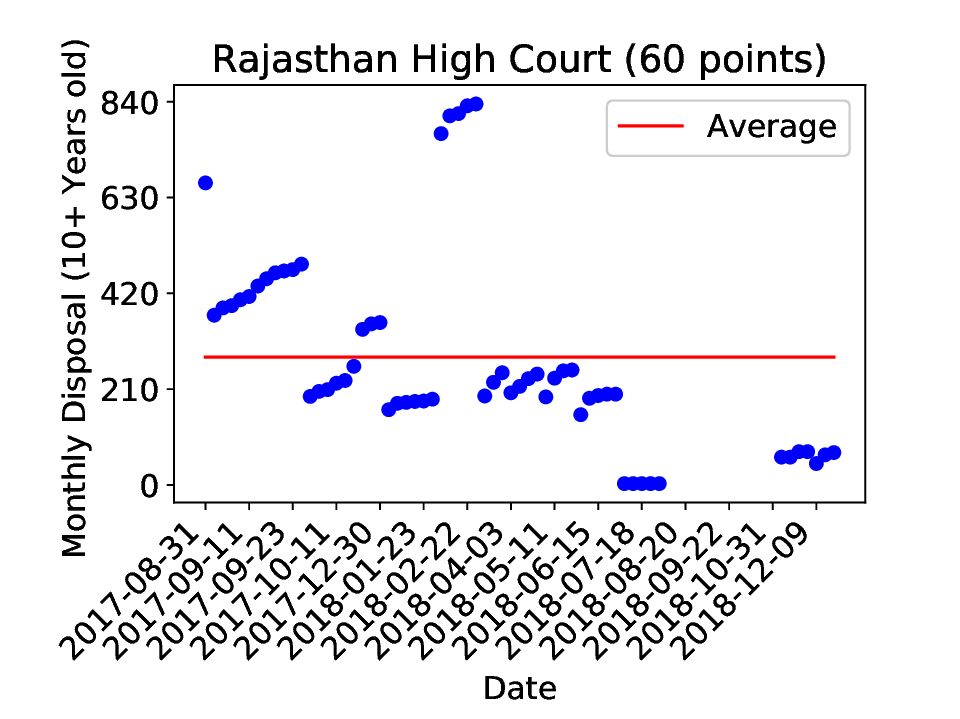}
\includegraphics[width=4.4cm]{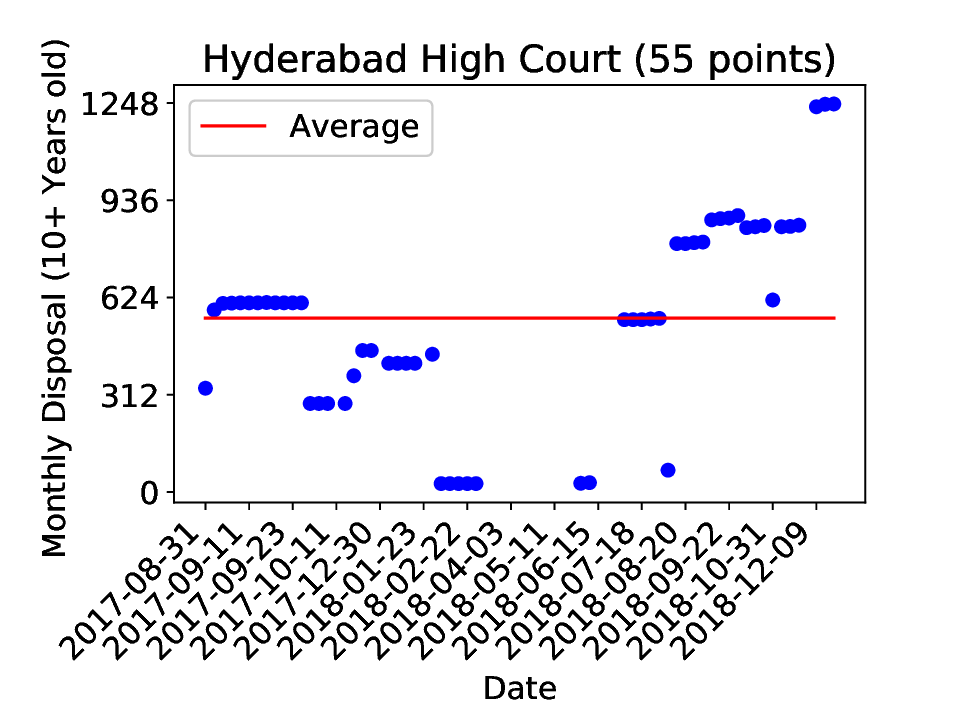}
\includegraphics[width=4.4cm]{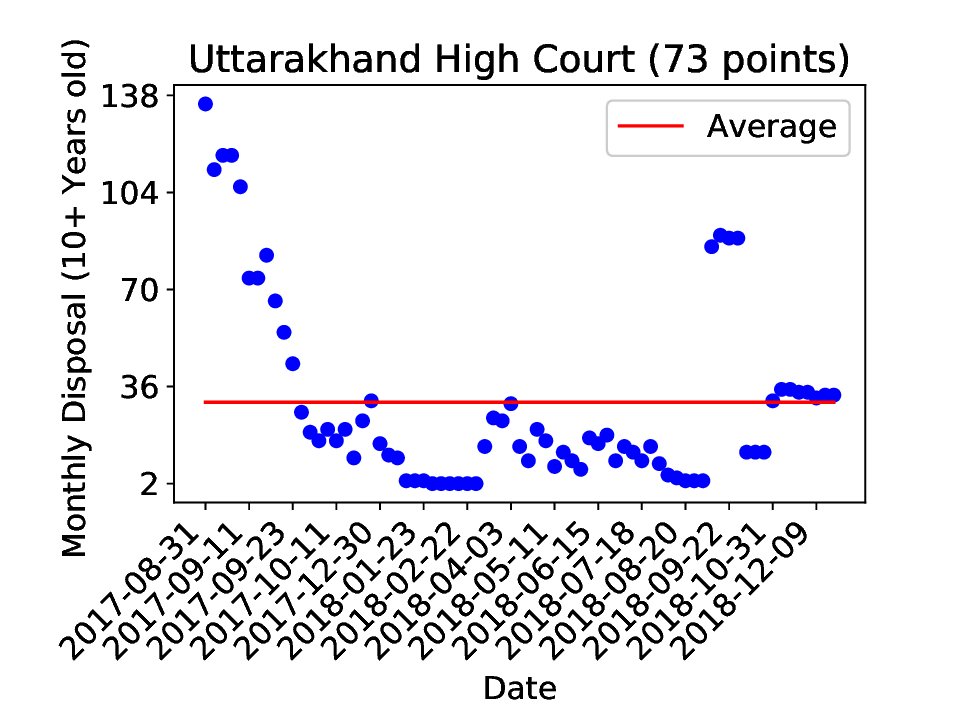}
\includegraphics[width=4.4cm]{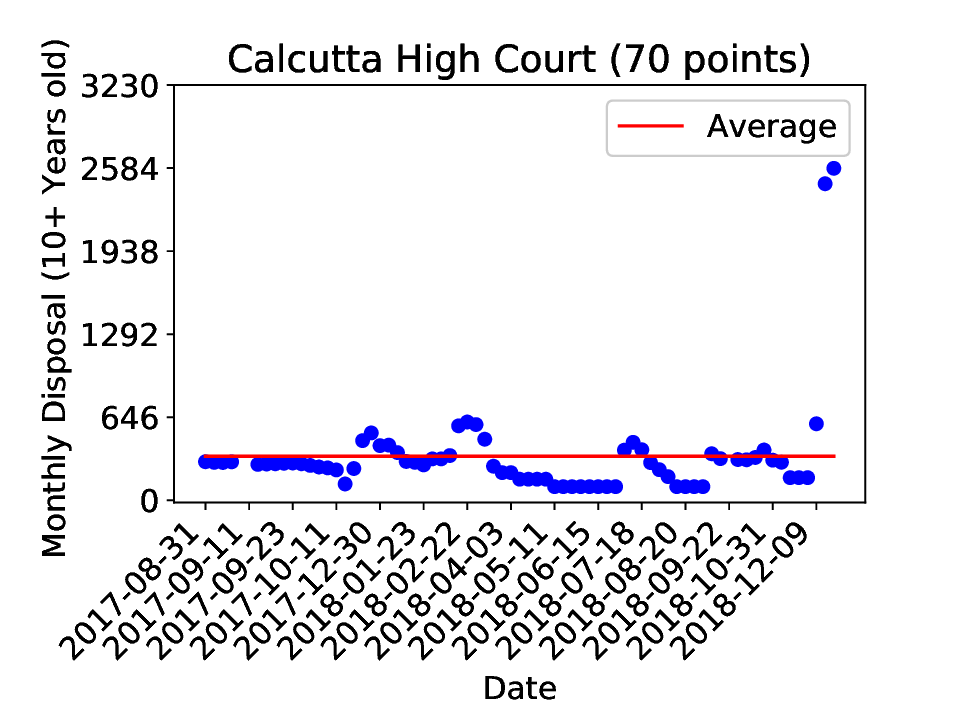}
\includegraphics[width=4.4cm]{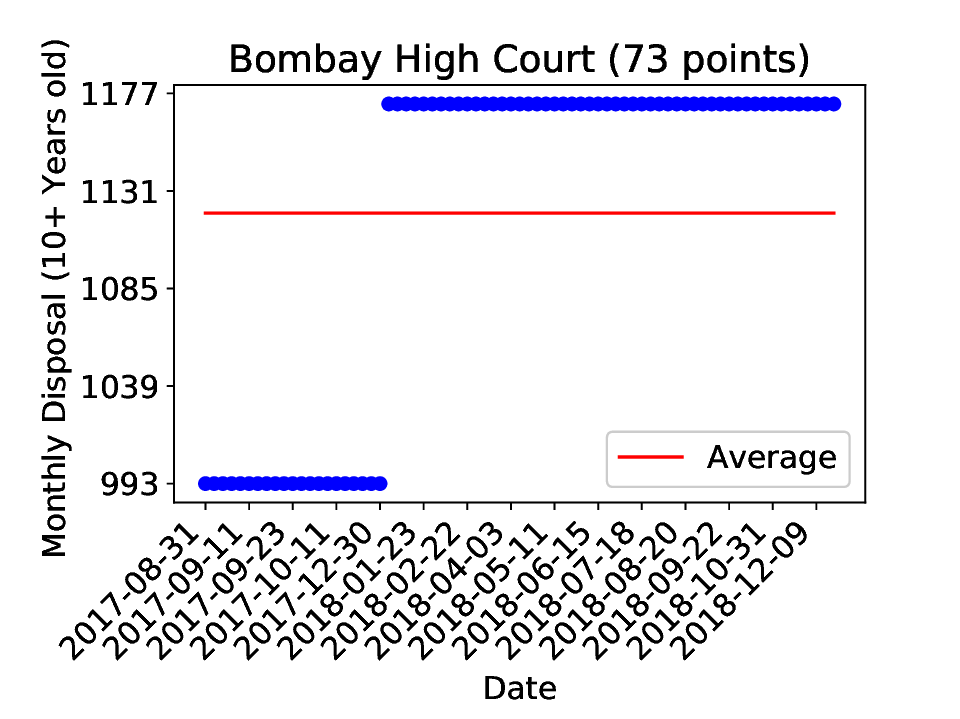}
\includegraphics[width=4.4cm]{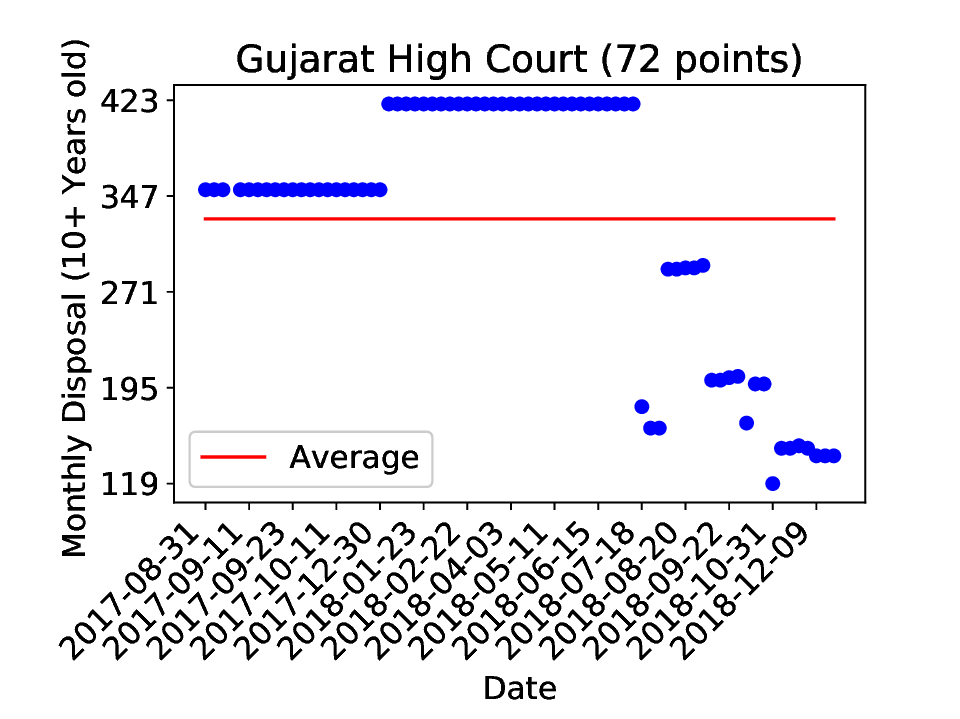}
\includegraphics[width=4.4cm]{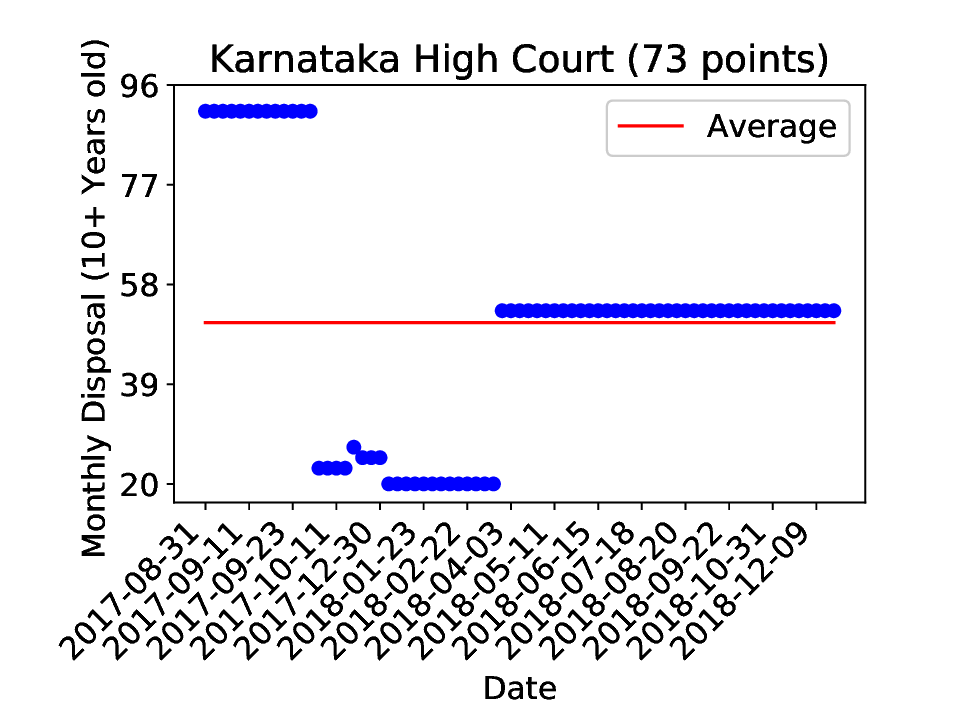}
\includegraphics[width=4.4cm]{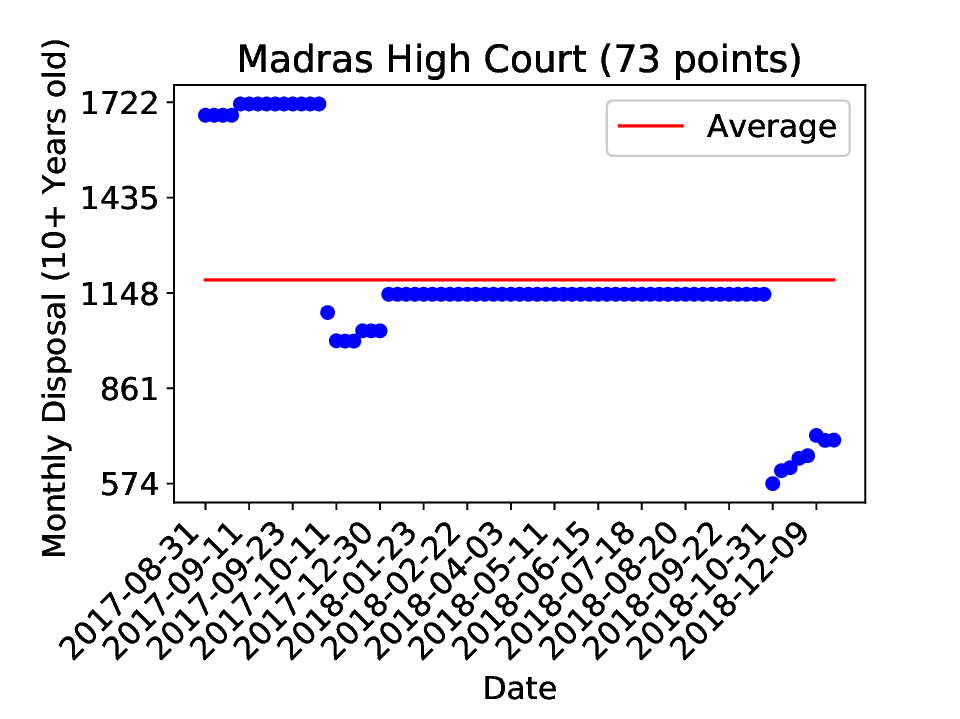}
\includegraphics[width=4.4cm]{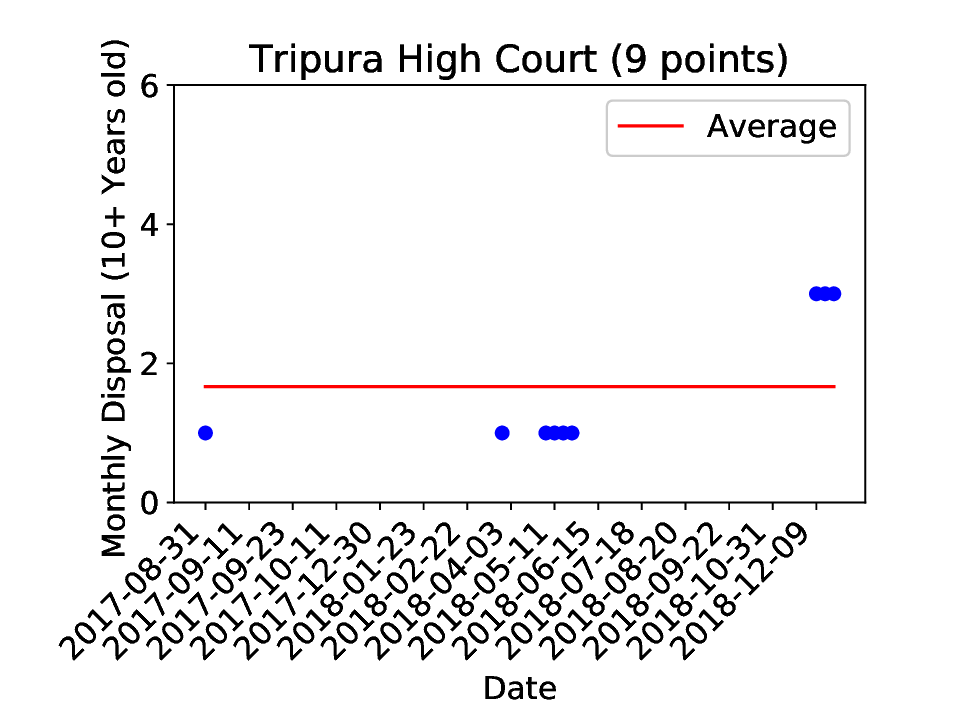}
\includegraphics[width=4.4cm]{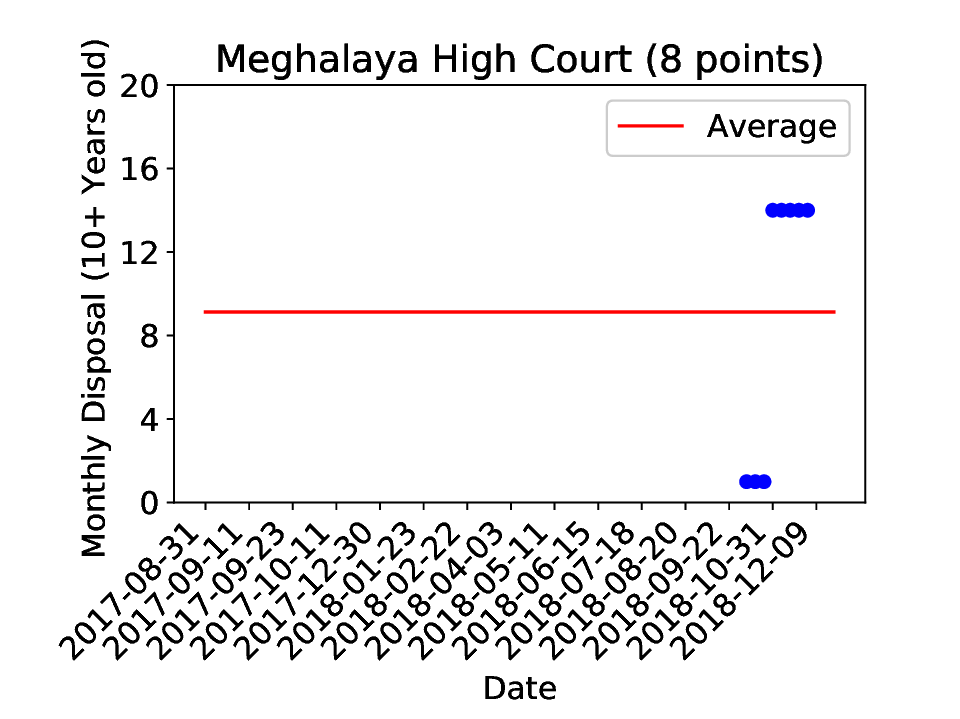}
\includegraphics[width=4.4cm]{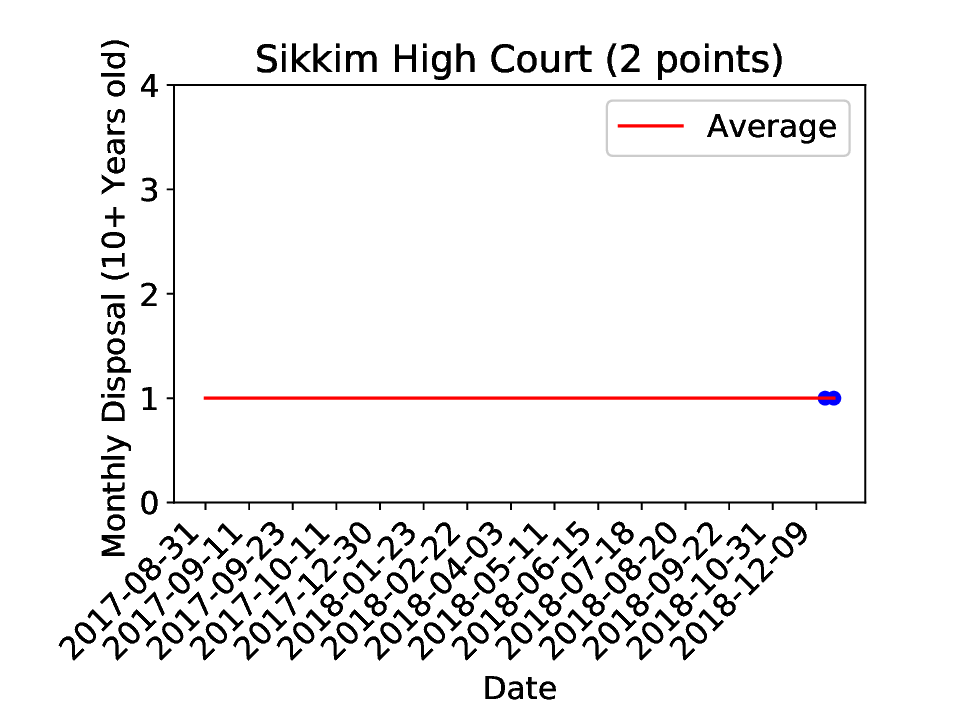}
\caption{Monthly disposed cases (10+ years old). The rate of disposal of 10+ years old cases seems to follow no trend as such, apart from Uttarakhand High Court that is disposing less than 30 cases (monthly), as per the HC-NJDG data. Ideally, it would be great to see the rate of disposal of such cases increasing so that the pendency of 10+ years can be quickly removed. There are few High Courts like Meghalaya, Sikkim and Tripura that have close to zero or zero 10+ years pending cases. Those high courts that show a long constant straight line, the data has not been updated for them. An important thing to note here is that all the high courts have data related to this metric.}
\label{fig:disp10_hc}
\end{figure*}

\fref{fig:disp10_hc} presents the disposal of those cases which are pending for more than ten years. We believe that such cases should be given priority while preparing causelists.

\subsection{Cases Under Objection and Pending Registrations}
\label{sec:obj_preg}

\begin{figure*}[h]
\includegraphics[width=4.4cm]{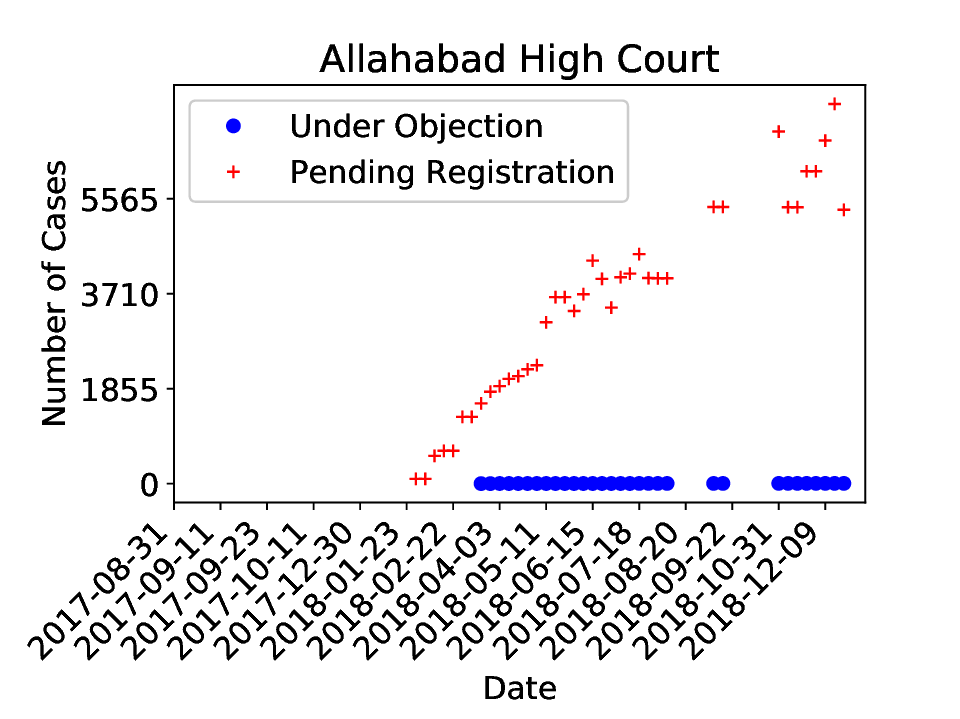}
\includegraphics[width=4.4cm]{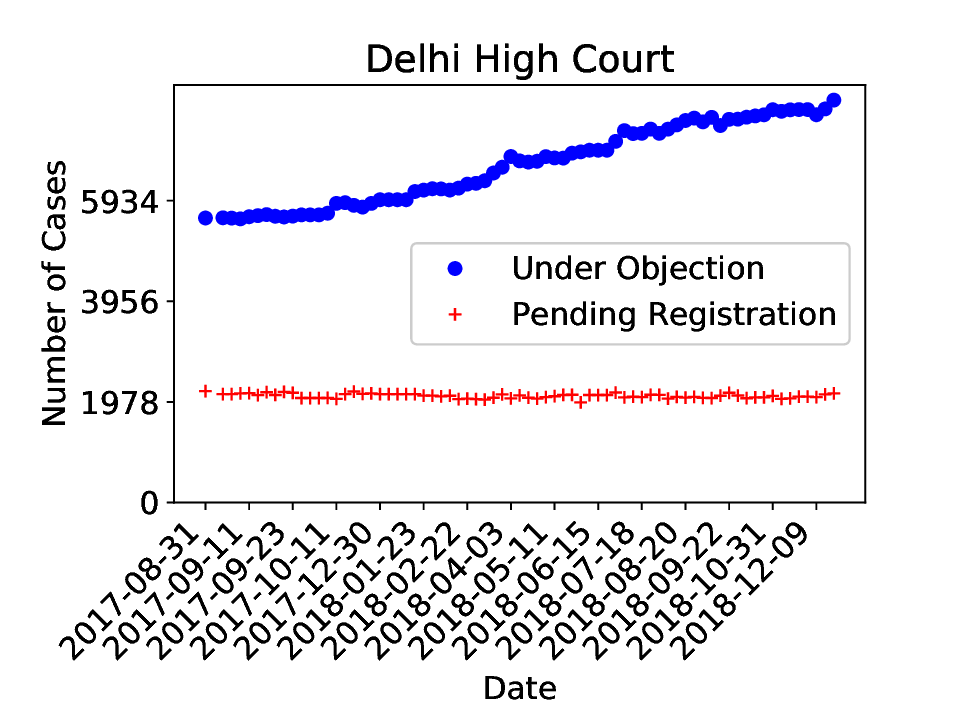}
\includegraphics[width=4.4cm]{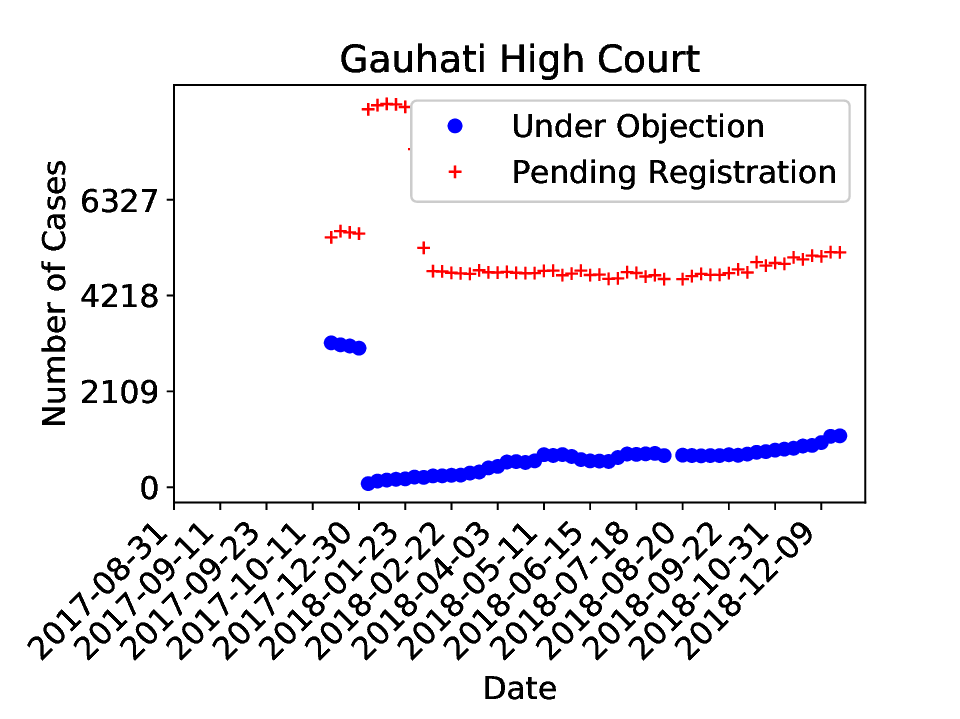}
\includegraphics[width=4.4cm]{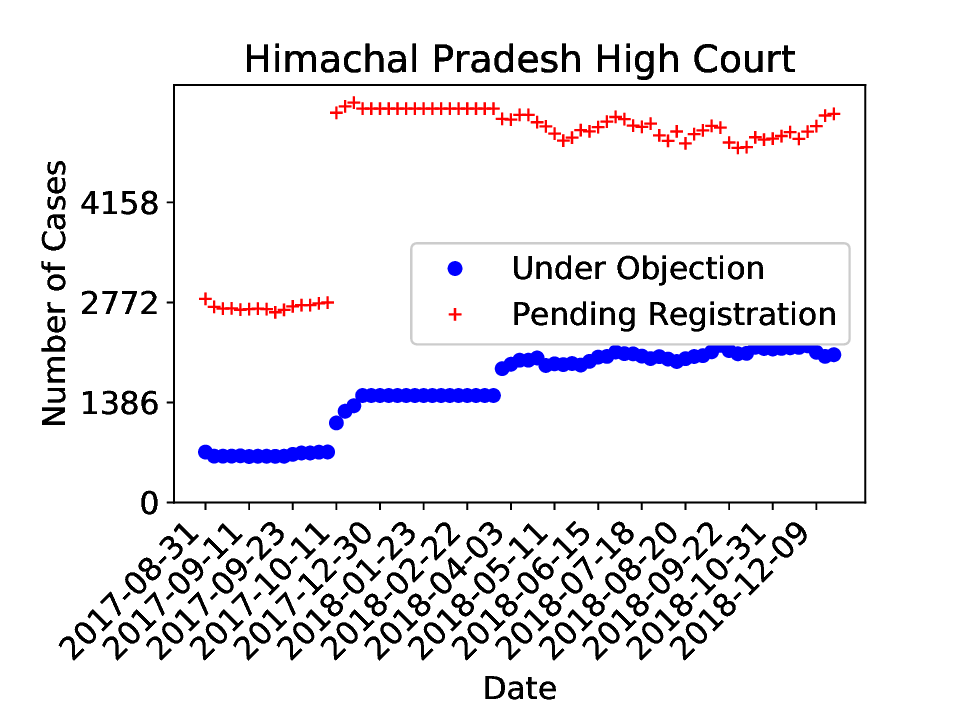}
\includegraphics[width=4.4cm]{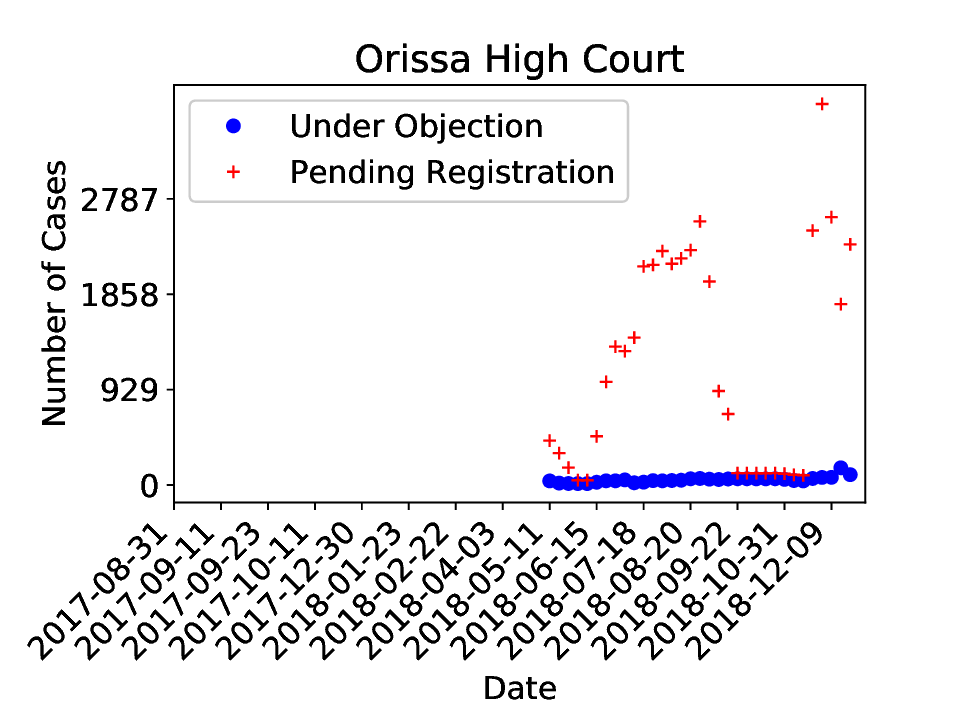}
\includegraphics[width=4.4cm]{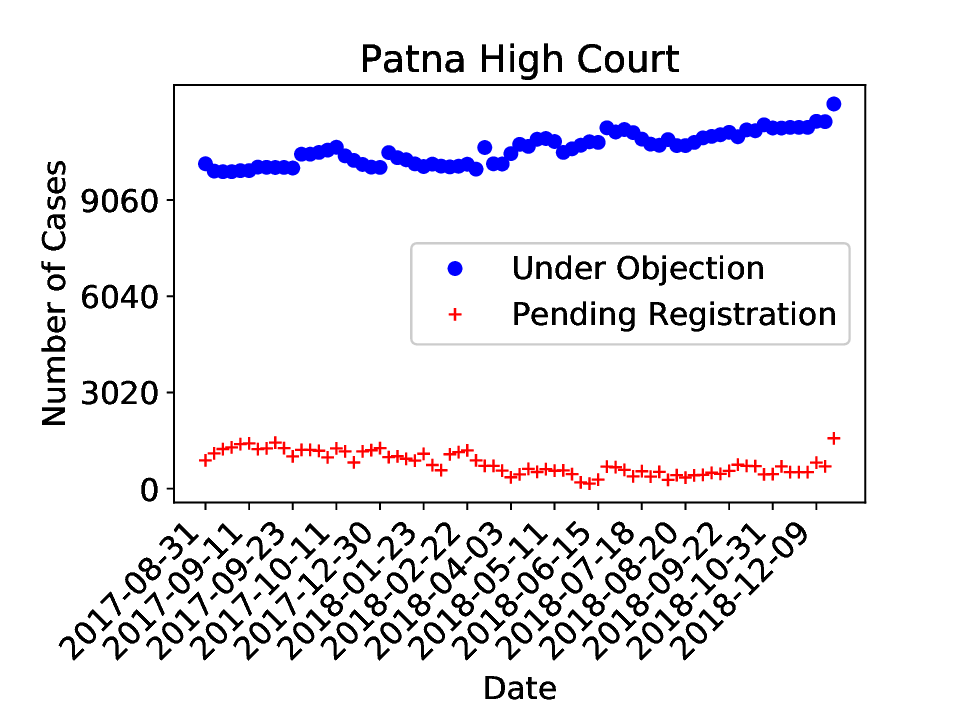}
\includegraphics[width=4.4cm]{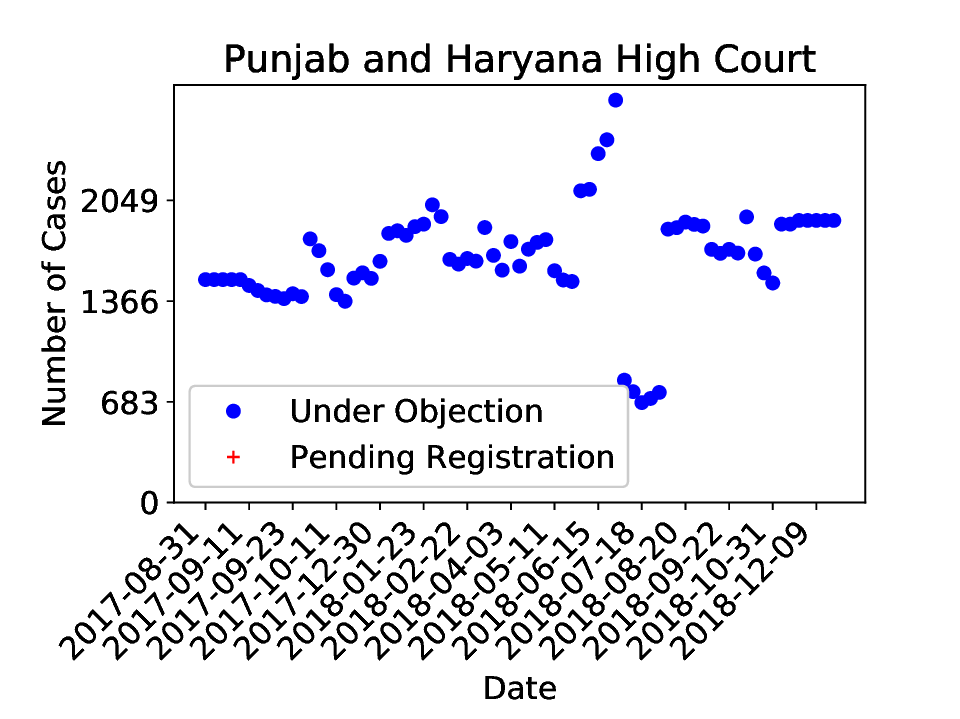}
\includegraphics[width=4.4cm]{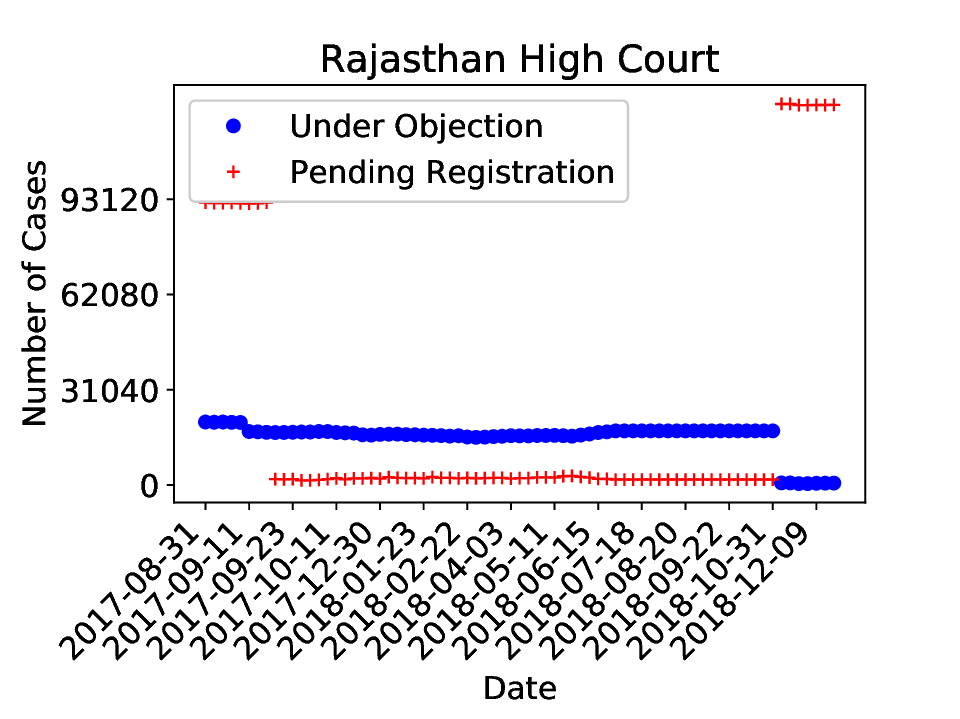}
\includegraphics[width=4.4cm]{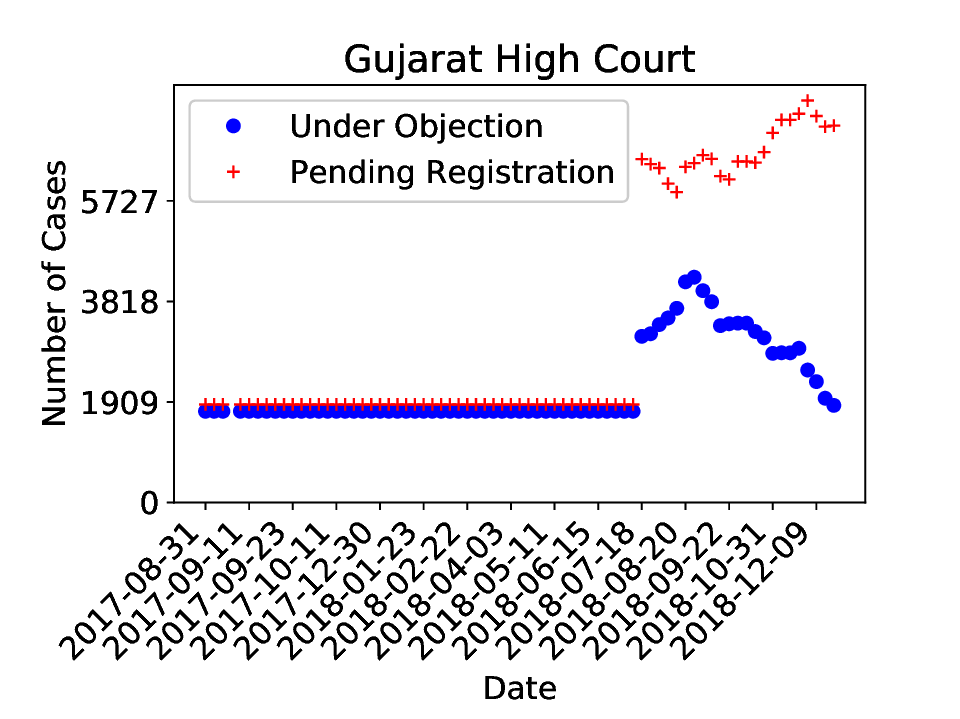}
\includegraphics[width=4.4cm]{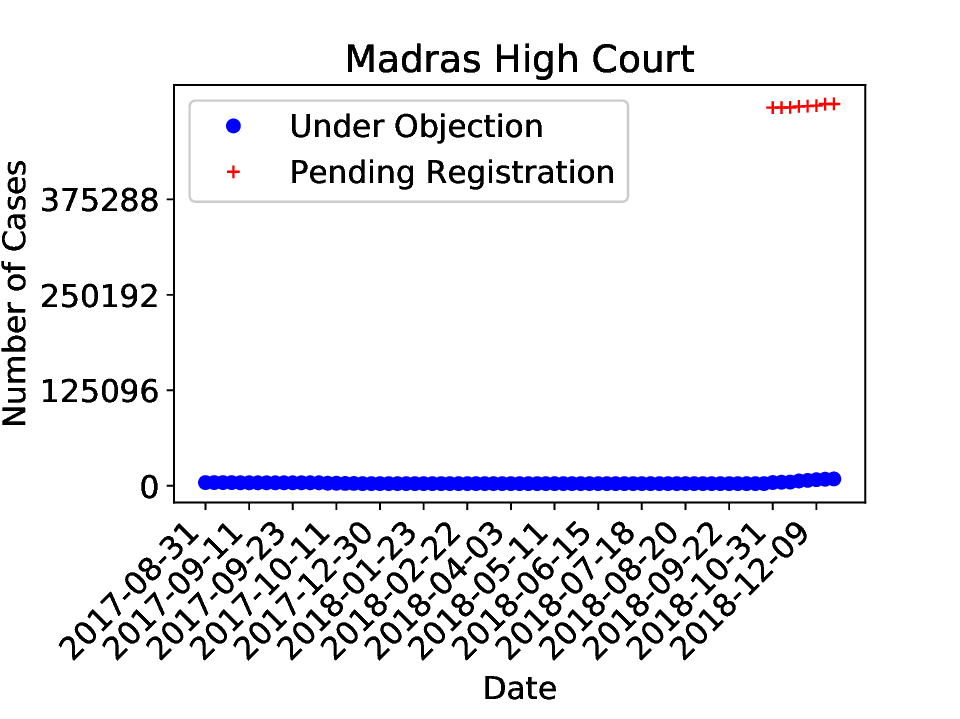}
\includegraphics[width=4.4cm]{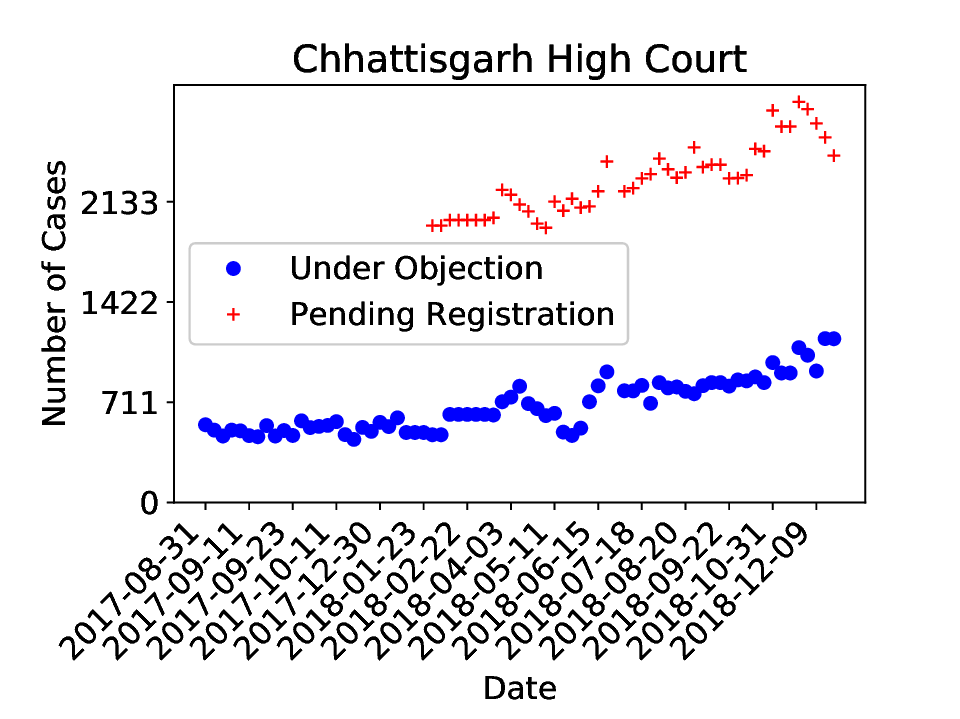}
\includegraphics[width=4.4cm]{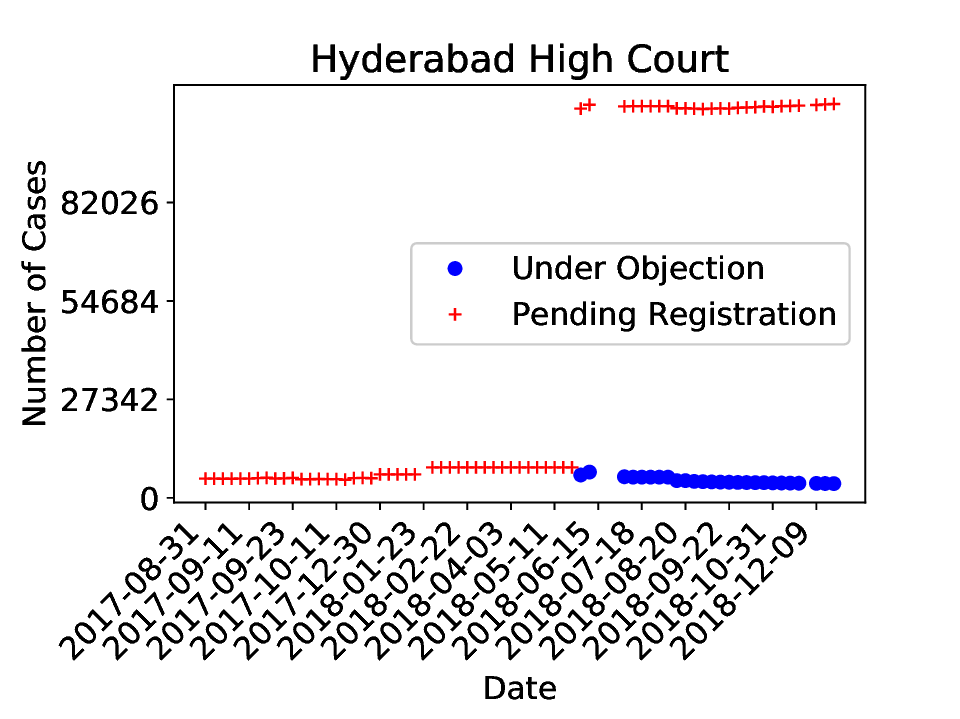}
\includegraphics[width=4.4cm]{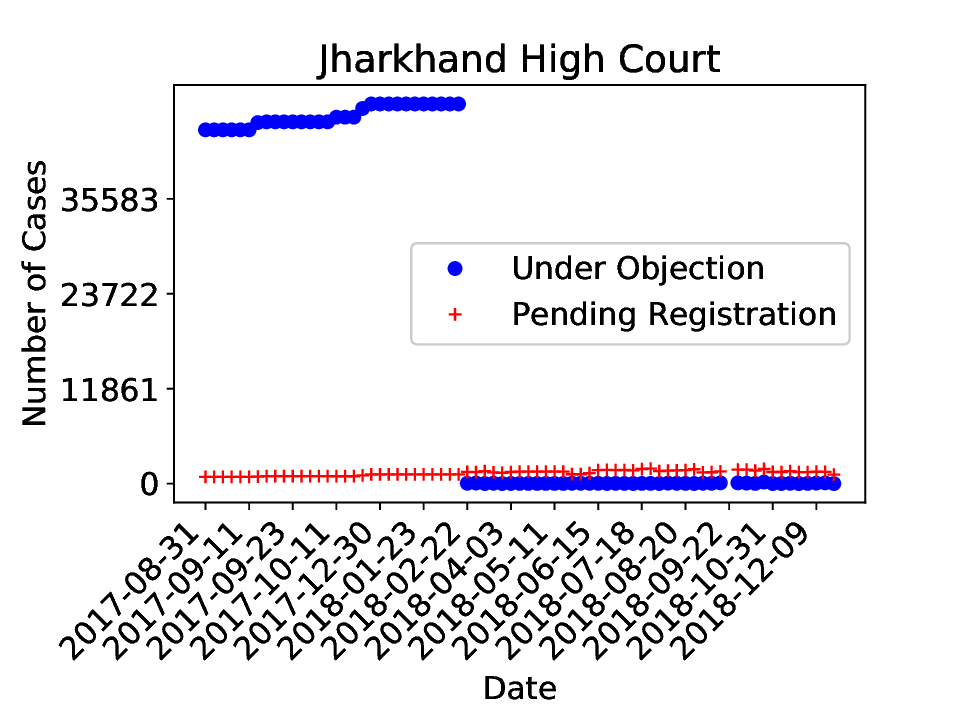}
\includegraphics[width=4.4cm]{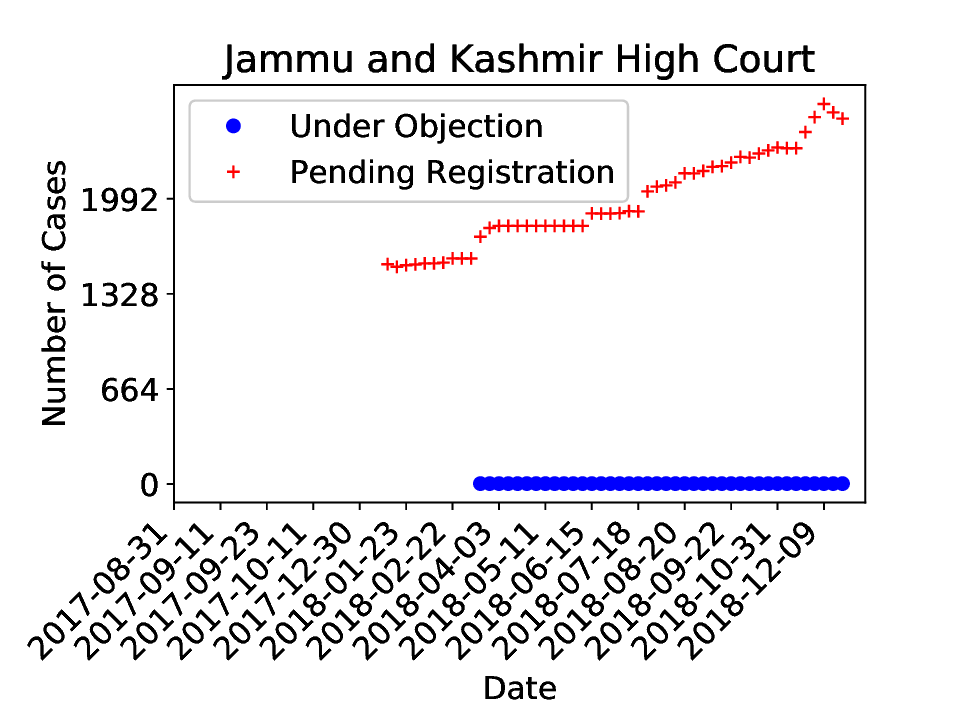}
\includegraphics[width=4.4cm]{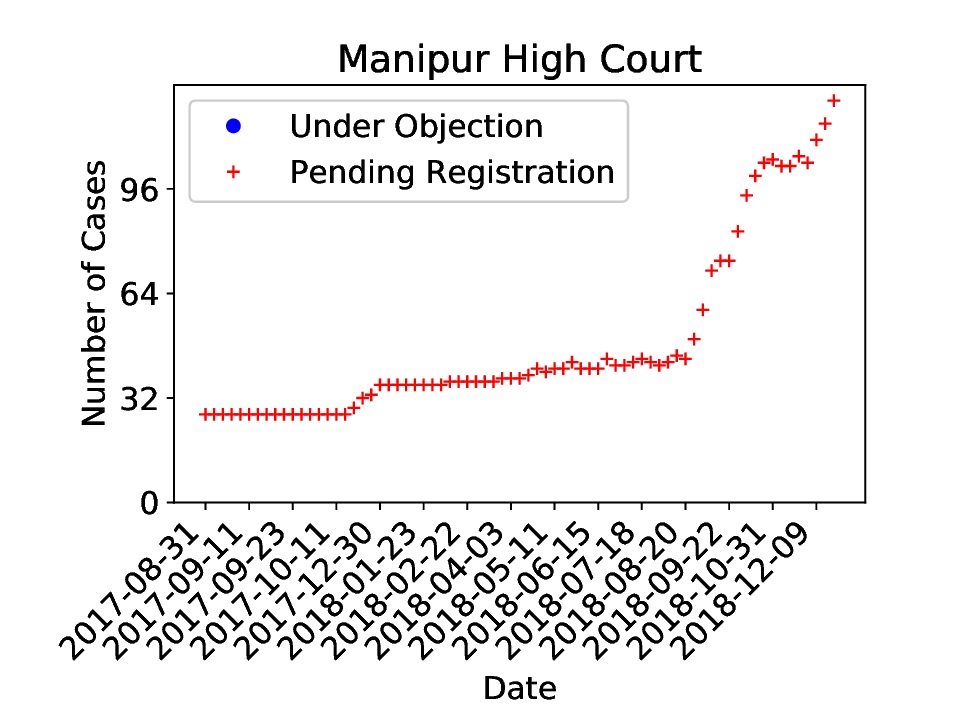}
\includegraphics[width=4.4cm]{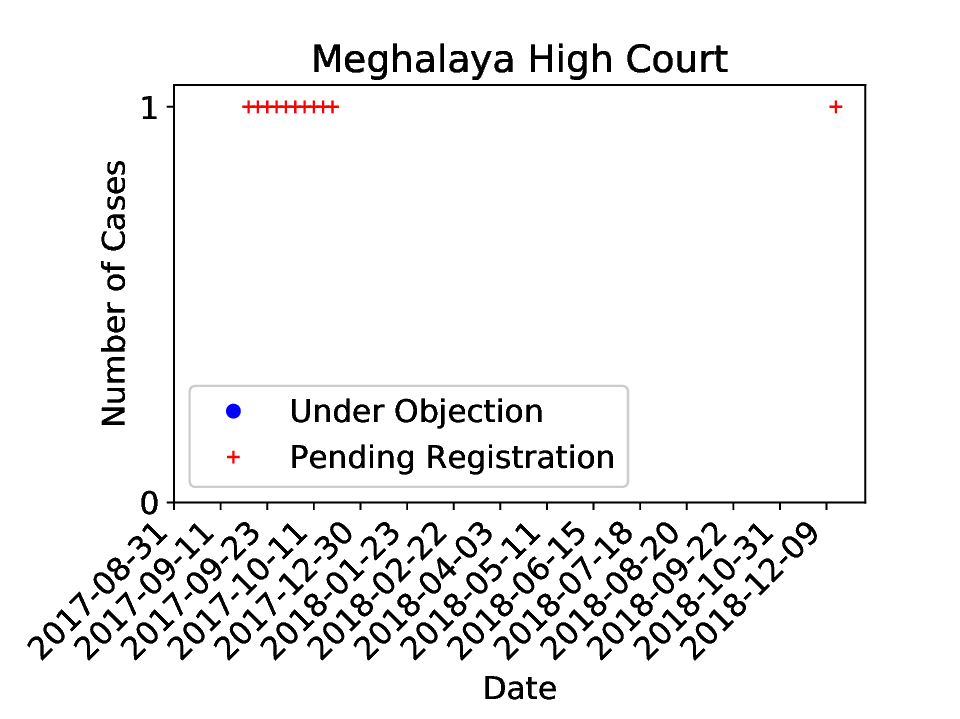}
\includegraphics[width=4.4cm]{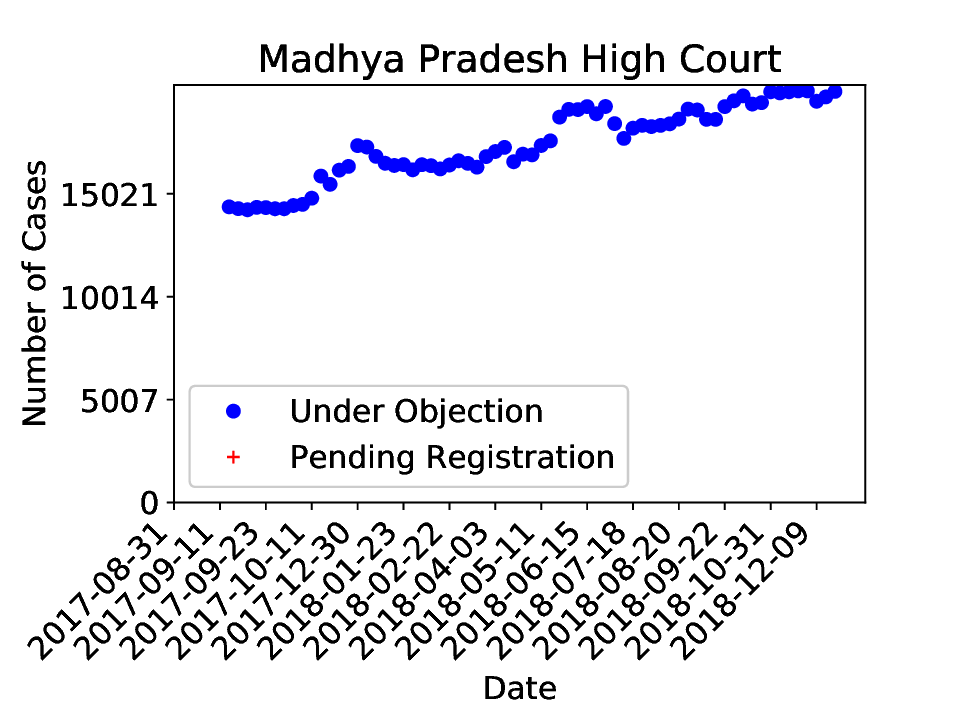}
\includegraphics[width=4.4cm]{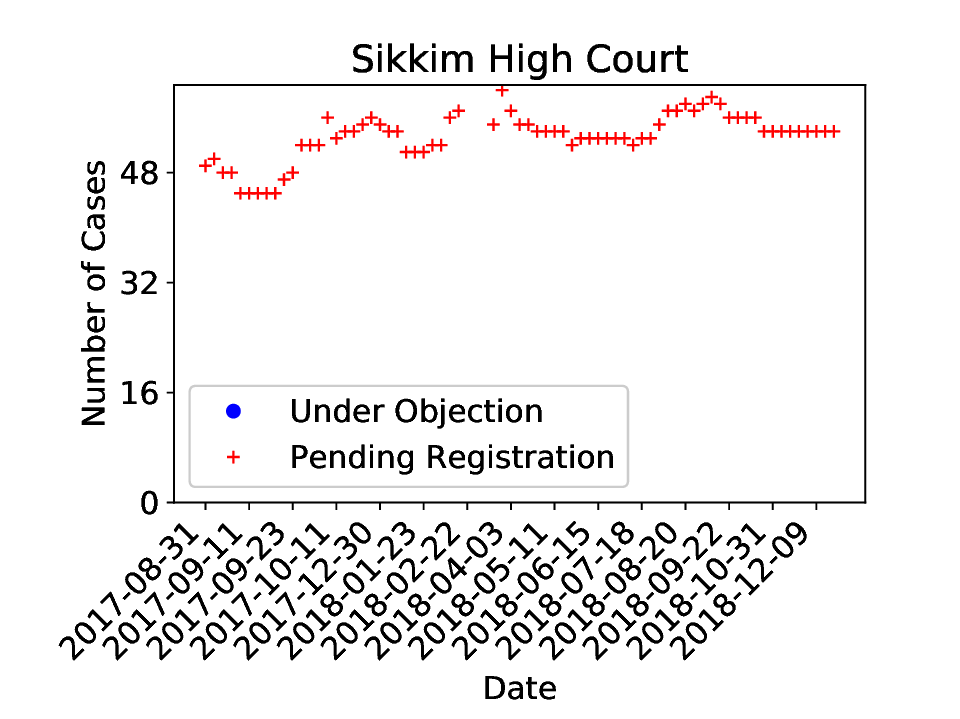}
\includegraphics[width=4.4cm]{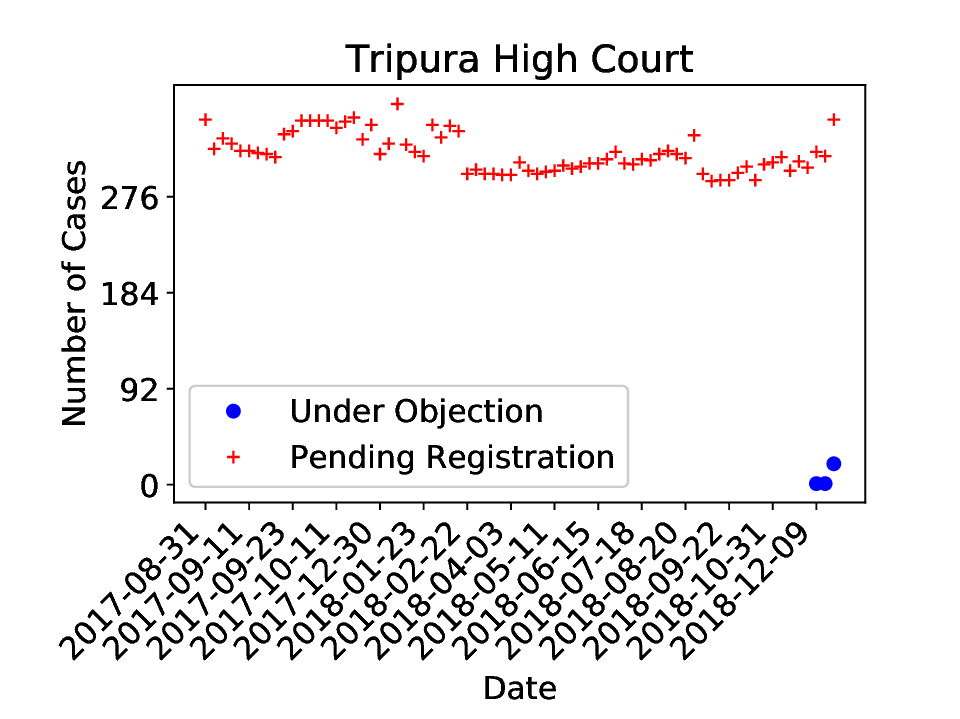}
\includegraphics[width=4.4cm]{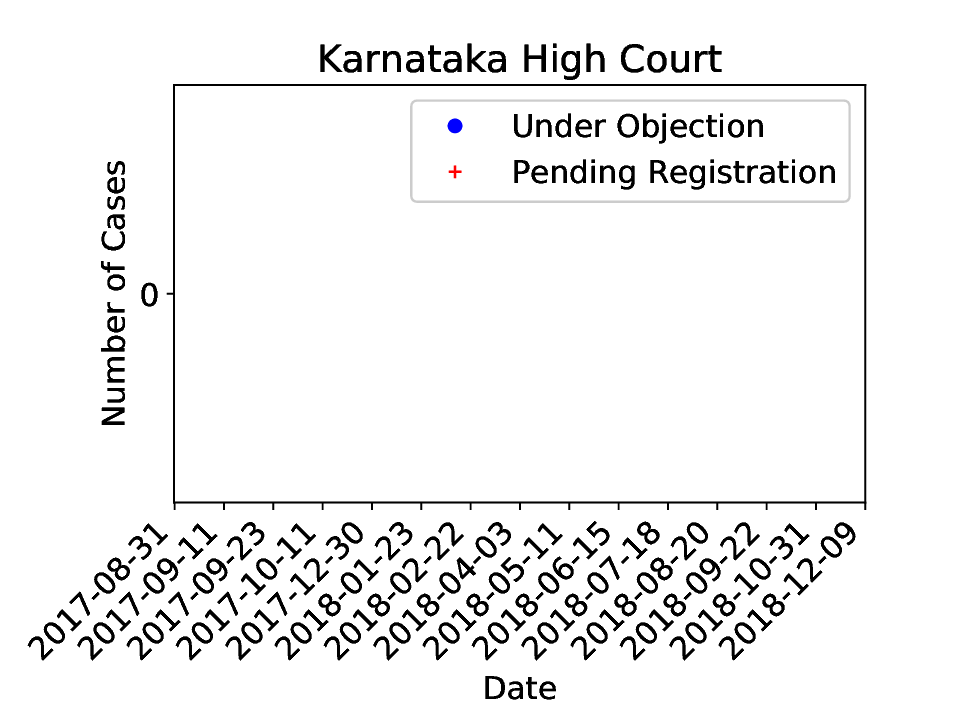}
\includegraphics[width=4.4cm]{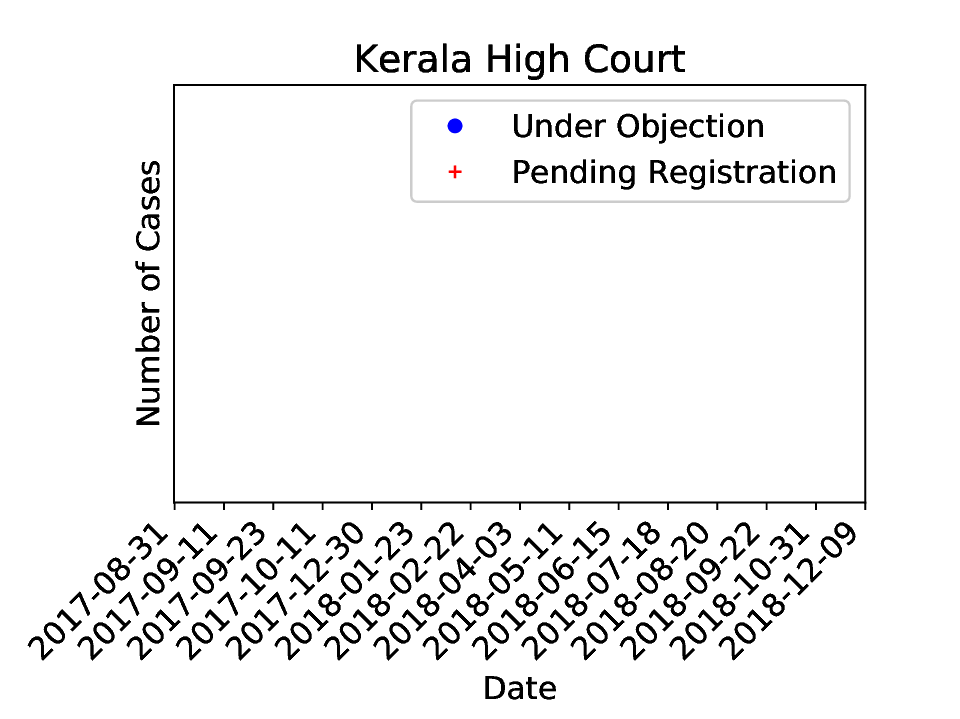}
\includegraphics[width=4.4cm]{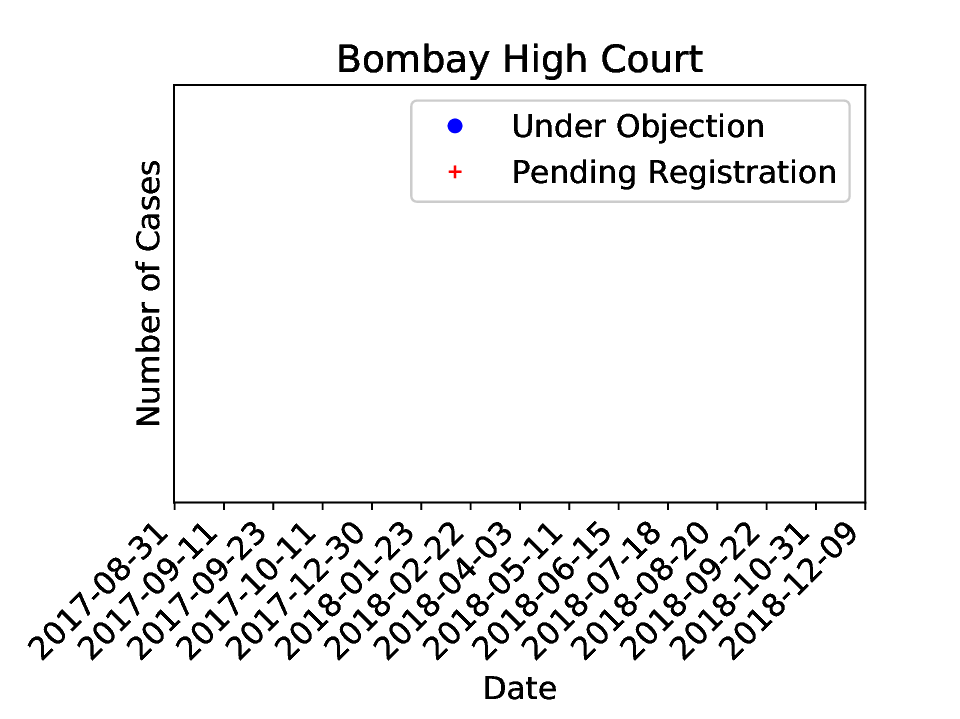}
\includegraphics[width=4.4cm]{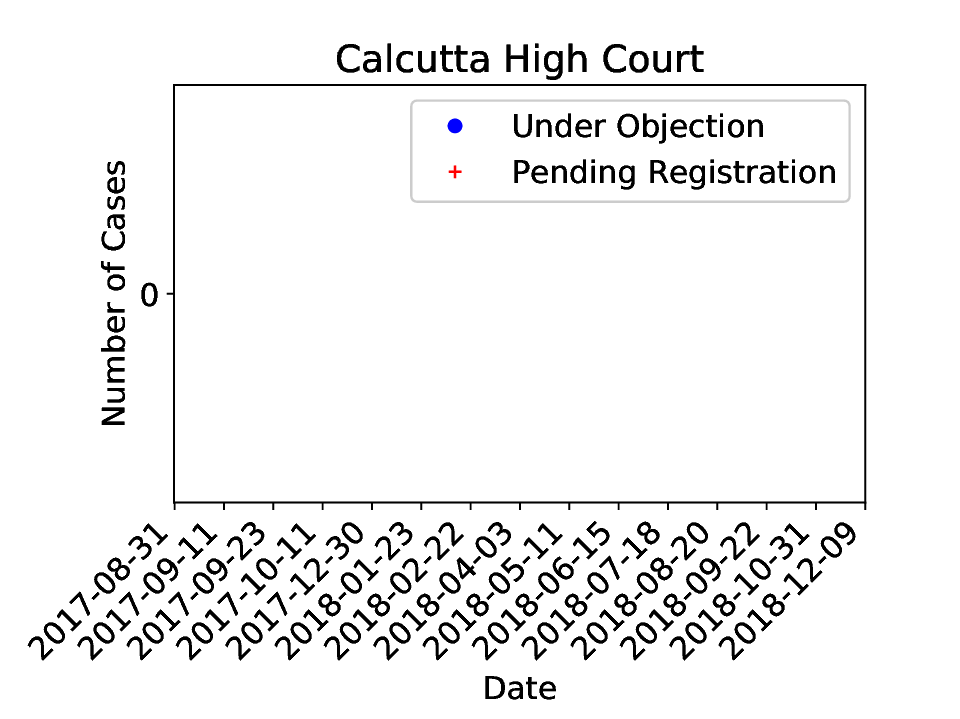}
\includegraphics[width=4.4cm]{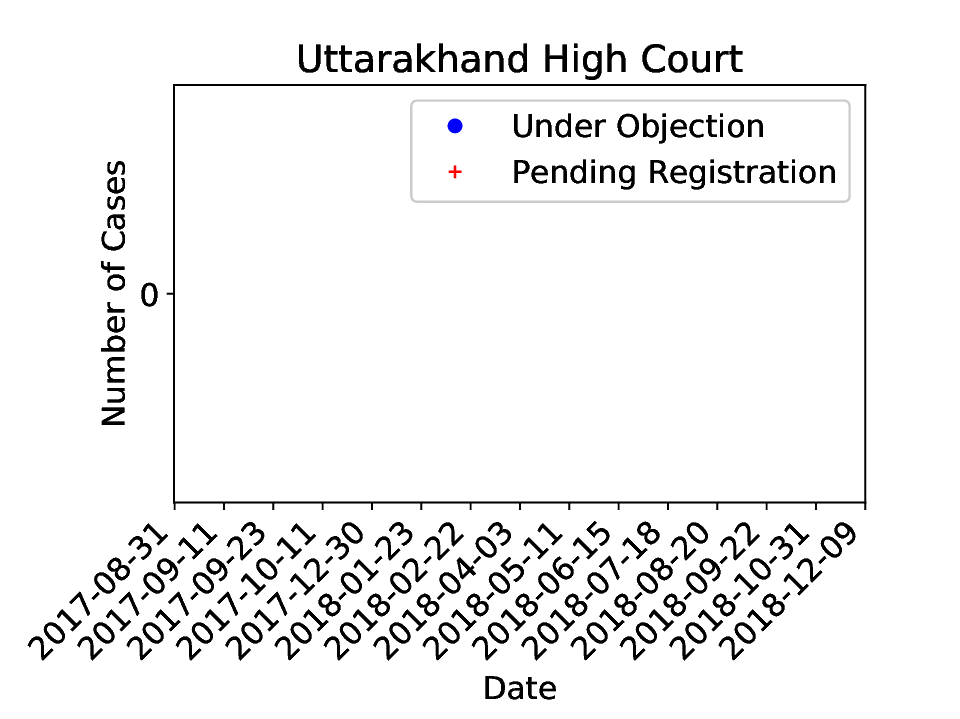}
\caption{Cases pending registration and cases under objection in various High Courts. These metrics are not defined properly. So we have decided not to make any inferences from this data, rather present the temporal graphs corresponding to the data.}
\label{fig:po_hc3}
\end{figure*}

We could not understand the metrics used here well enough to make some inferences from the data. We have still plotted the temporal graphs for each high court in \fref{fig:po_hc3}.

\section{Conclusion}
\label{sec:conclusion}

The paper details one of the fundamental problems that Indian Judiciary is faced with. The number of pending cases is well beyond the capacity of humans to solve. Hence, aid of ICT in judiciary is sought to reduce the backlog of millions of pending cases. We have analyzed sixteen month data sampled on 73 days as collected from the HC-NJDG portal. Our findings can be summarized as follows:

\begin{enumerate}
\item Maintaining NJDG flawlessly is a daunting task. Multiple levels of checks are required to ensure that the data provided on it is free from errors. 
\item Timely updates are an issue. Unless the updates on the portal are regular, it cannot be used for the envisaged purpose, which is to make it useful for reducing the pendency.
\item The gap between the number of disposed cases and the number of listed cases is huge. A reduction in the number of listed cases may help all the stakeholders without compromising on the quality of justice. On the contrary, it may improve the quality of life for all the stakeholders.
\item The number of cases filed by senior citizens and women are not at all proportional to their population. This is even more true when it comes to the criminal cases filed by women and senior citizens. Not many high courts are updating this field which may also result in the low numbers observed in the current data.
\item There are few undefined fields on the HC-NJDG portal. It will be easier for readers to interpret the data if the portal is backed up by a documentation. 
\item We have also estimated the number of years to elapse to nullify pendency provided that the working strength of each high court is same as its approved strength. As for the current working strength, pendency is only increasing.
\end{enumerate}

In the end, we hope that our work has served the role of bug reports for NJDG as well as helping in curbing pendency in high courts.

In future, we would like to strengthen our results by studying lower courts on the same scale. Our goal is to publish similar results related to each state and Union Territory that has its presence on NJDG. In order to improve upon the state of the art, relatively new computer science areas in artificial intelligence like deep learning, natural language processing, etc have to be applied to better utilize the ICT infrastructure procured by the courts. The fundamental problem that the judiciary in India is currently suffering from is the problem of scalability. Deep learning algorithms have proven to be very useful in improving scalability where human like tasks need to be done. Its applications to reducing pendency might be one such area.


\bibliographystyle{IEEEtran}
\bibliography{/home/kshitizv/Dropbox/law/LVI/2017/legal_informatics}

\end{document}